\documentclass[a4paper,12pt,oneoside]{report}

\usepackage[english,greek]{babel}
\usepackage[iso-8859-7]{inputenc}

\usepackage{graphicx}
\usepackage{mathrsfs}
\usepackage{amsfonts}
\usepackage{fancyhdr}
\usepackage{makeidx}
\usepackage{subfigure}
\usepackage{verbatim}
\usepackage{amssymb}

\def\hybrid{\topmargin -30pt    \oddsidemargin 0pt 
        \headheight 0pt \headsep 0pt
        \textwidth 6.25in       
        \textheight 9.5in       
        \marginparwidth .875in
        \parskip 5pt plus 1pt   \jot = 1.5ex}

\hybrid

\catcode`\@=11

\def\marginnote#1{}
\newcount\hour
\newcount\minute
\newtoks\amorpm
\hour=\time\divide\hour by60
\minute=\time{\multiply\hour by60 \global\advance\minute by-\hour}
\edef\standardtime{{\ifnum\hour<12 \global\amorpm={am}%
        \else\global\amorpm={pm}\advance\hour by-12 \fi
        \ifnum\hour=0 \hour=12 \fi
        \number\hour:\ifnum\minute<10 0\fi\number\minute\the\amorpm}}
\edef\militarytime{\number\hour:\ifnum\minute<10 0\fi\number\minute}

\def\draftlabel#1{{\@bsphack\if@filesw {\let\thepage\relax
   \xdef\@gtempa{\write\@auxout{\string
      \newlabel{#1}{{\@currentlabel}{\thepage}}}}}\@gtempa
   \if@nobreak \ifvmode\nobreak\fi\fi\fi\@esphack}
        \gdef\@eqnlabel{#1}}
\def\@eqnlabel{}
\def\@vacuum{}
\def\draftmarginnote#1{\marginpar{\raggedright\scriptsize\tt#1}}

\def\draft2{
        \def\@oddfoot{\sl \en preliminary draft \hfil
        \rm\thepage\hfil\sl\today\quad\militarytime}
        \let\@evenfoot\@oddfoot \overfullrule 3pt
        \let\label=\draftlabel
        \let\marginnote=\draftmarginnote
   \def\@eqnnum{(\theequation)\rlap{\kern\marginparsep\tt\@eqnlabel}%
\global\let\@eqnlabel\@vacuum}  }

\def\preprint{\twocolumn\sloppy\flushbottom\parindent 2em
        \leftmargini 2em\leftmarginv .5em\leftmarginvi .5em
        \oddsidemargin -.5in    \evensidemargin -.5in
        \columnsep .4in \footheight 0pt
        \textwidth 10.in        \topmargin  -.4in
        \headheight 12pt \topskip .4in
        \textheight 6.9in \footskip 0pt
        \def\@oddhead{\thepage\hfil\addtocounter{page}{1}\thepage}
        \let\@evenhead\@oddhead \def\@oddfoot{} \def\@evenfoot{} }

\def\numberbysection{\@addtoreset{equation}{section}
        \def\theequation{\thesection.\arabic{equation}}}

\def\underline#1{\relax\ifmmode\@@underline#1\else
        $\@@underline{\hbox{#1}}$\relax\fi}

\def\titlepage{\@restonecolfalse\if@twocolumn\@restonecoltrue\onecolumn
     \else \newpage \fi \thispagestyle{empty}\c@page\z@
        \def\thefootnote{\fnsymbol{footnote}} }

\def\endtitlepage{\if@restonecol\twocolumn \else \newpage \fi
        \def\thefootnote{\arabic{footnote}}
        \setcounter{footnote}{0}}  

\catcode`@=12
\relax

\def\figcap{\section*{Figure Captions\markboth
        {FIGURECAPTIONS}{FIGURECAPTIONS}}\list
        {Figure \arabic{enumi}:\hfill}{\settowidth\labelwidth{Figure
999:}
        \leftmargin\labelwidth
        \advance\leftmargin\labelsep\usecounter{enumi}}}
 \relax
\def\tablecap{\section*{Table Captions\markboth
        {TABLECAPTIONS}{TABLECAPTIONS}}\list
        {Table \arabic{enumi}:\hfill}{\settowidth\labelwidth{Table
999:}
        \leftmargin\labelwidth
        \advance\leftmargin\labelsep\usecounter{enumi}}}
 \relax
\def\reflist{\section*{References\markboth
        {REFLIST}{REFLIST}}\list
        {[\arabic{enumi}]\hfill}{\settowidth\labelwidth{[999]}
        \leftmargin\labelwidth
        \advance\leftmargin\labelsep\usecounter{enumi}}}
 \relax
\makeatletter
\newcounter{pubctr}
\def\publist{\@ifnextchar[{\@publist}{\@@publist}}
\def\@publist[#1]{\list
        {[\arabic{pubctr}]\hfill}{\settowidth\labelwidth{[999]}
        \leftmargin\labelwidth
        \advance\leftmargin\labelsep
        \@nmbrlisttrue\def\@listctr{pubctr}
        \setcounter{pubctr}{#1}\addtocounter{pubctr}{-1}}}
\def\@@publist{\list
        {[\arabic{pubctr}]\hfill}{\settowidth\labelwidth{[999]}
        \leftmargin\labelwidth
        \advance\leftmargin\labelsep
        \@nmbrlisttrue\def\@listctr{pubctr}}}
 \relax
\makeatother

\def\be{\begin{equation}}
\def\ee{\end{equation}}
\def\ba{\begin{eqnarray}}
\def\ea{\end{eqnarray}}

\def\del{\partial}

\def\k{\kappa}
\def\r{\rho}
\def\a{\alpha}
\def\A{\Alpha}
\def\b{\beta}
\def\B{\Beta}
\def\g{\gamma}
\def\G{\Gamma}
\def\d{\delta}
\def\D{\Delta}
\def\e{\epsilon}
\def\E{\Epsilon}
\def\p{\pi}
\def\P{\Pi}
\def\c{\chi}
\def\C{\Chi}
\def\th{\theta}
\def\Th{\Theta}
\def\m{\mu}
\def\n{\nu}
\def\om{\omega}
\def\Om{\Omega}
\def\l{\lambda}
\def\L{\Lambda}
\def\s{\sigma}
\def\S{\Sigma}
\def\vphi{\varphi}

\def\no{\noindent}

\def\qq{\qquad}

\begin{document}

\newcommand{\eqn}[1]{(\ref{#1})}

\linespread{1.3}

\newcommand{\en}{\selectlanguage{english}}
\newcommand{\gr}{\selectlanguage{greek}}
\def\be{\begin{equation}}
\def\ee{\end{equation}}
\def\ba{\begin{eqnarray}}
\def\ea{\end{eqnarray}}
\def\no{\noindent}
\def\del{\partial}

\def\k{\kappa}
\def\r{\rho}
\def\a{\alpha}
\def\A{\Alpha}
\def\b{\beta}
\def\B{\Beta}
\def\g{\gamma}
\def\G{\Gamma}
\def\d{\delta}
\def\D{\Delta}
\def\e{\epsilon}
\def\E{\Epsilon}
\def\p{\pi}
\def\P{\Pi}
\def\c{\chi}
\def\C{\Chi}
\def\th{\theta}
\def\Th{\Theta}
\def\m{\mu}
\def\n{\nu}
\def\om{\omega}
\def\Om{\Omega}
\def\l{\lambda}
\def\L{\Lambda}
\def\s{\sigma}
\def\S{\Sigma}
\def\vphi{\varphi}
\def\half{\frac{1}{2}}
\def \ov {\over}
\def\elF{{\bf F}}
\def\elK{{\bf K}}
\def\elPi{{\bf \Pi}}
\def\elE{{\bf E}}

\def\cA{{\cal A}}
\def\cB{{\cal B}}
\def\cD{{\cal D}}
\def\cF{{\cal F}}
\def\cG{{\cal G}}
\def\cH{{\cal H}}
\def\cL{{\cal L}}
\def\cM{{\cal M}}
\def\cN{{\cal N}}
\def\cO{{\cal O}}
\def\cP{{\cal P}}
\def\cQ{{\cal Q}}
\def\cR{{\cal R}}
\def\cV{{\cal V}}
\def\cY{{\cal Y}}

\hyphenation{δι-πλής  δι-δια-στα-της  δη-μι-ουρ-γί-ας  συ-μπα-γή δια-γρά-μμα-τος
 τα-λά-ντω-σης  συ-μπλη-ρώ-νε-ται   ε-πι-τυγ-χά-νε-ται  τα-χυ-ο-νι-κή  φερ-μι-ο-νι-κός
 συ-σχέ-τι-ση   α-πο-τε-λέ-σμα-τα   κα-τα-σκευ-ής   μι-κρο-σκο-πι-κή   γε-νί-κευ-ση
πα-ρου-σι-ά-ζου-με   ι-δι-ό-τη-τες   πα-ρα-κά-τω  συ-σχε-τί-ζει  κα-τάλ-λη-λους
τε-λε-στές  α-πει-κό-νι-ση   κί-νη-σης  προ-βλέ-πει  λο-γι-σμός  πα-ρου-σι-ά-στη-καν  π
α-ρα-τη-ρού-με  κα-τάλ-λη-λη ε-γκλω-βι-σμός διά-γρα-μμα ε-λεύ-θε-ρη υπερ-βα-ρύ-τη-τα
ε-νε-ργεί-α-κο αντι-κα-τά-στα-ση δια-φέ-ρει ε-γκλω-βι-σμού υ-πό-βα-θρο γε-νι-κευ-μέ-νη
υ-πο-λο-γι-σμέ-νες βα-ρύ-ο-νι-ων συ-μπε-ρι-λα-μβα-νο-μέ-νων δια-χω-ρι-στι-κού
ελα-χι-στο-ποιή-σου-με ελα-χι-στο-ποί-ηση κλα-σι-κή βα-ρυ-ο-νίων βρί-σκου-με
σφαι-ρι-κού συ-νά-ρτη-ση πα-ρα-πά-νω α-ντι-κα-τα-στή-σου-με ι-σο-τρο-πι-κός
προ-σα-να-το-λι-ζό-με-νη α-πο-στά-σεις χρω-μο-δυ-να-μι-κή στοι-χεί-α κβα-ντι-κή Για
υ-πο-λο-γι-σμό ε-πί-σης Κα-ραί-σκο δια-κλά-δω-σης υ-πο-λο-γί-σου-με διε-ξο-δι-κά μια
οι-κο-νο-μι-κή τμή-μα-τος ε-ξε-τά-σου-με δια-κυ-μά-νσε-ων υ-πε-ρί-ωδες προ-σέ-γγι-ζο-ντας
διά-στα-σης αδιά-στα-τες στην κα-νο-νι-κο-ποιή-σου-με πο-λυω-νυ-μι-κή
επα-να-κα-νο-νι-κο-ποιή-σι-μη ενε-ργει-ών σχε-τί-ζο-νται συ-νθή-κη αστα-θές
επα-να-λα-μβά-νο-ντας γνω-στό αρκε-τές έχει θεω-ρία ένα σύ-μμο-ρφη
χα-ρα-κτη-ρι-στι-κό υπά-ρχουν αντι-καθι-στώ-ντας δια-στά-σεις υπερ-φο-ρτία στο
πε-πλε-γμέ-νες πε-πλε-γμέ-νων AdS CFT σύ-μμο-ρφο Coulomb θω-ρα-κι-σμέ-νο
υπε-ριώ-δες προ-σε-γγι-στι-κά δυ-να-μι-κό φο-ρμα-λι-σμό δια-φο-ρε-τι-κές
συ-μπε-ρι-φο-ρά προ-σα-να-το-λι-σμού με-τρι-κής συ-ζευ-γμέ-νοι πα-ρα-μό-ρφω-σης
πα-ρου-σία πα-ρα-μο-ρφω-μέ-νης δια-κυ-μά-νσεις συ-μπε-ρά-σμα-τα αντι-στοι-χίας
απο-κλί-νου-σες εξι-σο-ρρό-πη-ση με-τα-στα-θής αντί-στοι-χη
κα-νο-νι-κο-ποιή-σει συ-νδε-δε-μέ-νο ακτί-νας κό-μβου δια-φο-ρε-τι-κών προ-σω-ρι-νά
μή-κους συ-νο-ψί-ζο-νται Εμμα-νου-ήλ βα-ρυο-νίων χα-ρα-κτη-ρι-στι-κά προ-πτυ-χια-κών
πα-ρα-δεί-γμα-τα γρά-ψου-με δια-τά-ξεις κου-άρκ υ-πο-στή-ρι-ξη δια-τα-ρα-κτι-κές
υπο-λο-γι-σμός ε-ργα-σιών α-ρι-θμού γρα-μμές δια-τα-ρα-κτι-κή}

\title{
\begin{figure}[!t]
\begin{center}
\includegraphics[scale=0.3]{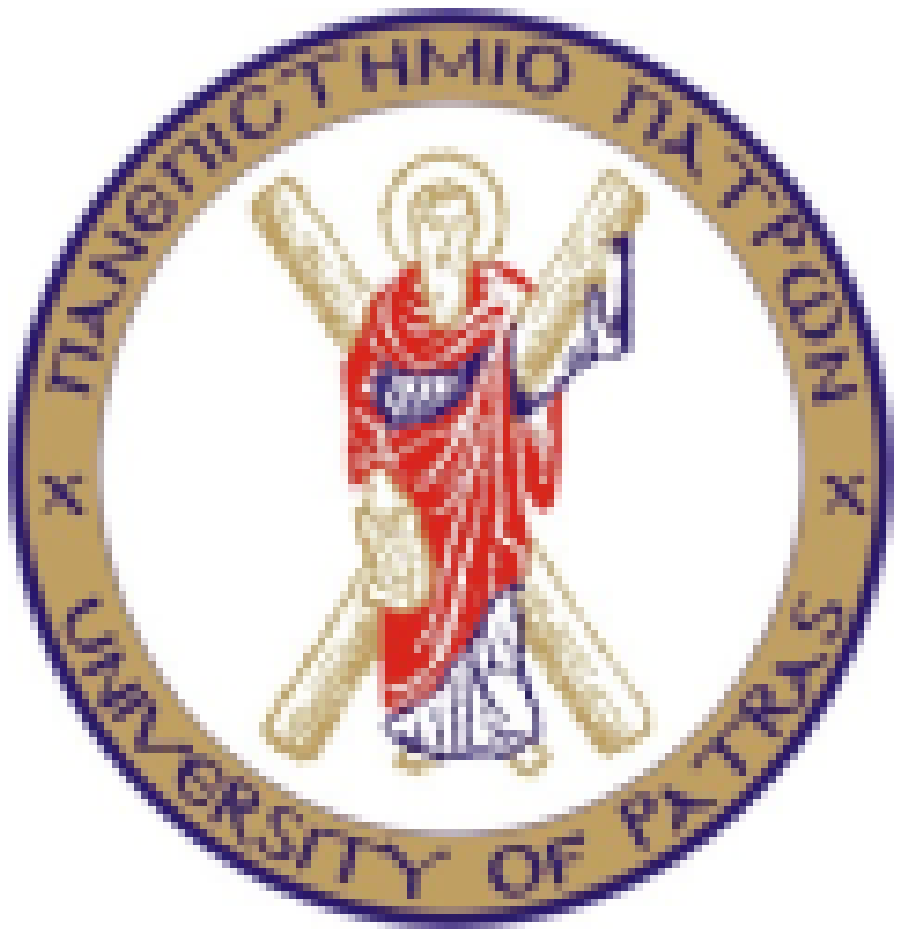}
\end{center}
\begin{center}
\LARGE{\gr Θεωρία χορδών και φυσικές εφαρμογές αυτής σε προβλήματα βαρύτητας και
θεωρίες βαθμίδας}
\end{center}
\label{St-Andreas}
\end{figure}}

\author{
    \Large{Διδακτορική Διατριβή}\\
}

\date{\vskip 0.5in
    Κωνσταντίνος Σιάμπος\\
    Γενικό Τμήμα, Tομέας Φυσικής\\
    Πανεπιστήμιο Πατρών\\
    \pagenumbering{roman}
}

\maketitle

\begin{titlepage}
\begin{center}

\vspace*{2cm}

{\Large Κωνσταντίνος Σιάμπος}

\vspace{2cm}

{\LARGE{Θεωρία χορδών και φυσικές εφαρμογές αυτής σε προβλήματα
βαρύτητας και θεωρίες βαθμίδας}}

\vspace{2cm}

\vspace{2cm}

\begin{tabular}{ll}
Επιβλέπων διατριβής:  Καθηγητής Γενικού Τμήματος Παν/μιου Πατρών Κωνσταντίνος Σφέτσος\\
\\
Καθηγητής Φυσικού Τμήματος Ε.Κ.Π.Α. Εμμανουήλ Φλωράτος\\
Καθηγητής Τμήματος Φυσικής Παν/μιου Πατρών Ιωάννης Μπάκας\\
Αναπληρωτή Καθηγητή Σ.Ε.Μ.Φ.Ε. Ε.Μ.Π. Αλέξανδρος Κεχαγιάς\\
Αναπληρωτής Καθηγητής Φυσικού Τμήματος Ε.Κ.Π.Α. Νικόλαος Τετράδης \\
Επίκουρος Καθηγητής, Τμήματος Φυσικής Παν/μιου Κρήτης Αναστάσιος Πέτκου\\
Λέκτορα Γενικού Τμήματος Παν/μιου Πατρών Αναστασία Δοίκου\\
\end{tabular}

\vspace{3cm}

Πάτρα, Σεπτέμβριος 2010
\end{center}
\end{titlepage}


\tableofcontents
\newpage
\newpage

\Huge

\addcontentsline{toc}{chapter}{Ευχαριστίες}
\textbf{Ευχαριστίες}

\normalsize

\vspace{2cm}
\no
Η παρούσα διδακτορική διατριβή εκπονήθηκε στο Γενικό Τμήμα του
Πανεπιστημίου Πατρών από τον Ιούνιο του 2005 έως τον Σεπτέμβριο του 2010
με επιβλέποντα τον Καθηγητή του Γενικού Τμήματος Κωνσταντίνο Σφέτσο
και βασίζεται στις ακόλουθες δημοσιεύσεις \cite{ASS1,ASS2,Sfetsos:2007nd,Sfetsos:2008yr}.
Στο παραπάνω χρονικό διάστημα εργάστηκα επίσης και
σε συμμετρίες δυισμού σε θεωρίες χορδών που οδήγησαν σε τρείς δημοσιεύσεις
\cite{Sfetsos:2009dj,Sfetsos:2009vt,Sfetsos:2010xa}.
Αυτές δεν εντάσσονται στο αντικείμενο αυτής της διδακτορικής διατριβής.
\no
Θα ήθελα να ευχαριστήσω τον Καθηγητή Κωνσταντίνο Σφέτσο για την υποστήριξη και
καθοδήγησή του κατά την διάρκεια του διδακτορικού όπως και κατά την διάρκεια της φοίτησης
μου στο διετές, μεταπτυχιακό πρόγραμμα του Φυσικού Τμήματος του Πανεπιστημίου Πατρών
(Σεπτέβριος 2003-Μάιος 2005).
Η υπόστηριξή του υπήρξε αμέριστη στα πλαίσια της επιστημονικής συνεργασίας αλλά και της
γενικότερης επικοινωνίας και συμπαράστασης.\\
\no
Ευχαριστώ τον Καθηγητή
Ιωάννη Μπάκα για την καθοδήγηση του κατά την διάρκεια των
σπουδών μου στο Φυσικό Τμήμα του Πανεπιστημίου Πατρών.
Οι οφειλές μου στην διατριβή αυτή συμπεριλαμβάνουν και τους Καθηγητές: Μάριο Πετρόπουλο
της \en Ec\'ole Polytechnique Paris,\ \gr Γεώργιο Παπαδόπουλο του Μαθηματικού Τμήματος
του \en King's College London \gr και Κωνσταντίνο Μπαχά της \en Ec\'ole Normale Superieure Paris
\gr για υποστήριξη τους και πολλές χρήσιμες συζητήσεις σε Φυσική.
Επίσης ευχαριστώ τους ιθύνοντες του
Μαθηματικού Τμήματος του \en King's College London, \gr και της \en Ec\'ole Polytechnique Paris
\gr για φιλοξενεία και για οικονομική υποστήριξη κατά το πρώτο εξάμηνο του 2008.
Συγκεκριμένα τα ευρωπαικά πρόγραμματα \en "Constituents, Fundamental Forces and
Symmetries of the Universe" \gr με συμβόλαιο \en MRTN-CT-2004-005104 \gr
και \en "European Superstring Theory Network" \gr με συμβόλαιο \en MCFH-2004-512194.\gr  \\
\no
Είμαι ευγνώμων στο Ίδρυμα Κρατικών Υποτροφιών (ΙΚΥ) για την οικονομική
ενίσχυση που μου παρείχε κατά την εκπόνηση της διδακτορικής μου διατριβής,
Νοέμβριος 2005 - Απρίλιος 2009. \\
\no
Τέλος ευχαριστώ τους μεταδιδακτορικούς ερευνητές Σπύρο Αβράμη και
Δημήτριο Ζωάκο  καθώς και τους διδακτορικούς φοιτητές Στέλιο Αλεξανδρή
και Νικόλαο Καραίσκο, για την συνεργασία τους καθώς και τους γονείς μου, την αδερφή
μου και τους φίλους μου για την
υποστήριξη τους κατά την διάρκεια εκπόνησης της διατριβής.

\vspace{0.0cm}
\begin{flushright}
\textit{Πάτρα, Σεπτέμβριος 2010}
\end{flushright}

\newpage
\Huge
\addcontentsline{toc}{chapter}{Πρόλογος}
\textbf{Πρόλογος}

\normalsize

\vspace{2cm}
\no
Η έως τώρα μικροσκοπική περιγραφή της ύλης και η οποία έχει επιβεβαιωθεί και
πειραματικά είναι μέσω της κβαντικής θεωρίας πεδίου. Τα στοιχειώδη σωματίδια
είναι σημειακά, αλληλεπιδρούν τοπικά με άλλα σωματίδια και είναι διεγέρσεις
κάποιου πεδίου. Παρότι, αυτή περιγράφει με μεγάλη ακρίβεια τη φύση
στις ενέργειες ή στις αποστάσεις που μελετούμε, υπάρχουν ενδείξεις για νέα
στοιχεία σε μεγαλύτερες ενέργειες (μικρότερες αποστάσεις) έως και την τάξη μεγέθους της μάζας
\en Plank. \gr Ο λόγος είναι ότι για αυτές τις  αποστάσεις (ή ενέργειες) η κβάντωση
της βαρύτητας είναι απαραίτητη. Ωστόσο, η κλασική βαρύτητα δεν είναι δυνατόν να κβαντωθεί με
τις συνήθεις διαταρακτικές μεθόδους, διότι είναι μη επανακανονικοποιήσιμη.
Η κβάντωση της βαρύτητας μπορεί να επιτευχθεί θεωρώντας
ότι τα θεμελιώδη αντικείμενα δεν είναι τα σημειακά σωματίδια,
αλλά μονοδιάστατα αντικείμενα, χορδές. Οι χορδές αυτές μπορούν να ταλαντώνονται και να περιστρέφονται,
συνεπώς έχουν ένα φάσμα ενεργειών
και στροφορμή. Οι ταλαντευόμενες αυτές χορδές είναι εντοπισμένες, και για μικρές
ενέργειες μπορούν να θεωρηθούν ως σωματιδιακές διεγέρσεις. Συνεπώς, μια
χορδή που ταλαντώνεται μπορεί να συμπεριφέρεται ως διάφορα σωματίδια, ανάλογα  με τον τρόπο ταλάντωσης
στον οποίο βρίσκεται. Στο φάσμα της περιέχει ένα σωμάτιο με μηδενική μάζα και
ιδιοστροφορμή δύο, το βαρυτόνιο, φορέα των βαρυτικών αλληλεπιδράσεων.
Από την δομή της και την απαίτηση να περιγράφει και φερμιονικές διεγέρσεις,
οδηγούμαστε σε ένα δεκαδιάστατο χώρο και στην ύπαρξη υπερσυμμετρίας \cite{Green:1984sg}.
Στην συνέχεια θεωρούμε τη θεωρία χορδών σε χώρο της μορφής $\mathbb{R}^{3,1}\times M^6$, όπου
ο χώρος $M^6$ είναι μια εξαδιάστατη συμπαγής πολλαπλότητα, η οποία καθορίζει και τις αλληλεπιδράσεις
για χαμηλές ενέργειες.\\
\no
Αυτός όμως, δεν ήταν ο πρωταρχικός λόγος ανάπτυξης της θεωρίας χορδών. Εφευρέθηκε την δεκαετία
του '60 ως ένα πρότυπο περιγραφής για τα μεσόνια και βαρυόνια.
Με την θεώρηση αυτή μπορούμε να περιγράψουμε κάποια από τα βασικά στοιχεία του φάσματος των αδρονίων.
Παραδείγματος χάριν, η μάζα του ελαφρότερου αδρονίου για δεδομένη ιδιοστροφορμή ακολουθεί
τις τροχιές \en Regge,
\gr $m^2\sim T s^2+\textrm{σταθερά}$, όπου $m$ η μάζα της χορδής, $T$ η τάση της και $s$ η ιδιοστροφορμή της.
Αργότερα βρέθηκε ότι τα αδρόνια αποτελούνται από κουάρκ και η περιγραφή τους δίνεται μέσα
απο την κβαντική χρωμοδυναμική (\en QCD) \gr και η θεωρία χορδών έπαψε να θεωρείται πρότυπο
περιγραφής των αδρονίων. \\
\no
Η \en QCD \gr είναι μια θεωρία βαθμίδας με ομάδα συμμετρίας $SU(3)$,
η οποία αντιστοιχεί στο γεγονός ότι τα κουάρκ έχουν τρία χρώματα. Η \en
QCD \gr είναι επίσης ασυμπτωτικά ελεύθερη \cite{bfunction}, δηλαδή η σταθερά ζεύξης είναι μικρή για μεγάλες
ενέργειες. Για χαμηλές ενέργειες είναι ισχυρά συζευγμένη και κατά συνέπεια
δεν μπορούμε να εφαρμόσουμε τις γνωστές διαταρακτικές μεθόδους. Αυτή είναι και η αιτία για την οποία
δεν παρατηρούμε ελεύθερα κουάρκ παρά μόνο εντός δέσμιων καταστάσεων. Η μελέτη της
για μικρές ενέργειες γίνεται μέσω αριθμητικών προσομοιώσεων στο πλέγμα.
Θεωρώντας ότι η ομάδα συμμετρίας ήταν η $SU(N)$, η θεωρία απλοποιείται
στο όριο του μεγάλου αριθμού χρωμάτων $N$ όπως προτάθηκε από τον
\en 't Hooft \cite{Hooft}, \gr και μπορεί να προσεγγιστεί από μια
ελεύθερη θεωρία κλειστών χορδών, της οποίας η σταθερά ζεύξης είναι το $1/Ν$. Το όριο
αυτό εξηγεί το λόγο για τον οποίο η θεωρία χορδών έδινε την αναμενόμενη εξάρτηση
μεταξύ μάζας και στροφορμής για τα αδρόνια. Το όριο του \en 't Hooft \gr όμως είναι πολύ
πιο γενικό και συνδέει διάφορα είδη θεωριών πεδίου και διάφορες θεωριές χορδών.
Αντικείμενο αυτής της διατριβής είναι η μελέτη της αντιστοιχίας μεταξύ θεωριών
βαθμίδας και θεωριών χορδών και οι εφαρμογές αυτής. Πιο συγκεκριμένα, η θεωρία
χορδών σε κάποια υπόβαθρα είναι δυική σε μια θεωρία πεδίου. Το αρχέτυπο αυτής της αντιστοιχίας
είναι η αναλογία μεταξύ της $\cN=4\ $\en  SYM \gr \cite{SYM} και της ΙΙΒ δεκαδιάστατης
υπερβαρύτητας σε έναν χώρο $AdS_5\times S^5$ \cite{adscft}(για μια
εισαγωγή δείτε την \cite{ads-intro}).\\
\no
Ένα από τα βασικά σημεία της διατριβής αυτής είναι η εφαρμογή της αντιστοιχίας
$ΑdS/CFT$ και των γενικεύσεων αυτής για την μελέτη δέσμιων καταστάσεων της
δυικής θεωρίας πεδίου, όπως μεσόνια, δυόνια και βαρυόνια. Στο πλαίσιο αυτής
της αναλογίας μπορούμε να κατασκευάσουμε δέσμιες καταστάσεις κουάρκ
στην θεωρία βαθμίδας και να υπολογίσουμε το δυναμικό αλληλεπίδρασης.
Για παράδειγμα, στην θεωρία πεδίου το δυναμικό αλληλεπίδρασης ενός μεσονίου
δίνεται από την αναμενόμενη τιμή ενός ορθογώνιου βρόχου \en Wilson \gr με μια
χωρική και μια χρονική ακμή. Στη βαρύτητα η μέση τιμή του βρόχου \en Wilson \gr
δίνεται από την ελαχιστοποίηση της δράσης \en Nambu--Goto \gr μιας χορδής που ταλαντεύεται
στο δυικό βαρυτικό υπόβαθρο και της οποίας τα άκρα βρίσκονται στις δύο χωρικές
ακμές του βρόχου \en Wilson, \gr εκεί που βρίσκονται τα κουάρκ. Αντίστοιχες
κατασκευές μέσω κλασικών λύσεων μπορούν να γίνουν για τον υπολογισμό των δέσμιων
καταστάσεων δυονίων και βαρυονίων.
Ο υπολογισμός των ενεργειών δέσμιων καταστάσεων στο πλαίσιο της αντιστοιχίας
$ΑdS/CFT$ οδηγεί συχνά σε συμπεριφορές οι οποίες διαφέρουν απο το αυτές που αναμένουμε
απ' την θεωρία πεδίου. Για να απομονώσουμε τις φυσικώς ενδιαφέρουσες
συμπεριφορές, θα πρέπει να εξετάσουμε την ευστάθεια των λύσεων χορδών οι οποίες
είναι δυικές σε αυτές τις δέσμιες καταστάσεις των κουάρκ. Η ανάλυση αυτή
θα στηριχθεί στην μελέτη των μικρών διακυμάνσεων περί την κλασική λύση.
Θα αποδείξουμε γενικά θεωρήματα που σχετίζονται με την διαταρακτική ευστάθεια των
λύσεων αυτών και θα εφαρμόσουμε τα
αποτελέσματα μας σε διάφορα υπόβαθρα και για τα τρία είδη δέσμιων καταστάσεων που περιγράφουν,
μεσόνια, δυόνια και βαρυόνια. Σε όλες τις περιπτώσεις, οι προβληματικές περιοχές
και συμπεριφορές είναι ασταθείς και κατά συνέπεια αντιστοιχούν σε
φυσικώς αδιάφορες περιοχές.\\
\no
Η διατριβή αυτή οργανώνεται ως εξής: Στο πρώτο μέρος, θα δώσουμε κάποιες
εισαγωγικές έννοιες απαραίτητες για την μελέτη μας, όπως μια ανασκόπηση
στις θεωρίες πεδίου βαθμίδας για μεγάλο $N$,
την σύμμορφη συμμετρία και τους χώρους $AdS$, όπως και την διατύπωση
της αναλογίας \en AdS/CFT.\ \gr Στο δεύτερο μέρος, θα συνοψίσουμε τα αποτελέσματα
των εργασιών μας \cite{ASS1,ASS2} για την περίπτωση των μεσονίων.
Συγκεκριμένα θα μελετήσουμε τις κλασικές λύσεις αυτών και θα εξετάσουμε την ευστάθεια
αυτών των λύσεων όπου αυτό κρίνεται
απαραίτητο, αποδεικνύοντας διάφορα θεωρήματα που σχετίζονται με την ευστάθεια αυτών.
Στην συνέχεια θα εφαρμόσουμε τα αποτελέσματα μας σε διάφορα φυσικώς ενδιαφέροντα
παραδείγματα και θα δείξουμε ότι τα αποτελέσματα μας είναι σε πλήρη συμφωνία με τα
αναμενόμενα αποτελέσματα από την θεωρία πεδίου.
Στο τρίτο μέρος,  θα συνοψίσουμε τα αποτελέσματα
της εργασίας μας \cite{Sfetsos:2007nd} για την περίπτωση των δυονίων.
Συγκεκριμένα, θα κατασκευάσουμε δυόνια
εντός της \en AdS/CFT \gr και θα μελετήσουμε την ευστάθεια των λύσεων όπου αυτή κρίνεται απαραίτητη.
Θα δείξουμε ότι τα αποτελέσματα από την ανάλυση της ευστάθειας των κλασικών λυσεων
διαφέρουν από αυτό που οδηγείται κανείς στηριζόμενος σε ενεργειακά επιχειρήματα.
Τέλος στο τέταρτο μέρος, θα συνοψίσουμε τα αποτελέσματα
της εργασίας μας \cite{Sfetsos:2008yr} για την περίπτωση των βαρυονίων.
Θα μελετήσουμε την κατασκευή των βαρυονίων εντός της αντιστοιχίας \en AdS/CFT \gr
και θα δείξουμε ότι μπορούμε να κατασκευάσουμε δέσμιες καταστάσεις με πλήθος κουάρκ
μικρότερο του $N$, όπου $N$ ο αριθμός των χρωμάτων.
Αυτό έρχεται σε αντίθεση με τα προσδοκώμενα αποτελέσματα απ' την
θεωρία πεδίου αλλά όπως θα δείξουμε ότι η ανάλυση της ευστάθειας βελτιώνει το κλασικό φράγμα.

\newpage

\newpage

\pagenumbering{arabic}

\part{Εισαγωγικές έννοιες}
\chapter{Θεωρίες βαθμίδας για μεγάλο $N$ ως θεωρίες χορδών}
\rm{Σε αυτό το κεφάλαιο θα συνοψίσουμε μερικά βασικά στοιχεία που
αφορούν στη σύνδεση των θεωριών πεδίου (βαθμίδας) και των θεωριών κλειστών χορδών.} \gr

\no
Η σύνδεση των θεωρίων χορδών και θεωριών βαθμίδας
είναι ένα ενδιαφέρον αντικείμενο έρευνας για περισσότερο από
σαράντα χρόνια. Η θεωρία χορδών αναπτύχθηκε αρχικά ως
μια υποψήφια θεωρία για την περιγραφή των ισχυρών αλληλεπιδράσεων, όπως ο
εκλωβισμός των κουάρκ και οι τροχιές \en Regge. \gr Αργότερα
βρέθηκε πως η περιγραφή των ισχυρών αλληλεπιδράσεων δίνεται μέσω μιας
θεωρίας βαθμίδας \en $SU(3)$ (QCD),\ \gr
η οποία είναι συνεπής με όλα τα πειραματικά δεδομένα έως σήμερα.
Παρόλα αυτά, ενώ οι θεωρίες βαθμίδας δίνουν μια πολύ καλή περιγραφή
των ισχυρών αλληλεπιδράσεων σε υψηλές ενέργειες, είναι πολύ δύσκολο να
μελετήσουμε φαινόμενα σε χαμηλές ενέργειες όπως ο εγκλωβισμός των κουάρκ
και το σπάσιμο της χειραλικής συμμετρίας. \\
\no
Τα τελευταία χρόνια έχουν βρεθεί πολλά παραδείγματα ενός φαινομένου το οποίο είναι γνωστό ως
δυισμός, σύμφωνα με τον οποίο μια θεωρία έχει τουλάχιστον συμπληρωματικές περιγραφές,
όπως για παράδειγμα στο ασθένες και στο ισχυρό όριο ζεύξης όπως προαναφέραμε.
Υπάρχουν αρκετές ενδείξεις ότι μια τέτοια δυική περιγραφή των ισχυρών αλληλεπιδράσεων
σε χαμηλές ενέργειες όπου έχουμε ισχυρή ζεύξη των θεωριών βαθμίδας, μπορεί να είναι η θεωρία χορδών.
Συγκεκριμένα στην \en QCD \gr υπάρχουν αντικείμενα παρεμφερή των χορδών
όπως οι γραμμές ροής ή γραμμές \en Wilson. \gr ’ν προσπαθήσουμε
να χωρίσουμε ένα ζεύγος κουάρκ και αντικουάρκ, μια γραμμή
ροής θα δημιουργηθεί μεταξύ αυτών των δύο. Οι γραμμές ροής συμπεριφέρονται όπως οι χορδές
και έχουν γίνει αρκετές προσπάθειες ώστε να γράψουμε μια θεωρία χορδών που να περιγράφει τις ισχυρές
αλληλεπιδράσεις και στην οποία οι γραμμές ροής θα είναι τα βασικά αντικείμενα.
Είναι προφανές ότι αυτή η περιγραφή έχει αρκετά ενδιαφέρουσες φαινομενολογικές
εφαρμογές. Η πιο ξεκάθαρη ένδειξη είναι ότι η θεωρία βαθμίδας μπορεί να περιγραφεί ως θεωρία χορδών
για μεγάλο $N$, το όριο \en 't Hooft \cite{Hooft}, \gr το οποίο ακολούθως θα περιγράψουμε.
Μία τετραδιάστατη $SU(N)$ θεωρία βαθμίδας \en Yang-Mills \gr έχει, πέρα από την αδιάστατη
σταθερά ζεύξης $g_{YM}$ (και την αντίστοιχη κλίμακα της \en QCD $\L_{QCD}$), \gr μια επιπρόσθετη παράμετρο,
που είναι ο αριθμός των χρωμάτων, $N$. Θέλουμε να κατανοήσουμε πως συμπεριφέρεται η σταθερά ζεύξης στο
όριο των μεγάλων $N$. Η συνάρτηση βήτα για την $SU(3)$ θεωρία \en Yang-Mills \gr υπολογίσθηκε
στην εργασία \cite{bfunction}.
Ενώ στην περίπτωση του $SU(N)$ (δείτε για παράδειγμα τις \cite{Zuber,Peskin}) είναι

\ba
\label{1-1}
\mu\frac{dg_{YM}}{d\mu}=-\frac{11}{3}N\frac{g_{YM}^3}{(4\pi)^2}+{\cal O}(g_{YM}^5)\ ,
\ea

\no
όπου έχουμε θεωρήσει ότι ο αριθμός χρωμάτων $N$ είναι πολύ μεγαλύτερος αριθμό γεύσεων $N_f$.
Παρατηρούμε ότι η σταθερά ζεύξης είναι μια φθίνουσα συνάρτηση της κλίμακας κάτι
το οποίο ερμηνεύεται ως ότι η θεωρία είναι ασυμπτωτικά ελέυθερη στο όριο των μεγάλων ενεργειών.
Μπορούμε να βρούμε ένα κατάλληλο όριο απαιτώντας ότι οι κυρίαρχοι όροι είναι της ιδίας τάξεως.
Αυτό μας υποδεικνύει ότι ο συνδυασμός
\ba
\label{1-2}
\l=g_{YM}^2N,
\ea
\no
πρέπει να είναι πεπερασμένος καθώς το $N\to\infty$. Αυτό είναι γνωστό ως όριο του \en 't Hooft
\gr και το $\l$ ονομάζεται σταθερά ζεύξης \en 't Hooft.
\gr Αντικαθιστώντας την Εξ. \eqn{1-2} στην Εξ. \eqn{1-1} βρίσκουμε την συνάρτηση βήτα για
την σταθερά σύζευξης $\l$

\ba
\label{1-3}
\mu\frac{d\l}{d\mu}=-\frac{11}{24\pi^2}\l^2\ +{\cal O}(\l^3)\ .
\ea
\no
\gr
Αξίζει να αναφέρουμε ότι η Εξ.\eqn{1-3}, ισχύει και στην περίπτωση ύπαρξης κατάλληλου
αριθμού άλλων σωματιδίων πέρα των γκλουονίων της \en Yang--Mills, \gr όπως φερμιονίων ή βαθμωτών πεδίων.
’ν η θεωρία είναι σύμμορφη όπως η ${\cal N}=4$ \en Super-Yang-Mills \gr τότε υπάρχουν κι άλλα όρια πέρα του
πεπερασμένου $\l$, όπως το $\l\to\infty$ τα οποία είναι δυνατά.\\
\no
Η θεωρία διαταραχών για μεγάλο $N$ και η σύνδεση της με μια θεωρία βαθμίδας είναι
χαρακτηριστικό κάθε γενική θεωρίας της οποίας τα πεδία μετασχηματίζονται στην προσαρτημένη (\en adjoint) \gr
αναπαράσταση της $SU(N)$. Η τελευταία έχει πεδία $X_i^\a$, όπου το $\a$ σχετίζεται με την
προσαρτημένη αναπαράσταση της $SU(N)$ και το $i$ περιγράφει
την ταυτότητα του σωματιδίου (δείκτης \en Lorentz \gr ή \en spin \gr ή γεύση, κ.λ.π.).
Σχηματικά η Λαγκρανζιανή αυτών των πεδίων είναι της μορφής

\ba
\label{1-4}
{\cal L}\sim Tr[dX_idX_i+g_{YM}^2d_{ijkl}X_iX_jX_kX_l]\ ,
\ea
\no
όπου $d_{ijkl}$ σταθερά. Κανονικοποιώντας τα πεδία $g_{YM}X_i\equiv\tilde{X}_i$,
η Λαγκρανζιανή γράφεται ως

\ba
\label{1-5}
{\cal L}\sim \frac{N}{\l}Tr[d\tilde{X}_id\tilde{X}_i
+d_{ijkl}\tilde{X}_i\tilde{X}_j\tilde{X}_k\tilde{X}_l]\ ,
\ea
\no
όπου η πλήρης εξάρτηση απο την σταθερά ζεύξης εμφανίζεται ως σταθερά αναλογίας στην δράση.
Τα διαγράμματα \en Feynman \gr της θεωρίας αυτής γράφονται με συμβολισμό διπλής γραμμής,
στον οποίο τα πεδία $X^\a$ που μετασχηματίζονται στην προσαρτημένη αναπαράσταση
συμβολίζονται ως ένα ευθύ γινόμενο ενός θεμελιώδους και ενός
αντι-θεμελιώδους πεδίου πεδίου, $X_j^i$. Οι διαδότες αυτών των $SU(N)$ πεδίων έχουν έναν
όρο ανάμειξης ο οποίος είναι διόρθωση σε πρώτη τάξης ως προς $1/N$, διότι

\ba
\label{1-6}
\langle \tilde{X}_j^i\tilde{X}_l^k\rangle \sim {\l\ov N}\left(\d_l^i\d_k^j-\frac{1}{N}\d_i^j\d_k^l\right)\ .
\ea

\no
Στο όριο των μεγάλων $N$ μπορούμε να αμελήσουμε αυτόν τον όρο. Για αυτό
κάθε διάγραμμα \en Feynman \gr των προσαρτημένων πεδίων μπορεί να γραφεί και
μέσω διπλών γραμμών. Σε έναν τέτοιο συμβολισμό κάθε διάγραμμα αντιστοιχεί σε μια
κλειστή προσανατολισμένη διδιάστατη επιφάνεια.\footnote{Η ύπαρξη του
προσανατολισμού σχετίζεται με το γεγονός ότι το κουάρκ και το αντικουάρκ μετασχηματίζονται
σε διαφορετικές αναπαραστάσεις της $SU(N)$. ’ν όμως,
η ομάδα συμμετρίας ήταν η $SO(N)$, στην οποία το κουάρκ και αντικουάρκ μετασχηματίζονται στην ίδια
αναπαράσταση, θα μπορούσαμε να έχουμε και μη προσανατολισμένες επιφάνειες,
για παράδειγμα το μπουκάλι του \en Klein.\gr}
Κάθε βρόχος είναι μια έδρα της επιφάνειας σε μια αποσυναρμολόγηση της,
κάθε διαδότης είναι μια ακμή αντίστοιχα, και κάθε κόμβος
αλληλεπίδρασης είναι μια κορυφή της επιφάνειας. Η επιφάνεια γίνεται συμπαγής
αν προσθέσουμε στο χώρο μας ένα σημείο στο άπειρο.
Για να υπολογίσουμε τους παράγοντες $N$ και $\l$ που συνδέονται με κάθε διάγραμμα,
παρατηρούμε ότι από τη μορφή της Εξ. \eqn{1-5} για κάθε κόμβο έχουμε έναν παράγοντα ανάλογο του
$N/\l$, ενώ οι διαδότες είναι ανάλογοι του $\l/N$. Επιπλέον δυνάμεις του $N$ προκύπτουν από άθροιση
δεικτών στους βρόχους, οι οποίοι δίνουν έναν παράγοντα $N$ σε κάθε βρόχο(κάθε δείκτης έχει $N$ δυνατές
τιμές στη θεμελιώδη αναπαράσταση). Οπότε, ένα διάγραμμα με $K$ κόμβους (κορυφές), $A$ διαδότες (ακμές) και
$E$ βρόχους (έδρες) έχει έναν συντελέστη ανάλογο του

\ba
\label{1-7}
N^{K+E-A}\l^{A-K}=N^\chi \l^{A-K}\ ,
\ea
όπου $\chi=K+E-A$ είναι ο αριθμός \en Euler \gr της διδιάστατης επιφανείας που αντιστοιχεί στο διάγραμμα.
Για μία κλειστή προσανατολισμένη επιφάνεια, $\chi=2-2g$, όπου $g$ το γένος της επιφάνειας.\gr
Για παράδειγμα η σφαίρα έχει $g=0$ και ο τόρος έχει $g=1$.

\begin{figure}[!t]
\begin{center}
(α) \includegraphics[scale=0.5]{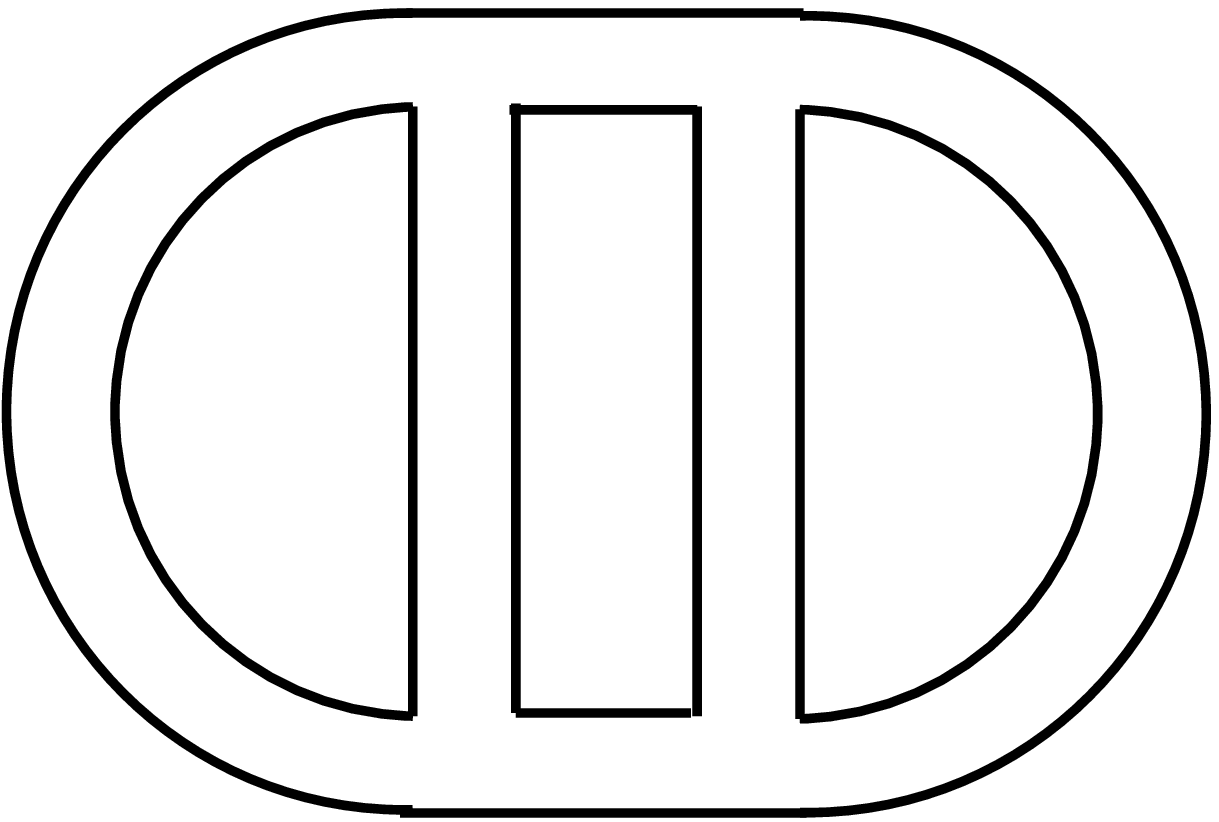}\qq
(β) \includegraphics[scale=0.5]{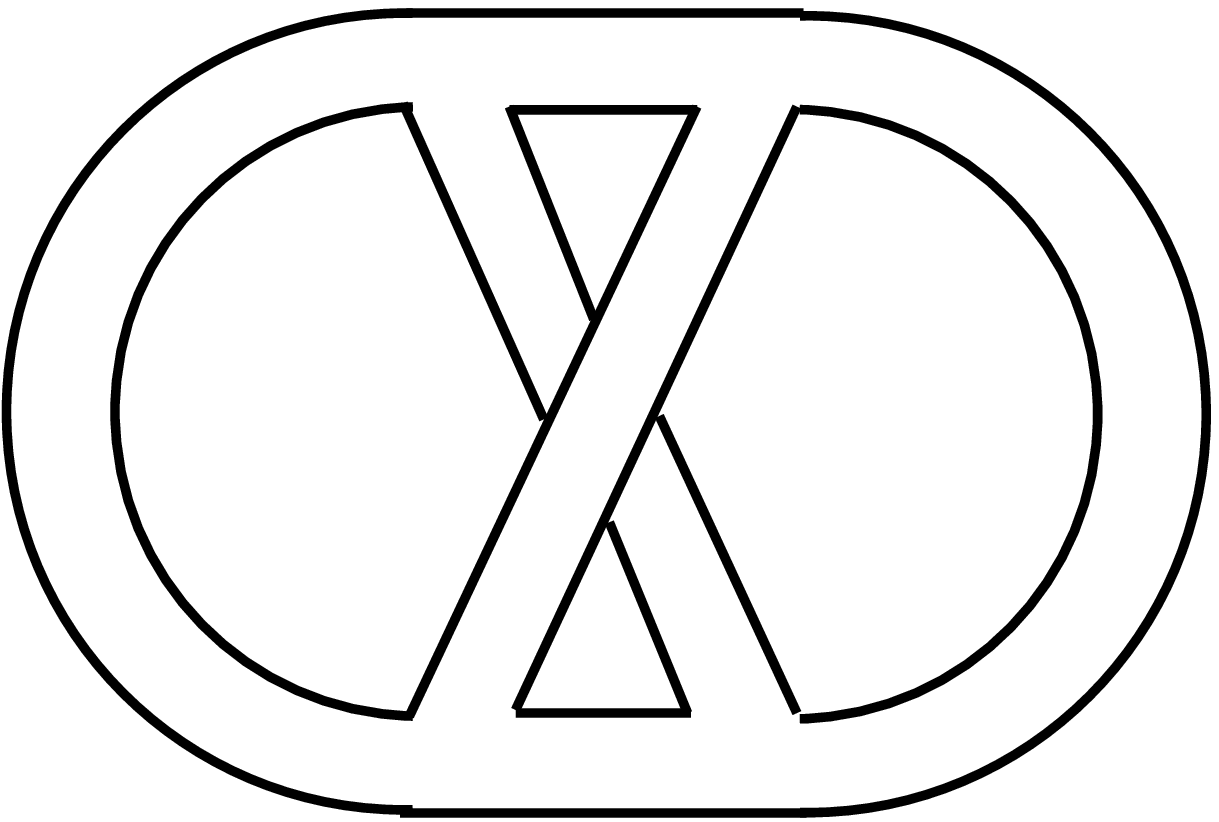}
\end{center}
\caption{ Διαγράμματα στην θεωρία πεδίου με πεδία στην προσαρτημένη αναπαράσταση
με το συμβολισμό της διπλής γραμμής. Το (α) διάγραμμα είναι επίπεδο και τάξεως $\l^2 N^2$, ενώ το
διάγραμμα (β) είναι τάξεως $\l^2$. Γιά πεπερασμένο $Ν$ τα διαγράμματα είναι ισοβαρή.
} \label{Meson}
\end{figure}
\no
\gr Οπότε, κάθε διαταρακτική σειρά ενός οποιουδήποτε διαγράμματος της θεωρίας αυτής μπορεί
να γραφεί σαν ένα διπλό ανάπτυγμα για μεγάλο $N$, (και μικρό $\l$) ως εξής:

\ba
\label{1-8}
\sum_{g=0}^\infty N^{2-2g}\sum_{i=0}^\infty c_{g,i}\l^i=\sum_{g=0}^\infty N^{2-2g}Z_g(\l)\ .
\ea
Στο όριο των μεγάλων $N$ το αποτέλεσμα έχει ως κυρίαρχο όρο επιφάνειες με μεγάλο $\chi$ ή μικρό $g$,
οι οποίες είναι επιφάνειες με την τοπολογία της σφαίρας. Αυτά τα διαγράμματα
ονομάζονται επίπεδα διαγράμματα διότι η σφαίρα είναι τοπολογικά ισοδύναμη με ένα επίπεδο.
Η μορφή του αναπτύγματος Εξ.\eqn{1-8} είναι ανάλογη με το αντίστοιχο ανάπτυγμα στην διαταρακτική
θεωρία των προσανατολισμένων κλειστών χορδών, αν θεωρήσουμε το $1/N$ ως την σταθερά ζεύξης της χορδής.
Η αναλογία της Εξ.\eqn{1-8} με την διαταρακτική θεωρία των χορδών  είναι μια ένδειξη ότι οι θεωρίες πεδίων
και θεωρίες χορδών συσχετίζονται και η σύνδεση αυτή είναι πιο εμφανής στο όριο των μεγάλων $N$ όπου
η δυική θεωρία χορδών είναι σε ασθενή ζεύξη. Όμως, η αναλογία αυτή στηρίχτηκε σε επιχειρήματα τα
οποία είναι βασισμένα σε θεωρία διαταραχων, και η οποία εν γένει δεν συγκλίνει. Ώς εκ τούτου τα παραπάνω
επιχειρήματα αποτελούν απλώς μια ένδειξη του δυισμού και απέχουν πολύ από το να θεωρηθούν απόδειξη.
Όπως θα δούμε η \en AdS/CFT \gr στηρίζεται στον παραπάνω δυισμό.


\chapter{Χώροι \en Anti-de-Sitter \gr και σύμμορφες θεωρίες πεδίου}
\rm{Σε αυτό το κεφάλαιο θα συνοψίσουμε τα βασικά χαρακτηριστικά του
χώρου \en Anti-de-Sitter (AdS) \gr και των σύμμορφων θεωριών πεδίου. \gr
Η γεωμετρία αυτού του χώρου έχει μελετηθεί εκτενώς στην θεωρία
χορδών και είναι ένα απο τα βασικά εργαλεία στην δυικότητα (\en duality)
AdS/CFT.}

\no
\gr Η κατανόηση μιας θεωρίας πεδίου στηρίζεται σε μεγάλο βαθμό
στην ύπαρξη συμμετριών, όπως είναι οι \en Lorentz \gr και \en Poincare.
\gr Μια ενδιαφέρουσα γενίκευση της συμμετρίας \en Poincare \gr
είναι η ύπαρξη συμμετρίας κάτω απο αλλαγές κλίμακας.
Όπως για παράδειγμα, η θεωρία \en Maxwell \gr σε τέσσερις διαστάσεις και
απουσία πηγών είναι αναλλοίωτη σε κλασικό επίπεδο κάτω από
αλλαγές κλίμακας. Όμως η συμμετρία αυτή παύει να
ισχύει σε κβαντικό επίπεδο, λόγω ύπαρξης μιας ενέργειας αποκοπής, η οποία εισάγεται
για λόγους επανακανονικοποίησης της θεωρίας. Οι θεωρίες αυτές όμως είναι εν γένει αναλλοίωτες
κάτω από την πλήρη σύμμορφη ομάδα, η οποία είναι γενίκευση της ομάδας συμμετρίας \en Poincare. \gr

\section{Η σύμμορφη ομάδα}

Η σύμμορφη ομάδα είναι η ομάδα των μετασχηματισμών που αφήνουν αναλλοίωτη την μορφή της
μετρικής πέραν της απόκτησης ενός παράγοντα κλίμακας, $g_{\mu\nu}\to\Om^2(x)g_{\mu\nu}$
όπου $\mu=0,\dots,d-1$.
Η σύμμορφη ομάδα του χώρου \en Minkowski \gr συμπεριλαμβάνει τους
μετασχηματισμούς \en Poincare, \gr την αλλαγή κλίμακας και τους ειδικούς σύμμορφους
 μετασχηματισμούς αντίστοιχα
\ba
\label{3-1}
&&x^\mu\to\l x^\mu\ , \nonumber \\
&&x^\mu\to\frac{x^\mu+\a^\mu x^2}{1+2x\cdot\a+\a^2x^2}\ .
\ea
Οι γεννήτορες της σύμμορφης ομάδας είναι οι $M_{\mu\nu}$ για τους μετασχηματισμούς
\en Lorentz \gr και τις στροφές, $P_\mu$  για τους μετασχηματισμούς μεταφοράς,
$D$ για τους μετασχηματισμούς κλίμακας
και $K_\mu$ για τους ειδικούς σύμμορφους μετασχηματισμούς. Οι γεννήτορες αυτοί κλείνουν στην
σύμμορφη άλγεβρα
\ba
\label{3-2}
&&[M_{\mu\nu},P_\rho]=-i(\eta_{\mu\rho}P_\nu-\eta_{\nu\rho}P_\mu); \quad
[M_{\mu,nu},K_\rho]=-i(\eta_{\mu\rho}K_\nu-\eta_{\nu\rho}K_\mu); \nonumber \\
&&[M_{\mu\nu},M_{\rho\s}]=-i\eta_{\mu\rho}M_{\nu\s}\pm \textrm{ εναλλαγές}; \quad
[M_{\mu\nu},D]=0; \quad [D,K_\mu]=iK_\mu; \\
&&[D,P_\mu]=-iP_\mu; \quad [P_\mu,K_n]=2iM_{\mu\nu}-2i\eta_{\mu\nu}D\ . \nonumber
\ea
\no
Η άλγεβρα αυτή είναι ισόμορφη με την άλγεβρα $SO(d,2)$(με υπογραφή
$(-,+,+,\cdots,+,-)$) με γεννήτορες $J_{ab}$ ($a,b=0,1\dots,d+1$) που ορίζονται ως
\ba
\label{3-3}
J_{\mu\nu}=M_{\mu\nu};\quad J_{\mu d}=\half(K_\mu-P_\mu);
\quad J_\mu(d+1)=\half(K_\mu+P_\mu);\quad J_{(d+1)d}=D\ .
\ea
\no
Αξίζει να σημειώσουμε ότι στην ειδική περίπτωση που το $d=2$; η
σύμμορφη ομάδα είναι απειροδιάσταση.

\section{Στοιχειώδεις ιδιότητες των χώρων \en AdS}

Ένα απο τα βασικά χαρακτηριστικά της δυικότητας \en AdS/CFT \gr
είναι η ταυτοποίηση της ομάδας ισομετρίας του $AdS_{d+1}$ και της
σύμμορφης συμμετρίας του επίπεδου χώρου \en Minkowskι $\mathbb{R}^{1,d-1}$.
\gr Ο χώρος \en AdS \gr διάστασης $d+1$ μπορεί να παρασταθεί ως ένα υπερβολοειδές

\ba
\label{3-4}
X_0^2+X_{d+1}^2-\sum_{i=1}^d X_i^2=R^2\ ,
\ea
σε ένα επίπεδο χώρο δίαστασης $d+2$ με μετρική
\ba
\label{3-5}
ds^2=-dX_0^2-dX_{d+1}^2+\sum_{i=1}^d dX_i^2\ .
\ea
Εκ κατασκευής, ο χώρος
αυτός έχει ομάδα ισομετρίας $SO(d,2)$, είναι ομογενής
και ισοτροπικός. Η εξίσωση \eqn{3-4} μπορεί να λυθεί θέτοντας
\ba
\label{3-6}
&&X_0=R\cosh\rho\cos\tau\ , \quad X_{d+1}=R\cosh\rho\sin\tau\ , \nonumber \\
&&X_i=R\sinh\rho\ \Om_i\ , i=1,2\dots,d; \quad \sum_{i=1}^d\Om_i^2=1\ .
\ea
Αντικαθιστώντας την Εξ.\eqn{3-6} στην Eξ.\eqn{3-5} βρίσκουμε την μετρική του $AdS_{d+1}$
\ba
\label{3-7}
ds^2=R^2(-\cosh^2\rho d\tau^2+d\rho^2+\sinh^2\rho d\Om^2)\ .
\ea
\no
Αξίζει να σημειώσουμε ότι η παραμετροποίηση της Εξ.\eqn{3-6}
για $\rho\geqslant0$ και $0\leqslant\tau\leqslant 2\pi$
καλύπτει όλο το υπερβολοειδές και οι συντεταγμένες αυτές λέγονται ολικές (\en global).
\gr Πέρα από την παραμετροποίηση της Εξ.\eqn{3-6} του
$AdS$, υπάρχει ένα άλλο σύνολο συντεταγμένων $(u,t,\vec{x})$
($u>0,\vec{x}\in\mathbb{R}^d$) το οποίο θα μας χρειαστεί αργότερα
\ba
\label{3-8}
&&X_0=\frac{1}{2u}(1+u^2(R^2+\vec{x}^2-t^2))\ , \quad X_{d+1}=Rut,\nonumber \\
&&X_i=Rux_i\ , \quad i=1,\dots,d-1\ , \nonumber \\
&&X_d=\frac{1}{2u}(1-u^2(R^2-\vec{x}^2+t^2))\ .
\ea
\no
Σε αυτές τις συντεταγμένες καλύπτουμε το μισό υπερβολοειδές Εξ.\eqn{3-4}.
Αντικαθιστώντας τις συντεταγμένες αυτές στην Εξ.\eqn{3-5}, βρίσκουμε μια
άλλη μορφή της μετρικής του χώρου $AdS_{d+1}$
\ba
\label{3-9}
ds^2=R^2\left(\frac{du^2}{u^2}+u^2(-dt^2+d\vec{x}^2)\right)\ .
\ea
\no
Οι συντεταγμένες $(u,t,\vec{x})$, ονομάζονται συντεταγμένες \en Poincare.
\gr Στην μορφή αυτή της μετρικής, οι υποομάδες $ISO(1,d-1)$ και $SO(1,1)$ της
ισομετρίας $SO(d,2)$ της αρχικής μετρικής είναι εμφανείς, όπου $ISO(1,d-1)$
είναι οι \en Poincare \gr μετασχηματισμοί στο $(t,\vec{x})$  και $SO(1,1)$ είναι:
\ba
\label{3-10}
(u,t,\vec{x})\rightarrow(c^{-1}u,ct,c\vec{x}), \quad c>0.
\ea
Στην \en AdS/CFT, \gr η συμμετρία αυτή αντιστοιχεί στον τελεστή κλίμακας $D$ της σύμμορφης
ομάδας του $\mathbb{R}^{1,d-1}$. Ο χώρος \en AdS \gr αποτελεί
λύση των εξισώσεων \en Einstein \gr με αρνητική κοσμολογική σταθερά.

\chapter{Μελανές \en Dp-\gr Βράνες}
\rm{Σε αυτήν την ενότητα θα αναφέρουμε μερικά στοιχεία για τις $Dp$-βράνες. Οι
$Dp$-βράνες είναι φορτισμένα αντικείμενα ως προς μια $p+1$ μορφή
(\en form) \gr η οποία έχει επίσης φορτίο διαστελονίου και ενέργεια.
Όπως θα δούμε, το αντικείμενο αυτό μπορεί να θεωρηθεί ως ημικλασική
λύση της υπερβαρύτητας.}

\section{Λύσεις υπερβαρύτητας}

Για να βρούμε τα πεδία μεγάλης εμβέλειας των $D$-βρανών κρατάμε τους όρους
δύο παραγώγων της ενεργού (\en effective) \gr δράσης της θεωρίας ΙΙΑ/ΙΙΒ
στο σύστημα της χορδής (\en string frame) \gr \cite{Horowitz:1991cd}

\ba
\label{4-1}
{\cal S}=\frac{1}{(2\pi)^7\ell_s^8g_s^2}\int d^{10}x\sqrt{-g}
\left(e^{-2\Phi}(R+4(\nabla\Phi)^2)-\frac{1}{2(p+2)!}F_{p+2}^2\right)\ ,
\ea

\no
όπου $\ell_s$ είναι το μήκος της χορδής, το οποίο συνδέεται με την τάση
της χορδής $(2\pi\a')^{-1}$ ως $\a'=\ell_s^2$ και $F_{p+2}$ είναι ο τανυστής πεδίου της
$p+1$ μορφής, $F_{p+2}=dC_{p+1}$.

\no
Από λογισμό μεταβολών για την δράση \eqn{4-1} βρίσκουμε ότι

\ba
\label{4-2}
&&R_{\mu\nu}+2\nabla_\mu\nabla_\nu\Phi=\frac{e^{2\Phi}}{2(p+1)!}
\left(F_{\mu\nu}^2-\frac{g_{\mu\nu}}{2(p+2)}F^2\right)\ , \\
&&d\star F_{p+2}=0\ , \qq R=4(\nabla\Phi)^2-4\Box\Phi\ . \nonumber
\ea

\no
Θα υποθέσουμε μια μορφή για την μετρική στο σύστημα της χορδής

\ba
\label{4-3}
ds_{10}^2=\frac{-f(r)dt^2+d\vec{x}\cdot d\vec{x}}{\sqrt{H_p(r)}}+
\sqrt{H_p(r)}\left(\frac{dr^2}{f(r)}+r^2 d\Om^2_{8-p}\right)\ ,
\ea

\no
όπου $d\Om^2_{8-p}$ είναι το στοιχείο στερεάς γωνίας της μοναδιαίας σφαίρας
διάστασης $(8-p)$. Οι συντεταγμένες του χωροχρόνου είναι $t,x^i,\ i=1,2,\dots p$,
και είναι κάθετες στην συντεταγμένη $r$ και τις συντεταγμένες της σφαίρας. Λύνoντας τις Εξ.
\eqn{4-2} με την μορφή της Εξ. \eqn{4-3} βρίσκουμε ότι το διαστελόνιο ισούται με

\ba
\label{4-4}
e^{2\Phi}=g_s^2 H_p^{(3-p)/2}\ ,
\ea

\no
όπου $H_p,f$ είναι αρμονικές συναρτήσεις στον εγκάρσιο χώρο.
Παρατηρούμε επίσης ότι το διαστελόνιο είναι σταθερό για $p=3$.
Για σφαιρικά συμμετρικές λύσεις έχουμε ότι

\ba
\label{4-5}
H_p(r)=1+\frac{L^{7-p}}{r^{7-p}}\ , \qq f(r)=1-\frac{r_0^{7-p}}{r^{7-p}}\ ,
\ea

\no
όπου τις έχουμε νορμαλίσει στην μονάδα όταν το $r\to\infty$.
Η μετρική αυτή, είναι ασυμπτωτικά επίπεδη στο άπειρο του εγκάρσιου χώρου, $r\to\infty$.
Η $R-R$ μορφή ισούται με

\ba
\label{4-6}
C_{012\dots p}(r)=\sqrt{1+\frac{r_0^{7-p}}{r^{7-p}}}\ \frac{H_p(r)-1}{H_p(r)}\ .
\ea

\no
Όλες οι άλλες συνιστώσες μηδενίζονται, με εξαίρεση την ειδική περίπτωση όπου $p=3$
και στην οποία η συνθήκη αυτο-συζυγίας

\ba
\label{4-7}
F_{\mu_1\dots\mu_5}=\star F_{\mu_1\dots\mu_5}\equiv\frac{1}{5!\sqrt{g}}\
\epsilon_{\mu_1\dots\mu_5}\ ^{\nu_1\dots\nu_5}
F_{\nu_1\dots\nu_5}\ ,
\ea

\no
επιβάλει ύπαρξη μη μηδενικών μορφών στις εγκάρσιες διευθύνσεις.
Μπορούμε επίσης να υπολογίσουμε  το φορτίο \en Ramond-Ramond \gr ολοκληρώνοντας
το $\displaystyle{\frac{1}{(2\pi)^7\ell_s^8g_s^2}\star F_{p+2}}$ πάνω σε μια σφαίρα $S^{8-p}$
και θέτωντας το ίσο με εναν ακέραιο $N$ επί της τάση μιας $p$-βράνης, $T_p$

\ba
\label{4-8}
&&N=\frac{A(7-p)\Om_{7-p}}{T_p}L^{(7-p)/2}\sqrt{r_0^{7-p}+L^{7-p}}\ , \\
&&A=\frac{1}{(2\pi)^7\ell_s^8g_s^2}\ , \quad T_p=\frac{1}{(2\pi)^p\ell^{p+1}g_s}\ ,
\quad \Om_{n-1}=\frac{2\pi^{n/2}}{\G(n/2)}\ . \nonumber
\ea

\no
Μπορούμε επίσης να υπολογίσουμε με την μέθοδο \en ADM\ \gr την μάζα της λύσης από
την $g_{00}$ συνιστώσα της μετρικής

\ba
\label{4-9}
M=\frac{\Om_{8-p}V_p}{2\pi)^7\ell_s^8g_s^2}[(8-p)r_0^{7-p}+(7-p)L^{7-p}]\ ,
\ea

\no
όπου $V_p$ είναι ο όγκος μιας $Dp$ επίπεδης βράνης.

\section{Ορίζοντες και ανωμαλίες της μετρικής}

Για να κατανοήσουμε τα ειδικά σημεία $r=0$ και $r=r_0$ πρέπει να αλλάξουμε συντεταγμένες
\ba
\label{4-10}
\rho^{7-p}=L^{7-p}+r^{7-p}\ , \quad r_-=L\ , \quad r_+^{7-p}=r_0^{7-p}+L^{7-p}\ .
\ea

\no
Η μετρική Εξ. \eqn{4-3} γράφεται ως

\ba
\label{4-11}
&&ds^2=-\frac{f_+(\rho)}{\sqrt{f_-(\rho)}}\ dt^2+\sqrt{f_-(\rho)}d\vec{x}\cdot d\vec{x}
+f_-(\rho)^{-\half-\frac{5-p}{7-p}}
\left(\frac{d\rho^2}{f_+(\rho)}+\rho^2f_-(\rho)\right) , \\
&&f_\pm=1-\left(\frac{r_\pm}{\rho}\right)\ . \nonumber
\ea

\no
Από την μορφή της μετρικής παρατηρούμε ότι υπάρχει
ένας εσωτερικός ορίζοντας στο $r_-$ και ένας εξωτερικός στο $r_+$.
Υπολογίζοντας την βαθμωτή ποσότητα \en Ricci \gr στο σύστημα της χορδής
παρατηρούμε ότι

\ba
\label{4-12}
R\simeq\frac{(p+1)(3-p)(p-7)^2}{4r^{(p-3)/2}L^{(7-p)/2}}[1+{\cal O}(r^{7-p})]\ .
\ea

\no
Όμως για να βρούμε τυχόν ανώμαλιες της μετρικής θα πρέπει να πάμε στο σύστημα \en Einstein \gr
ως εξής:
\ba
\label{4-13}
g_{\mu\nu}^E=e^{-\frac{\Phi}{2}}g_{\mu\nu}^\s\ , \qq R^E\simeq r^{-(p-3)^2/8}\ ,
\ea

\no
όπου παρατηρούμε ότι η μετρική έχει πάντα έναν απειρισμό στο $r=0$ ή $\rho=L$
εκτός και αν $p=3$. Για φυσικά αποδεκτή λύση θα πρέπει να έχουμε $r_+\geq r_-$ ή
ισοδύναμα $r_0\geq0$ αλλιώς έχουμε μία γυμνή ανωμαλία.\\
\no
Με αυτή την παραδοχή και χρησιμοποιώντας τις εξισώσεις \eqn{4-8}
και \eqn{4-9} παρατηρούμε ότι

\ba
\label{4-14}
\frac{M}{V_p}\geq T_p N.
\ea

\no
Στην ειδική περίπτωση όπου $r_0=0$, έχουμε ότι η βράνη ικανοποιεί το
όριο \en Bogomolny ($M=NV_pT_p$) \gr και η οποία λύση διατηρεί
το μισό της υπερσυμμετρίας του χωροχρόνου. Η περιοχή γώρω απο το $r=0$ λέγεται
λαιμός (\en throat) \gr της λύσης. Το μέγεθος $L$ του λαιμού μπορεί να βρεθεί απο
το φορτίο Εξ. \eqn{4-8}

\ba
\label{4-15}
\left(\frac{L}{2\pi\ell_s}\right)^{7-p}=\frac{g_sN}{7-p}\frac{\G(\frac{9-p}{2})}{2\pi^{(9-p)/2}}\ .
\ea

\chapter{Η αντιστοιχία \en AdS/CFT}
\rm{Σε αυτήν την ενότητα θα διατυπώσουμε ένα επιχείρημα το οποίο συνδέει
μια θεωρία χορδών τύπου $IIB$ με μια θεωρία ${\cal N}=4$ \en
Super-Yang-Mills (SYM) \cite{SYM}. \gr Αρχικά θα συνοψίσουμε τα βασικά στοιχεία μιας
θεωρίας ${\cal N}=4$ \en SYM \gr και στην συνέχεια θα
διατυπώσουμε την αναλογία $AdS/CFT$ \cite{adscft}.}

\section{${\cal N}=4$ \en Super-Yang-Mills}

Η θεωρία ${\cal N}=4$ \en SYM \gr είναι η μέγιστα υπερσυμμετρική θεωρία βαθμίδας
σε τέσσερις διαστάσεις χωρίς βαρυτόνια. Η θεωρία αυτή είναι μια πολύ ειδική
περίπτωση τετραδιάστατης θεωρίας βαθμίδας διότι είναι σύμμορφη.
Τα βασικά συστατικά της ${\cal N}=4$ θεωρίας βαθμίδας είναι: έξι βαθμωτά πεδία
$\Phi_m$, τέσσερα φερμιονικά
$\Psi_{\a a},\dot{\Psi}^a_{\dot{\a}}$ και το διανυσματικό πεδίο $A_\mu$.
Ορίζουμε την συναλοίωτη παράγωγο $D_\mu$ και την δράση της πάνω στα πεδία

\ba
\label{5-1}
D_\mu=\del_\mu-ig A_\mu\ , \qq D_\mu W=[D_\mu,W]=\del_\mu W+ig[W,A_\mu]\ ,
\ea
\no
όπου το $A_\mu$ είναι το διανυσματικό πεδίο και $W$ είναι το σύνολο των πεδίων

\ba
\label{5-2}
W=(D_\mu,\Psi_{\a a},\dot{\Psi}^a_{\dot{\a}},\Phi_m)\ .
\ea

\no
Οι δείκτες $\mu,\nu$ ανήκουν στην άλγεβρα \en Lorentz $SO(3,1)$,\ \gr
με $\mu,\nu=1,\dots,4$ και $\a,\b,\dot{\a},\dot{\b}=1,2$. Οι λατινικοί δείκτες
αντιστοιχούν στην συμμετρία $R$ της άλγεβρας $SO(6)=SU(4)$. Τα βαθμωτά πεδία ανήκουν
στην θεμελιώδη αναπαράσταση της $SO(6)$ και συνεπώς τα $m,n=1,\dots,6$. Ενώ τα
φερμιόνια ανήκουν στην σπινοριακή αναπαράσταση και οι δείκτες παίρνουν τιμές $a,b=1,\dots,4$.

\no
Η Λαγκρανζιανή της ${\cal N}=4$ \en SYM \gr είναι η ακόλουθη:

\ba
\label{5-3}
{\cal L} &=& Tr\left(\frac{1}{4}F^{\mu\nu}F_{\mu\nu}+\half D^\mu\Phi^\nu D_\mu\Phi_\nu-
\frac{1}{4}g^2[\Phi^m,\Phi^n][\Phi_m,\Phi_n]\right. \nonumber \\
&&\left. +\dot{\Psi}_{\dot{\a}}^a\s_\mu^{\dot{a}\b}D^\mu\Psi_{\b a}-
\half ig\Psi_{\a a}\s_m^{ab}\varepsilon_{\a\b}[\Phi^m,\Psi_{\b b}]-
\half ig\dot{\Psi}^a_{\dot{\a}}\s_{ab}^m\varepsilon^{\dot{\a}\dot{\b}}[\Phi_m,\dot{\Psi}^b_{\dot{\b}}]
\right)\ ,
\ea

\no
όπου $F_{\mu\nu}$ είναι ο τανυστής πεδίου του πεδίου $A_\mu$ και ορίζεται ως
$F_{\mu\nu}\equiv ig^{-1}[D_\mu,D_\nu]$ και $\s^\mu,\s^m$ είναι οι πίνακες
γάμμα της χειραλικής συμμετρίας.

\section{Το όριο αποσύζευξης}

Έστω $N$ παράλληλες συμπίπτουσες $D3$ βράνες εντός ενός δεκαδιάστατου χωρόχρονου.
Οι βράνες αυτές, εκτείνονται σε ένα υπερεπίπεδο διάστασης ($3+1$) και βρίσκονται στο ίδιο σημείο ως προς
τον εγκάρσιο χώρο. Σε αυτό το υπόβαθρο, έχουμε δύο ειδών διεγέρσεις: τις ανοιχτές και τις
κλειστές χορδές. Οι ανοιχτές χορδές είναι οι διεγέρσεις των $D3$ βρανών, ενώ οι κλειστές
χορδές είναι οι διεγέρσεις του δεκαδιάστατου υποβάθρου.
Η περιγραφή των κλειστών χορδών στο όριο των χαμηλών ενεργειών δίνεται απο
την υπερβαρύτητα τύπου $IIB$. Αντίστοιχα, η περιγραφή των ανοιχτών χορδών στο όριο
των χαμηλών ενεργειών δίνεται απο την θεωρία ${\cal N}=4$ \en SYM. \gr \\
\no
Μπορούμε να γράψουμε την δράση του συστήματος στο όριο
των χαμηλών ενεργειών ως

\ba
\label{5-4}
{\cal S}={\cal S}_{\textrm{υποβάθρου}}+{\cal S}_{\textrm{βράνης}}+{\cal S}_{\textrm{αλληλεπιδράσεων}}\ ,
\ea
όπου ο πρώτος και ο δεύτερος όρος περιέχουν τις αλληλεπιδράσεις των κλειστών
και ανοιχτών χορδών, ενώ ο τρίτος περιέχει τις αλληλεπιδράσεις μεταξύ των
κλειστών και ανοιχτών χορδών. Πιο συγκεκριμένα, ο πρώτος όρος είναι η
δράση της δεκαδιάστατης υπερβαρύτητας, και διορθώσεις ανωτέρων παραγώγων.
Ο δεύτερος ορίζεται στον χωρόχρονο της $3+1$ βράνης, και περιέχει την
${\cal N}=4$ \en SYM \gr και κάποιες διορθώσεις ανωτέρων παραγώγων.
Ο τρίτος όρος περιέχει τις αλληλεπιδράσεις μεταξύ των τρόπων ταλάντωσης (\en modes)
\gr της βράνης και του υποβάθρου.\\
Μπορούμε να αναπτύξουμε την δράση υποβάθρου γύρω απο το ελεύθερο
σημείο, σε δυνάμεις της βαρυτικής ζεύξης $\k$. Θεωρώντας την μετρική
υποβάθρου $g_{\mu\nu}$, και κάνοντας ανάπτυξη γύρω απο τον χώρο \en
Minkowski, $g_{\mu\nu}=\eta_{\mu\nu}+\k\ h_{\mu\nu}$ \gr βρίσκουμε
ότι

\ba
\label{5-5}
{\cal S}_{\textrm{υποβάθρου}}\sim
\frac{1}{2\k^2}\int\sqrt{g}R\ \sim\ \int (\del h)^2+\k (\del h)^2h+\dots\ ,
\ea

\no
όπου η σταθερά $\k$ σχετίζεται με την σταθερά του Νεύτωνα
σε δέκα διαστάσεις μέσω της σχέσης $G_{10}=8\pi^6g_s^2\ell_s^4=2\k^2$ \cite{adscft}.
Στην ανάπτυξη αυτή συμπεριλάβαμε μόνο το βαρυτόνιο, αλλά υπάρχουν
αντίστοιχα αναπτύγματα και για τους άλλους όρους.
Παρατηρούμε ότι όλοι οι όροι αλληλεπίδρασης είναι ανάλογοι
θετικών δυνάμεων του $\k$. Συνεπώς, σε χαμηλές ενέργειες, όλες αυτές
οι αλληλεπιδράσεις είναι αμελητέες. Αυτό είναι ισοδύναμο με
το ότι η βαρύτητα είναι ελεύθερη στο υπέρυθρο.
Συνεπώς καταλήγουμε ότι σε χαμηλές ενέργειες οι αλληλεπιδράσεις
αυτές είναι αμελητέες. 'Η ισοδύναμα με το να θεωρήσουμε ότι η βαρύτητα
είναι αποσυζευγμένη στο υπόβαθρο.

\section{Γεωμετρία του ορίζοντα}

Στην ειδική περίπτωση όπου έχουμε $N$ $D3$ βράνες, από την Εξ.\eqn{4-3}
για $p=3$, προκύπτει

\ba
\label{5-6}
&&ds^2=H^{-1/2}(-dt^2+d\vec{x}\cdot d\vec{x})+H^{1/2}(dr^2+r^2d\Om_5^2)\ , \\
&&F_5=(1+\star)dH^{-1}\wedge dt\wedge dx_1\wedge dx_2\wedge dx_3\ ,
\quad H=1+\frac{L^4}{r^4}, \nonumber \\
&&L^4=4\pi g_s\ell_s^4N, \quad \Phi=\textrm{Σταθερό}, \nonumber
\ea

\no
όπου χρησιμοποιήσαμε επίσης την Εξ.\eqn{4-15} για $p=3$. Παρατηρώντας
ότι το στοιχείο της μετρικής $g_{tt}$ δεν είναι σταθερό, συμπεραίνουμε ότι η ενέργεια
$E_p$ ενός αντικειμένου όπως μετριέται απο έναν παρατηρητή σε απόσταση $r$
και η ενέργεια $E$ όπως μετριέται απο έναν παρατηρητή στο άπειρο συσχετίζονται
με μια μετατόπιση στο ερυθρό

\ba
\label{5-7}
E=H^{-1/4}E_p\ .
\ea

\no
θεωρώντας ότι η ενέργεια στο άπειρο είναι πεπερασμένη,
ένα αντικείμενο το οποίο κινείται προς το $r=0$ φαίνεται ότι έχει
όλο και μικρότερη ενέργεια. Ακολούθως, θα θεωρήσουμε
το ασθενές ενεργειακό όριο της Εξ.\eqn{5-7}. Υπάρχουν δύο ειδών ασθενείς ενεργειακές
διεγέρσεις για έναν παρατηρητή ο οποίος βρίσκεται στο άπειρο. Συγκεκριμένα, μπορούμε να έχουμε
άμαζα σωμάτια τα οποία διαδίδονται στον χώρο με μεγάλα μήκη κύματος, ή οποιουδήποτε
είδους διεγέρσεις κοντά στο $r=0$. Στο ασθενές ενεργειακό όριο αυτές οι διεγέρσεις
είναι αποσυζευγμένες μεταξύ τους. Επιπλέον τα άμαζα σωμάτια είναι αποσυζευγμένα στην περιοχή
περί του ορίζοντα (γύρω απο το $r=0$), διότι το μήκος κύματος του σωματιδίου είναι πολύ
μεγαλύτερο απο το εύρος της βαρυτικής δύναμης της βράνης. Ομοίως, οι διεγέρσεις περί του
$r=0$ δεν μπορούν να ξεφύγουν στο άπειρο λόγω του βαρυτικού δυναμικού. Συνεπώς,
το ασθενές ενεργειακό όριο αποτελείται απο δύο αποσυζευγμένα μέρη, το ένα να είναι
η ελεύθερη υπερβαρύτητα στο κενό και το δεύτερο είναι η περιοχή περί του ορίζοντα.
Στην περιοχή κοντά στον ορίζοντα, $r\ll L$, η μετρική της Εξ. \eqn{5-6} προσεγγίζεται ως

\ba
\label{5-8}
ds^2=\frac{r^2}{L^2}(-dt^2+d\vec{x}\cdot d\vec{x})+L^2\frac{dr^2}{r^2}+L^2d\Om^2_5\ ,
\ea

\no
ή την ισοδύναμη έκφραση με την αντικατάσταση $u=L^2/r$,

\ba
\label{5-9}
ds^2=\frac{L^2}{u^2}\ (du^2-dt^2+d\vec{x}\cdot d\vec{x})+L^2d\Om^2_5\ .
\ea

\no
Το πρώτο μέρος περιγράφει όπως έχουμε δει τον πενταδιάστατο χώρο \en Anti-de Sitter \gr ($AdS_5$),
\gr ενώ το δεύτερο περιγράφει μια πενταδιάστατη σφαίρα $S^5$.
Το σύνορο του $AdS_5$ σε αυτές τις συντεταγμένες είναι το $u=0$.\\
\no
Και στις δυο περιγραφές, βρήκαμε ότι στο όριο χαμηλών ενεργειών το σύστημα
αποτελείται από δύο μη-αλληλεπιδρώντα συστήματα. Αυτή η παρατήρηση οδηγεί
στην εικασία της \en AdS/CFT \gr ότι η ${\cal N}=4$ \en SYM \gr είναι
δυική με την υπερβαρύτητα τύπου $IIB$ \cite{adscft}.

\section{Στοιχεία της αντιστοιχίας}

Στο όριο όπου $\ell_s\to 0$ και για πεπερασμένες ενέργειες θεωρούμε ότι η ποσότητα
$E_r\ell_s$ είναι πεπερασμένη. Από την Εξ.\eqn{5-7} βρίσκουμε ότι
\ba
\label{5-10}
E_\infty\sim E_r\frac{r}{\ell_s}=(E_r\ell_s)\frac{r}{\ell_s^2}\ ,
\ea

\no
όπου θέλουμε η $E_\infty$ όπως αυτή θα μετρηθεί στην θεωρία πεδίου,
να είναι πεπερασμένη. Ως εκ τούτου πρέπει να ορίσουμε μια νέα
μεταβλητή $U=\frac{r}{\ell_s^2}$ η οποία θα πρέπει να είναι πεπερασμένη καθώς
το $r\to 0$. Σε αυτό το σύστημα συντεταγμένων, στο οποίο η $U$ έχει διαστάσεις
ενέργειας, ενώ η μετρική, Εξ.\eqn{5-6}, περί του ορίζοντα δίνεται από

\ba
\label{5-11}
ds^2=\ell_s^2\left(\frac{U^2}{\sqrt{4\pi g_sN}}(-dt^2+d\vec{x}\cdot d\vec{x})+
\sqrt{4\pi g_sN}\frac{dU^2}{U^2}+\sqrt{4\pi g_sN}d\Om_5^2\right)\ .
\ea

\no
Θα πρέπει να συγκρίνουμε τις ολικές συμμετρίες του χώρου $AdS_5\times S^5$
και της θεωρίας ${\cal N}=4$ \en SYM.\ \gr Οι ολικές μποζονικές συμμετρίες της
τετραδιάστατης σύμμορφης θεωρίας πεδίου έχουν γεννήτορες στην σύμμορφη ομάδα
$SO(4,2)$ και της συμμετρίας $R$, $SO(6)_R$. Αυτές είναι ακριβώς και οι συμμετρίες
του χώρου $AdS_5\times S^5$. Πέρα όμως απο τις μποζονικές συμμετρίες έχουμε και τις
φερμιονικές συμμετρίες. Η θεωρία ${\cal N}=4$ \en SYM \gr έχει 16 υπερφορτία, δηλαδή την
υπερσυμμετρία που διατηρείται απο τις $N\ D3$ βράνες. Όμως η υπερσυμμετρική σύμμορφη ομάδα
έχει άλλα 16 φορτία, οπότε στο σύνολο έχουμε 32 υπερφορτία. Αυτή η ενισχυμένη υπερσυμμετρία
εμφανίζεται και στην θεωρία χορδών στο υπόβαθρο $AdS_5\times S^5$. Οπότε, η θεωρία βαθμίδας και
η θεωρία χορδών, έχουν τις ίδιες (υπερ)συμμετρίες.
Θα πρέπει να εξετάσουμε τα όρια ισχύος των δύο θεωριών. Στο επίπεδο της θεωρίας πεδίου
η ανάλυση σε διαγράμματα \en Feynman \gr της θεωρίας \en Yang--Mills \gr
στηρίζεται στο ότι η σταθερά ζεύξης του \en 't Hooft \gr πρέπει να
είναι

\ba
\label{5-12}
\l=g_{YM}^2N\sim g_sN\sim\frac{L^4}{l_s^4}\ll 1\ , \quad g_s\sim g_{YM}^2\ .
\ea

\no
Από την άλλη πλευρά, για να είναι η κλασική υπερβαρύτητα αξιόπιστη
θα πρέπει η ακτίνα $L$ του χώρου $AdS_5$ και της $S^5$ να είναι πολύ
μεγαλύτερη απο το μήκος της χορδής

\ba
\label{5-13}
\frac{L^4}{l_s^4}\sim g_sN\sim g_{YM}^2N=\l\gg 1\ .
\ea

\no
Από τις εξισώσεις \eqn{5-12} και \eqn{5-13} παρατηρούμε ότι
τα όρια της βαρύτητας και θεωρίας πεδίου είναι εκ διαμέτρου
αντίθετα. Αυτό είναι ο λόγος γιατί η αντιστοιχία \en AdS/CFT \gr
ονομάζεται δυισμός. Πρέπει να αναφέρουμε ότι τα επιχειρήματα που δώσαμε δεν αποτελούν σε καμμία
περίπτωση απόδειξη της αντιστοιχίας απλά μας οδηγούν στην εικασία
της ύπαρξης της. Η εικασία αυτή έχει επεκταθεί και για
υπόβαθρα τα οποία είναι ασυμπτωτικά στον $AdS_5\times S^5$ όπως και
σε περιπτώσεις με λιγότερη υπερσυμμετρία \cite{LM,frolov,LS}.

\part{Μεσόνια}

\chapter{Εισαγωγή}

\no
\gr Ο υπολογισμός ενός
στατικού βρόχου \en Wilson \gr για το ζεύγος κουάρκ-αντικουάρκ
 σε μια ${\cal N}=4$ \en SYM \gr απεικονίζεται στην
ελαχιστοποίηση της δράσης για μια χορδή, που συνδέει το ζεύγος κουάρκ-αντικουάρκ
που βρίσκεται στο σύνορο του χώρου $AdS_5$, και η οποία
εκτείνεται στην ακτινική διεύθυνση. Η προσέγγιση αυτή εφαρμόστηκε για πρώτη φορά
στην εργασία  \cite{maldaloop} για την σύμμορφη περίπτωση και γενικεύτηκε
στις εργασίες  \cite{wilsonloopTemp,bs} για μη σύμμορφες περιπτώσεις.\\
\no
Μολονότι στην σύμμορφη περίπτωση οι υπολογισμοί δίδουν την
προσδωκόμενη συμπεριφορά \en Coulomb \gr του δυναμικού,
στις μη σύμμορφες γενικεύσεις η συμπεριφορά του δυναμικού διαφέρει
με την προσδοκώμενη απο την θεωρία πεδίου και το πείραμα.
Για παράδειγμα, στην θεωρία ${\cal N}=4$ \en SYM, \gr το δυναμικό του
κουάρκ-αντικουάρκ έχει πλειότιμη συμπεριφορά όπως και συμπεριφορά
ανάλογη με μετατροπή φάσεως, ενώ στον κλάδο \en Coulomb \gr της ${\cal N}=4$ \en SYM
\gr βρίσκουμε ότι υπάρχει εγκλωβισμός σε ορισμένες περιοχές του παραμετρικού
χώρου. Για να ελένξουμε αν οι λύσεις αυτές όντως προκύπτουν απο την αντιστοιχία
βαθμίδας/βαρύτητας, πρέπει να είναι συνεπείς κάτω απο ορισμένους
ελέγχους, ένας απο αυτούς είναι η ευστάθεια τους κάτω από μικρές διακυμάνσεις.

\no
Για την σύμμορφη περίπτωση, οι εξισώσεις κίνησης για τις διαταραχές έχουν
βρεθεί στο \cite{cg,kmt}, και το φάσμα των διαταραχών ήταν θετικό ως αναμενόταν.
Στην περίπτωση της πεπερασμένης θερμοκρασίας, όπου η λύση έχει δύο κλάδους,
το θέμα της ευστάθειας μελετήθηκε στην εργασία \cite{michalogiorgakis,ASS1} για την περίπτωση κινούμενου
ζεύγους κουάρκ-αντικουάρκ. Η ανάλυση των μικρών διακυμάνσεων στο πρόβλημα αυτό
οδήγησε σε ένα σύστημα πεπλεγμένων διαφορικών εξισώσεων, και εφαρμόζοντας αριθμητικές μεθόδους
\cite{michalogiorgakis} είτε αναλυτικές \cite{ASS1}, βρέθηκε ότι ο άνω κλάδος είναι ασταθής κάτω απο μικρές διαταραχές στις διαμήκεις διευθύνσεις.\\
\no
Αρχικά, θα μελετήσουμε την ευστάθεια στατικών διατάξεων (\en configurations)\ \gr
στηριζόμενοι σε αναλυτικές μεθόδους. Θα μελετήσουμε διαταραχές σε όλες τις συντεταγμένες
συμπεριλαμβανομένων και των γωνιακών και θα μελετήσουμε την ${\cal N}=4$ \en SYM \gr σε πεπερασμένη
θερμοκρασία και στον κλάδο \en Coulomb. \gr Σε αυτά τα υπόβαθρα, οι εξισώσεις είναι μη πεπλεγμένες
και το πρόβλημα επιλύεται με αναλυτικές μεθόδους. Το πλάνο μας είναι να γράψουμε τις εξισώσεις κίνησεις
που ικανοποιούν οι διακυμάνσεις υπό μορφή \en Sturm--Liouville \gr και \en Schr\"odinger \gr
και θα ανάγουμε την μελέτη του φάσματος συχνοτήτων των διακυμάνσεων είτε σε μηδενικούς τρόπους
(\en zero mode)\gr ενός \en Sturm-Liouville \gr προβλήματος, είτε σε μελέτη δυναμικών \en Schr\"odinger,
\gr χρησιμοποιώντας αναλυτικές και αριθμητικές μεθόδους.\\
\no
Η ενότητα των μεσονίων είναι βασισμένη στις εργασίες \cite{ASS1,ASS2} οργανωμένη ως εξής.
Στο 6ο κεφάλαιο, θα περιγράψουμε τον υπολογισμό
ενός βρόχου \en Wilson \gr στο πλαίσιο της \en AdS/CFT \gr μέσω διατάξεων χορδών σε υπόβαθρα
$D3$ βρανών και θα αναλύσουμε τα γενικά χαρακτηριστικά των δυναμικών κουάρκ-αντικουάρκ.
Στο 7ο κεφάλαιο, θα θεωρήσουμε μικρές διακυμάνσεις περί των κλασικών λύσεων,
και θα ανάγουμε το πρόβλημα σε ένα ισοδύναμο μονοδιάστατο πρόβλημα \en Schr\"odinger,
\gr και θα βρούμε κάποια γενικά αποτελέσματα για αναλυτικό και προσεγγιστικό
υπολογισμό των ενεργειακών ιδιοτιμών. Στο 8ο κεφάλαιο, θα εφαρμόσουμε τα γενικά
αποτελέσματα του κεφαλαίου 6ο στις περιπτώσεις των μελανών (\en non-extremal) $D3$ \gr βρανών
και των πολυκεντρικών $D3$ βρανών και θα προσδιορίσουμε ορισμένες περιοχές στον παραμετρικό χώρο
στις οποίες τα δυναμικά είναι σε αντίθεση με τα προσδοκώμενα φυσικά αποτελέσματα.
Στο 9ο κεφάλαιο, θα εφαρμόσουμε την θεωρία των μικρών διακυμάνσεων για αυτές τις διατάξεις
και θα αποδείξουμε ότι όλες αυτές οι προβληματικές περιοχές είναι διαταρακτικά ασταθείς.
Στο 10ο κεφάλαιο θα μελετήσουμε υπόβαθρα δυικά σε ${\cal N}=1$ \en SYM. \gr Όπως θα δούμε
ακόμα και η ευστάθεια της σύμμορφης περίπτωσης σε οριακά παραμορφωμένα (\en marginal) \gr
υπόβαθρα απαιτεί το φανταστικό μέρος $\s$ της παραμέτρου παραμόρφωσης να είναι μικρότερο
από μια τιμή, ενώ σε κάποιες περιοχές του κλάδου \en Coulomb \gr επιτρέπεται γραμμικός εγκλωβισμός
σε ένα διάστημα τιμών της παραμέτρου $\s$.

\chapter{Κλασικές Λύσεις}
θα υπολογίσουμε μέσω της $AdS/CFT$ το στατικό δυναμικό ενός βαρεού ζεύγους κουάρκ-αντικουάρκ
σύμφωνα με την μεθοδολογία του \cite{maldaloop} για την σύμμορφη
περίπτωση και τις γενικεύσεις της πέρα του σύμμορφου ορίου
\cite{wilsonloopTemp,bs}. Στην θεωρία πεδίου βαθμίδας, αυτό το
δυναμικό υπολογίζεται από την αναμενόμενη τιμή ενός βρόχου \en Wilson \gr με μια χρονική και μια
χωρική ακμή. Στην βαρύτητα, η αναμενόμενη τιμή του βρόχου \en Wilson
\gr δίνεται απο την ελαχιστοποίηση της δράσης \en Nambu--Goto \gr
για μια χορδή η οποία διαδίδεται στο δυικό
υπερβαρυτικό υπόβαθρο, της οποίας τα άκρα είναι
δεσμευμένα στις δύο άκρες του χωρικού μέρους του βρόχου.

\section{Βρόχοι \en Wilson \gr και κουάρκ}
Στην η ενότητα θα εισάγουμε την έννοια του βρόχου \en
Wilson \gr ο οποίος θα μας χρειαστεί αργότερα για τον
υπολογισμό ενεργειών δέσμιων καταστάσεων. Τα κουάρκ στην \en QCD \gr
χωρίζονται στα θεμελιώδη (ελαφριά κουάρκ) και στις εξωτερικές προσθήκες
(απείρως βαρέα) και τα οποία δεν υπάρχουν στην δράση.
\no
Στη \en QCD, \gr τα ελαφριά κουάρκ δεν βρίσκονται
ποτέ ελεύθερα στο κενό, αλλά πάντα σε δέσμιες καταστάσεις.
Για παράδειγμα, σχηματίζοντας ζεύγη με αντικουάρκ.
\no
Αφού τα εξωτερικά κουάρκ είναι πολύ βαριά, η απόσταση μεταξύ του ζεύγους
κουάρκ-αντικουάρκ δεν παραμείνει σταθερή με το χρόνο. Θέλουμε να υπολογίσουμε
το δυναμικό αλληλεπίδρασης μεταξύ των δύο κουάρκ, $V_{q\bar{q}}(L)$ και για αυτό
χρειαζόμαστε κάποιο φυσικό μέγεθος που δεν θα εξαρτάται απο την βαθμίδα.
Μια σημαντική ποσότητα η οποία δεν εξαρτάται απο την βαθμίδα στις θεωρίες
\en Yang-Mills \gr είναι ο βρόχος \en Wilson \cite{Peskin} \gr
\ba
\label{7-1}
W(C;R)=Tr_R[P e^{i\oint_C A}],
\ea
ο οποίος δίνεται απο ένα εκθετικό διατασσόμενης διαδρομής (\en path ordered) \gr
πάνω στο πεδίο βαθμίδας, κατά μήκος μιας κλειστής τροχιάς $C$ ενώ το ίχνος έχει παρθεί
πάνω σε μια τυχαία αναπαράσταση $R$ του πεδίου βαθμίδας. Αποδεικνύεται ότι η ποσότητα αυτή
είναι ανεξάρτητη απο την επιλογή βαθμίδας \cite{ads-intro}.
Από την αναμενόμενη τιμή του βρόχου \en Wilson $W(C;R)$ \gr μπορούμε να υπολογίσουμε
το δυναμικό αλληλεπίδρασης. Για το σκοπό αυτό θεωρούμε ένα ορθογώνιο παραλληλόγραμμο
στο ευκλείδιο χώρο $t-x$, με μήκος $T$ κατά μήκος του χρόνου $t$ και μήκος $L$
κατά μήκος του χώρου $x$. Η Εξ.\eqn{7-1} ισούται με ένα άθροισμα παραγόντων
\en Boltzmann \gr σε όλες τις ενέργειες και για μεγάλα $T$ ($Τ\gg L$) η αναμενόμενη τιμή
συμπεριφέρεται ως $e^{-TE}$, όπου $E$ είναι η ελάχιστη ενέργεια του ζεύγους
κουάρκ και αντικουάρκ. Οπότε, βρίσκουμε ότι μια εναλλακτική
σχέση για την αναμενόμενη τιμή του βρόχου \en Wilson \gr
\ba
\label{7-2}
\langle W(c)\rangle\ \simeq e^{-TV(L)}\ ,
\ea
\no
όπου $V(L)$ είναι το δυναμικό κουάρκ και αντικουάρκ.
Όπως θα δούμε αργότερα για μεγάλα $N$ και μεγάλες τιμές του $g_{YM}^2N$,
μέσω της αναλογίας \en AdS/CFT \gr ο υπολογισμός του $\langle W(c)\rangle$ της \en CFT
\gr γίνεται από την ελαχιστοποίηση μιας επιφάνειας στον χώρο $AdS$.

\section{Υπολογισμός κλασικής λύσης}

\no
\gr Έστω η γενική διαγώνια μετρική της μορφής

\ba
\label{7-3}
ds^2=G_{tt}dt^2+G_{yy}dy^2+G_{uu}du^2+G_{xx}dx^2+G_{\th\th}d\th^2+\ldots\ ,
\ea

\no
με υπογραφή \en Lorentz \gr και η οποία στο σύμμορφο όριο ανάγεται σε μετρική του χώρου
$AdS_5\times S^5$. Η αγνοήσιμη (κυκλική) συντεταγμένη $y$
είναι κατά μήκος της χωρικής πλευράς του βρόχου \en Wilson, \gr $u$ είναι η ακτινική
συντεταγμένη που αντικατοπτρίζει την κλίμακα ενεργειών στην δυική θεωρία βαθμίδας
και εκτείνεται από το υπεριώδες ($u\to\infty$) μέχρι το υπέρυθρο σε κάποια
ελάχιστη τιμή $u_{min}$ η οποία καθορίζεται απο την γεωμετρία. Η συντεταγμένη
$x$ είναι μια γενική αγνοήσιμη συντεταγμένη και η $\th$ είναι μια γενική μη αγνοήσιμη συντεταγμένη
από την οποία εξαρτώνται τα στοιχεία της μετρικής, αλλά που μπορεί να λάβει μια
σταθερή τιμή συμβατή με τις εξισώσεις κίνησης όπως θα δούμε. Οι λοιπές συντεταγμένες δεν
έχουν συμπεριληφθεί θεωρώντας ότι ανήκουν σε μια από τελευταίες δύο κατηγορίες.
Είναι βολικό για την ανάλυση μας, να ορίσουμε τις συναρτήσεις

\be
\label{7-4}
g(u,\th) = - G_{tt} G_{uu}\ ,\qq f_y(u,\th) = - G_{tt} G_{yy}\ ,
\ee

\no
καθώς και τις

\be
\label{7-5}
f_x(u,\th) = - G_{tt} G_{xx}\ ,\qq f_\th(u,\th) = - G_{tt} G_{\th\th}\ ,
\qq h(u,\th) = G_{yy} G_{uu}\ .
\ee

\no
Είναι χρήσιμο να βρούμε τις ασυμπτωτικές συμπεριφορές των συναρτήσεων αυτών στο σύμμορφο
όριο. Όπως προανεφέρθει θεωρούμε ότι η μετρική Εξ.\eqn{7-3} στο σύμμορφο όριο ($u\to\infty$)
ανάγεται σε μετρική του χώρου $AdS_5\times S^5$, οπότε έχουμε

\be
\label{7-6}
-G_{tt} \simeq G_{xx} \simeq G_{yy} \simeq {u^2 \ov R^2} \ ,\quad G_{uu}
\simeq {R^2\ov u^{2}} \ , \quad G_{\th\th} \simeq R^2\ ,\qq {\textrm{όταν}} \quad
u\to \infty\ ,
\ee

\no
όπου $R=(4\pi g_sN)^{1/4}$ είναι η ακτίνα του χώρου $AdS_5\times S^5$
στο σύστημα μονάδων της χορδής και $g_s$ είναι η σταθερά ζεύξης της χορδής.
Αντικαθιστώντας την Εξ.\eqn{7-6} στην Εξ.\eqn{7-5} βρίσκουμε ότι

\be
\label{7-7}
g \simeq h \simeq 1\ ,\quad f_x \simeq f_y \simeq u^4\ ,\quad
f_\th \simeq u^2 \ , \qq {\textrm{όταν}} \quad u\to \infty\ .
\ee

\no
Στο πλαίσιο της αντιστοιχίας \en AdS/CFT, \gr το δυναμικό αλληλεπίδρασης του ζεύγους
κουάρκ-αντικουάρκ δίνεται απο τη σχέση

\be
\label{7-8}
e^{-{\rm i} E T} = \langle W(C) \rangle = e^{{\rm i} S[C]}\ ,
\ee

\no
όπου

\be
\label{7-9}
S[C] = - {1 \ov 2 \pi} \int_C d \tau d \sigma \sqrt{- \det g_{\a \b} }\ ,
\qq g_{\a\b} = G_{\mu\nu}
\partial_\alpha x^\mu \partial_\b x^\nu \ ,
\ee

\no
είναι η δράση \en Nambu--Goto \gr για μια χορδή που διαδίδεται στο δυικό υπερβαρυτικό
υπόβαθρο και της οποίας τα άκρα βρίσκονται στον βρόχο $C$. Στην συνέχεια
καθορίζουμε την διδιάστατη παραμετροποίηση επιλέγοντας την βαθμίδα

\be
\label{7-10}
t=\tau \ ,\qq u=\s \ .
\ee

\no
Υποθέτοντας ότι δεν έχουμε εξάρτηση απο τον χρόνο $t$, θεωρούμε
την παρακάτω εμβάπτιση

\be
\label{7-11}
y = y(u)\ ,\qq x =0\ ,\qq \th = \th_0 = \textrm{Σταθερά} \ ,\qq \hbox{Υπόλοιπες} = \textrm{Σταθερές}\ .
\ee
\no
Οι συντεταγμένες υπόκεινται στις συνοριακές συνθήκες
\be
\label{7-12}
u \left(\pm {L \ov 2} \right) = \infty\ ,
\ee
\begin{figure}[!t]
\begin{center}
\includegraphics[scale=0.4]{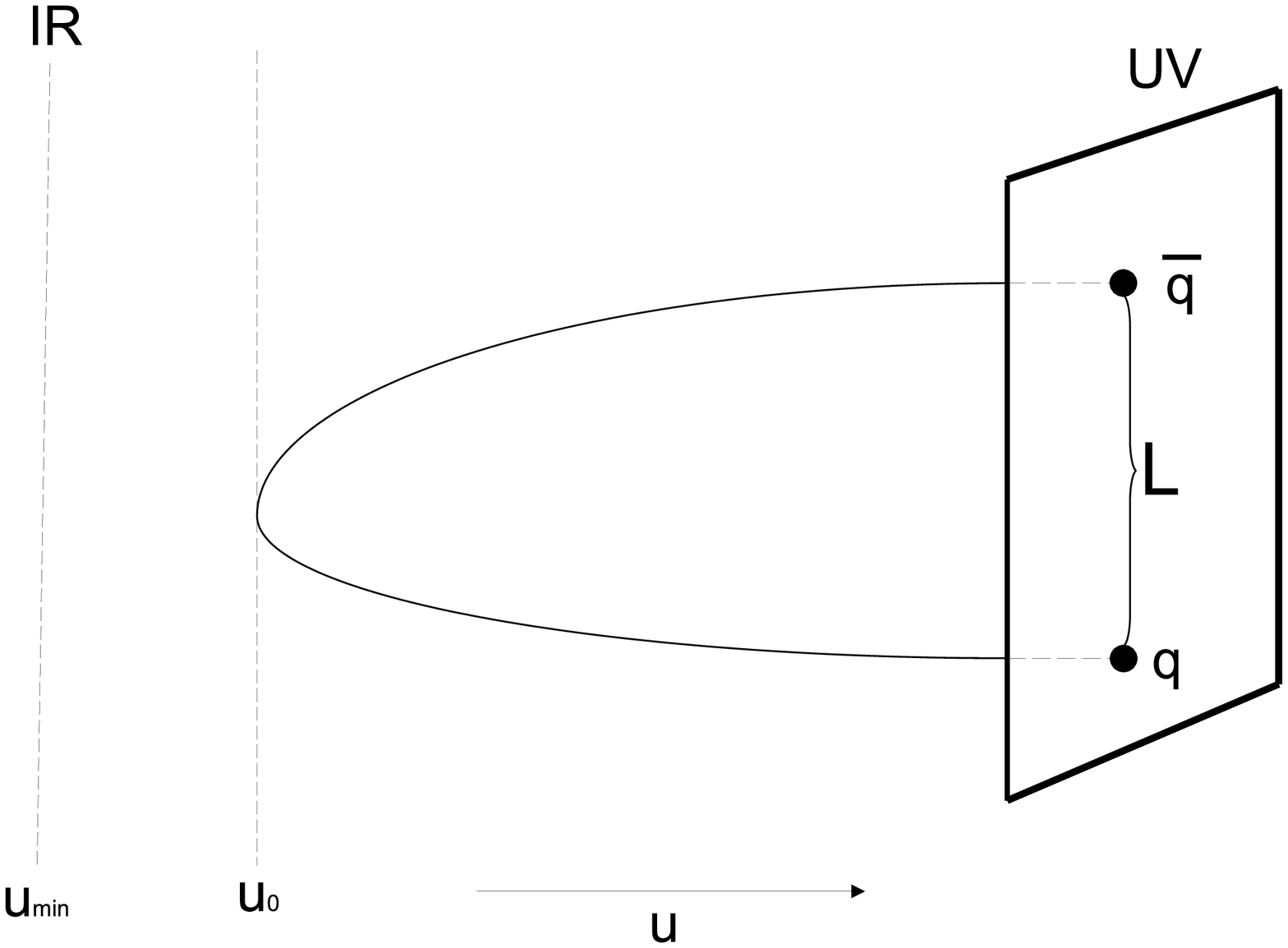}
\end{center}
\caption{Το σημείο στροφής της χορδής είναι το $u_0$ και το ελάχιστο είναι το $u_{\rm min}$.
} \label{Meson}
\end{figure}

\no
οι οποίες είναι κατάλληλες για ένα κουάρκ στο $y=-L/2$ και ένα αντικουάρκ
στο $y=L/2$, Σχήμα 7.1. Στην υπόθεση της \eqn{7-11}, η σταθερή τιμή $\th_0$ της μη αγνοήσιμης μεταβλητής
$\th$ πρέπει να είναι συνεπής με τις εξισώσεις κίνησης. Όπως θα δούμε αργότερα, αυτό επιβάλει
την αναγκαία συνθήκη

\be
\label{7-13}
\partial_\th g(u,\th) |_{\th=\th_0} = \partial_\th f_y(u,\th) |_{\th=\th_0} = 0\ ,
\ee

\no
Με αυτές τις υποθέσεις, η δράση \en Nambu--Goto \gr ισούται
με

\be
\label{7-14}
S = - {T \ov 2 \pi} \int d u \sqrt{ g(u) + f_y(u) y^{\prime 2}}\ ,
\ee

\no
όπου $T$, είναι η χρονική ακμή του βρόχου \en Wilson,\ \gr ο τόνος
συμβολίζει παραγώγιση ως προς το $u$ ενώ $g(u) \equiv
g(u,\th_0)$ και $f_y(u) \equiv f_y(u,\th_0)$ είναι οι συναρτήσεις
της Εξ.\eqn{7-4} υπολογισμένες σε μια σταθερά τιμή $\th_0$ του $\th$.
Η Λαγκραζιανή του προβλήματος είναι ανεξάρτητη της μεταβλητής
$y$, οπότε η αντίστοιχη γενικευμένη ορμή διατηρείται

\be
\label{7-15} {f_y y_{\rm cl}^\prime \ov \sqrt{ g + f_y y_{\rm
cl}^{\prime 2}}} = \pm f_{y0}^{1/2}\qq \Longrightarrow \qq y_{\rm
cl}^\prime = \pm {\sqrt{f_{y0} F}\ov f_y}\ ,
\ee

\no
όπου $u_0$ είναι η τιμή του $u$ στο σημείο στρέψης της χορδής,
$f_{y0}\equiv f_y(u_0)$, $y_{\rm cl}$ είναι η κλασική λύση με τα
δύο πρόσημα να αντιστοιχούν στους δύο συμμετρικούς κλάδους της χορδής
περί το σημείο στρέψης, και $F$ είναι η ποσότητα

\be
\label{7-16} F = {g f_y \ov f_y - f_{y0}}\ .
\ee

\no
Ολοκληρώνοντας την Εξ.\eqn{7-15}, μπορούμε
να εκφράσουμε την απόσταση του ζεύγους ως

\be
\label{7-17}
L = 2 f_{y0}^{1/2} \int_{u_0}^{\infty} d u {\sqrt{F} \ov f_y}\ .
\ee

\no
Τελικά, αντικαθιστώντας την λύση για την $y_{\rm cl}^\prime$ στην
Εξ.\eqn{7-14} και αφαιρώντας τις αποκλίνουσες μάζες
των κουάρκ και αντικουάρκ, γράφουμε την ενέργεια αλληλεπίδρασης
ως

\be
\label{7-18}
E = {1 \ov \pi} \int_{u_0}^\infty d u \sqrt{F} - {1 \ov \pi}
\int_{u_{\rm min}}^\infty d u \sqrt{g}\ .
\ee

\no
Στην ιδανική περίπτωση, θα θέλαμε να υπολογίσουμε αναλυτικά τα ολοκληρώματα
των Εξ.\eqn{7-17} και \eqn{7-18}, και στην συνέχεια να λύσουμε την Εξ.\eqn{7-17}
ως προς $u_0$, να αντικαταστήσουμε στην Εξ.\eqn{7-18} και να έκφράσουμε την ενέργεια
αλληλεπίδρασης σαν συνάρτηση της απόστασης $L$ του ζεύγους. Ωστόσο, αυτό εν γένει
δεν μπορεί να γίνει, εκτός από πολύ απλές περιπτώσεις, όπως η σύμμορφη περίπτωση,
οπότε οι Εξ. \eqn{7-17} και \eqn{7-18} μπορούν να θεωρηθούν ως παραμετρικές σχέσεις
των $L$ και $E$ με παραμέτρο το $u_0$. Το προκύπτον δυναμικό του ζεύγους $E(L)$
πρέπει να ικανοποιεί την συνθήκη κοιλότητας \cite{concavity}

\be
\label{7-19}
{d E \ov d L} > 0\ ,\qq {d^2 E \ov d L^2} \leqslant 0\ ,
\ee

\no
η οποία ισχύει για οποιαδήποτε θεωρία βαθμίδας, ανεξαρτήτως της ομάδας βαθμίδας
και των πεδίων ύλης, και το οποίο υποδηλώνει ότι η δύναμη είναι ελκτική
και φθίνουσα συνάρτηση του μήκους του ζεύγους. Στην περίπτωση μας,
έχει δειχθεί ότι \cite{bs}

\be
\label{7-20}
{dE\ov dL}= {1 \ov 2 \pi} f_{y0}^{1/2} > 0 \ ,\qq {d^2E\ov dL^2}=
{1\ov 4\pi} {f^\prime_{y0} \ov f^{1/2}_{y0}} {1 \ov
L^\prime(u_0)}\ ,
\ee

\no
και σε όλα τα γνωστά παραδείγματα έχουμε ότι $f^\prime_{y0} > 0$.
Η συνθήκη κοιλότητας επιβάλλει στην παράμετρο ολοκλήρωσης $u_0$
να ικανοποιεί την ανισότητα

\be
\label{7-21}
L^\prime(u_0) \leqslant 0\ .
\ee

\no
Στην σύμμορφη περίπτωση \cite{maldaloop},
η ενέργεια αλληλεπίδρασης συναρτήσει του μήκους έχει συμπεριφορά
\en Coulomb, \gr μια γνησίως φθίνουσα υπερβολή με σταθερά αναλογίας
$\sqrt{g_{YM}^2N}$ αντί για $g_{YM}^2N$, το οποίο είναι αποτέλεσμα
του μη διαταρακτικού υπολογισμού. Η συμπεριφορά τύπου \en Coulomb \gr υπάρχει
στο υπεριώδες σε όλα τα ασυμπωτικά υποβάθρα στον χώρο $AdS_5\times S^5$ όταν το $u\to\infty$.
Ωστόσο, στο υπέρυθρο, η συμπεριφορά είναι διαφορετική και εξαρτάται από
τις λεπτομέρειες της λύσης της υπερβαρύτητας. Συγκεκριμένα, στην ${\cal N}=4$ \en SYM
\gr σε πεπερασμένη θερμοκρασία, δυική σε μελανές $D3$ βράνες \cite{wilsonloopTemp},
όπως και στον κλάδο \en Coulomb \gr της ${\cal N}=4$ \en SYM, \gr δυική σε πολυκεντρικές
κατανομές $D3$ βράνων \cite{bs} υπάρχουν τριών ειδών συμπεριφορές, τις οποίες θα
περιγράψουμε παρακάτω.

\begin{itemize}

\item Μόνο για μήκη $L$ μικρότερο απο μία μέγιστη τιμή
$L_{\rm c}=L(u_{0\rm c})$ υπάρχει λύση για το $u_0=u_0(L)$
και είναι μια δίτιμη συνάρτητη του $L$, Σχήμα 7.2(α). Αυτό υποδηλώνει
την ύπαρξη δυο κλασικών χορδών οι οποίες ικανοποιούν τις ίδιες οριακές συνθήκες, η
μικρού μήκους χορδή για $u_0>u_{0c}$ και η μεγάλου μήκους χορδή $u_0<u_{0c}$, που ικανοποιούν
$L^\prime(u_0)<0$ και $L^\prime(u_0)>0$ αντίστοιχα. Ώς εκ τούτου η ενέργεια
είναι μια δίτιμη συνάρτηση του μήκους με άνω και κάτω κλάδο. Παρότι ο άνω κλάδος έχει μεγαλύτερη
ενέργεια απο τον κάτω, η κλασική λύση δεν μας ξεκαθαρίζει αν ο άνω είναι διαταρακτικά ευσταθής
ή μετασταθής και επομένως αν είναι φυσικά αποδεκτή. Επιπρόσθετα ο άνω κλάδος παραβιάζει την
συνθήκη κυρτότητας Εξ.\eqn{7-19}. Η συμπεριφορά αυτή συναντάται στη θεωρία ${\cal N}=4$ \en SYM \gr
σε πεπερασμένη θερμοκρασία και σε κάποιες περιοχές του κλάδου \en Coulomb. \gr

\item Για μήκη $L$ μικρότερα απο μία μέγιστη τιμή $L_c$ υπάρχει λύση για το $u_0=u_0(L)$ και
είναι μονότιμη συνάρτηση του $L$. Η $E(L)$ είναι μια μονότιμη συνάρτηση και περιγράφει ένα θωρακισμένο
δυναμικό \en Coulomb, \gr Σχήμα 7.2(β). Ωστόσο, το μήκος θωράκισης έχει έντονη εξάρτηση απο τον προσανατολισμό της
χορδής σε σχέση με την κατανομή των βρανών.

\item Για όλα τα μήκη $L$, υπάρχει λύση για το $u_0=u_0(L)$. Επιπλέον η $E(L)$ είναι μια μονότιμη
συνάρτηση του $L$ η οποία ταλαντεύεται μεταξύ μιας συμπεριφοράς \en Coulomb \gr και ενός γραμμικού
δυναμικού εγκλωβισμού, Σχήμα 7.2(γ). Ωστόσο, ο εγκλωβισμός δεν αναμένεται στις δυικές θεωρίες πεδίου που μελετούμε
λόγω της σύμμορφης συμμετρίας σε συνδυασμό με την μέγιστη υπερσυμμετρία.\\
\no
Τα δύο τελευταία είδη συμπεριμπεριφοράς συναντούνται σε ορισμένες περιοχές του κλάδου
\en Coulomb \gr στην ${\cal N}=4$ \en SYM. \gr Σημειώνουμε ότι στις περιπτώσεις αυτές δεν
παραβιάζεται η συνθήκη κοιλότητας Εξ.\eqn{7-19}. Αυτές οι διαφορές εμφανίζονται λόγω του ότι
η θεωρία ${\cal N}=4$ \en SYM \gr σε μηδενική θερμοκρασία δεν αναμένεται να προβλέπει
συμπεριφορά εγκλωβισμού και ότι το μήκος θωράκισης δεν θα πρέπει να έχει μια έντονη εξαρτηση
στην συγκεκριμένη τροχιά της χορδής που χρησιμοποιούμε για να υπολογίσουμε το δυναμικό του
ζεύγους.

\end{itemize}

\no
Και στις τρείς αυτές περιπτώσεις, η συμπεριφορά της ενέργειας αλληλεπίδρασης ως συνάρτηση
του μήκους είναι αντίθετη με τα προσδοκώμενα για την δυική θεωρία πεδίου και ως εκ τούτου απαιτείται
μια πιο προσεκτική ερμηνεία. Για την πρώτη περίπτωση, είναι εμφανές ότι η ανάλυση των μικρών διακυμάνσεων
θα καθορίσει αν η μεγάλου μήκους χορδή (άνω κλαδος) είναι διαταρακτικά ασταθής. Στις άλλες δύο
περιπτώσεις δεν υπάρχουν ενδείξεις αστάθειας στηριζόμενοι σε ενεργειακά επιχειρήματα.
Όπως θα δούμε, οι παραμετρικές περιοχές που δίνουν έντονη εξάρτηση του μήκους θωράκισης απο τον
προσανατολισμό της χορδής (δεύτερη περίπτωση) και η γραμμική συμπεριφορά εγκλωβισμού (τρίτη περίπτωση)
θα αποδειχθούν ότι είναι ασταθείς κάτω απο μικρές διακυμάνσεις.

\begin{figure}[!t]
\begin{center}
\begin{tabular}{ccc}
\includegraphics[height=3.2cm]{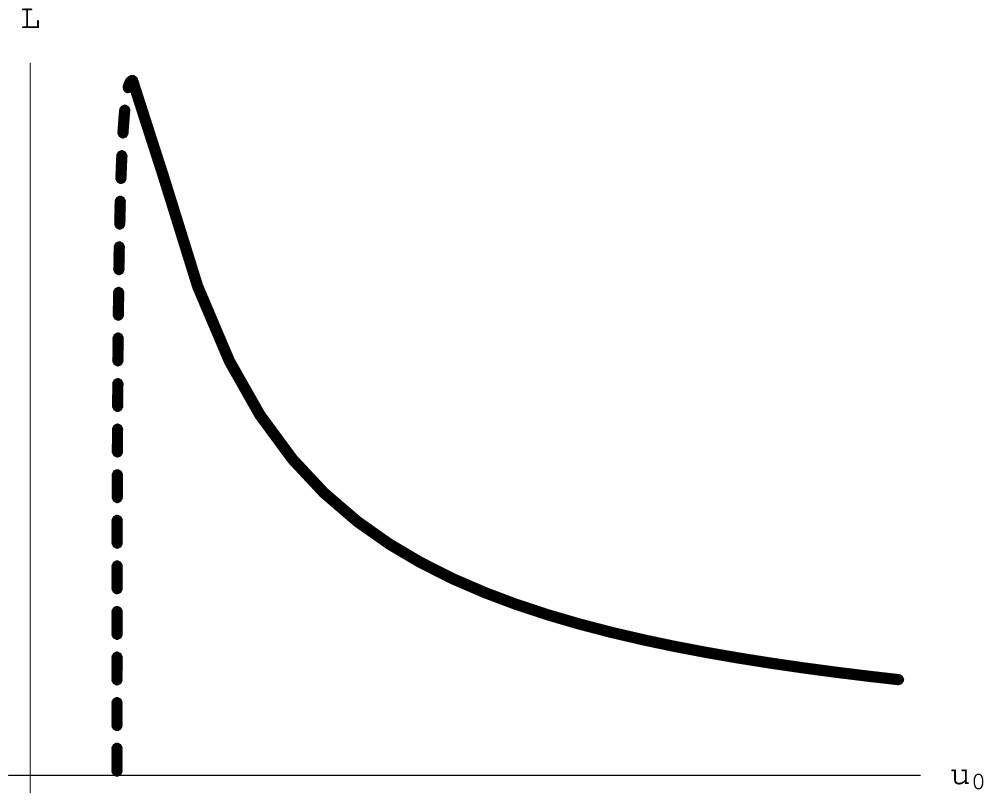}
&\includegraphics[height=3.2cm]{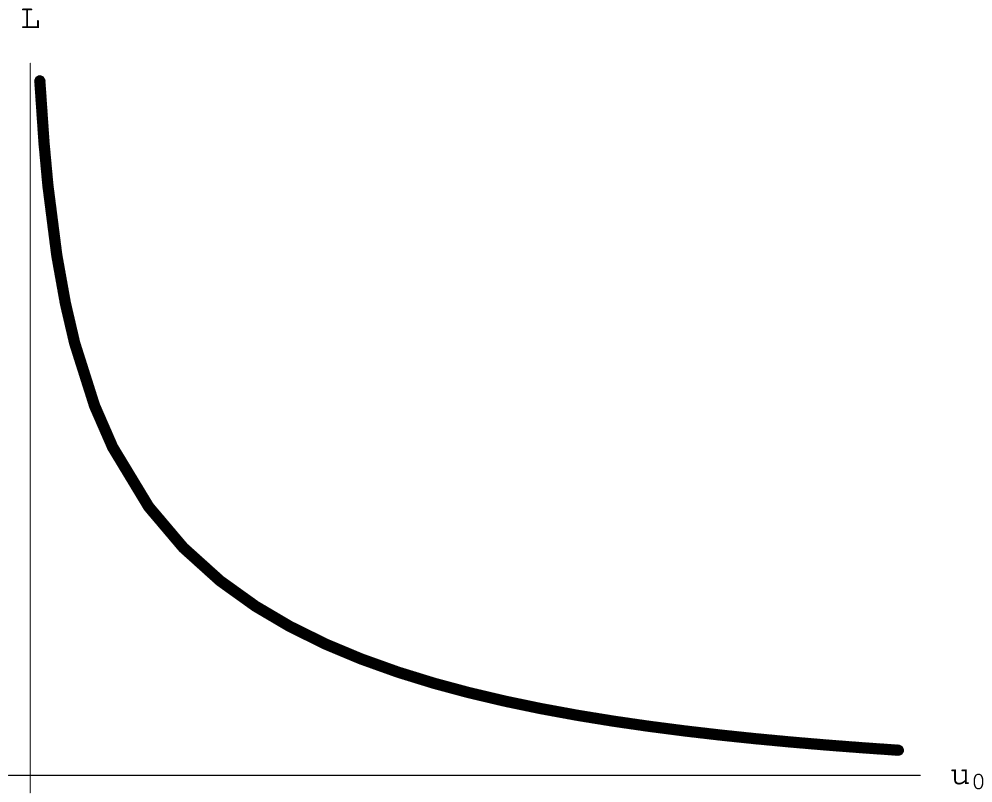}
&\includegraphics[height=3.2cm]{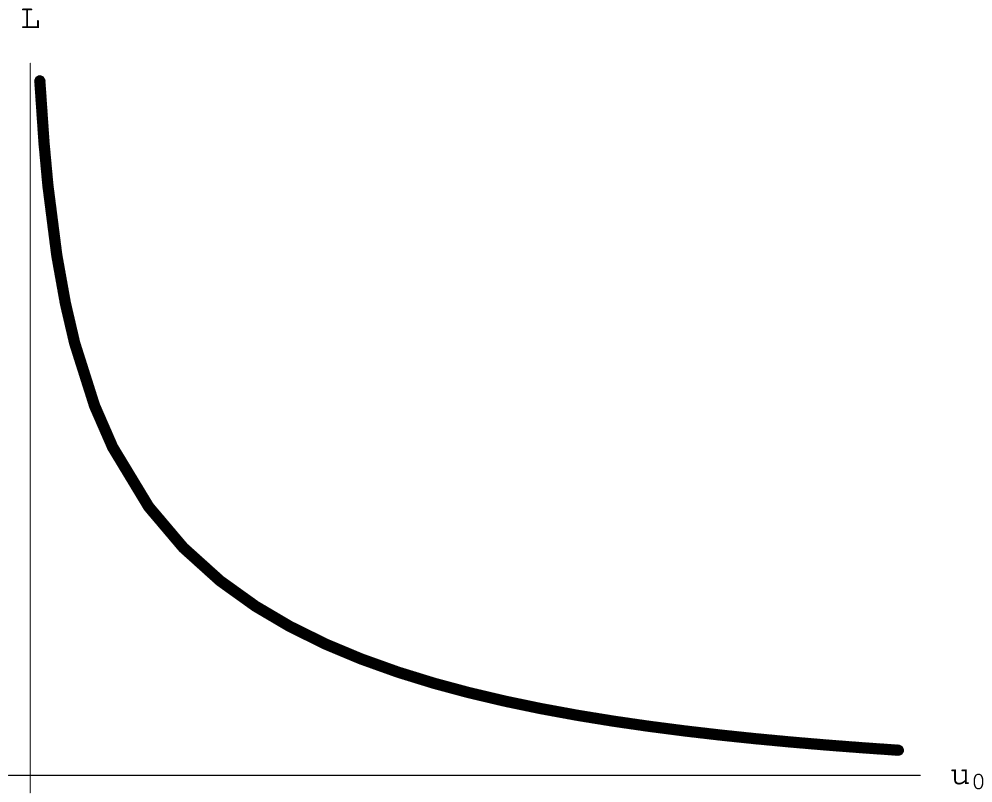}\\
\includegraphics[height=3.2cm]{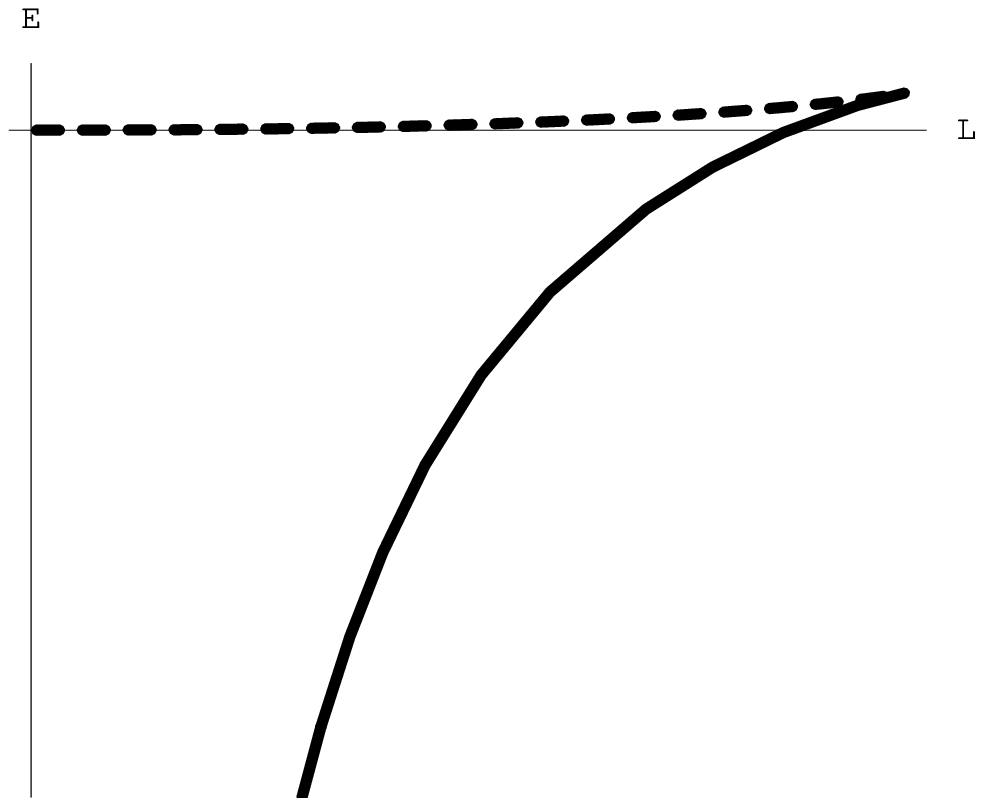}
&\includegraphics[height=3.2cm]{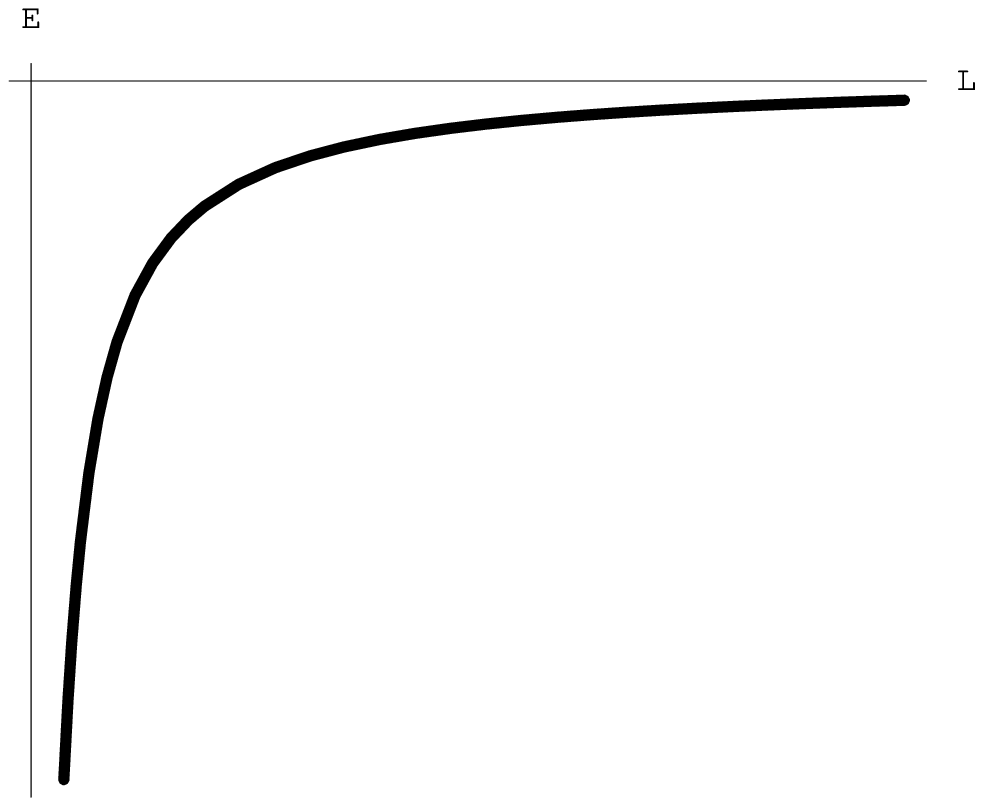}
&\includegraphics[height=3.2cm]{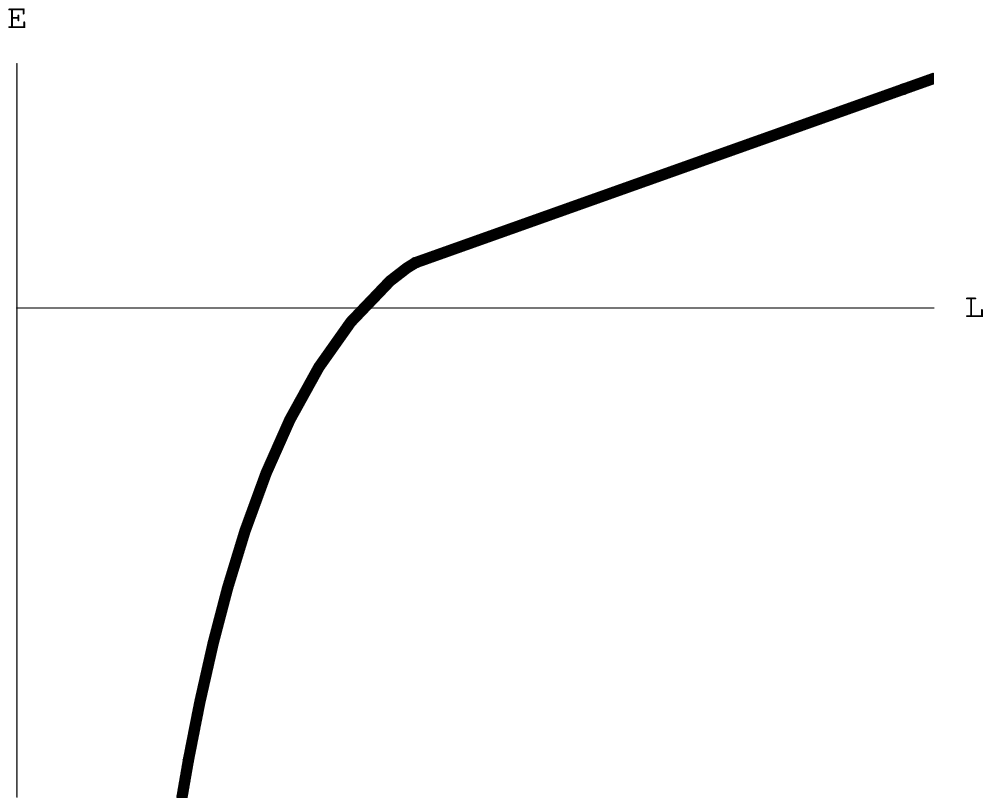}
\\
(α) & (β) & (γ)
\end{tabular}
\end{center}
\caption{Τα τρία γενικά είδη συμπεριφορών των $L(u_0)$ και $E(L)$
που απαντώνται σε υπολογισμούς βρόχων \en Wilson \gr στα πλαίσια της \en AdS/CFT , \gr
και αντιστοιχούν σε (α) ένα ακρότατο, δίτιμο δυναμικό,
(β) μη ύπαρξη ακροτάτων, θωρακισμένο ή μη δυναμικό \en Coulomb, \gr
(γ) μη ύπαρξη ακροτάτων, σε δυναμικό \en Coulomb \gr και εγκλωβισμού.
} \label{Mesonplots}
\end{figure}
\no

\chapter{Ανάλυση Ευστάθειας}
Έχοντας περιγράψει τα βασικά χαρακτηριστικά των υπό μελέτη
κλασικών χορδών, θα εξετάσουμε την ευστάθεια αυτών των διατάξεων,
με σκοπό να απομονώσουμε τις φυσικά ενδιαφέρουσες περιοχές στις
οποίες η αντιστοιχία \en AdS/CFT \gr είναι αξιόπιστη.
Σε αυτό το κεφάλαιο θα δώσουμε μια γενική περιγραφή της ανάλυσης των μικρών διακυμάνσεων
για την μετρική της μορφής Εξ.\eqn{7-1} και θα εξάγουμε διάφορα αποτελέσματα
τα οποία θα μας επιτρέψουν να αναγνωρίσουμε τις ευσταθείς και ασταθείς περιοχές
με έναν συνδυασμό αναλυτικών και προσεγγιστικών μεθόδων \cite{ASS1}.

\section{Μικρές Διακυμάνσεις}

Για να μελετήσουμε την ευστάθεια των διατάξεων, θα θεωρήσουμε μικρές διακυμάνσεις
γύρω απο τις κλασικές λύσεις που μελετήσαμε πριν. Συγκεκριμένα, θα μελετήσουμε τριών
ειδών διακυμάνσεις, (1) εγκάρσιες διακυμάνσεις, που αντιστοιχούν στις αγνοήσιμες
συντεταγμένες κάθετες στον άξονα του ζεύγους όπως $x$, (2) διαμήκεις διακυμάνσεις,
που αντιστοιχούν στην αγνοήσιμη συντεταγμένη $y$ κατά μήκος του άξονα του ζεύγους και
(3) γωνιακές διακυμάνσεις, που αντιστοιχούν στην μη αγνοήσιμη συντεταγμένη $\th$.
Θεωρούμε την παρακάτω διαταραχή στην εμβάπτιση

\be
\label{8-1}
x = \d x (t,u)\ ,\qq y = y_{\rm cl}(u) + \d y (t,u)\ ,\qq \th = \th_0 + \d \th(t,u)\ .
\ee
\newpage
\no κρατώντας σταθερή την επιλογή της βαθμίδας Εξ.\eqn{7-10} λόγω
του αναλλοίωτου κάτω απο διδιάστατες παραμετροποιήσεις.\footnote{
Εναλλακτικά, θα μπορούσαμε να διαταράξουμε την βαθμίδα θέτοντας $u =
\s + \d u(t,\s)$, κρατώντας το $y$ ως $y_{\rm cl}(\s)$ όπως στο
\cite{michalogiorgakis}. Το πλεονέκτημα της επιλογής μας είναι ότι η
διαφορική εξίσωση για το $\d y$ είναι πιο απλή απο αυτή για το $\d
u$. Ωστόσο, όπως θα δούμε στην παράγραφο 8.2, για να καθορίσουμε τις
συνοριακές συνθήκες πρέπει να χρησιμοποιήσουμε το $\d u$ αντί για το
$\d y$. Για λεπτομέρειες για τις διάφορες δυνατές επιλογές βαθμίδας,
δείτε \cite{kinar}.} Στην συνέχεια υπολογίζουμε την δράση \en
Nambu-Goto \gr για αυτήν την υπόθεση και την αναπτύσουμε σε δυνάμεις
των διακυμάνσεων. Η τελική μορφή της ανάπτυξης γράφεται ως εξής

\be
\label{8-2}
S = S_0 + S_1 + S_2 + \ldots\ ,
\ee

\no
όπου οι δείκτες στους διάφορους όρους αντιστοιχούν στην τάξη του αναπτύγματος. Ο όρος
μηδενικής τάξεως ισούται με την κλασική δράση.
Με χρήση των εξισώσεων κίνησης για τις διαμήκεις διακυμάνσεις
ο όρος πρώτης τάξεως μπορεί να δειχθεί ότι είναι ίσος με
\be
\label{8-3} S_1 = - {1 \ov 2\pi} \int dt du \left[ \sqrt{f_{y0}}\
\d y^\prime + \left( {1 \ov 2 F^{1/2}} \partial_\th g + {f_{y0}
F^{1/2}
 \ov 2 f_y^2} \partial_\th f_y \right) \d \th \right]\ ,
\ee

\no
όπου $\partial_\th g$ και $\partial_\th f_y$ είναι οι παράγωγοι ως προς
$\th$ των συναρτήσεων $g(u,\th)$ και $f_y(u,\th)$ αντίστοιχα.
Ο πρώτος όρος είναι επιφανειακός όρος και μπορεί να απαλειφθεί αν προσθέσουμε
στην κλασική δράση έναν επιφανειακό όρο που δεν θα επηρεάσει τις εξισώσεις κίνησης.
Στο δεύτερο όρο, ο συντελεστής του $\d\th$ είναι οι εξισώσεις κίνησης για το $\th$ και
μια αναγκαία συνθήκη μηδενισμού του είναι η Εξ.\eqn{7-13}. Χρησιμοποιώντας
αυτή την συνθήκη και μετά απο αλγεβρικές πράξεις, βρίσκουμε ότι η δευτέρας τάξεως
συνεισφορά μπορεί να γραφεί ως
\ba
\label{8-4} \!\!\!\!\!\!\!\!\!\!\!\! S_2 &=& - {1 \ov 2\pi} \int
dt du \biggl[ {f_x \ov 2 F^{1/2}}
\d x^{\prime 2} - {h f_x F^{1/2} \ov 2 g f_y} \d \dot{x}^2 \nonumber\\
 && \qq\qq\quad\; + {g f_y \ov 2 F^{3/2}} \d y^{\prime 2} - {h \ov 2 F^{1/2}} \d \dot{y}^2 \\
&& \qq\qq\quad\; + {f_\th \ov 2 F^{1/2}} \d \th^{\prime 2} - {h
f_\th F^{1/2} \ov 2 g f_y} \d \dot{\th}^2 + \left( {1 \ov 4
F^{1/2}}
\partial_\th^2 g
+ {f_{y0} F^{1/2} \ov 4 f_y^2} \partial_\th^2 f_y \right) \d \th^2
\biggr]\ ,\nonumber
\ea

\no
όπου όλες οι συναρτήσεις καθώς και οι παράγωγοι αυτών ως προς $\th$ είναι υπολογισμένες
στο $\th=\th_0$. Παρατηρούμε ότι, με την χρήση της Εξ. \eqn{7-13}, όλα τα είδη
των διακυμάνσεων είναι αποσυζευγμένα μεταξύ τους.
Γράφοντας τις εξισώσεις κίνησης αυτής της δράσης και χρησιμοποιώντας
το ότι οι συναρτήσεις δεν εξαρτώνται από το χρόνο $t$,
αναπτύσουμε σε συνιστώσες \en Fourier \gr

\be
\label{8-5}
\d x^\m (t,u) = \d x^\m (u) e^{-{\rm i} \omega t}\ ,
\ee

\no
και βρίσκουμε τις ακόλουθες συνήθεις γραμμικές διαφορικές εξισώσεις
δευτέρας τάξεως

\ba
\label{8-6}
&&\left[ {d \ov du} \left({f_x \ov F^{1/2}} {d \ov du} \right)
+ \omega^2 {h f_x F^{1/2} \ov g f_y} \right] \d x = 0\ ,
\nonumber\\
&&\left[ {d \ov du} \left( {g f_y \ov F^{3/2}} {d \ov du} \right)
+ \omega^2 {h \ov F^{1/2}} \right] \d y = 0\ ,
\\
&&\left[ {d \ov du} \left({f_\th \ov F^{1/2}} {d \ov du} \right)
 + \left( \omega^2 {h f_\th F^{1/2} \ov g f_y}
- {1 \ov 2 F^{1/2}} \partial_\th^2 g - {f_{y0} F^{1/2} \ov 2
f_y^2} \partial_\th^2 f_y \right) \right] \d \th = 0\ . \nonumber
\ea

\no
Ως εκ τούτου, το πρόβλημα καθορισμού της ευστάθειας των διατάξεων των χορδών
ανάγεται στο καθιερωμένο πρόβλημα ιδιοτιμών για τους διαφορικούς τελεστές
των τριών διακυμάνσεων. Συγκεκριμένα, το πρόβλημα ανάγεται σε μελέτη
ενός προβλήματος \en Sturm--Liouville \gr
\be
\label{8-7} \left\{ - {d \ov du} \left[ p(u;u_0) {d \ov du}
\right] - r(u;u_0)
 \right\} \Phi(u) = \omega^2 q(u;u_0) \Phi(u)\ ,\quad u_{\rm min}\leqslant u_0\leqslant u
< \infty\ ,
\ee

\no
όπου οι συναρτήσεις $p(u;u_0)$, $q(u;u_0)$ και $r(u;u_0)$ δίδονται στην Εξ.\eqn{8-6}
για την εκάστοτε διακύμανση και εξαρτώνται απο την παράμετρο $u_0$ μέσω της συνάρτησης
$F$ στην Εξ.\eqn{7-16}. Το πρόβλημα συνεπώς ισοδυναμεί με τον καθορισμό των τιμών της παραμέτρου
$u_0$ για τις οποίες η παράμετρος $\om^2$ είναι αρνητική, οδηγώντας έτσι σε αστάθεια της κλασικής
λύσης.\\
\no
Παρότι στις περισσότερες περιπτώσεις η περιγραφή \en Sturm--Liouville \gr είναι αρκετή,
σε άλλες περιπτώσεις είναι πιο κομψή η περιγραφή του προβλήματος μετασχηματίζοντας
το πρόβλημα μας σε τύπου \en Schr\"odinger.\ \gr Για το λόγο αυτό, πρέπει να αλλάξουμε μεταβλητές

\be
\label{8-8} x = \int_u^\infty du \sqrt{q\ov p}\ = \int_u^\infty
du \sqrt{h\ov f_y-f_{y0}} \ ,\qq \Phi = (p q)^{-1/4} \Psi\ ,
\ee

\no
όπου η έκφραση αυτή για την μεταβλητή $x$ ισχύει και για τα
τρία είδη διακυμάνσεων. Με την βοήθεια της Εξ.\eqn{8-8} η Εξ.\eqn{8-7}
γράφεται ως εξίσωση τύπου \en Schr\"odinger \gr

\be
\label{8-9} \left[ -{d^2 \ov dx^2} + V(x;u_0) \right] \Psi(x) =
\omega^2 \Psi(x)\ ,
\ee
με δυναμικό
\be
\label{8-10}
V = -{r \ov q} + {p^{1/4} \ov q^{3/4}} {d \ov du}
\left[ \left( {p \ov q} \right)^{1/2} {d \ov du}
 (p q)^{1/4} \right] =  -{r \ov q} + (p q)^{-1/4} {d^2 \ov dx^2} (p q)^{1/4}\ ,
\ee

\no
που εκφράζεται σαν συνάρτηση $V(u;u_0)$ της μεταβλητής $u$ στην πρώτη σχέση
και σαν συνάρτηση $V(x;u_0)$ της μεταβλητής $x$ στην δεύτερη. Το ισοδύναμο
πρόβλημα \en Schr\"odinger \gr ορίζεται στην περιοχή $x \in [0,x_0]$
όπου η παράμετρος $x_0$ δίδεται από την

\be
\label{8-11}
x_0 = \int_{u_0}^\infty du \sqrt{q\ov p} \ =
\int_{u_0}^\infty du \sqrt{h \ov f_y-f_{y0}} \ ,
\ee

\no
και είναι πεπερασμένη. Αυτό μπορεί να επιβεβαιωθεί απο τις ασυμπτωτικές εκφράσεις
Εξ.\eqn{7-7} και το γεγονός ότι $p/q \sim u-u_0$ όταν $u\to u_0^+$. Η ακριβής εξάρτηση
της παραμέτρου $x_0$ συναρτήσει της παραμέτρου $u_0$ εξαρτάται από τις λεπτομέρειες
του υποβάθρου Εξ.\eqn{7-3}. Ωστόσο, χρησιμοποιώντας τις ασυμπτωτικές εκφράσεις
Εξ.\eqn{7-7} βρίσκουμε ότι

\be
\label{8-12}
x(u_0)\simeq {\G(1/4)^2\ov 4 \sqrt{2 \pi} \ u_0} \ ,
\qq {\textrm{όταν}}\quad u_0\to \infty\ .
\ee

\no
Από το γεγονός ότι το μέγεθος του διαστήματος είναι πεπερασμένο συνεπάγεται
ότι το φάσμα των διακυμάνσεων είναι κβαντισμένο και το γεγονός ότι είναι
όλο και στενότερο στο υπεριώδες υποδηλώνει ότι δεν υπάρχει καμμία αστάθεια
στο σύμμορφο όριο της θεωρίας. Με την βοήθεια του μετασχηματισμού Εξ.\eqn{8-8}
οι περιοχές του υπεριώδους και του υπέρυθρου απεικονίζονται στις περιοχές στο
$x=0$ και στο $x=x_0$ αντίστοιχα. Στην τύπου \en Schr\"odinger \gr περιγραφή,
ο σκοπός μας είναι να καθορίσουμε το εύρος της παραμέτρου $u_0$ για την οποία η
θεμελιώδης κατάσταση του δυναμικού Εξ.\eqn{8-10} έχει αρνητική ενέργεια.  Αξίζει να
σημειώσουμε, ότι εν γένει, δεν μπορούμε να βρούμε την μεταβλητή $u$ σε κλειστή
έκφραση της μεταβλητής $x$ και για αυτό δεν είναι δυνατό να εκφράσουμε το δυναμικό σαν
συνάρτηση του $x$. Παρόλα αυτά, σε πολλές περιπτώσεις η έκφραση του δυναμικού σαν συνάρτηση
της μεταβλητής $u$ δίνει χρήσιμες πληροφορίες. Για παράδειγμα, αν το δυναμικό $V(u;u_0)$ είναι
θετικό τότε οι ιδιοτιμές $\om^2$ είναι πάντα θετικές και συνεπώς δεν υπάρχει αστάθεια.
Η δε περιγραφή του μέσω της μεταβλητής $x$ είναι χρήσιμη στην εφαρμογή διαταρακτικών μεθόδων.

\section{Συνοριακές συνθήκες}

Για να καθορίσουμε πλήρως το πρόβλημα ιδιοτιμών, πρέπει να εφαρμόσουμε
κατάλληλες συνοριακές συνθήκες για τις διακυμάνσεις στο υπέρυθρο
$u\to\infty$ ($x=0$) και στο υπεριώδες $u=u_0$ ($x=x_0$). Ξεκινώντας
από την περιγραφή \en Sturm--Liuville, \gr η συνοριακή συνθήκη στο υπεριώδες
μπορεί να βρεθεί εύκολα απο την συμπεριφορά των διαφορικών εξισώσεων Εξ. \eqn{8-6}
στο $u\to\infty$. Όντως, σε αυτό το όριο βρίσκουμε ότι

\be
\label{8-13}
\Phi^{\prime\prime} + {4\ov u} \Phi^\prime = {\cal O}(u^{-4}) \
,\qq {\textrm{όταν}} \quad u\to \infty\ ,
\ee

\no
το οποίο υποδηλώνει ότι το όριο, $u\to\infty$ είναι ένα απαλείψιμο
ανώμαλο σημείο και ότι οι δύο ανεξάρτητες λύσεις (Βροσκιανή διάφορη του μηδενός)
της Εξ.\eqn{8-13} έχουν την μορφή

\ba
\label{8-14}
\Phi_1 &= & \sum_{n=0} c_n u^{-n}\ ,
\nonumber\\
\Phi_2 & = & d \ln u + {1\ov u^3} \sum_{n=0}^\infty d_n u^{-n}\ .
\ea

\no
Στην πρώτη λύση, η οποία είναι ομαλή στο άπειρο, έχουμε
την ελευθερία να θέσουμε την προσθετική σταθερά $c_0$ στο μηδέν.
Για την δεύτερη λύση, η οποία απειρίζεται στο άπειρο,
επιλέγουμε το $d=0$. Οπότε υποθέτουμε την συνοριακή συνθήκη

\be
\label{8-15}
\Phi(u) = 0\ ,\qq {\textrm{όταν}}\quad u\to \infty \ .
\ee

\no
Με άλλα λόγια υποθέτουμε ότι η διακύμανση δεν θα επηρεάσει τα άκρα
της χορδής στα οποία βρίσκονται τα κουάρκ και αντικουάρκ.
\gr Όσον αναφορά την συνοριακή συνθήκη στο υπέρυθρο, αυτή απαιτεί
ιδιαίτερη προσοχή, λόγου του ότι πρέπει να συρράψουμε τον άνω και κάτω
κλάδο της κλασικής λύσης $y_{\rm cl}$, που αντιστοιχεί στα δυο διαφορετικά πρόσημα
της Εξ.\eqn{7-15}. Το σημείο $u=u_0$ είναι επίσης ένα απαλείψιμο ανώμαλο σημείο
και οι δύο ανεξάρτητες λύσεις μπορούν να αναπτυχθούν ως εξής\footnote{Υποθέτουμε ότι
$u_0> u_{\rm min}$. Στην περίπτωση της ισότητας η δομή των ανωμαλιών είναι εντελώς
διαφορετική και θα εξεταστεί ενδελεχώς σε συγκεκριμένα παραδείγματα που θα ακολουθήσουν.}

\ba
\label{8-16}
\Phi_{1} & = & (u-u_0)^{\r} \left[ c_0 + \sum_{n=1}^\infty c_n (u-u_0)^n\right]\ ,
\nonumber\\
\Phi_{2}  & = & d_0 + \sum_{n=1}^\infty d_n (u-u_0)^n\ ,
\ea

\no
όπου $\rho=1/2$ για τις διακυμάνσεις $x,\th$ και $\rho=-1/2$ για τις $y$.
Απαιτούμε ότι, στις δυο πλευρές της κλασικής λύσης, οι διακυμάνσεις και οι
πρώτες παράγωγοι αυτών ως προς την κλασική λύση πρέπει να είναι ίσες. Ωστόσο,
αυτό δεν μπορεί να γίνει χρησιμοποιώντας την κλασική λύση $y_{\rm cl}$, αλλά
αυτή σε συνδυασμό με την Εξ.\eqn{8-1}. Ισοδύναμα, αν ορίσουμε μια νέα μεταβλητή
$\bar{u}$ ως

\be
\label{8-17}
u=\bar u + \d u(t,u) \ ,\qq \d u(t,u) = - {\d y(t,u)\ov
y^\prime_{\rm cl}(u)}\ ,
\ee

\no
η κλασική λύση δεν διαταράσσεται καθόλου. Αυτή η αλλαγή μεταβλητής δεν επηρεάζει
τις $x$ και $\th$ διακυμάνσεις διότι έχουν τετριμμένη κλασική λύση και επειδή κρατάμε
όρους γραμμικούς στην διακύμανση. Ωστόσο, εφόσον $y_{\rm cl}^\prime \sim (u-u_0)^{-1/2}$
κοντά στο $u=u_0$, οι διακυμάνσεις $\d u$ έχουν ένα ανάπτυγμα όπως στην Εξ.\eqn{8-16} αλλά
με $\rho=1/2$. Εφόσον, γύρω απο το σημείο συρραφής, οι δύο κλάδοι της κλασικής λύσης
είναι συμμετρικοί και διαφέρουν κατά ένα πρόσημο, πρέπει να έχουμε

\be
\label{8-18}
{d\Phi\ov d y_{\rm cl}}\bigg |_{u=u_0} = 0\ ,
\ee

\no
όπου το $\Phi$ αναφέρεται στις $\d x,\d\th$ και $\d u$ διακυμάνσεις. Χρησιμοποιώντας και πάλι
τις αρχικές μεταβλητές στις διακυμάνσεις βρίσκουμε, στα αναπτύγματα Εξ.\eqn{8-16}, όσον αφορά τους
συντελεστές $c_0,d_0$  ότι $c_0=0$ για τις $\d x,\d\th$ διακυμάνσεις και $d_0=0$ για τις $y$. Ισοδύναμα

\ba
\label{8-19} \d x,\d \th  : & & \phantom{xxx}
(u-u_0)^{1/2}\Phi^\prime = 0 \  ,\qq {\textrm{όταν}}\quad u\to u_0^+ ,
\nonumber\\
\d y  : & & \phantom{xxx} \Phi + 2 (u-u_0) \Phi^\prime = 0 \ ,\qq
{\textrm{όταν}}\quad u\to u_0^+ \ .
\ea

\no
Οι συνοριακές αυτές συνθήκες για την συνάρτηση $\Phi$ στο πρόβλημα τύπου \en Sturm--Liouville \gr
μπορούν να μεταφραστούν ως συνοριακές συνθήκες για την συνάρτηση $\Psi$ στο πρόβλημα τύπου
\en Schr\"odinger \gr με χρήση των ασυμπτωτικών σχέσεων $x\sim 1/u$ όταν το $u\to\infty$
και $u-u_0 \sim (x_0-x)^2$ όταν $u \to u_0^+$ και $x \to x_0^-$, οι οποίες απορρέουν
απο την Εξ.\eqn{8-8} και τις συναρτήσεις $p$ και $q$. Μετά απο κάποιες
αλγεβρικές πράξεις, καταλήγουμε ότι για όλα τα είδη των διακυμάνσεων πρέπει να έχουμε

\be
\label{8-20} \Psi(0) = 0 \ ,\qq \Psi^\prime(x_0)= 0 \ ,
\ee

\no
οι οποίες αντιστοιχούν σε συνοριακές συνθήκες τύπου \en Dirichlet \gr και \en Neumann,
\gr στο υπεριώδες και υπέρυθρο αντίστοιχα.

\section{Μηδενικοί τρόποι}

Η μέθοδος που περιγράψαμε, καθορίζει εν γένει τις περιοχές ευστάθειας των λύσεων,
υπό την προυπόθεση ότι τα προβλήματα \en Sturm--Liouville \gr και \en Schr\"odinger \gr
μπορούν να λυθούν επακριβώς. Ωστόσο,  σε πολλές περιπτώσεις, τα πρόβληματα αυτά
είναι αρκετά πεπλεγμένα και το φάσμα είναι αδύνατο να καθοριστεί αναλυτικά. Από την άλλη πλευρά,
μπορούμε να πάρουμε χρήσιμες πληροφορίες μελετώντας ένα απλούστερο
πρόβλημα, το πρόβλημα των μηδενικών τρόπων των αντίστοιχων διαφορικών τελεστών.
Σε ότι ακολουθεί, αποδεικνύουμε ότι (1) δεν υπάρχουν μηδενικοί τρόποι για τις εγκάρσιες
διακυμάνσεις, (2) οι διαμήκεις μηδενικοί τρόποι είναι σε αναλογία ένα προς ένα με τα
υποψήφια ακρότατα του μήκους $L(u_0)$, και (3) οι γωνιακοί μηδενικοί τρόποι μπορούν να
βρεθούν με πολύ καλή ακρίβεια προσεγγίζοντας το αντίστοιχο πρόβλημα \en Schr\"odinger \gr
με ένα απειρόβαθρο πηγάδι. Τα αποτελέσματα αυτά θα είναι χρήσιμα για την ανάλυση του
9ου κεφαλαίου.

\subsection{Εγκάρσιοι μηδενικοί τρόποι}

Αρχικά θα θεωρήσουμε την περίπτωση των εγκάρσιων διακυμάνσεων. Ο μηδενικός τρόπος
που ικανοποιεί την Εξ.\eqn{8-15} είναι:

\ba
\label{8-21}
\d x &  \sim & \int^{\infty}_u {du\ov f_x} \sqrt{gf_y\ov f_y-f_{y0}}
\nonumber\\
& = & -2 {\sqrt{g f_y}\ov f_x f_y^\prime} \sqrt{f_y-f_{y0}} - 2
\int_u^\infty du \sqrt{f_y-f_{y0}}\ \del_u \left(\sqrt{g f_y}\ov
f_x f_y^\prime \right)
\\
& = &  - 2 \int_{u_0}^\infty du \sqrt{f_y-f_{y0}} \ \del_u
\left(\sqrt{g f_y}\ov f_x f_y^\prime \right)
  - {2\ov f_{x0}}\ \sqrt{g_0 f_{y0}\ov f^\prime_{y0}} (u-u_0)^{1/2} + {\cal O}(u-u_0)\ ,
\nonumber
\ea

\no
όπου στο δεύτερο βήμα κάναμε ολοκλήρωση κατά παράγοντες. Η ύπαρξη του μηδενικού τρόπου στηρίζεται
στο αν μπορούμε να ικανοποιήσουμε την συνοριακή συνθήκη Εξ.\eqn{8-19}, δηλαδή να μηδενίζεται
ο συντελεστής του όρου $(u-u_0)^{1/2}$ για κάποια τιμή της παραμέτρου $u_0$. Παρατηρούμε ότι
αυτό δεν είναι δυνατόν και ως εκ τούτου δεν υπάρχουν μηδενικοί τρόποι για τις εγκάρσιες
διακυμάνσεις. Για αυτό, αν η χαμηλότερη τιμή της ιδιοτιμής του τελεστή \en Schr\"odinger \gr
που αντιστοιχεί στις εγκάρσιες διακυμάνσεις είναι θετική για μια τιμή της παραμέτρου $u_0$,
θα παραμείνει θετική. Όντως, όπως θα δούμε, σε όλα τα παραδείγματα που θα μελετήσουμε, οι
κλασικές λύσεις είναι ευσταθείς κάτω απο εγκάρσιες διακυμάνσεις.

\subsection{Διαμήκεις μηδενικοί τρόποι}

Στην συνέχεια θα μελετήσουμε τις διαμηκείς διακυμάνσεις, για τις οποίες θα αποδείξουμε ότι
οι επανακανονικοποιήσιμοι μηδενικοί τρόποι υπάρχουν μόνο για τιμές της παραμέτρου $u_0$ για τα υποψήφια
ακρότατα της συνάρτησης $L(u_0)$, δηλαδή όπου $L^\prime(u_0) = 0$. Η σπουδαιότητα αυτού του
αποτελέσματος στηρίζεται στο ότι η εύρεση υποψήφιων ακροτάτων της συνάρτησης $L(u_0)$ για κάποιες
τιμές της παραμέτρου $u_0$ ισοδυναμεί με την ύπαρξη μηδενικών τρόπων για τις διαμήκεις διακυμάνσεις.
Για να κατανοήσουμε αυτήν την σύνδεση μεταξύ των μηδενικών τρόπων για τις διαμήκεις διακυμάνσεις και
των υποψήφιων ακροτάτων του μήκους, πρέπει να βρούμε την λύση του μηδενικού τρόπου των διαμήκων
διακυμάνσεων που ικανοποιούν την Εξ.\eqn{8-13}

\ba
\label{8-22}
\d y &  = & \int^{\infty}_u du {\sqrt{gf_y}\ov (f_y-f_{y0})^{3/2}}
\nonumber\\ & = & 2 {1\ov f^\prime_y } \sqrt{g f_y\ov f_y-f_{y0}}
+ 2 \int_u^\infty {du\ov \sqrt{f_y-f_{y0}}}\ \del_u \left(\sqrt{g
f_y}\ov f_y^\prime \right)
\\
& = &   2 \sqrt{g_0 f_{y0}\ov f_{y0}^{\prime 3}}\ (u-u_0)^{-1/2}
  +   2 \int_{u_0}^\infty { du\ov \sqrt{f_y-f_{y0}}}
\ \del_u \left(\sqrt{g f_y}\ov f_y^\prime \right)
 + {\cal O}\left((u-u_0)^{1/2}\right)\ .
\nonumber
\ea

\no
Οπότε, απο την συνοριακή συνθήκη Εξ.\eqn{8-19} συνεπάγεται ότι
υπάρχει αυτός ο τρόπος αν

\be
\label{8-23}
\int_{u_0}^\infty { du\ov \sqrt{f_y-f_{y0}}} \ \del_u
\left(\sqrt{g f_y}\ov f_y^\prime \right) = 0  \ .
\ee

\no
Στην συνέχεια παραγωγίζουμε την συνάρτηση του μήκους Εξ.\eqn{7-17}
ως προς την παράμετρο $u_0$. Το αποτέλεσμα ισούται με

\ba
\label{8-24} L^\prime(u_0) & = & {f_{y0}^\prime \ov \sqrt{f_{y0}}}
\int_{u_0}^\infty du {\sqrt{gf_y}\ov (f_y-f_{y0})^{3/2}}
 - 2 f_{y0}^{1/2} {F^{1/2} \ov f_y} \biggr|_{u = u_0}
\nonumber\\
&=&  2 {f_{y0}^\prime \ov \sqrt{f_{y0}}} \int_{u_0}^\infty { du\ov
\sqrt{f_y-f_{y0}}} \ \del_u \left(\sqrt{g f_y}\ov f_y^\prime
\right)
\\
&& +
{2 \ov f_{y0}^{1/2}} \lim_{u \to u_0} \left[ F^{1/2}
\left( {\partial_{u_0} f_{y0} \ov \partial_u f_y}
- {f_{y0} \ov f_y} \right) \right]\ ,
\nonumber
\ea

\no
όπου όπως και πριν κάναμε μια ολοκλήρωση κατά παράγοντες.
Η τελευταία γραμμή είναι μηδέν, οπότε

\be
\label{8-25}
L^\prime (u_0) =  2 {f_{y0}^\prime \ov \sqrt{f_{y0}}}
\int_{u_0}^\infty { du\ov \sqrt{f_y-f_{y0}}} \ \del_u
\left(\sqrt{g f_y}\ov f_y^\prime \right)\ ,
\ee

\no
το οποίο είναι πεπερασμένο, σε αντίθεση με τους δύο όρους
της πρώτης γραμμή της Εξ.\eqn{8-25} που είναι αποκλίνοντες.
Συγκρίνοντας τις Εξ.\eqn{8-23} και \eqn{8-25} βρίσκουμε
ότι τα πιθανά ακρότατα της συνάρτησης $L(u_0)$ είναι οι μηδενικοί τρόποι.
Συνεπώς, μπορούμε να
βρούμε το πρόσημο της χαμηλότερης ιδιοτιμής $\om^2$ σε όλες τις
περιοχές αναπτύσοντας το δυναμικό \en Schr\"odinger \gr στο $u_0$
γύρω από κάθε κρίσιμο σημείο $u_{0\rm c}$ και βρίσκοντας αν η
ιδιοτιμή $\om^2$ αλλάζει πρόσημο εκεί.

\no
Εκτός από την παραπάνω ανάλυση, υπάρχει ένας εναλλακτικός τρόπος
για να κατανοήσουμε την εμφάνιση ενός κρίσιμου σημείου $u_{0 \rm c}$.
Δεν είναι δύσκολο να αποδείξουμε ότι το δυναμικό \en Schr\"odinger \gr
έχει την ακόλουθη συμπεριφορά στις οριακές τιμές της μεταβλητής $u$,

\be
\label{8-26} V(u;u_0)  =   2 u^2 \ , \qq {\textrm{όταν}} \quad u\to
\infty\ ,
\ee
και
\be
\label{8-27} V_0(u_0)=V(u_0;u_0)  =  {f_{y0}^{\prime\prime}\ov 2
h_0} + {h_0^\prime f_{y0}^\prime \ov 8 h_0^2} -{3\ov 8}
{f_{y0}^\prime \ov h_0}\left({g_0^\prime \ov g_0}+{f_{y0}^\prime
\ov f_{y0}} \right) \ .
\ee

\no
Εφόσον το δυναμικό αυξάνει από μια ελάχιστη τιμή στο άπειρο στο
πεπερασμένο διάστημα $x\in [0,x_0]$, όπου το $x_0$ δίνεται απο την \eqn{8-11},
το φάσμα των διακυμάνσεων είναι διάκριτο. Επιπρόσθετα,  αν η ελάχιστη τιμή
$V_0$ του δυναμικού είναι αρκετά αρνητική το δυναμικό μπορεί να υποστηρίξει
δέσμιες καταστάσεις με αρνητική ενέργεια και αυτό συμβαίνει αν η Εξ.\eqn{8-23} έχει
λύση για κάποιες τιμές της παραμέτρου $u_0$. Η μεγαλύτερη τιμή απο αυτές της παραμέτρους είναι
η $u_{0 \rm c}$ για την οποία η μικρότερη ιδιοτιμή της ενέργειας γίνεται μηδενική.

\subsection{Γωνιακοί μηδενικοί τρόποι}

Τέλος θα θεωρήσουμε τους γωνιακούς μηδενικούς τρόπους οι οποίοι,
σε αντίθεση με τους διαμήκεις και εγκάρσιους δύο δεν μπορούν να εκφραστούν μέσω ενός ολοκληρώματος
λόγω του όρου μάζας που υπάρχει στην αντίστοιχη εξίσωση \en Sturm--Liouville, \gr
στην τρίτη γραμμή της Εξ.\eqn{8-6}. Από την άλλη πλευρά, στην περιγραφή
\en Schr\"odinger, \gr μπορούμε προσεγγίζοντας το αντίστοιχο δυναμικό,
να βρούμε το πλήρες φάσμα των διακυμάνσεων με πολύ μεγάλη ακρίβεια.
Ακολούθως, θα περιγράψουμε αυτήν την διαδικασία.\\
\no
Αρχικά δίνουμε την συμπεριφορά των γωνιακών δυναμικών
\en Schr\"odinger \gr στα όρια $u\to\infty$ και $u=u_0$
για τα παραδείγματα που θα μελετήσουμε στην συνέχεια

\be
\label{8-28}
V_\infty (u_0) \equiv V(\infty;u_0) = \half \lim_{u\to \infty}
u^2\del_\th^2 g\  ,
\ee
και
\be
\label{8-29}
V_0(u_0) \equiv V(u_0;u_0)  = {1\ov 8} {g_0 f_{y0} f_{y0}^\prime
\ov h_0^2 f_{\th 0}^2}\ \del_{u_0} \left(h_0 f_{\th 0}^2\ov g_0
f_{y0}\right) + \half {g_0\ov h_0 f_{\th 0}} \del^2_\th f_{y0}\ .
\ee
όπου το όριο της πρώτης εξίσωσης είναι πεπερασμένο. Εφόσον το
δυναμικό ξεκινάει απ'ο μια ελάχιστη τιμή στο πεπερασμένο διάστημα
$x\in [0,x_0]$, όπου $x_0$ δίδεται από την \eqn{8-11}, μπορούμε
να το προσεγγίσουμε με ένα απειρόβαθρο πηγάδι της μορφής

\be
\label{8-30}
V_{\textrm{προσεγ}} =
\left\{ \begin{array}{l l} \half (V_0+V_{\infty}) , \quad 0\leqslant x \leqslant x_0\\
\infty\  ,\quad {\textrm{οπουδήποτε αλλού}} \end{array}\right\}\ .
\ee

\no
Με την βοήθεια των συνοριακών συνθηκών της Εξ.\eqn{8-20} βρίσκουμε ότι το ενεγειακό
φάσμα δίδεται από την

\be
\label{8-31}
\omega^2_n(u_0) = {\pi^2 (2 n + 1)^2\ov 4
[x_0(u_0)]^2} + \half \left[V_0(u_0)+V_{\infty}(u_0)\right] \ ,\qq n=0,1,\dots \ .
\ee

\no
Παρατηρούμε ότι έχουμε χρησιμοποιήσει την μέση τιμή του δυναμικού για τις δύο
ακραίες τιμές της μεταβλητής $u$. Ικανή συνθήκη (αλλά όχι αναγκαία) για να υπάρχει λύση,
είναι η παραπάνω μέση τιμή να είναι αρνητική σε κάποιο πεπερασμένο εύρος των
τιμών της παραμέτρου $u_0$. Η κρίσιμη παράμετρος $u_{0 \rm c}$ μπορεί να βρεθεί λύνοντας
αριθμητικά την εξίσωση $\omega^2_0(u_0)=0$. Σημειώνουμε ότι διαφορετικές επιλογές του
$V_{\textrm{προσεγ}}$ είναι δυνατές (όπως για παράδειγμα η μέση τιμή του δυναμικού) αλλά
εννοιολογικά, δεν προσφέρουν κάτι νέο.\\
\no
Το πλεονέκτημα της μεθόδου που μόλις περιγράψαμε είναι ότι μπορεί να δώσει προσεγγιστικά το
πλήρες φάσμα των διακυμάνσεων και την εξάρτηση αυτών από την παράμετρο $u_0$ χωρίς να
καταφύγουμε σε προχωρημένες αριθμητικές μεθόδους.
Συγκεκριμένα, μπορούμε να σχεδιάσουμε τη χαμηλότερη ιδιοτιμή
$\om_0^2$ ως συνάρτηση της παραμέτρου $u_0$ και να ελέγξουμε αν είναι μονοτόνως αύξουσα,
όπως είναι σε όλα τα παραδείγματα μας. Επιπρόσθετα, όπως θα δούμε, σε μια περίπτωση η προσέγγιση
του απειρόβαθρου πηγαδιού είναι και ακριβής. Τέλος, θα δείξουμε ότι υπάρχει μια ιεραρχία στις
διάφορες κρίσιμες τιμές $u_{0{\rm c},n}$, $n=0,1,\dots$, με $u_{0 \rm c} \equiv u_{0 {\rm
c},0}$ και $u_{0 {\rm c},n}>u_{0 {\rm c},n+1}$, για τις οποίες οι αντίστοιχες ιδιοτιμές $\om_n^2$
είναι μηδενικές και γίνοται χαμηλότερα αρνητικές.\\
\no
Σε αυτό το σημείο θα μπορούσαμε να σκεφτούμε ότι η προσέγγιση του απειρόβαθρου πηγαδιού
θα μπορούσε να εφαρμοστεί στις διαμήκεις διακυμάνσεις, χρησιμοποιώντας για παράδειγμα την Εξ.\eqn{8-27}
ως την τιμή για τον πυθμένα του πηγαδιού, και έτσι να βρούμε προσεγγιστικά
το πλήρες φάσμα των διακυμάνσεων. Στα παραδείγματα μας, όπως θα δείξουμε, μια
τέτοια επιλογή δίνει με καλή προσέγγιση το αποτέλεσμα της Εξ.\eqn{8-23} για το $u_{0 \rm c}$,
δίνει όμως και ένα πλήθος από αρνητικές ιδιοτιμές όσο χαμηλώνουμε το $u_0$, το οποίο έρχεται σε
αντίθεση με τα γενικά συμπεράσματα που ακολούθησαν μετά την Εξ.\eqn{8-27}. Αυτή η αντίφαση
οφείλεται στο ότι το πλήρες δυναμικό μεταβάλεται κατά πολύ στο διάστημα $[0,x_0]$ και για αυτό
υποστηρίζει λιγότερες αρνητικές καταστάσεις από ότι η προσέγγιση του απειρόβαθρου
πηγαδιού που χρησιμοποιήσαμε.

\section{Θεωρία διαταραχών}

Όπως προανεφέραμε, η ευστάθεια των λύσεων έναντι των διαμήκων διακυμάνσεων θα
καθοριστεί απο την εύρεση των κρίσιμων σημείων του $L(u_0)$ και στην συνέχεια θα λύσουμε
την εξίσωση \en Schr\"odinger \gr για μικρές αποκλίσεις του $u_0$ από κάθε
κρίσιμο σημείο $u_{0 \rm c}$ μέσω θεωρίας διαταραχών. Κάνοντας αυτό, πρέπει να
παρατηρήσουμε ότι η παράμετρος $u_0$ υπεισέρχεται στο πρόβλημα όχι μόνο
μέσα στο δυναμικό $V(x;u_0)$, αλλά και στο μέγεθος του διαστήματος $x_0$
στο οποίο ορίζεται το πρόβλημα. Σε ότι ακολουθεί, θα παρουσιάσουμε
τις σχέσεις της διαταρακτικής θεωρίας που είναι κατάλληλες σε αυτήν την περίπτωση.\\
\no
Θεωρούμε το πιο γενικό πρόβλημα \en Schr\"odinger \gr που ορίζεται στο διάστημα
$[0,x_0]$ με δυναμικό που εξαρτάται από μια παράμετρο $u_0$, για την οποία γνωρίζουμε
ότι για ορισμένα $u_0=u_{0 \rm c}$ και $x_0=x_{0 \rm c}$ η εξίσωση \en Schr\"odinger \gr
\eqn{8-9} έχει μοναδική λύση με δεδομένη ιδιοτιμή $\om^2$ (ίση με το μηδέν στην περίπτωση μας).
Θα θέλαμε να καθορίσουμε την διόρθωση στην ιδιοτιμή της ενέργειας όταν οι παράμετροι $u_0$ και $x_0$
παρεκλίνουν από $u_{0 \rm c}$ και $x_{0 \rm c}$ αντίστοιχα. Εύκολα μπορούμε να δείξουμε ότι

\ba
\label{8-32}
V(x;u_0) & = & V(x;u_{0 \rm c})+ \d V(x)\ ,
\nonumber\\
 \d V(x)& = & {\d x_0 \ov x_{0 \rm c}}\Big[2 V(x;u_{0 \rm c})+ x \del_x V(x;u_{0 \rm c})\Big]
+ \d u_0 \del_{u_{0 \rm c}}V(x;u_{0 \rm c}) + \dots \ ,
\ea
\no
όπου $x\in[0,x_{0 \rm c}]$ ενώ $\d u_0 = u_0 - u_{0 \rm c}$ και
$\d x_0 = x_0 - x_{0 \rm c}$. Στην συνέχεια, ένας προσεκτικός υπολογισμός,
κρατώντας του επιφανειακούς όρους, δίνει την μετατόπιση της ενέργειας
(σε πρώτη τάξη)

\ba
\label{8-33}
\d \omega^2 & = & \int_{0}^{x_{0 \rm c}} dx\
|\Psi(x)|^2 \d V(x)
\nonumber\\
& = & {\d x_0 \ov x_{0 \rm c}}\ \left\{ 2 \omega^2 + \left[ \half
(\Psi^* \Psi^\prime + \Psi \Psi^{*\prime}) -x\Big[ |\Psi^\prime|^2
+ (\omega^2-V) |\Psi|^2\Big]\right]_{0}^{x_{0 \rm c}} \right\}
\nonumber\\
&& +\ \d u_0\ \int_{0}^{x_{0 \rm c}} dx\ |\Psi(x)|^2\ \del_{u_{0
\rm c}} V(x;u_{0 \rm c})\ ,
\ea

\no
όπου $(\omega^2-V) |\Psi|^2$ μπορεί να γραφεί εναλλακτικά
με χρήση της εξίσωσης \en Schr\"odinger \gr ως $-\half \left(\Psi^{\prime\prime} \Psi^* +
\Psi^{*\prime\prime} \Psi\right)$. Για να το δείξουμε αυτό,
έχουμε υποθέσει ότι η Χαμιλτονιανή είναι ερμιτιανή, το οποίο ισχύει
όταν οι συνοριακές συνθήκες είναι \en Dirichlet, Neumann \gr ή γραμμικός συνδυασμός αυτών.\\
\no
Για να εφαρμόσουμε αυτό στην περίπτωση μας, θέτουμε $\om^2=0$ και παρατηρούμε
ότι οι παραμέτροι $x_0$ και $u_0$ συσχετίζονται μέσω της Εξ.\eqn{8-11}. Οπότε, η μετατόπιση
της ενέργειας είναι:

\be
\label{8-34}
\d \omega^2 = \d u_0 \left[ x^\prime_0(u_{0 \rm c}) V(x_{0 \rm
c};u_{0 \rm c})|\Psi(x_{0 \rm c})|^2 + \int_{0}^{x_{0 \rm c}} dx\
|\Psi(x)|^2\ \del_{u_{0 \rm c}} V(x;u_{0 \rm c}) \right]\ ,
\ee
με $x_{0 \rm c}=x_0(u_{0 \rm c})$ και όπου έχουμε χρησιμοποιήσει τις συνοριακές συνθήκες
Εξ.\eqn{8-20}. Όσον αναφορά στον πρώτο όρο, από την παραγώγιση της Εξ.\eqn{8-11}
ως προς την παράμετρο $u_0$, βρίσκουμε

\be
\label{8-35}
x^\prime_0(u_0) = f^\prime_{y0} \int_{u_0}^\infty du\
{\del_u\left(h^{1/2} {f^\prime_y}^{-1}\right) \ov
\sqrt{f_y-f_{y0}}}\ .
\ee

\no
Όσον αναφορά στον δεύτερο όρο, παρατηρούμε ότι η
παράγωγος ως προς την παράμετρο $u_0$ δρα σε ένα δυναμικό $V(x;u_0)$ του οποίου
η μορφή δεν είναι εν γένει γνωστή. Για να μετατρέψουμε αυτήν την
έκφραση σε μια που να περιέχει το δυναμικό $V(u;u_0)$, πρέπει να λάβουμε υπόψιν μας ότι
μεταβάλλουμε την παράμετρο $u_0$ κρατώντας την $x$ σταθερή και συνεπώς
η $u$ μεταβάλλεται. Με άλλα λόγια, η $\del_{u_0}V(x;u_0)$ δίνεται από
την σχέση:

\be
\label{8-36}
\del_{u_0} V(x;u_0) = \del_{u_0} V(u;u_0) + {\del u\ov \del
{u_0}}\ \del_u V(u;u_0)\ ,
\ee

\no
όπου ο δεύτερος όρος μπορεί να υπολογιστεί με χρήση της έκφρασης
\be
\label{8-37}
{\del u\ov \del u_0}= {f^\prime_{y0}\ov f^\prime_y} +
f^\prime_{y0} \sqrt{f_y-f_{y0}\ov h} \int_{u}^\infty du\
{\del_u\left(h^{1/2} {f^\prime_y}^{-1}\right) \ov
\sqrt{f_y-f_{y0}}}\ ,
\ee
όπως προκύπτει από παραγώγιση της Εξ.\eqn{8-8}, ως προς $u_0$.
Αξίζει να σημειώσουμε, ότι αν $\d\om^2=0$, τότε πρέπει να προχωρήσουμε
σε θεωρία διαταραχών ανώτερης τάξεως (μέχρι να βρούμε την πρώτη μη-μηδενική
διόρθωση) αλλά τέτοια συμπεριφορά δεν συναντάται στα παραδείγματα μας.

\chapter{\en D3-\gr υπόβαθρα: Κλασικές λύσεις}
Σε αυτή την ενότητα, θα συνοψίσουμε τα δυναμικά κουάρκ-αντικουάρκ
 τα οποία εμφανίζονται σε υπολογισμούς βρόχων \en Wilson \gr σε
μελανά και πολυκεντρικά υπόβαθρα $D3$ βρανών. Ο σκοπός αυτής της
σύνοψης είναι να προσδιορίσουμε τα τρία είδη προβληματικής
συμπεριφοράς που αναφέρθηκαν στο τέλος του 6ου κεφαλαίου, και τα
οποία μας οδήγησαν στην μέλετη ευστάθειας των λύσεων.
Πέρα από τις εκφράσεις της μετρικής, το μοναδικό μη τετριμμένο
πεδίο στο υπόβαθρο μας είναι η αυτοδυική μορφή τάξεως πέντε.
Εφόσον θα εργαστούμε με την δράση \en Nambu--Goto, \gr θα
χρειαστούμε μόνο τις εκφράσεις της μετρικής.

\section{Μελανές $D3$ βράνες}

Έστω ένα υπόβαθρο το οποίο περιγράφει μια δέσμη από $N$ μελανές
$D3$ βράνες. Στο όριο της θεωρίας πεδίου, η μετρική γράφεται ως

\be
\label{9-1}
ds^2 = {u^2 \ov R^2} \left[ - \left( 1 - {\m^4 \ov
u^4} \right)
 dt^2 + d \vec{x}_3^2 \right]
+ R^2 \left( {u^2 \ov u^4 - \m^4}\ du^2 + d \Omega_5^2 \right)\ ,
\ee

\no
όπου ο ορίζοντας της μετρικής βρίσκεται στο $u=\mu$ και η θερμοκρασία \en Hawking
\gr είναι $T={\m \ov \pi R^2}$. Η μετρική αυτή είναι ένα ευθύ γινόμενο του χώρου
\en ${\rm AdS}_5$--Schwarzschild \gr με ${\rm S}^5$ και είναι δυική με την
$\cN=4$\ \en SYM \gr σε πεπερασμένη θερμοκρασία. Είναι πιο βολικό για τους υπολογισμούς που θα
ακολουθήσουν να ορίσουμε αδιάστατες παραμέτρους αλλάζοντας την κλίμακα όλων των ποσοτήτων
μέσω της παραμέτρου $\mu$. Θέτοντας $u \to\m u$ και $u_0 \to \m u_0$ και εισάγοντας
αδιάστατες παραμέτρους για το μήκος και την ενέργεια ως εξής,

\be
\label{9-2} L \to {R^2 \ov \m} L\ ,\qq E \to {\m \ov \pi} E\ ,
\ee
βλέπουμε ότι δεν υπάρχει εξάρτηση στις παραμέτρους $\mu$ και $R$ και συνεπώς
μπορούμε να θέσουμε $\mu\to 1$ και $R\to 1$ σε ότι ακολουθεί. Οι συναρτήσεις των
Εξ.\eqn{7-2} και \eqn{7-3} εξαρτώνται μόνο απο το $u$ (αντανακλώντας το γεγονός ότι όλες οι
γωνίες είναι ισοδύναμες) και δίδονται από τις σχέσεις:

\be
\label{9-3}
g(u) = 1\ ,\qq f_y(u) = u^4-1\ ,
\ee
και
\be
\label{9-4}
f_x(u) = u^4-1\ ,\qq f_\th(u) = {u^4-1 \ov u^2}\ ,
\qq h(u) = {u^4 \ov u^4 -1}\ ,
\ee
αντίστοιχα. Οπότε βρίσκουμε ότι οι εξισώσεις κίνησης \eqn{7-11} για την γωνία $\th$
ικανοποιούνται εκ ταυτότητας για όλες τις τιμές της παραμέτρου $\th_0$.

\no
Στην συνέχεια, θα υπολογίσουμε το δυναμικό του κουάρκ-αντικουάρκ σύμφωνα με τον
φορμαλισμό του 7ου κεφαλαίου. Τα ολοκληρώματα του μήκους και της ενέργειας δίνονται από
\cite{wilsonloopTemp,bs}
\begin{figure}[!t]
\begin{center}
\begin{tabular}{cc}
\includegraphics[height=5.2cm]{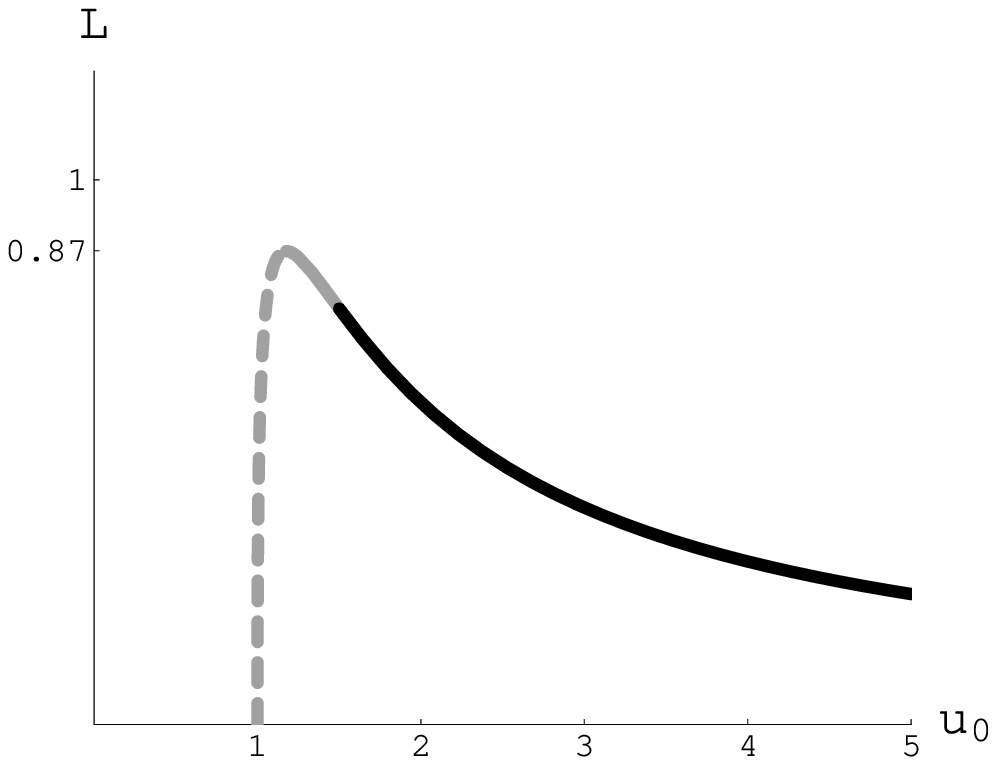}
&\includegraphics[height=5.2cm]{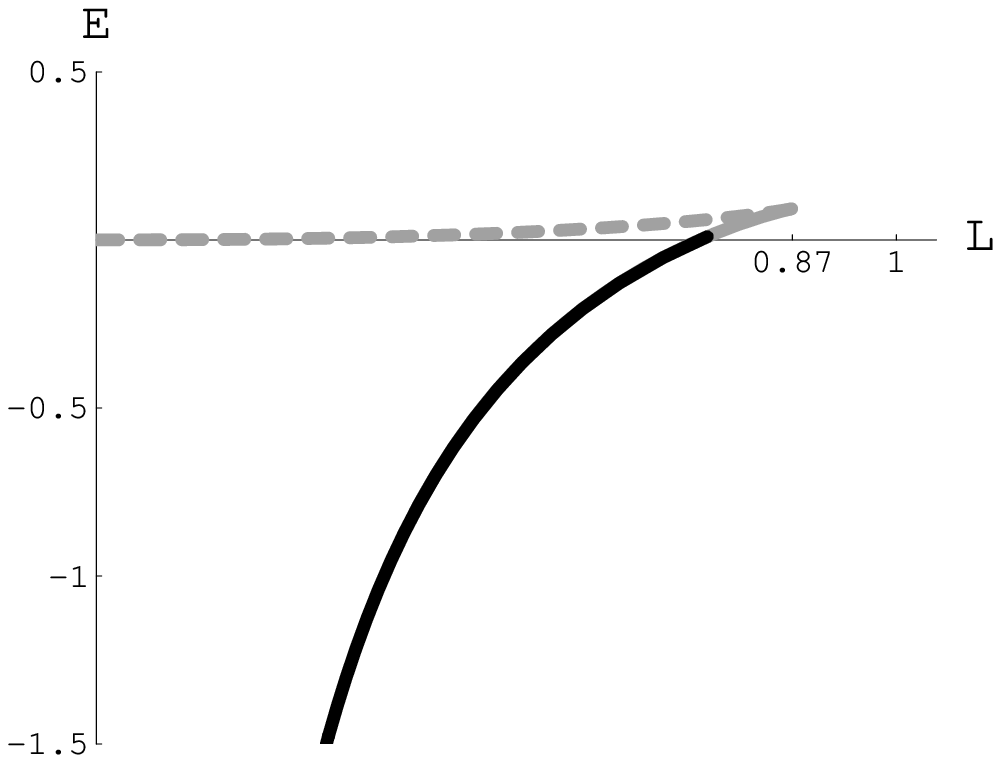}\\
(a) & (b)
\end{tabular}
\end{center}
\vskip -.5 cm \caption{Γραφικές παραστάσεις του $L(u_0)$
 και $E(L)$ των μελανών $D3$-βρανών.
Σε αυτά και στα σχήματα που θα ακολουθήσουν,
τα δίαφορα είδη γραμμών αντιστοιχούν σε ευσταθείς (μαύρες),
μετασταθείς (γκρί) και ασταθείς (διακεκομμένες γκρί) διατάξεις,
όπου η ευστάθεια θα καθοριστεί απο την ανάλυση του 9ου κεφαλαίου.}
\label{fig1.mesons}
\end{figure}
\be
\label{9-5}
L = 2\sqrt{u_0^4-1} \int_{u_0}^\infty {du \ov
 \sqrt{(u^4-1)(u^4-u_0^4)}} = {2 \sqrt{2} \pi^{3/2}
\ov \G(1/4)^2}{\sqrt{u_0^4-1} \ov u_0^3}
 {}_2F_1\left( \half,{3 \ov 4},{5 \ov 4};{1 \ov u_0^4} \right)\ ,
\ee
και
\be
\label{9-6} E = \int_{u_0}^\infty du \left( \sqrt{{u^4-1 \ov
u^4-u_0^4}} -1 \right)  - (u_0-1)= - {\sqrt{2} \pi^{3/2} \ov
\G(1/4)^2} u_0\,  {}_2F_1\left( - \half,- {1 \ov 4},{1 \ov 4};{1 \ov
u_0^4} \right) + 1\ ,
\ee

\no
όπου ${}_2F_1(a,b,c;x)$ είναι η υπεργεωμετρική συνάρτηση και $u_0\geq 1$. Για
$u_0\gg 1$ έχουμε συμπεριφορά τύπου \en Coulomb  \gr ενώ στο αντίθετο όριο,
$u_0\to 1$, έχουμε την ακόλουθη ασυμπτωτική λύση

\be
\label{9-7}
 L \simeq \sqrt{u_0-1} \left( \ln {8 \ov u_0-1} -
 {\pi \ov 2} \right)\ ,\qq E \simeq {u_0-1 \ov 2}
\left( \ln {8 \ov u_0-1} - 1 - {\pi \ov 2} \right)\ .
\ee
Η συνάρτηση $L(u_0)$ έχει ένα ολικό μέγιστο, το οποίο μαζί με τη
αντίστοιχη μέγιστη ενέργεια δίνονται απο
\cite{bs}
\be
\label{9-8}
 u_{\rm c} \simeq 1.177\ , \qq  L_{\rm c} \simeq  0.869\ ,\qq
E_c \simeq 0.093\ .
\ee
Για $L>L_c$, μόνο η αποκομμένη λύση υπάρχει.
Για $L<L_c$, η Εξ.\eqn{9-5} για δεδομένη τιμή του
μήκους έχει δύο λύσεις για την παράμετρο $u_0$,
που αντιστοιχούν στην μικρή και μεγάλου μήκους χορδή αντίστοιχα και
η ενέργεια είναι μια δίτιμη συνάρτηση του μήκους. Επιπλέον, υπάρχει μια
επιπλέον τιμή του μήκους, που στην περίπτωση μας είναι $\tilde{L}_{\rm c} \simeq 0.754$,
πάνω απο την οποία η αποκομμένη διάταξη είναι ενεργειακά
προτιμητέα και στην οποία η μικρού μήκους χορδή γίνεται μετασταθής,
το οποίο θα μπορούσε να είναι ένας εναλλακτικός τρόπος ορισμού του
μήκους θωράκισης. Η συμπεριφορά που περιγράψαμε παραπάνω, εμφανίζεται στα Σχήματα 8.1
και, με δεδομένο ότι ο άνω κλάδος του $E(L)$ απορρίπτεται απο φυσικής απόψεως,
αντιστοιχεί σε ένα θωρακισμένο δυναμικό \en Coulomb. \gr

\section{Πολυκεντρικές $D3$ βράνες}

Στην συνέχεια θα μελετήσουμε την περίπτωση των πολυκεντρικών κατανομών $D3$ βρανών.
Αυτές ανακαλύφτηκαν ως οριακές περιπτώσεις λύσεων που περιγράφουν περιστρεφόμενες $D3$ βρανές
\cite{cy,rs} στο \cite{trivedi,sfet1} και ανήκουν σε μία μεγάλη κατατηγορία απο
συνεχείς κατανομές βρανών στην θεωρία $M$ και θεωρία χορδών πάνω σε ελλειψοειδή
ανώτερης διάστασης \cite{Basfe2}. Οι κατανομές αυτές έχουν χρησιμοποιηθεί στο πλαίσιο
της αντιστοιχίας \en AdS/CFT \gr, ξεκινώντας απο τις εργασίες \cite{warn,bs}.
Σε αυτήν την ενότητα θα μελετήσουμε τις περιπτώσεις ομοιόμορφων κατανομών
$D3$ βρανών σε δίσκο και τριδιάστατη σφαίρα.

\subsection{Ο δίσκος}

Στο όριο της θεωρίας πεδίου η μετρική των $N\ D3$ βρανών που είναι
ομοιόμορφα κατανεμημένες σε έναν δίσκο ακτίνας $r_0$ γράφεται ως

\ba
\label{9-9}
ds^2 &=& H^{-1/2} (- dt^2 + d \vec{x}_3^2 ) + H^{1/2}
{u^2+r_0^2 \cos^2\th \ov u^2+r_0^2}\ du^2
\nonumber\\
&+& H^{1/2}\left[(u^2+r_0^2\cos^2\th )d\th^2
+ r^2 \cos^2\th d\Om_3^2 + (u^2+r_0^2)\sin^2\th d \phi_1^2 \right]\ ,
\ea
όπου
\be
\label{9-10}
H = {R^4 \ov u^2 (u^2 + r_0^2 \cos^2 \th )}\ ,
\ee
ενώ το διαφορικό στοιχείο $d\Om_3^2$ είναι η μετρική της τριδιάστατης σφαίρας $S^3$.
Εφόσον η μόνη παράμετρος η οποία υπεισέρχεται στην λύση της υπερβαρύτητας είναι
το $r_0$ είναι βολικό να ορίσουμε αδιάστατες παραμέτρους αλλάζοντας την κλίμακα των ποσοτήτων
μέσω της παραμέτρου $r_0$. Θέτοντας $u \to r_0 u$ και $u_0 \to r_0 u_0$ και ορίζοντας
αδιάστατες παραμέτρους για το μήκος και την ενέργεια με
\be
\label{9-11}
L \to {R^2 \ov r_0} L\ ,\qq E \to {r_0 \ov \pi} E\ .
\ee
βλέπουμε ότι δεν υπάρχει εξάρτηση απο τις παραμέτρους $r_0$ και $R$ se ότι ακολουθεί.
Οι συναρτήσεις Εξ.\eqn{7-4} και \eqn{7-5} εξαρτώνται απο την γωνία $\th$ και δίδονται απο
\be
\label{9-12}
g(u,\th) = {u^2 + \cos^2\th \ov u^2 +1}\ ,\qq f_y(u,\th) = u^2(u^2+\cos^2\th)\ ,
\ee
και
\be
\label{9-13}
f_x(u,\th) = u^2(u^2+\cos^2\th)\ ,\qq f_\th(u,\th)
= u^2+\cos^2\th\ ,\qq h(u,\th) = {u^2 + \cos^2\th \ov u^2 +1}\ ,
\ee
αντίστοιχα. Οι εξισώσεις κίνησης \eqn{7-13} για την γωνία $\th$ ικανοποιούνται μόνο για
$\th_0=0$ και $\th_0=\pi/2$, οι οποίες αντιστοιχούν σε ορθογώνιες και παράλληλες προς το
δίσκο τροχιές. Για να υπολογίσουμε το δυναμικό του κουάρκ και αντικουάρκ,
\gr θα εξετάσουμε τις δύο αυτές τροχιές.

\no $\bullet$ $\th_0=0$. Σε αυτήν την περίπτωση, τα ολοκληρώματα που αντιστοιχούν για τις αδιάστατες
παραμέτρους του μήκους και της ενέργειας είναι\cite{bs}
\ba
\label{9-14}
L &=& 2 u_0 \sqrt{u_0^2 + 1} \int_{u_0}^\infty
 {du \ov u \sqrt{(u^2+1)(u^2-u_0^2)(u^2+u_0^2+1)}} \nonumber\\
&=& {2 u_0 k^{\prime} \ov u_0^2 +1} \left[ \elPi (k^{\prime 2},k) - \elK(k) \right]
\ea
και
\ba
\label{9-15}
E &=& \int_{u_0}^\infty du
\left[ u \sqrt{u^2+1 \ov (u^2-u_0^2)(u^2+u_0^2+1)} - 1 \right] - u_0 \nonumber\\
&=& \sqrt{2u_0^2+1} \left[ k^{\prime 2} \elK (k) - \elE(k) \right] \ ,
\ea
όπου $\elK(k)$, $\elE(k)$ και $\elPi(\a,k)$ είναι τα πλήρη ελλειπτικά ολοκληρώματα
πρώτου, δευτέρου και τρίτου είδους αντίστοιχα και
\be
\label{9-16}
k={u_0 \ov \sqrt{2u_0^2+1}}\ , \qq k^{\prime}=\sqrt{1-k^2}\ ,
\ee
\begin{figure}[!t]
\begin{center}
\begin{tabular}{cc}
\includegraphics[height=5.2cm]{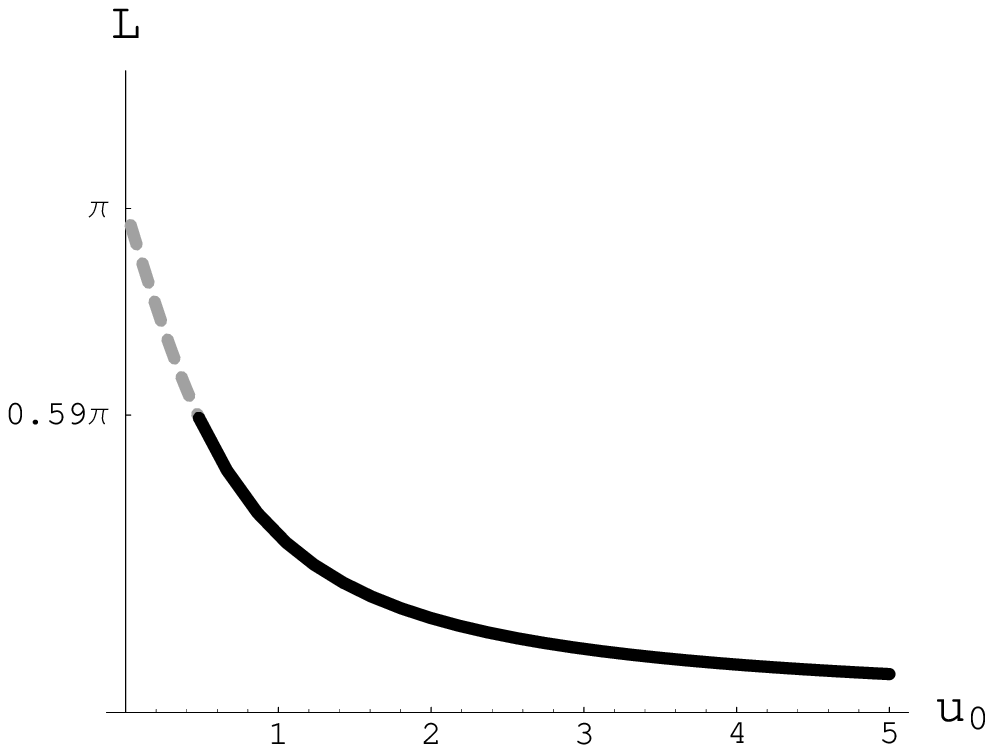}
&\includegraphics[height=5.2cm]{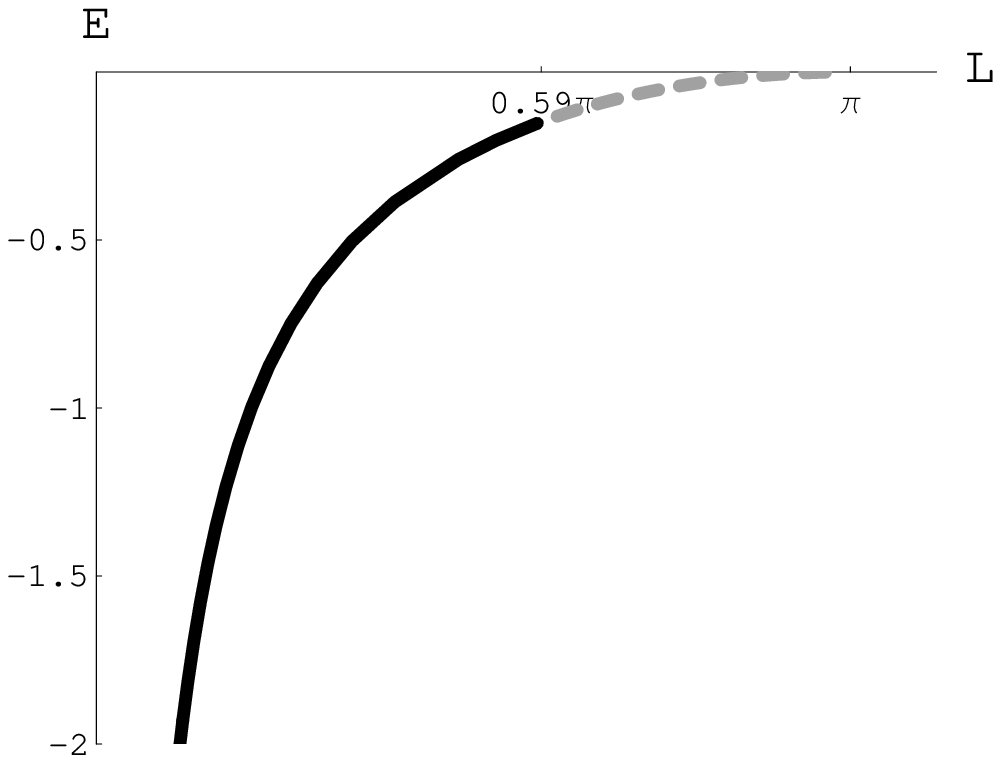}\\
(a) & (b)
\end{tabular}
\end{center}
\vskip -.5 cm \caption{Γραφικές παραστάσεις του $L(u_0)$ και $E(L)$ στον δίσκο
για $\th_0=0$.} \label{fig2.mesons}
\end{figure}
είναι το όρισμα (\en modulus)\gr και το συμπληρωματικό όρισμα των ελλειπτικών ολοκληρωμάτων.
Για $u_0\gg 1$, έχουμε συμπεριφορά τύπου \en Coulomb \gr και στο αντίθετο όριο,
$u_0\to 0$, έχουμε την ασυμπτωτική συμπεριφορά \cite{bs}
\be
\label{9-17} L \simeq \pi(1-u_0)\ ,\qq E \simeq - {\pi \ov 4}
u_0^2\ ,
\ee
Σε αυτήν την περίπτωση, η συνάρτηση του μήκους $L(u_0)$ είναι γνησίως φθίνουσα
συνάρτηση, της οποίας το μέγιστο είναι στο $u_{0 {\rm c}} = 0$ και είναι ίσο με
$L_{\rm c} =\pi$. Για $L > L_{\rm c}$, υπάρχει μόνο η αποκομμένη λύση, ενώ αν
$L < L_{\rm c}$, η Εξ.\eqn{9-14} έχει μοναδική λύση για δεδομένη τιμή της παραμέτρου
$u_0$ και η ενέργεια είναι μονότιμη συνάρτηση του μήκους. Η συμπεριφορά που περιγράψαμε παραπάνω,
εμφανίζεται στα Σχήματα 8.2, και αντιστοιχεί σε θωρακισμένο δυναμικό
\en Coulomb.\gr

\no $\bullet$ $\th_0=\pi/2$. Σε αυτή την περίπτωση, τα ολοκληρώματα
για τις αδιάστατες ποσότητες που αντιστοιχούν στο μήκος και την ενέργεια είναι
\ba
\label{9-18}
L &=& 2 u_0^2 \int_{u_0}^\infty {du \ov u \sqrt{(u^2+1)(u^4-u_0^4)}} \nonumber\\
&=& {2 u_<^2 \ov \sqrt{u_0^2+u_>^2}}\left[ {\bf \Pi}
\left({u_>^2\ov u_0^2+u_>^2},k\right)-{\bf K}(k)\right ]
\ea
και
\ba
\label{9-19}
E &=& \int_{u_0}^\infty du u
\left[ {u^2 \ov \sqrt{(u^2+1)(u^4-u_0^4)}}
- {1 \ov \sqrt{u^2+1}} \right] - \int_0^{u_0} {du u \ov \sqrt{u^2+1}} \nonumber\\
&=&{u_0^2\ov \sqrt{u_0^2+u_>^2}} \ {\bf K}(k) - \sqrt{u_0^2+u_>^2} \ {\bf E}(k) + 1\ ,
\ea
όπου τώρα
\be
\label{9-20}
k^2 = {u_>^2 -u_<^2 \ov u_0^2+u_>^2}\ , \qq k^{\prime}=\sqrt{1-k^2}\
\ee
\begin{figure}[!t]
\begin{center}
\begin{tabular}{cc}
\includegraphics[height=5.2cm]{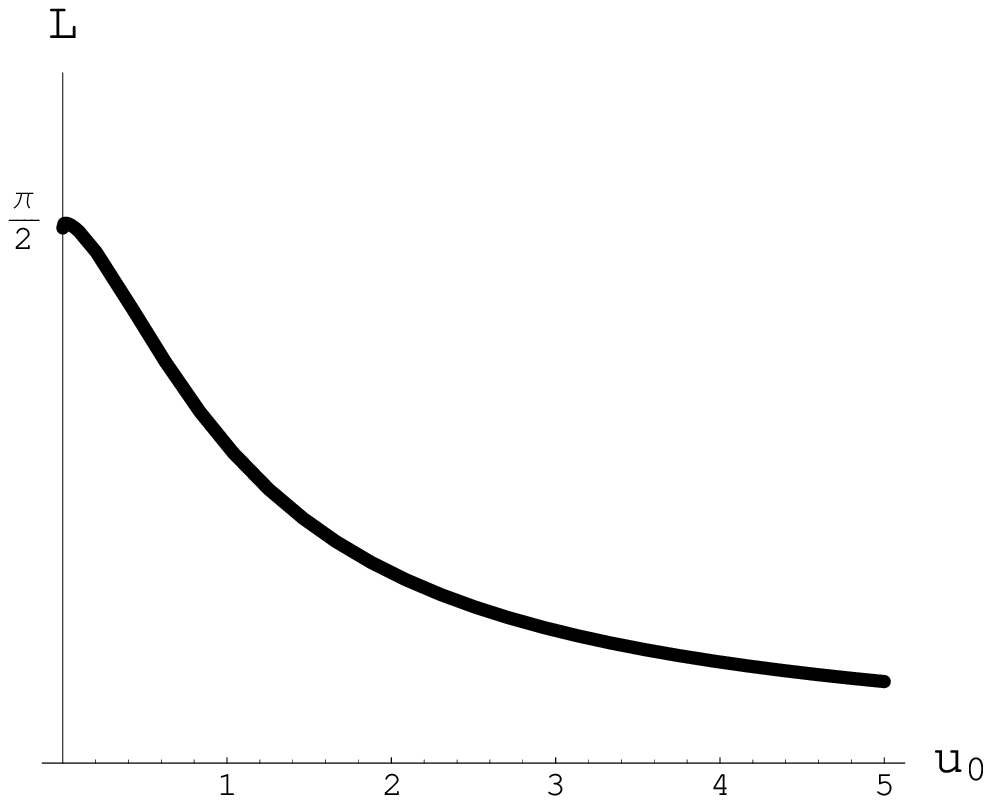}
&\includegraphics[height=5.2cm]{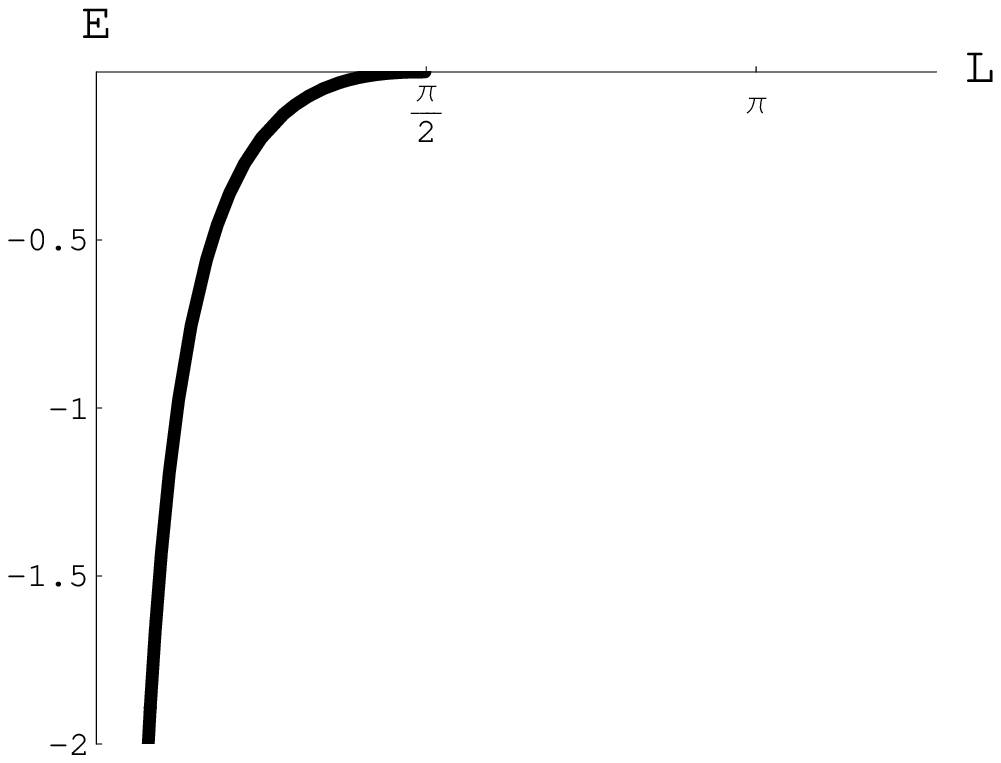}\\
(a) & (b)
\end{tabular}
\end{center}
\vskip -.5 cm \caption{Γραφικές παραστάσεις του $L(u_0)$ και $E(L)$ στον δίσκο
για $\th_0=\pi/2$. Αναφέρουμε ότι το μήκος θωράκισης είναι μικρότερο απο το αντίστοιχο
στην περίπτωση $\th_0=0$ κατά ένα παράγοντα δύο.} \label{fig3.mesons}
\end{figure}
και $u_>$ ($u_<$) συμβολίζουν το μεγαλύτερο (μικρότερο) μεταξύ του $u_0$ και
$1$. Στο όριο $u_0 \to 0$, έχουμε την ασυμπτωτική συμπεριφορά \cite{bs}
\be
\label{9-21}
L \simeq {\pi\ov 2}\left[1- u_0^2 \left(\ln{8 \ov u_0^2}
-1\right)\right]\ , \qq E \simeq - {1\ov 8} u_0^4 \left(\ln {8 \ov
u_0^2}-{3\ov 2}\right)\ .
\ee
Αυτή η συμπεριφορά είναι ποιοτικά η ίδια με την προηγούμενη περίπτωση.
\be
\label{9-22}
L_{\rm c} = {\pi \ov 2}\ .
\ee
Συγκρίνοντας τις περιπτώσεις $\th_0=0$ και $\th_0=\pi/2$, μολονότι
οι συμπεριφορές είναι  ανάλογες, οι εκφράσεις για το μήκος θωράκισης
διαφέρουν κατά έναν παράγοντα δύο. Ο παράγοντας αυτός, είναι αρκετά μεγάλος
μιάς και δεν αναμένεται ο προσανατολισμός της χορδής να επηρεάζει
σε τέτοιο βαθμό τα φυσικά μεγέθη της θεωρίας βαθμίδας. Αυτή η ασυμφωνία θα επιλυθεί
από την ανάλυση της ευστάθειας της κλασικής λύσης.

\subsection{Η τριδιάστατη σφαίρα}

Το όριο της θεωρία πεδίου για την μετρική $N\ D3$ βρανών ομοιόμορφα κατανεμημένων
πάνω σε μια τριδιάστατη σφαίρα ακτίνας $r_0$ ισούται με
\ba
\label{9-23}
ds^2 &=& H^{-1/2} (- dt^2 + d \vec{x}_3^2 ) + H^{1/2}
{u^2-r_0^2 \cos^2\th \ov u^2-r_0^2}\ du^2
\nonumber\\
&+& H^{1/2}\left[(u^2-r_0^2\cos^2\th )d\th^2 + u^2 \cos^2\th d\Om_3^2
+ (u^2-r_0^2)\sin^2\th d \phi_1^2 \right]\ ,
\ea
όπου
\be
\label{9-24}
H = {R^4 \ov u^2 (u^2 - r_0^2 \cos^2 \th )}\ .
\ee
Παρατηρούμε ότι μπορούμε να πάρουμε αυτήν την λύση από την μετρική του δίσκου
Εξ.\eqn{9-9} για $r_0^2\to-r_0^2$. Κάνοντας τις ίδιες κανονικοποιήσεις όπως πριν,
βρίσκουμε ότι οι συναρτήσεις Εξ.\eqn{7-4} και \eqn{7-5} εξαρτώνται από την γωνία
$\th$ και δίδονται απο
\be
\label{9-25}
g(u,\th) = {u^2 - \cos^2\th \ov u^2 - 1}\ ,\qq f_y(u,\th) = u^2(u^2-\cos^2\th)\
\ee
και
\be
\label{9-26}
f_x(u,\th) = u^2(u^2-\cos^2\th)\ ,\qq f_\th(u,\th)
= u^2-\cos^2\th\ ,\qq h(u,\th) = {u^2 - \cos^2\th \ov u^2 - 1}\ ,
\ee
αντίστοιχα, και οι εξισώσεις κίνησης \eqn{7-13} για την γωνία $\th$,
ικανοποιούνται για $\th=0$ και $\th=\pi/2$. Στην συνέχεια θα
εξετάσουμε με λεπτομέρεια αυτές τις δύο περιπτώσεις.

\no $\bullet$ $\th_0=0$. Για αυτή την περίπτωση, οι αδιάστατες
ποσότητες που αντιστοιχούν στο μήκος και την ενέργεια δίδονται από \cite{bs}

\ba
\label{9-27}
L &=& 2 u_0 \sqrt{u_0^2 - 1} \int_{u_0}^\infty {du \ov u \sqrt{(u^2-1)(u^2-u_0^2)(u^2+u_0^2-1)}} \nonumber\\
&=& {2 u_0 k^{\prime} \ov u_0^2-1} \left[ \elPi (k^{\prime 2},k) - \elK(k) \right]
\ea
και
\ba
\label{9-28}
E &=& \int_{u_0}^\infty du
\left[ u \sqrt{u^2-1 \ov (u^2-u_0^2)(u^2+u_0^2-1)} - 1 \right] - (u_0-1) \nonumber\\
&=& \sqrt{2u_0^2-1} \left[ k^{\prime 2} \elK (k) - \elE(k) \right] + 1\ ,
\ea
όπου
\be
\label{9-29}
k={u_0 \ov \sqrt{2u_0^2-1}}\ ,\qq k^{\prime}=\sqrt{1-k^2}\ .
\ee
\begin{figure}[!t]
\begin{center}
\begin{tabular}{cc}
\includegraphics[height=5.2cm]{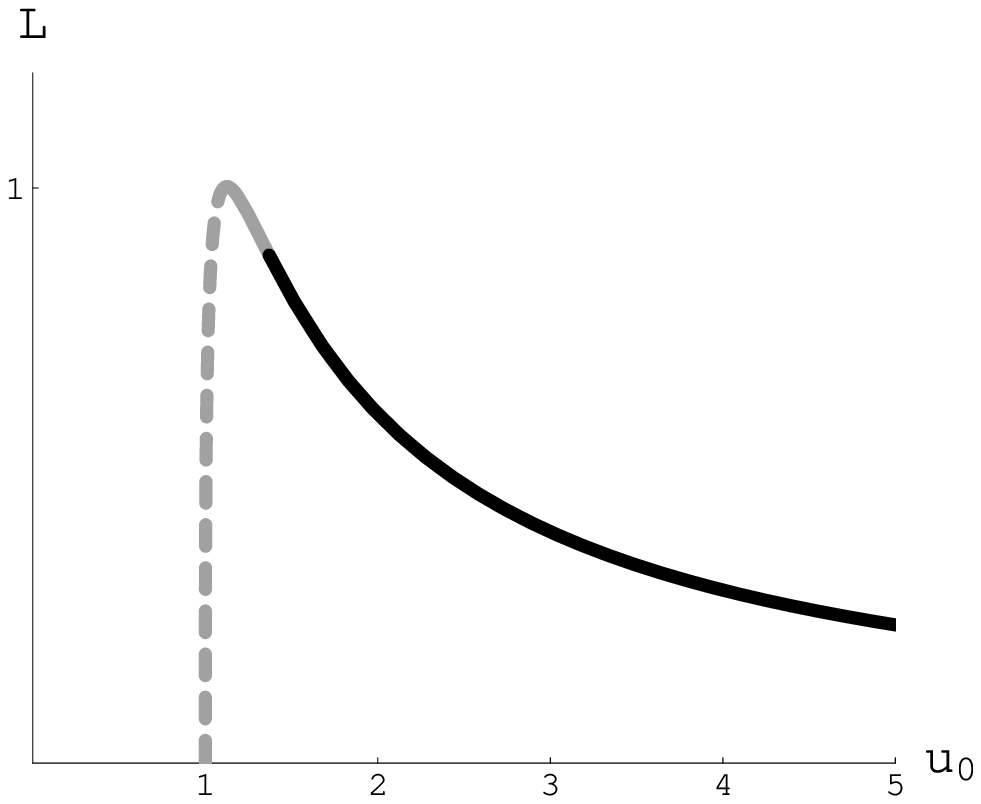}
&\includegraphics[height=5.2cm]{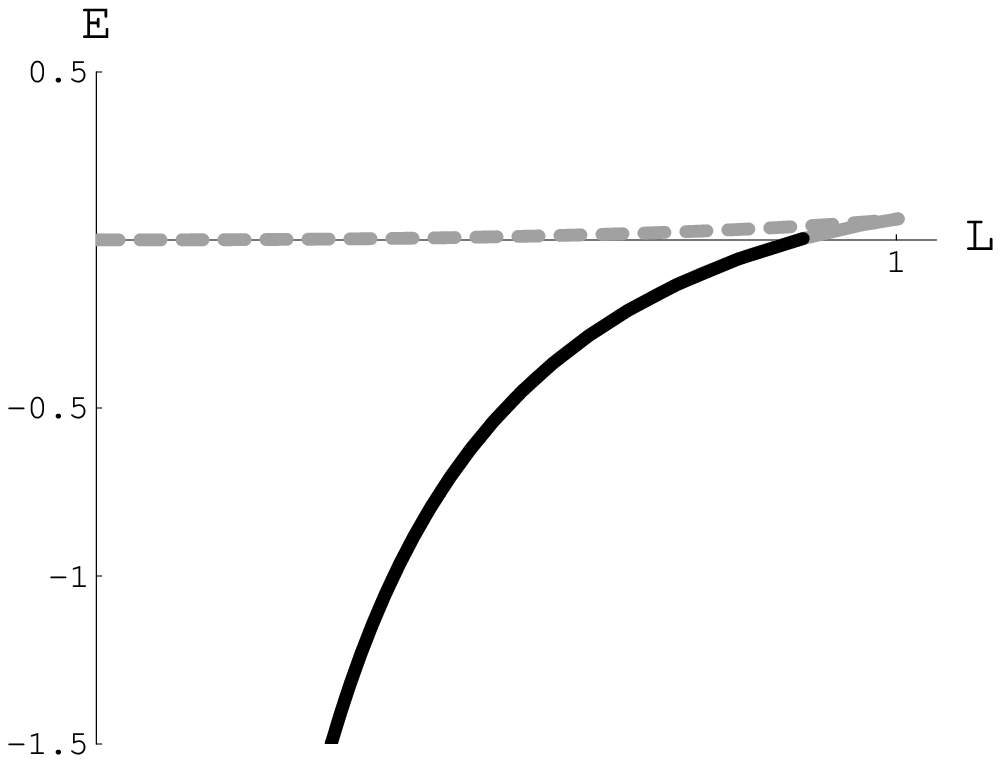}\\
(a) & (b)
\end{tabular}
\end{center}
\vskip -.5 cm \caption{Γραφικές παραστάσεις του $L(u_0)$ και $E(L)$ στην σφαίρα για
$\th_0=0$. Σημειώνουμε την εμφάνιση δύο κλαδων στην παράσταση $E(L)$.} \label{fig4.mesons}
\end{figure}
Για $u_0 \gg 1$, η συμπεριφορά είναι τύπου \en Coulomb, \gr ενώ στο αντίθετο όριο, δηλαδή $u_0\to 1$,
έχουμε την ασυμπτωτική συμπεριφορά \cite{bs}
\be
\label{9-30}
L \simeq \sqrt{2(u_0-1)}
 \left[ \ln \left({8 \ov u_0-1}\right) - 2 \right]\ ,
\qq E \simeq {u_0-1 \ov 2} \left[ \ln \left({8 \ov u_0-1}\right) - 3 \right]\ .
\ee
Η συνάρτηση του μήκους $L(u_0)$ έχει ένα ολικό μέγιστο. Η θέση, η τιμή του μήκους
και της αντίστοιχης ενέργειας είναι \cite{bs}
\be
\label{9-31}
u_{0 \rm c} \simeq 1.125\ , \qq  L_{\rm c} \simeq 1.002\ ,\qq E_c
\simeq 0.063\ .
\ee
Για $L > L_{\rm c}$, υπάρχει μόνο η αποκομμένη λύση. Για $L < L_{\rm c}$,
η Εξ.\eqn{9-27} για δεδομένη τιμή του μήκους έχει δύο λύσεις για την παράμετρο $u_0$
και η ενέργεια είναι δίτιμη συνάρτηση του μήκους. Η συμπεριφορά που περιγράψαμε παραπάνω,
εμφανίζεται στα Σχήματα 4 και με δεδομένο ότι ο άνω κλάδος απορρίπτεται απο φυσικής απόψεως,
συνεπάγεται ότι αντιστοιχεί σε θωρακισμένο δυναμικό \en Coulomb.\gr

\no $\bullet$ $\th_0=\pi/2$. Σε αυτή την περίπτωση, τα ολοκληρώματα που αντιστοιχούν στις αδιάστατες
παραμέτρους του μήκους και της ενέργειας είναι\cite{bs}

\be
\label{9-32}
L = 2 u_0^2 \int_{u_0}^\infty {du \ov u \sqrt{(u^2-1)
(u^4-u_0^4)}} = {\sqrt{2} \ov u_0} \left[ \elPi \left( \half,k \right) -\elK(k) \right]\ ,
\ee
και
\be
\label{9-33}
E = \int_{u_0}^\infty {du u \ov \sqrt{u^2-1}}
\left( {u^2 \ov \sqrt{u^4-u_0^4}} - 1 \right) - \int_1^{u_0}
{du u \ov \sqrt{u^2-1}} = {u_0 \ov \sqrt{2}} \left[ \elK(k)- 2 \elE(k) \right]\ .
\ee
όπου
\be
\label{9-34}
k=\sqrt{{u_0^2+1 \ov 2 u_0^2}}\ ,\qq k^{\prime}=\sqrt{1-k^2}\ .
\ee
\begin{figure}[!t]
\begin{center}
\begin{tabular}{cc}
\includegraphics[height=5.2cm]{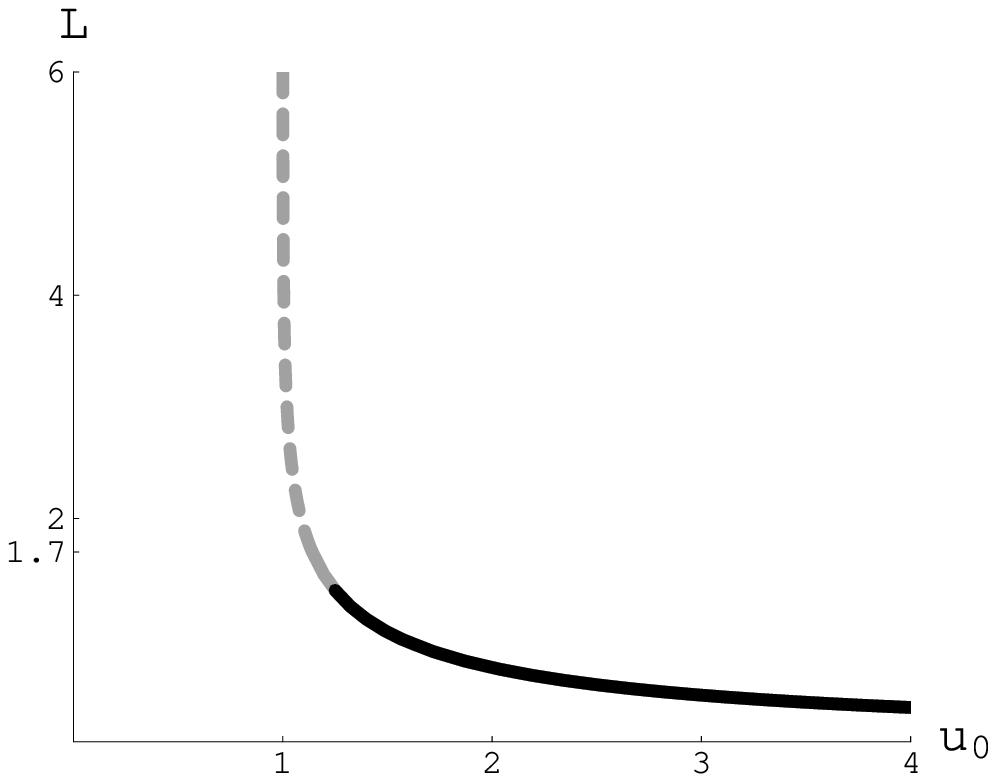}
&\includegraphics[height=5.2cm]{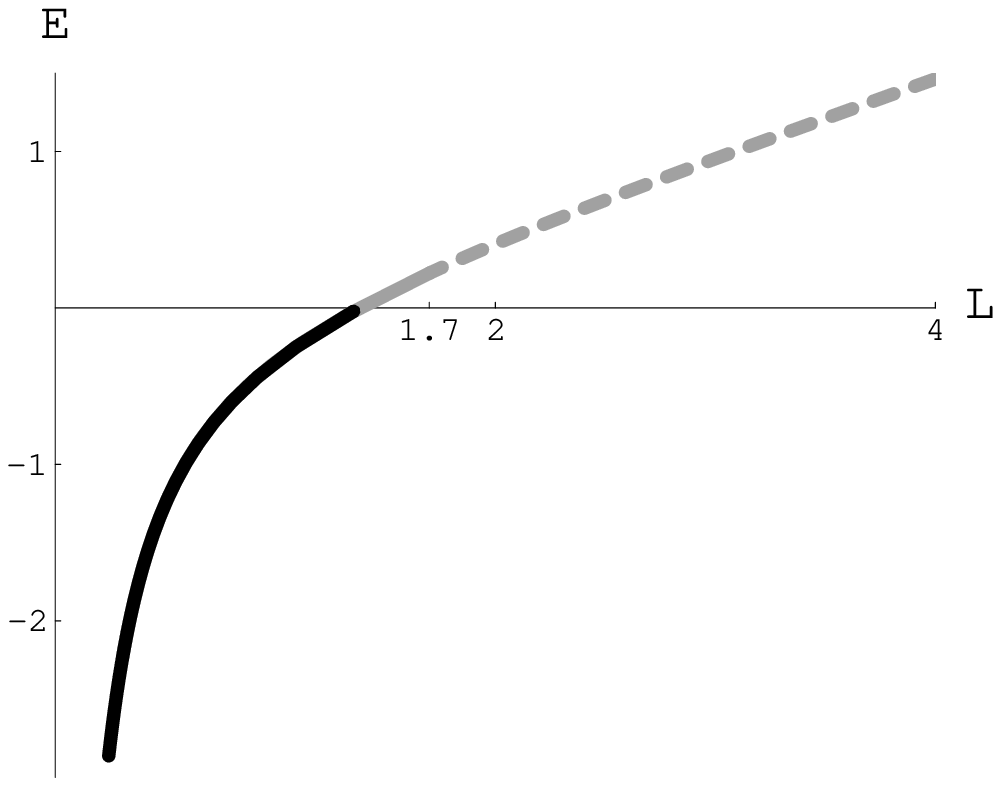}\\
(a) & (b)
\end{tabular}
\end{center}
\vskip -.5 cm \caption{Γραφικές παραστάσεις του $L(u_0)$ και $E(L)$ στην σφαίρα
για $\th_0=\pi/2$.
Σημειώνουμε την ύπαρξη ενός δυναμικού εγκλωβισμού για $L\gg 1$.}
\label{fig5.mesons}
\end{figure}
Για $u_0 \gg 1$, έχουμε συμπεριφορά τύπου \en Coulomb, \gr
ενώ στο αντίθετο όριο, $u_0\to 1$, έχουμε την ασυμπτωτική
συμπεριφορά \cite{bs}
\be
\label{9-35}
L \simeq {1 \ov \sqrt{2}} \left[ \ln \left({16 \ov
u_0-1}\right) - 2\sqrt{2} \ln(1+\sqrt{2}) \right]\ , \qq E \simeq
{1 \ov 2 \sqrt{2}} \left[ \ln \left({16 \ov u_0-1}\right) - 4
\right]\ .
\ee
Η συνάρτηση του μήκους $L(u_0)$ είναι μία γνησίως φθίνουσα
συνάρτηση που απειρίζεται όταν η παράμετρος $u_0\to 1$ και μηδενίζεται όταν
 $u_0\to\infty$ και συνεπώς δεν υπάρχει μέγιστο μήκος, ενώ η εξίσωση
\eqn{9-32} έχει μοναδική λύση για δεδομένη τιμή του $u_0$, και η $E$
είναι μια μονότιμη συνάρτηση του $L$. Οπότε, φαίνεται ότι δεν
υφίσταται φαινόμενο θωράκισης αλλά, σύμφωνα με την ακόλουθη εξίσωση
ένα δυναμικό εγκλωβισμού
\be
\label{9-36} E \simeq  {L \ov 2}\ ,\qq {\textrm{για}} \quad L \gg 1 \ ,
\ee
το οποίο δεν αναμενόταν, δεδομένου ότι η δυική θεωρία βαθμίδας είναι
η ${\cal N}=4$ \en SYM \cite{Seiberg}. \gr\\
\no
Συγκρίνοντας τις περιπτώσεις για $\th_0=0$ και $\th_0=\pi/2$, σημειώνουμε
ότι η ποιοτική συμπεριφορά του δυναμικού φαίνεται να είναι τελείως διαφορετική:
στη πρώτη περίπτωση το δυναμικό είναι μια δίτιμη συνάρτηση του μήκους $L$ και
εκτείνεται μέχρι ένα μέγιστο μήκος, ενώ στην δεύτερη περίπτωση το δυναμικό
υπάρχει για κάθε τιμή του μήκους και "ταλαντώνεται" μεταξύ μιας συμπεριφοράς
\en Coulomb \gr και ενός δυναμικού εγκλωβισμού.
Ωστόσο, τέτοιες διαφορετικές ποιοτικές συμπεριφορές δεν αναμένονται
από μια απλή αλλαγή προσανατολισμού της χορδής και, επιπρόσθετα,
η ύπαρξη ενός δυναμικού εγκλωβισμού είναι απροσδόκητη. Και τα δύο
αυτά ζητήματα θα επιλυθούν από την ανάλυση ευστάθειας.

\chapter{\en D3-\gr υπόβαθρα: Ανάλυση ευστάθειας}
Σε αυτό το κεφάλαιο θα εφαρμόσουμε την ανάλυση της ευστάθειας που
αναπτύξαμε στο 7ο κεφάλαιο για τις κλασικές λύσεις που συνοψίσαμε
στο 6ο κεφάλαιο. Θα βρούμε ότι οι μελανές $D3$ βράνες και η
σφαίρα για $\th_0=0$ θα έχουν μια αστάθεια που οφείλεται στις
διαμήκεις διακυμάνσεις, και η αστάθεια αυτή βρίσκεται στον άνω κλάδο
της καμπύλης της ενέργειας ως συνάρτηση του μήκους, ενώ ο δίσκος
και η σφαίρα για $\th_0=\pi/2$ έχουν αστάθεια στο υπέρυθρο που οφείλεται στις γωνιακές
διακυμάνσεις, παρότι το δυναμικό εκεί είναι μια μονότιμη συνάρτηση.

\section{Σύμμορφη περίπτωση}

Ως ένα απλό παράδειγμα, και σαν έλεγχος συνέπειας,
θα μελετήσουμε πρώτα την σύμμορφη περίπτωση, η
οποία αντιστοιχεί στο όριο $\mu\to 0$ ή $r_0\to 0$ των παραπάνω λύσεων.
Σε αυτήν την περίπτωση, τα δυναμικά \en Schr\"odinger \gr είναι
\ba
\label{10-1}
V_x(u;u_0) &=& 2 u^2 \ ,
\nonumber\\
V_\th(u;u_0) &=& 0\ ,\\
V_y(u;u_0) &=& 2 {u^{4} - u_0^4 \ov u^2}\ ,
\nonumber
\ea
όπου ο δείκτης αντιστοιχεί στην αντίστοιχη διακύμανση. Η αλλαγή μεταβλητών
της Εξ.\eqn{8-8} δίνει

\be
\label{10-2}
x={1\ov u}\ {}_2F_1\left(\half,{1\ov 4},{5\ov 4};{u_0^4\ov
u^4}\right)\ ,
\ee
όπου, $x\in [0,x_0]$ με
\be
\label{10-3}
x_0 = {\G(1/4)^2\ov 4 \sqrt{2 \pi}}\ {1\ov u_0}\ .
\ee
Εφόσον όλες οι εξισώσεις \en Schr\"odinger \gr ορίζονται
σε ένα πεπερασμένο διάστημα με θετικά δυναμικά,
οι αντίστοιχες ενεργειακές ιδιοτιμές είναι θετικές
οπότε δεν υπάρχει αστάθεια. Οι εξίσωσεις που αντιστοιχούν
στις σχέσεις \eqn{8-6} για τις εγκάρσιες ($\d x$) και τις διαμήκεις
($\d y$) διακυμάνσεις δίδονται από τις Εξ.(7) στο \cite{cg} και
της Εξ.(16) στο \cite{kmt} αντίστοιχα.
Η σύγκριση για τις $\d x$ είναι
εμφανής αν κάνουμε αλλαγή συντεταγμένων
$u=1/z$ και επανανομάσουμε $z_m=1/u_0$, όπου η μεταβλητή $z$
και η παράμετρος $z_m$ είναι αυτές που χρησιμοποιούνται στο \cite{cg}.
Για τις $\d y$, η σύγκριση γίνεται με την ίδια αλλαγή μεταβλητής και
$\d y = u_0^2 u^2 (u^4-u_0^4)^{-1/2} \d\bar y$, όπου $\d \bar{y}$ είναι
η μεταβλητή $u$ στο \cite{kmt}. Επιπρόσθετα, εφόσον η ανεξάρτητη μεταβλητή
στην Εξ. (16) του \cite{kmt} είναι η $y$($x$ στον συμβολισμό τους) πρέπει
επίσης να χρησιμοποιήσουμε την σχέση των διαφορικών $dy,du$
($dy = u_0^2/u^2 (u^4-u_0^4)^{-1/2}du$) όπως αυτή προκύπτει
απο τις κλασικές εξισώσεις κίνησης \eqn{7-15}.

\section{Μελανές \en D3 \gr βράνες}

Στην συνέχεια εξετάζουμε την ευστάθεια των μελανών $D3$ βρανών,
όπου η ενέργεια είναι μια δίτιμη συνάρτηση του μήκους του ζεύγους.
Τα δυναμικά \en Schr\"odinger \gr για τα τρία είδη των διακυμάνσεων
δίνονται απο
\ba
\label{10-4}
V_x(u;u_0) &=& 2 {u^8 - u_0^4 \ov u^6}\ ,\nonumber\\
V_\th(u;u_0) &=& 0\ ,\\
V_y(u;u_0) &=& 2 {u^{12} - u_0^4 u^8 - (4u_0^4 -3 ) u^4 + u_0^4
\ov u^6 (u^4-1)}\ .
\nonumber
\ea
Από την Εξ.\eqn{8-11} βρίσκουμε την τιμή του άκρου $x_0$
\be
\label{10-5}
x_0 = {\G(1/4)^2\ov 4 \sqrt{2 \pi}}\ {1\ov u_0}\
{}_2F_1\left(\half,{1\ov 4},{3\ov 4};{1\ov u_0^4}\right)\ ,
\ee
με τις ασυμπτωτικές συμπεριφoρές να δίνονται απο την Εξ.\eqn{8-12}
και
\ba
\label{10-6}
x_0(u_0)\simeq -{1\ov 4} \ln(u_0-1) +{\pi +\ln 64\ov 8} + {\cal
O}(u_0-1)\ .
\ea
Το δυναμικό $V_x$ είναι θετικό για όλες τις τιμές του $u_0$, και συνεπώς η λύση
είναι ευσταθής κάτω απο εγκάρσιες διαταραχές, σε συμφωνία με τα
γενικά αποτελέσματα της ενότητας (7.3.1). Επίσης, το δυναμικό
$V_\th$ είναι εκ ταυτότητας μηδέν, το οποίο σημαίνει, ότι για πεπερασμένο $x_0$,
το φάσμα είναι θετικό και η λύση είναι επίσης ευσταθής κάτω απο γωνιακές διακυμάνσεις.
Αντίθετα, το δυναμικό $V_y$ ξεκινάει από μια αρνητική τιμή στο σημείο $u=u_0$
η οποία δίνεται από την σχέση
\ba
\label{10-7}
V_{y 0} = -{8\ov u_0^2} \ ,\qq -8 \leqslant V_{y 0} < 0\
\ea
και συμπεριφέρεται όπως στην Εξ.\eqn{8-27} για $u\to\infty$. Για να εξετάσουμε
τις αστάθειες κάτω από διαμήκεις διακυμάνσεις, θα εφαρμόσουμε τα αποτελέσματα
της παραγράφου (7.3.2) και τις εξισώσεις της θεωρίας διαταραχών της
παραγράφου (7.4) στο πρόβλημα μας. Βρίσκουμε ότι υπάρχει μηδενικός τρόπος για
τις διαμήκεις διακυμάνσεις στο σημείο $u_0=u_{0 \rm c}$ όπου το $u_{0 \rm c}$
δίδεται λύνοντας αριθμητικά την Εξ.\eqn{8-23}, η οποία στην περίπτωση μας,
είναι
\ba
\label{10-8}
{9 \ov 5 u_0^4} \ {}_2F_1\left(\half ,{7\ov 4},{9\ov 4};{1\ov
u_0^4}\right) = {}_2F_1\left(\half ,{3\ov 4},{5\ov 4};{1\ov
u_0^4}\right) \ .
\ea
Η εξίσωση αυτή έχει μια λύση και δίνεται στην Εξ.\eqn{9-8}, $u_{0\rm c}\simeq 1.177$.
Προκειμένου να βρούμε την αλλαγή στην ιδιοτιμή της ενέργειας $\om^2$ για μικρές
διακυμάνσεις της παραμέτρου $u_0$ γύρω από το $u_{0\rm c}$, αρχικά
χρησιμοποιούμε την πρώτη γραμμή της Εξ.\eqn{8-22} ώστε να βρούμε την ακριβή
έκφραση για τον μηδενικό τρόπο που αντιστοιχεί στις διαμήκεις διακυμάνσεις
\ba
\label{10-9}
\d y = {1.463\ov u^3}\ F{}_1\left({3\ov 4}; {3\ov 2} , -{1\ov 2} ,
{7\ov 4} ;{u_{0\rm c}^4\ov u^4}, {1\ov u_{0\rm c}^4}\right)\ ,
\ea
όπου $F{}_1(a\ ;b_1,b_2,c\ ;x,y)$ είναι η υπεργεωμετρική συνάρτηση \en Appell
\gr της οποίας η κανονικοποίηση έγινε μέσω της αντίστοιχης κυματοσυνάρτησης
\en Schr\"odinger \gr
\be
\label{10-10}
\Psi = {1.463 \ov u^2} \sqrt{u^4-u_{0\rm c}^4\ov u^4-1} \
F{}_1\left({3\ov 4}; {3\ov 2} , -{1\ov 2} , {7\ov 4} ;{u_{0\rm
c}^4\ov u^4}, {1\ov u_{0\rm c}^4}\right)\ .
\ee
Οπότε, χρησιμοποιώντας την Εξ.\eqn{8-34}, βρίσκουμε την εξής μετατόπιση στην ενέργεια
\be
\label{10-11}
\d \omega^2 \simeq ( 51.9 + 29.6 - 20.1 )\ \d u_0 = 61.4\ \d u_0 \
,
\ee
όπου ο κάθε όρος μέσα στην παρένθεση αντιπροσωπεύει την συνεισφορά του αντίστοιχου
όρου  της Εξ.\eqn{8-34}. Οπότε, η χορδή μεγάλου μήκους (άνω κλάδος) ($u_0 < u_{0\rm c}$)
είναι ασταθής ($\d \omega^2 < 0$) και η χορδή μικρού μήκους (κάτω κλάδος) είναι ευσταθής ($\d \omega^2 > 0$).
Συνεπώς, η ανάλυση της ευστάθειας της κλασική λύσης είναι σε συμφωνία
με τα ενεργειακά επιχειρήματα, όπως αυτό φαίνεται στο Σχήμα 8.1(β).\\
\no
Τέλος, αξίζει να αναφέρουμε ότι η παραπάνω συμπεριφορά της μελανής $D3$ βράνης
εμφανίζεται σε κάθε μελανή οπή. Πιο συγκεκριμένα, κάθε μελανή οπή μπορεί
να προσεγγιστεί ως ένας χώρος \en Rindler \gr γύρω από την περιοχή
του ορίζοντα. Όπως δείξαμε αναλυτικά για τον χώρο αυτό στο παράρτημα $C$ του \cite{ASS2}, η ενέργεια
αλληλεπίδρασης ενός μεσονίου είναι μια δίκλαδη συνάρτηση του μήκους και ο
άνω κλάδος είναι διαταρακτικά ασταθής.

\section{Πολυκεντρικές $D3$ βράνες}

Τέλος, θα μελετήσουμε την περίπτωση των πολυκεντρικών $D3$ βρανών,
των οποίων οι προβληματικές περιοχές έχουν αναφερθεί στο τέλος του 6ου κεφαλαίου.
Συγκεκριμένα, θα παρουσιάσουμε τα αποτελέσματα της ανάλυσης ευστάθειας όλων των
περιπτώσεων μελετήσαμε στην ενότητα 8.2.

\subsection{Ο δίσκος}

Για την κατανομή του δίσκου, οι δύο επιτρεπτοί προσανατολισμοί της χορδής
αντιστοιχούσαν σε θωρακισμένο δυναμικό \en Coulomb, \gr και τα μήκη θωράκισης τους διαφέρανε
κατά έναν παράγοντα 2. Τα αποτελέσματα της ανάλυσης της ευστάθειας της κλασικής λύσης
για τους δύο προσανατολισμούς είναι

\no
$\bullet$ $\th_0=0$. Σε αυτή την περίπτωση, έχουμε τα δυναμικά \en Schr\"odinger \gr
\ba
\label{10-12}
V_x(u;u_0) &=& {8 u^8 + 18 u^6 + 11 u^4 - [2 u_0^2(u_0^2+1)-1]
 u^2 + u_0^2(u_0^2+1) \ov 4 u^2(u^2+1)^2}\ ,\nonumber\\
V_\th(u;u_0) &=& - {2 u^6 + u^4 + [6 u_0^2(u_0^2+1)-1] u^2 + 3
 u_0^2(u_0^2+1) \ov 4 u^2(u^2+1)^2}\ ,\\
V_y(u;u_0) &=& {8 u^8 + 18 u^6 - [ 8 u_0^2 (u_0^2+1) - 11] u^4
- [6 u_0^2(u_0^2+1)-1] u^2 -3 u_0^2(u_0^2+1) \ov 4 u^2(u^2+1)^2}\ ,
\nonumber
\ea
και η αντίστοιχη τιμή του άκρου $x_0$ είναι
\ba
\label{10-13}
x_0 = \sqrt{2 {k'}^2 -1} \ {\bf K}(k')\ ,
\ea
όπου $k'$ είναι το συμπληρωματικό όρισμα που ορίσαμε στην Εξ.\eqn{9-16}.
Οι ασυμπτωτικές συμπεριφορές του $x_0$ δίνονται στην Εξ.\eqn{8-12} και
\ba
\label{10-14}
x_0(u_0)\simeq -\ln {u_0\ov 4} + {\cal O}(u_0^2\ln u_0)\ .
\ea
Εφόσον τα δυναμικά $V_x$ και $V_y$ είναι θετικά για όλες τις τιμές
της παραμέτρου $u_0$, η λύση είναι ευσταθής κάτω από εγκάρσιες
και διαμήκεις διακυμάνσεις, σύμφωνα με τα γενικά αποτελέσματα της ενότητας
7.3 λόγω απουσίας ακρότατου στην συνάρτηση του μήκους. Από την άλλη
πλευρά, το δυναμικό $V_\th$ είναι αρνητικό σε όλο το εύρος της παραμέτρου
$u_0$, με τιμές για $u=u_0$ και $u\to\infty$ να δίνονται από
\ba
\label{10-15}
V_{\th 0} &=& -2 +{3\ov 2(u_0^2+1)} \ ,\qq -2 < V_{\th 0}
\leqslant -\half \ ,
\nonumber\\
V_{\th\infty} &=& -\half\ .
\ea
Για να εξετάσουμε την ευστάθεια της λύσης, θα χρηισμοποιήσουμε την προσέγγιση του
απειρόβαθρου πηγαδιού Εξ.\eqn{8-30} και θα μελετήσουμε την συμπεριφορά της μικρότερης
ιδιοτιμής $\om_0^2$ σαν συνάρτηση της παραμέτρου $u_0$. Βρίσκουμε ότι είναι μια γνησίως αύξουσα
συνάρτηση η οποία ξεκινά από αρνητικές τιμές για $u_0=0$ και αλλάζει πρόσημο σε μια κρίσιμη
τιμή $u_{0\rm c}$. Η τιμή αυτή, και το αντίστοιχο μήκος και ενέργεια, δίνονται
\ba
\label{10-16}
u_{0\rm c}\simeq 0.48\ ,\qq L_{\rm c} \simeq 0.59 \pi\ ,\qq E_{\rm
c} \simeq  -0.15 \ .
\ea
Ως εκ τούτου, όταν η απόσταση μεταξύ κουάρκ και αντικουάρκ είναι
μεγαλύτερη απο $L_c$, οι μικρές διακυμάνσεις της γωνίας $\th$ αποσταθεροποιούν την
κλασική λύση και κατά συνέπεια το δυναμικό του ζεύγους είναι αναξιόπιστο.
Συνεπώς, το πραγματικό μήκος θωράκισης δίνεται από την δεύτερη σχέση της Εξ.\eqn{10-16}
και είναι συγκρίσιμο με το αντίστοιχο μήκος στην $\th_0=\pi/2$ περίπτωση.

\no $\bullet$ $\th_0=\pi/2$. Σε αυτήν την περίπτωση, τα δυναμικά \en Schr\"odinger \gr
δίνονται από τις
\ba
\label{10-17}
V_x(u;u_0) &=& {2 u^6 + u^4 + u_0^4 \ov u^4}\ ,
\nonumber\\
V_\th(u;u_0) &=& 1\ ,\\
V_y(u;u_0) &=& {2 u^6 + u^4 - 2 u_0^4 u^2 - 3u_0^4 \ov u^4}\ ,
\nonumber
\ea
και η τιμή του άκρου $x_0$ ισούται με
\ba
\label{10-18}
x_0 = {1\ov \sqrt{u_0^2 +u_>^2}}\ {\bf K}(k)\ ,
\ea
όπου $k$ είναι το όρισμα που ορίστηκε στην Εξ.\eqn{9-20}.
Οι ασυμπτωτικές τιμές του δίνονται από την Εξ.\eqn{8-12}
και
\be
\label{10-19}
x_0(u_0)\simeq - \ln {u_0 \ov \sqrt{8}} + {\cal O}(u_0^4 \ln u_0)\
.
\ee
Εφόσον, τα δυναμικά $V_x$ και $V_\th$ είναι θετικά, η λύση είναι ευσταθής κάτω από
εγκάρσιες και γωνιακές διαταραχές. Επίσης, μολονότι το $V_y$ έχει ένα αρνητικό μέρος
για κάποιο διάστημα της μεταβλητής $u$, το γεγονός ότι η συνάρτηση του μήκους
δεν έχει ακρότατα σημαίνει ότι η λύση είναι επίσης ευσταθής κάτω από διαμήκεις διακυμάνσεις.\\
\no
Ο στόχος αυτής της ανάλυσης είναι να δείξουμε ότι τα δύο μήκη θωράκισης για τις
δύο διαφορετικές οριακές γωνίες στις οποίες το δυναμικό του ζεύγους μπορεί να
υπολογιστεί είναι στην πράξη το ίδιο. Συνεπώς, αν το σύστημα ξεκινήσει από $\th_0=0$
και $L_c < L < \pi$, οι μικρές διακυμάνσεις θα οδηγήσουν το σύστημα τελικά στο $\th=\pi/2$.

\subsection{Η τριδιάστατη σφαίρα}

Για την σφαιρική κατανομή, οι δύο επιτρεπτοί προσανατολισμοί της χορδής οδήγησαν σε τελείως
διαφορετικές συμπεριφορές και σε ένα δυναμικό εγκλωβισμού για $\th_0=\pi/2$.

\no
$\bullet$ $\th_0=0$. Σε αυτήν την περίπτωση, έχουμε τα δυναμικά \en Schr\"odinger \gr
\ba
\label{10-20}
V_x(u;u_0) &=& {8 u^8 - 18 u^6 + 11 u^4 + [2 u_0^2(u_0^2-1)-1]
 u^2 + u_0^2(u_0^2-1) \ov 4 u^2(u^2-1)^2}\ ,\nonumber\\
V_\th(u;u_0) &=& {2 u^6 - u^4 + [6 u_0^2(u_0^2-1)-1] u^2 - 3
u_0^2(u_0^2-1) \ov 4 u^2(u^2-1)^2}\ ,\\
V_y(u;u_0) &=& {8 u^8 - 18 u^6 - [ 8 u_0^2 (u_0^2-1) - 11] u^4
 + [6 u_0^2(u_0^2-1)-1] u^2 -3 u_0^2(u_0^2-1) \ov 4 u^2(u^2-1)^2}\ ,\nonumber
\ea
και την τιμή του άκρου $x_0$
\ba
\label{10-21}
x_0 = \sqrt{1-2 {k'}^2}\ {\bf K}(k')\ ,
\ea
όπου $k'$ είναι το συμπληρωματικό όρισμα που ορίστηκε
στην Εξ.\eqn{9-29}. Οι ασυμπτωτικές συμπεριφορές του $x_0$
δίνονται απο την Εξ.\eqn{8-12} και
\ba
\label{10-22}
x_0(u_0)\simeq {\pi\ov 2} -{3\pi\ov 4} (u_0-1) + {\cal O}(u_0-1)^2\ .
\ea
Εφόσον τα δυναμικά $V_x$ και $V_\th$ είναι θετικά για όλες τις τιμές της
παραμέτρου $u_0$, η λύση είναι ευσταθής κάτω από εγκάρσιες και γωνιακές διακυμάνσεις.
Από την άλλη πλευρά, το δυναμικό $V_y$ ξεκινά από αρνητικές τιμές και στην συνέχεια
γίνεται θετικό. Επαναλαμβάνοντας την ανάλυση της ενότητας 9.2, βρίσκουμε ότι υπάρχει
αστάθεια στις διαμήκεις διακυμάνσεις για τιμές της παραμέτρου $u_0$ μικρότερες από
την κρίσιμη τιμή $u_{0\rm c}$ που δίνεται στην Εξ.\eqn{9-31}.

\no $\bullet$ $\th_0=\pi/2$. Σε αυτή την περιπτωση, τα δυναμικά \en Schr\"odinger \gr
δίνονται από την
\ba
\label{10-23}
V_x(u;u_0) &=& {2 u^6 - u^4 - u_0^4 \ov u^4}\ ,\nonumber\\
V_\th(u;u_0) &=& -1\ ,\\
V_y(u;u_0) &=& {2 u^6 - u^4 - 2 u_0^4 u^2 + 3u_0^4 \ov u^4}\ .
\nonumber
\ea
και η τιμή του άκρου $x_0$ είναι
\ba
\label{10-24}
x_0 = \sqrt{2 k^2-1\ov 2}\ {\bf K}(k)\ ,
\ea
όπου το όρισμα $k$ ορίστηκε στην Εξ.\eqn{9-34}.  Οι ασυμπτωτικές
τιμές του $x_0$ δίνονται από την Εξ.\eqn{8-12}
\ba
\label{10-25}
x_0(u_0)\simeq -{1\ov 2\sqrt{2}}\ln {u_0-1\ov 16}  + {\cal
O}\left((u_0-1)\ln(u_0-1)\right)\ .
\ea
Εφόσον τα δυναμικά $V_x$ και $V_y$ είναι θετικά για όλες τις τιμές τις παραμέτρου
$u_0$. η λύση είναι ευσταθής κάτω από εγκάρσιες και διαμήκεις διαταραχές,
σύμφωνα με τα γενικά αποτελέσματα της ενότητας (7.3). Από την άλλη πλευρά,
το δυναμικό $V_\th$ έχει μια σταθερή αρνητική τιμή και κατά συνέπεια η προσέγγιση
του απειρόβαθρου πηγαδιού είναι ακριβής. Εξετάζοντας την συμπεριφορά της μικρότερης ιδιοτιμής
$\om_0^2$, βρίσκουμε ότι είναι μια γνησίως αύξουσα συνάρτηση ξεκινώντας από αρνητικές τιμές
για $u_0=0$ και αλλάζοντας πρόσημο για μια κρίσιμη τιμή $u_{0\rm c}$ με

\ba
\label{10-26}
u_{0\rm c}\simeq 1.14\ ,\qq L_{\rm c} \simeq 1.7\ ,\qq E_{\rm c} \simeq
0.22 \ .
\ea
Οπότε, αν το μηκός του ζεύγους είναι μεγαλύτερο από $L_{\rm c}$,
οι μικρές διακυμάνσεις της γωνίας $\th$ θα αποσταθεροποιήσουν την κλασική λύση
και το δυναμικό του ζεύγους θα είναι αναξιόπιστο. Εφόσον το γραμμικό δυναμικό εγκλωβισμού
αντιστοιχεί σε μήκη μεγαλύτερα από το $L_{\rm c}$ της Εξ.\eqn{10-26} για το οποίο
έχουμε αστάθεια, καταλήγουμε σε ένα θωρακισμένο δυναμικό τύπου \en Coulomb \gr με μέγιστο μήκος
το $L_{\rm c}$. Ο στόχος αυτής της ανάλυσης είναι να δείξουμε ότι η γραμμική συμπεριφορά εγκλωβισμού
είναι ασταθής και ότι και τα δύο είναι θωρακισμένα δυναμικά τύπου \en Coulomb,\ \gr
με συγκρίσιμα μήκη θωράκισης.

\chapter{Οριακές παραμορφώσεις \en D3 \gr βρανών}
Σε αυτό το κεφάλαιο, θα θεωρήσουμε τις οριακές διαμορφώσεις \en Lunin--Maldacena \cite{LM}
\gr των λύσεων $D3$ βρανών, οι οποίες είναι δυικές στις \gr οριακές παραμορφώσεις
\en Leigh--Strassler \cite{LS} \gr της \en ${\cal N}=4$ SYM.\ \gr Κάτω από αυτές τις
οριακές διαμορφώσεις η \en ${\cal N}=4$ SYM \gr έχει σπάσει σε ${\cal N}=1$
\en SYM.\gr  Αυτά τα υπόβαθρα χαρακτηρίζονται
από μια μιγαδική παράμετρο $\b = \g + \tau \s$ (όπου $\tau$ είναι το αξιόνιο-διαστελόνιο
(\en axion-dilaton) \gr της θεωρίας χορδών ΙΙΒ και $\g$, $\s$ είναι πραγματικές παράμετροι)
οι οποία στην θεωρία βαθμίδας, αντιπροσωπεύει την μιγαδική φάση που υπάρχει στο υπερδυναμικό
\en Leigh--Strassler (\gr όπου $\tau$ είναι η μιγαδική σταθερά ζεύξης). Συγκεκριμένα,
θα μελετήσουμε οριακές διαμορφώσεις του σύμμορφου υποβάθρου $AdS_5\times S^5$ \cite{LM} και
των πολυκεντρικών υποβάθρων που αντιστοιχούν σε $D3$ βράνες οι οποίες κατανέμονται πάνω σε μία
σφαίρα και σε ένα δίσκο \cite{asz1}. Για την ανάπτυξη αυτών των παραδειγμάτων θα χρειαστούμε
γενικεύσεις των αποτελεσμάτων που παρουσιάστηκαν στα Κεφάλαια 6 και 7 και οι οποίες αναπτύχθηκαν
διεξοδικά στις ενότητες 2 και 3 του \cite{ASS2}.

\section{Σύμμορφη περίπτωση}

Ξεκινούμε από την παραμόρφωση του σύμμορφου υποβάθρου $AdS_5\times S^5$,
αλλάζοντας την κλίμακα των παραμέτρων παραμόρφωσης ως $(\b,\g,\s) \to {2 \ov R^2} (\b,\g,g_{\rm s}\s)$
και γράφουμε την μετρική ως
\be
\label{11-1}
ds^2 = {\cal H}^{1/2} \left[ {u^2\ov R^2} \left( -  dt^2 +
dx^2 + dy^2 + dz^2 \right) + R^2 \left(  {du^2 \ov u^2} +
d\Omega_{5,\b}^2 \right) \right]\ ,
\ee
όπου $d\Om_{5,\b}^2$ είναι η μετρική πάνω στην παραμορφωμένη πενταδιάστατη σφαίρα,
η οποία δίνεται από
\ba
\label{11-2}
d\Omega_{5,\b}^2 &=& d \th^2 + {\cal G} \sin^2 \th d
\phi_1^2 + \cos^2 \th [ d \psi^2  + {\cal G} ( \sin^2 \psi d\phi_2^2 +
\cos^2 \psi d \phi_3^2 ) ] \nonumber\\
&+& {\cal G} |\b|^2 \cos^4 \th \sin^2 \th \sin^2 2 \psi
\left(d\phi_1+d\phi_2+d\phi_3 \right)^2\ ,
\ea
και οι συναρτήσεις ${\cal G}$ και ${\cal H}$ δίνονται από
\ba
\label{11-3}
{\cal G}^{-1} &=& 1 + 4 |\b|^2 \cos^2 \th (\sin^2 \th +
\cos^2 \th \cos^2 \psi \sin^2 \psi)\ , \nonumber\\ {\cal H} &=& 1 +
4\s^2 \cos^2 \th (\sin^2 \th + \cos^2 \th \cos^2 \psi \sin^2
\psi)\ .
\ea
Η λύση περιέχει και ένα μη μηδενικό πεδίο \en Kalb--Ramond \gr
το οποίο, στην περίπτωση που η παράμετρος παραμόρφωσης $\g$
μηδενίζεται, ισούται με
\be
\label{11-4}
B_2 = {\s \ov 2}
\cos^4\th\sin{2\psi}(d\phi_1+d\phi_2+d\phi_3)\wedge d\psi\ ,
\ee
και επομένως είναι της μορφής (2.2) του \cite{ASS2}. Για την ανάλυση ευστάθειας
αυτής της μετρικής, οι εγκάρσιες συντεταγμένες είναι οι
$(x,z,\phi_{1,2,3})$, με τις $\phi_i$ να είναι συζευγμένες μεταξύ τους
και όλες να είναι αποσυζευγμένες από την διαμήκη συντεταγμένη $y$.
Οι γωνιακές συντεταγμένες είναι οι $(\th,\psi)$ και είναι αποσυζευγμένες
η μια από την άλλη. Η εξίσωση κίνησης για τις γωνιακές συντεταγμένες \eqn{7-13}
(για τις $(\th,\psi)$) καθορίζει τις επιτρεπτές γωνιακές διευθύνσεις της χορδής σε
$(\th_0,\psi_0)=(0,\pi/4)$, $(\th_0,\psi_0)=(0,0 \hbox{ ή }\pi/2)$ (οι δύο επιλογές
για το $\psi_0$ οδηγούν σε ισοδύναμα αποτελέσματα), $(\th_0,\psi_0)=(\pi/2,{\textrm{οποιαδήποτε}})$, και
$(\th_0,\psi_0)=(\sin^{-1}(1 / \sqrt{3}),\pi/4)$. Σημειώνουμε επίσης ότι για τις τροχιές με
$\th_0=\pi/2$, η μεταβλητή $\psi$ είναι μια αποσυζευγμένη εγκάρσια συντεταγμένη και επομένως οι
διακυμάνσεις της είναι ευσταθείς. Επίσης, η παρουσία του μη μηδενικού πεδίου \en Kalb--Ramond
\gr στην Εξ.\eqn{11-2}
οδηγεί στο ότι για τις τροχιές με $(\th_0,\psi_0)=(0,\pi/4)$ και
$(\th_0,\psi_0)=(\sin^{-1}(1 / \sqrt{3}),\pi/4)$ η γωνιακή διακύμανση $\d\psi$ και οι
εγκάρσεις $\d\phi_i$ είναι συζευγμένες. Ωστόσο, οι συζευγμένοι όροι είναι ανάλογοι της ιδιοτιμής
$\om$ (δες Εξ.(3.6) του \cite{ASS2}) και για αυτό δεν επηρεάζουν τα απολέσματα
της ανάλυσης μας η οποία στηρίζεται στους μηδενικούς τρόπους. Για $(\th_0,\psi_0)=(0,0\hbox{ ή }
\pi/2),(\pi/2,\textrm{οποιαδήποτε})$, οι διακυμάνσεις $\d \psi$ και $\d \phi_i$ είναι αποσυζευγμένες.\\
\no
Στην περίπτωση μας, η ενέργεια $E$ μπορεί να δοθεί ως συνάρτηση του δυναμικού σε κλειστή
μορφή
\ba
\label{11-5}
E(L) = k(\s) \left(- {4\pi^2R^2\ov\Gamma(1/4)^4}{1\ov
L} \right)\ ,
\ea
όπου $k(\s)$ είναι ένας παράγοντας που εξαρτάται απο τις γωνίες
και την παράμετρο $\s$, και η μορφή του θα δοθεί στα παραδείγματα που θα ακολουθήσουν
(γίνεται ίσος με ένα για $\s=0$). Ο παράγοντας εντός παρενθέσεως είναι
το αποτέλεσμα στην σύμμορφη περίπτωση \cite{maldaloop},
δίνοντας την αναμενόμενη συμπεριφορά \en Coulomb \gr λόγω ύπαρξης της συγκεκριμένης συμμετρίας.
Σημειώνουμε ότι η ύπαρξη $\g$ παραμορφώσεων αφήνει αυτόν παράγοντα
ως έχει, εκτός και αν τα κουάρκ και αντικουάρκ
βρίσκονται σε απόσταση ως προς την παραμορφωμένη (\en deformed)\gr εσωτερική σφαίρα \cite{hsz}.\\
\no
Όσον αφορά την ευστάθεια, οι εγκάρσιες και διαμήκεις διακυμάνσεις είναι ευσταθείς,
ενώ όλα τα δυναμικά \en Schr\"odinger \gr για τις γωνιακές ταλαντώσεις έχουνε
την μορφή
\be
\label{11-6}
 V_{\theta,\psi}= a_{\theta,\psi}(\s) u^2\ ,
\ee
όπου $a_{\th,\psi}(\s)$ είναι συναρτήσεις που εξαρτώνται απο την γωνία και
την παράμετρο $\s$ και των οποίων την μορφή θα δοθεί στα παραδείγματα που θα ακολουθήσουν.
Η έκφραση για την μεταβλητή \en Schr\"odinger $x$ \gr ως συνάρτηση της μεταβλητής $u$ είναι
\ba
\label{11-7}
x={1\ov u}\ {}_2F_1\left({1 \ov 4},{1 \ov 2},{5 \ov
4},{u_0^4 \ov u^4} \right)
\ea
και, αντίστοιχα, η τιμή του άκρου $x_0$ είναι
\ba
\label{11-8}
x_0={\Gamma(1/4)^2\ov 4 \sqrt{2\pi}}{1\ov u_0}\ .
\ea
Αστάθειες μπορούν να συμβούν μόνο εάν κάποιο από τα $a_{\theta,\psi}(\s)$
είναι αρνητικό. Μολονότι η Εξ.\eqn{11-7} δεν μπορεί να αντιστραφεί σε κλειστή
μορφή ως προς την μεταβλητή $u$ για να βρούμε μια αναλυτική λύση στο πρόβλημα
\en Schr\"odinger, \gr μπορούμε να βρούμε ένα κάτω φράγμα για το $\s$ πάνω από
το οποίο έχουμε αστάθειες. Πρώτα αναφέρουμε ότι δεν ισχύει
η προσέγγιση του απειρόβαθρου πηγαδιού, διότι το δυναμικό ικανοποιεί την
Εξ.\eqn{11-6} και ως εκ τούτου δεν μπορεί να θεωρηθεί σταθερά εντός του διαστήματος
$[0,x_0]$. Για να βρούμε ένα φράγμα για το $\s$, παρατηρούμε ότι στο σύμμορφο όριο
$u/u_0\gg 1$, το δυναμικό προσεγγίζεται ως $V_{\th,\psi} \simeq a_{\th,\psi}(\s) / x^2$
όπου $x\simeq 1/u$. Τότε, εφαρμόζοντας γνωστές τεχνικές από την κβαντομηχανική \cite{LLQM}
βρίσκουμε ότι αν $a_{\th,\psi}<-1/4$, το δυναμικό έχει ένα άπειρο πλήθος καταστάσεων αρνητική
ενέργειας με αποτέλεσμα, η λύση γίνεται ασταθής για κάθε τιμή της παραμέτρου $u_0$.
Η παρατήρηση αυτή μας δίνει ένα φράγμα για την παράμετρο $\s$. Τα αποτελέσματα για τις
επιτρεπτές τροχιές είναι

\no
$\bullet$
$(\th_0,\psi_0)=(0,\pi/4)$. Σε αυτήν την περίπτωση έχουμε
\be
\label{11-9}
k(\s) = \sqrt{1+\s^2}\ ,\qq a_\th(\s) = {2\s^2 \ov
1+\s^2}\ ,\qq a_\psi(\s) = - {4\s^2 \ov 1+\s^2}\ .
\ee
\no
Υπάρχουν αστάθειες από τις διακυμάνσεις $\d\psi$ για
$\s> 1 / \sqrt{15} \simeq 0.258$. Εφόσον στο υπεριώδες
όλα τα οριακά παραμορφώσιμα υπόβαθρα είναι ασυμπτωτικά
του υποβάθρου που δόθηκε στις Εξ.\eqn{11-1},\eqn{11-2}, το άνω φράγμα
είναι καθολικό εφόσον η αντίστοιχη τροχιά υπάρχει.
Όπως θα δούμε η τροχιά αυτή υπάρχει επίσης στην περίπτωση
της δίσκου και της τριδιάστατης σφαίρας.

\no $\bullet$ $(\theta_0,\psi_0)=(0,0 \hbox{ ή } \pi/2)$. Όπου,
έχουμε
\be
\label{11-10}
k(\s) = 1\ ,\qq a_\th(\s) = a_\psi(\s) = 4\s^2\ ,
\ee
και συνεπώς δεν υπάρχουν γωνιακές αστάθειες.

\no $\bullet$ $(\th_0,\psi_0)=(\pi/2,{\textrm{οποιαδήποτε}})$. Όπου, έχουμε
\be
\label{11-11}
k(\s) = 1\ ,\qq a_\th(\s) = 4 \s^2\ ,
\ee
και συνεπώς επίσης δεν υπάρχουν γωνιακές αστάθειες.

\no $\bullet$ $(\th_0,\psi_0)=(\sin^{-1}(1/\sqrt{3}),\pi/4)$.
Όπου, έχουμε
\be
\label{11-12}
k(\s) = \sqrt{1+{4\s^2 \ov 3}}\ ,\qq a_\th(\s) =
a_\psi(\s) = - {8 \s^2 \ov 3+4\s^2}\ .
\ee
Υπάρχουν αστάθειες για τις διακυμάνσεις $\d \th$ και $\d \psi$ για
$\s > \sqrt{3/28} \simeq 0.327$. Αυτό το φράγμα δεν υπάρχει στην περίπτωση
των πολυκεντρικών κατανομών $D3$ βρανών διότι δεν υπάρχει η αντίστοιχη τροχιά.

\section{Η τριδιάστατη σφαίρα}

Στην συνέχεια θα θεωρήσουμε την παραμόρφωση του υποβάθρου που αντιστοιχεί
σε μια κατανομή $D3$  βρανών πάνω σε μια τριδιάστατη σφαίρα ακτίνας $r_0$.
Χρησιμοποιώντας αδιάστατες μονάδες ($\displaystyle{u \to r_0 u\ ,u_0 \to r_0 u_0, L \to {R^2
\ov r_0} L \ ,E \to {r_0 \ov \pi}E}$), γράφουμε την μετρική ως \cite{asz1}

\ba
\label{11-13} &&\!\!\!\!\!\!\!\!\!\!\!\!\!\! ds^2 = \cH^{1/2}
\Biggl\{ H^{-1/2} \left( - dt^2 + dx^2 + dy^2 + dz^2 \right) +
H^{1/2} {u^2 - r_0^2 \cos^2 \th \ov u^2 - r_0^2} du^2
\nonumber\\
&&\!\!\!\!\!\!\!\!\!\!\!\!\!\!\ \qq\qq\quad + \:\: H^{1/2}
\left[(u^2 - r_0^2 \cos^2 \th) d \th^2 + \cG (u^2 - r_0^2) \sin^2
\th d \phi_1^2 \right]
\nonumber\\
&&\!\!\!\!\!\!\!\!\!\!\!\!\!\!\ \qq\qq\quad + \:\: H^{1/2} u^2
\cos^2 \th \left[ d \psi^2 + \cG ( \sin^2 \psi d\phi_2^2 + \cos^2
\psi d \phi_3^2 ) \right]
\nonumber\\
&&\!\!\!\!\!\!\!\!\!\!\!\!\!\!\ \qq\qq\quad + \:\: H^{1/2} {|\b|^2
\cG u^2 (u^2-r_0^2) \cos^4 \th \sin^2 \th \sin^2 2 \psi \ov u^2 -
r_0^2 \cos^2 \th} \left(d\phi_1+d\phi_2+d\phi_3 \right)^2
\Biggr\}\ ,
\ea
με
\ba
\label{11-14} \cG^{-1} &=& 1 + 4|\b|^2 \cos^2 \th\ { ( u^2 - r_0^2
) \sin^2 \th + u^2 \cos^2 \th \cos^2 \psi \sin^2 \psi  \ov u^2 -
r_0^2 \cos^2 \th}\ , \nonumber\\ \cH &=& 1 + 4\s^2 \cos^2 \th\ { (
u^2 - r_0^2 ) \sin^2 \th + u^2 \cos^2 \th \cos^2 \psi \sin^2 \psi
\ov u^2 - r_0^2 \cos^2 \th}\ ,\nonumber \\
H&=&\frac{R^4}{u^2(u^2-r_0^2\cos^2\th)}\ .
\ea
Αναφέρουμε επίσης ότι στην περίπτωση μας υπάρχει μη μηδενικό πεδίο
\en Kalb--Ramond \gr το οποίο για $\g=0$ δίνεται από την Εξ.\eqn{11-4}.
Σε αυτήν την περίπτωση η εξίσωση κίνησης για τις γωνιακές συντεταγμένες
επιτρέπει μόνο τις τρείς πρώτες τροχιές που προαναφέρθησαν στην σύμμορφη περίπτωση,
$(\th_0,\psi_0)=(0,\pi/4)$, $(\th_0,\psi_0)=(0,0 \hbox{ ή
}\pi/2)$ και $(\th_0,\psi_0)=(\pi/2,{\textrm{οποιαδήποτε}})$.\\
\no
Στην συνέχεια θα μελετήσουμε αυτές τις περιπτώσεις:

\be
\label{11-15} L(u_0,\s) = { 2 u_0 \ov \sqrt{(u_0^2-{1 \ov
1+\s^2})(2 u_0^2-{1 \ov 1+\s^2})} } \left[ \elPi(a^2,k) - \elK(k)
\right]\ ,
\ee
και
\be
\label{11-16} E(u_0,\s) = \sqrt{1+\s^2} \left\{ \sqrt{2 u_0^2-{1
\ov 1+\s^2}} \left[ a^2 \elK(k) - \elE(k) \right] + \elE(c)-{\s^2
\ov 1+\s^2}\elK(c) \right\}\ ,
\ee
όπου
\be
\label{11-17} k = \sqrt{{u_0^2 + {\s^2 \ov 1+\s^2} \ov 2 u_0^2 - {1
\ov 1+\s^2}}}\ ,\qq a = \sqrt{{u_0^2 - {1 \ov 1+\s^2} \ov 2 u_0^2
- {1 \ov 1+\s^2}}}\ , \qq c={1 \ov \sqrt{1 + \s ^2}}\ .
\ee
Σημειώνουμε ότι τα αποτελέσματα αυτά δεν εξαρτώνται από την παράμετρο $\g$.
Ο λόγος είναι ότι, όπως προανεφέρθει στην σύμμορφη περίπτωση, θεωρούμε πως
τα κουάρκ και αντικουάρκ δεν βρίσκονται σε απόσταση ως προς
την παραμορφωμένη εσωτερική σφαίρα. Η χρήση του βρόχου \en Wilson \gr για τον
υπολογισμό δυναμικών στην περίπτωση αυτή, αλλά με μηδενικό $\s$, έγιναν στο \cite{hsz}.\\
\no
Για $\s=0$, η συμπεριφορά του μήκους και της ενέργειας είναι ίδια με
την περίπτωση που δεν έχουμε παραμόρφωση (Σχήμα~10.1γ). Όσο αυξάνουμε το $\s$, οι καμπύλες του μήκους και
της ενέργειας μοιάζουν με τις ισόθερμες καμπύλες \en van der Waals \gr για ένα στατιστικό
σύστημα με $u_0,L$ και $E$ να αντιστοιχούν στον όγκο, την πίεση και το δυναμικό \en Gibbs \gr
(δείτε για παράδειγμα \cite{Callen}). Συγκεκριμένα, υπάρχει μια κρίσιμη τιμή του $\s$, η οποία
δίνεται από $\s_{\rm cr}^{(y)}\simeq 0.209$ (το οποίο βρέθηκε αναλυτικά στο \cite{asz2}),
κάτω από την οποία το σύστημα συμπεριφέρεται σαν ένα στατιστικό σύστημα για $T < T_{cr}$ (Σχήμα~10.1δ)
και πάνω από την οποία το σύστημα συμπεριφέρεται σαν ένα στατιστικό σύστημα για $T > T_{cr}$ (Σχήμα~10.1β).
Για μη μηδενικά $\s$, έχουμε την συνήθη συμπεριφορά \en Coulomb \gr για μεγάλα $u_0$ (μικρά μήκη),
ενώ στο αντίθετο όριο, $u_0\to 1$, δηλαδή μεγάλα μήκη, οι ασυμπτωτικές συμπεριφορές των Εξ.\eqn{11-15}
και \eqn{11-16} οδηγούν στο γραμμικό δυναμικό
\be
\label{11-18}
E \simeq {\s \ov 2} L\ .
\ee
Για να εξετάσουμε την σημασία των αποτελεσμάτων,
είναι σημαντικό να εξετάσουμε την ευστάθεια των αντίστοιχων
διατάξεων χορδών.\\
\no
Τα γενικά μας αποτελέσματα δείχνουν ότι η λύση είναι ευσταθής κάτω από
εγκάρσιες διακυμάνσεις. Για τις διαμήκεις διακυμάνσεις, το γεγονός
ότι το $L(u_0)$ έχει δύο ακρότατα $u_{0 \rm c}^{(l1)}(\s)$ και
$u_{0 \rm c}^{(l2)}(\s)$ για $0<\s<\s^{(y)}_{\rm cr}$ υποδηλώνει
την ύπαρξη ασταθειών στις διαμήκεις διακυμάνσεις στην περιοχή
$u_{0 \rm c}^{(l1)}(\s) < u_0 < u_{0 \rm c}^{(l2)}(\s)$ (Σχήμα~10.1δ)
στην πρώτη περίπτωση, και στην μη ύπαρξη ασταθειών (Σχήμα~10.1β) στην δεύτερη περίπτωση.

\begin{figure}[!t]
\begin{center}
\begin{tabular}{cccc}
 \includegraphics[height=2.8cm]{fig1a-1.mesons.eps}
&\includegraphics[height=2.8cm]{fig1b-1.mesons.eps}
&\includegraphics[height=2.8cm]{fig1c-1.mesons.eps}
&\includegraphics[height=2.8cm]{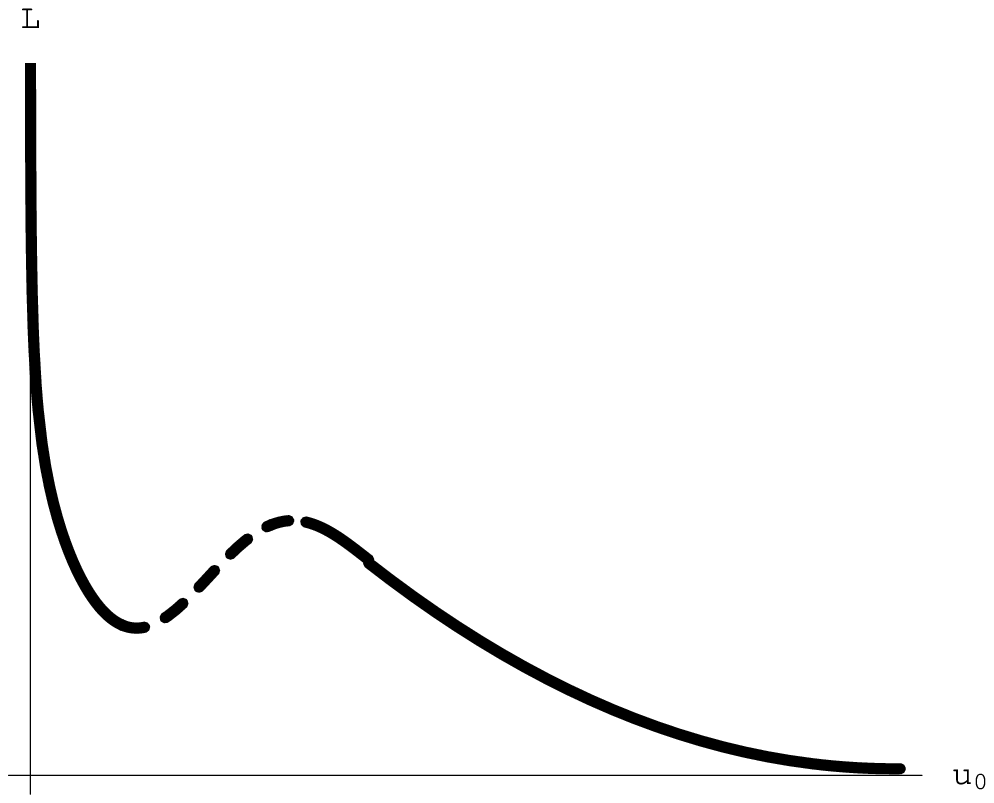} \\
 \includegraphics[height=2.8cm]{fig1a-2.mesons.eps}
&\includegraphics[height=2.8cm]{fig1b-2.mesons.eps}
&\includegraphics[height=2.8cm]{fig1c-2.mesons.eps}
&\includegraphics[height=2.8cm]{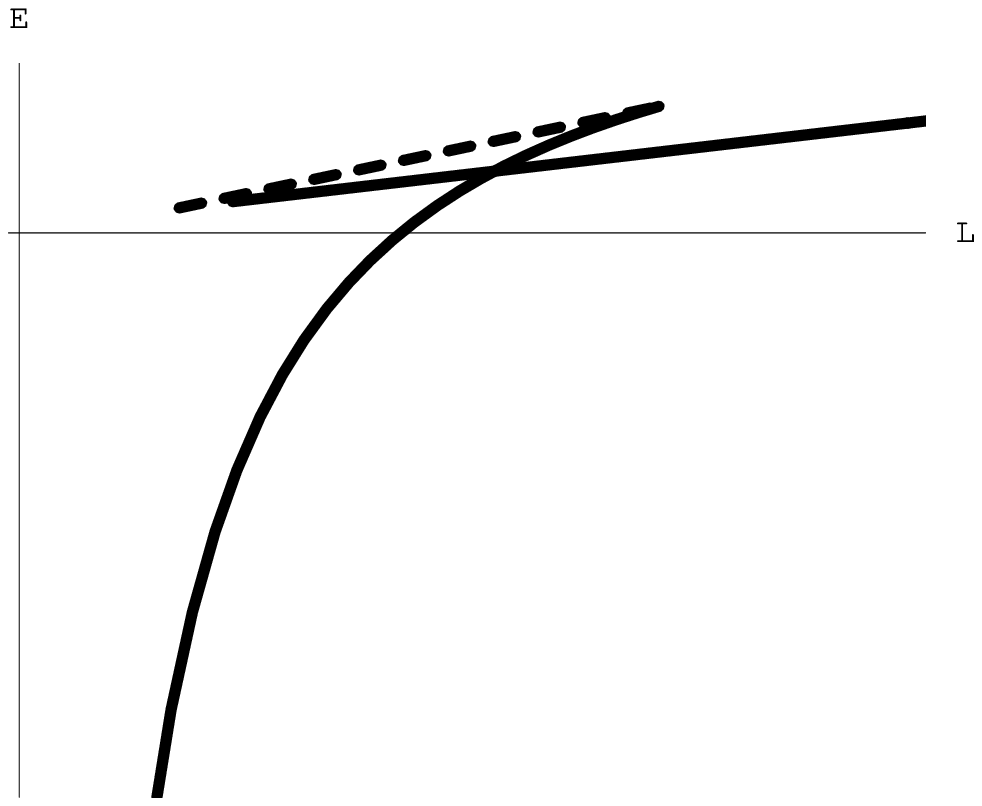} \\
(α) & (β) & (γ) & (δ)
\end{tabular}
\end{center}
\caption{Τα τέσσερα γενικά είδη συμπεριφορών των $L(u_0)$ και $E(L)$
που απαντώνται σε υπολογισμούς βρόχων \en Wilson \gr στα πλαίσια της \en AdS/CFT , \gr
και αντιστοιχούν σε (α) μη ύπαρξη ακροτάτων, θωρακισμένο ή μη δυναμικό \en Coulomb, \gr
(β) μη ύπαρξη ακροτάτων, σε δυναμικό \en Coulomb \gr και εγκλωβισμού, (γ) ένα ακρότατο,
 δίτιμο δυναμικό και (δ) δύο ακρότατα,
πλειότιμα δυναμικά. Τα διακεκομμένα τμήματα των καμπυλών
συμβολίζουν τις περιοχές που είναι ασταθής κάτω από διαμήκεις διακημάνσεις.
} \label{fig4.Deformed}
\end{figure}
\no
Για να ελένξουμε αν όντως η μικρότερη ιδιοτιμή $\om_0^2$ αλλάζει πρόσημο
σε αυτές τις τιμές, πραγματοποιήσαμε αριθμητική ανάλυση της οποίας τα αποτελέσματα φαίνονται
στο Σχήμα~10.2β και η οποία αποδίδει την αναμενόμενη συμπεριφορά. Τέλος, για τις
γωνιακές διακυμάνσεις, τα αντίστοιχα δυναμικά \en Schr\"odinger \gr δίνονται από
περίπλοκες εκφράσεις, βλέπε Εξ.(A.13) στο παράρτημα Α του \cite{ASS2}.
Ξεκινώντας απο το δυναμικό $V_\th$, βρίσκουμε ότι είναι θετικό για $\s < 0.71$,
ενώ εμφανίζει αρνητικό μέρος για $\s > 0.71$ και ως εκ τούτου μπορεί να
εμφανίσει δέσμια κατάσταση αρνητικής ενέργειας. Η εύρεση της κρίσιμης τιμής
$\s^{(\th)}_{\rm cr}$ πάνω από την οποία μπορεί να υπάρξουν δέσμιες καταστάσεις
γίνεται μέσω της προσέγγισης του απειρόβαθρου πηγαδιού και ισούται με
$\s^{(\th)}_{\rm cr} \simeq 0.78$, και η οποία είναι αρκετά κοντά στην ακριβή τιμή $\s^{(\th)}_{\rm cr}
\simeq 0.805$ η οποία βρέθηκε από αριθμητική ανάλυση. Καθώς το $\s$ αυξάνει πάνω από την τιμή αυτή,
η αντίστοιχή κρίσιμη τιμή της παραμέτρου $u_{0\rm c}^{(\th)}$ πλησιάζει την 1.097 όταν η παράμετρος
$\s$ γίνεται πολύ μεγάλη. Μελετώντας τις διακυμάνσεις $\d\psi$, για τον ίδιο  λόγο που προανεφέρθει στην
σύμμορφη περίπτωση για την ίδια τροχιά, υπάρχει μια κρίσιμη τιμή $\s^{(\psi)}_{\rm cr} = 1/\sqrt{15}$
πάνω από την οποία υπάρχει αστάθεια για κάθε τιμή της παραμέτρου $u_0$. Η ταύτιση των τιμών
οφείλεται στο ότι στο υπεριώδες το υπόβαθρο αυτό έχει την παθογένεια της σύμμορφης περίπτωσης.
Τα παραπάνω αποτελέσματα συνοψίζονται στο Σχήμα~10.2.

\begin{figure}[!t]
\begin{center}
\begin{tabular}{ccc}
 \includegraphics[height=4.4cm]{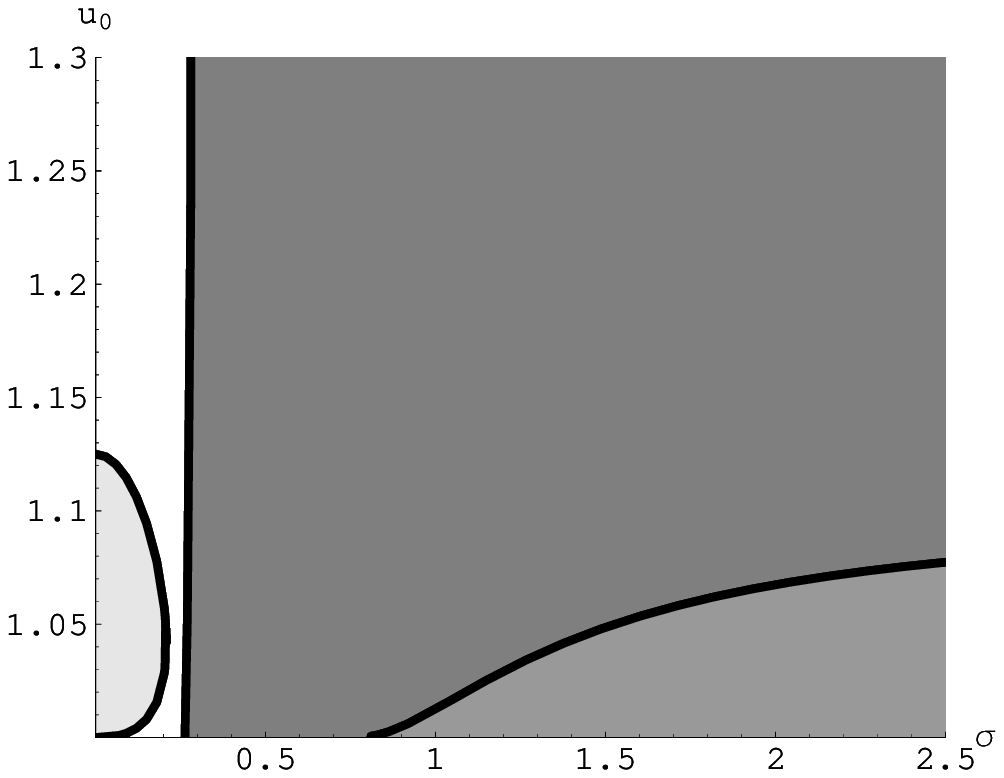}
&\includegraphics[height=4.4cm]{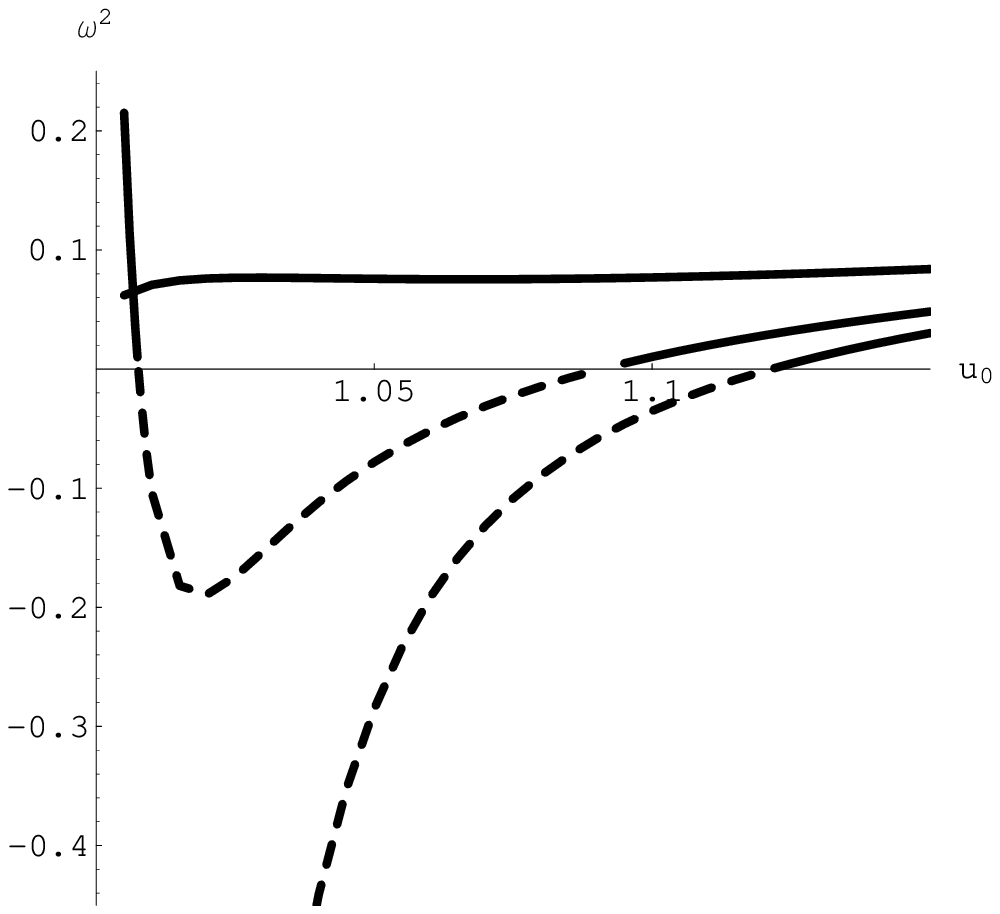}
&\includegraphics[height=4.4cm]{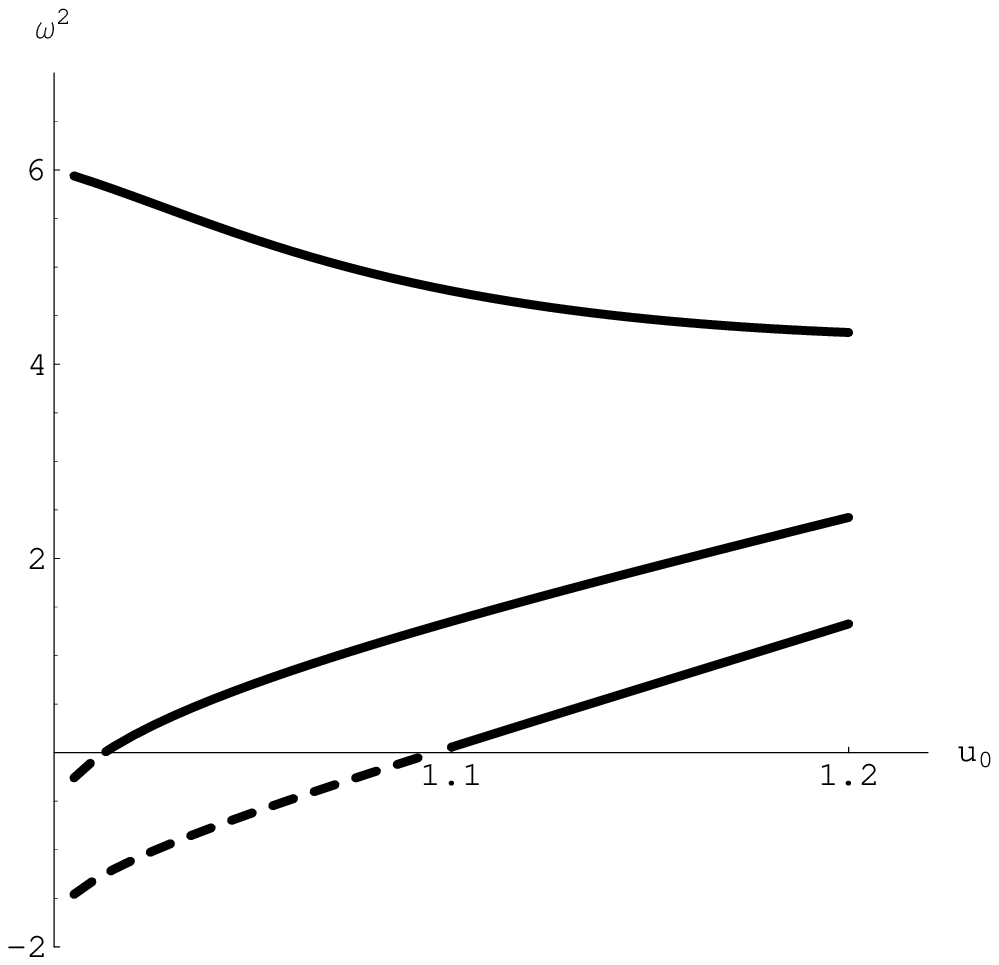}\\
(α) & (β) & (γ)
\end{tabular}
\end{center}
\vskip -.5 cm \caption{(α) Διάγραμμα ευστάθειας στο επίπεδο $(\s,u_0)$
για την παραμορφωμένη σφαίρα και την τροχιά
$(\th_0,\psi_0)=(0,\pi/4)$. Οι τρείς σκιασμένες περιοχές αντιστοιχούν
σε αστάθειες που οφείλονται στις διαμήκεις,
$\d \th$ και $\d \psi$
διακυμάνσεις (φωτεινότερο στο σκοτεινότερο). (β)
Γραφική παράσταση της χαμηλότερης ιδιοτιμής των διαμήκων διακυμάνσεων
σαν συνάρτηση της παραμέτρου $u_0$, για $\s=0$, $\s^{(y)}_{\rm
cr}> 0.15$ και $\s^{(y)}_{\rm cr}< 0.5$ (από κάτω προς τα πάνω). (γ) Το
αντίστοιχο για τις γωνιακές διακυμάνσεις $\d \th$, για $\s=0<\s^{(\th)}_{\rm cr}$,
$\s^{(\th)}_{\rm cr}< 1$ και $\s^{(\th)}_{\rm cr}\ll 10$ (από πάνω προς
τα κάτω).} \label{fig5.deformed}
\end{figure}

\no
Συνοψίζοντας, βρήκαμε ότι για ένα μικρό έυρος τιμών της παραμέτρου παραμόρφωσης,
συγκεκριμένα για
$0.209 \simeq \s^{(y)}_{\rm cr} < \s  < \s^{(\psi)}_{\rm cr} \simeq 0.258$,
η γραμμική συμπεριφορά εγκλωβισμού για μεγάλες αποστάσεις είναι ευσταθής,
έξω από το οποίο εύρος αποσταθεροποιείται απο τις γωνιακές διακυμάνσεις $\d\psi$ και
στην συνέχεια από τις $\d\th$. Επιπλέον, ακόμα και για μικρές τιμές του $\s$,
συγκεκριμένα για $0< \s< \s^{(y)}_{\rm cr}\simeq 0.209 $, η συμπεριφορά εγκλωβισμού
του δυναμικού είναι ευσταθής.

\no $\bullet$ $(\th_0=0,\psi_0=0 \hbox{ ή } \pi/2)$. Η συνάρτηση
του μήκους όπως και η ενέγεια έχουν τις αυτές εκφράσεις
με την περίπτωση που δεν έχουμε παραμόρφωση \cite{bs}
[Εξ.\eqn{9-27},\eqn{9-28},\eqn{9-29}],
\be
\label{11-19}
L(u_0) = {2 u_0 k^{\prime} \ov u_0^2-1} \left[ \elPi
(k^{\prime 2},k) - \elK(k) \right]\ ,
\ee
και
\be
\label{11-20} E(u_0) = \sqrt{2u_0^2-1} \left[ k^{\prime 2} \elK (k) -
\elE(k) \right] + 1\ ,
\ee
όπου
\be
\label{11-21} k={u_0 \ov \sqrt{2u_0^2-1}}\ ,\qq
k^{\prime}=\sqrt{1-k^2}\ .
\ee
Η συμπεριφορά τουσ είναι ανάλογη με το Σχήμα~10.1γ.\\
\no
Από τα γενικά μας αποτελέσματα, υπάρχει αστάθεια στις διαμήκεις διακυμάνσεις
που συμβαίνει για τιμές της παραμέτρου $u_0$ μικρότερες απο την κρίσιμη τιμή
για μη ύπαρξη παραμόρφωσης, δηλαδή $u_{0 \rm c}\simeq 1.125$. Για τις γωνιακές
διακυμάνσεις $\d\th$ και $\d\psi$, τα δυναμικά \en Schr\"odinger \gr δίνονται
από περίπλοκες εκφράσεις, βλέπε Εξ.(A.14) στο παράρτημα Α του \cite{ASS2}
και διαφέρουν από τα αντίστοιχα δυναμικά στην περίπτωση που δεν έχουμε παραμόρφωση,
κατά έναν προσθετικό παράγοντα $4\s^2u^2$. Τα παραπάνω αποτελέσματα συνιστούν ότι, εφόσον οι γωνιακές
διακυμάνσεις $\d\th$ και $\d\psi$ είναι ευσταθείς στην περίπτωση που δεν έχουμε παραμόρφωση,
είναι ευσταθείς και στην παρούσα περίπτωση.

\no $\bullet$ $(\th_0,\psi_0)=(\pi/2,{\textrm{οποιαδήποτε}})$. Σε αυτή την περίπτωση,
το μήκος και το δυναμικό είναι τα ίδια και πάλι με την περίπτωση που δεν έχουμε διαμόρφωση
και δίνονται από \cite{bs}
\be
\label{11-22} L(u_0) = {\sqrt{2} \ov u_0} \left[ \elPi \left( \half,k
\right) -\elK(k) \right]\ ,
\ee
και
\be
\label{11-23} E(u_0) = {u_0 \ov \sqrt{2}} \left[ \elK(k)- 2 \elE(k)
\right]\ ,
\ee
όπου
\be
\label{11-24} k = \sqrt{{u_0^2+1 \ov 2 u_0^2}}\ ,\qq
k^{\prime}=\sqrt{1-k^2}\ .
\ee
Η συμπεριφορά τουσ είναι ανάλογη με το Σχήμα~10.1β. Για $u_0\gg 1$ (μικρά $L$), η συμπεριφορά
είναι τύπου \en Coulomb, \gr ενώ στο αντίθετο όριο, $u_0\to 1$ (μεγάλα $L$), οι ασυμπτωτικές
συμπεριφορές των Εξ.\eqn{11-22} και \eqn{11-23} οδηγούν σε ένα γραμμικό δυναμικό
εγκλωβισμού
\be
\label{11-25}
E \simeq  {L \ov 2}\ ,\qq \textrm{για} \quad L \gg 1 \ .
\ee
Στην περίπτωση που δεν έχουμε παραμόρφωση,
η εμφάνιση ενός γραμμικού δυναμικού
εγκλωβισμού δεν αναμένεται στην δυική θεωρία βαθμίδας
(${\cN}=4$ \en SYM)\gr.
Το παράδοξο αυτό επιλύθηκε στο \cite{ASS1}
(αναφέρθηκε αναλυτικά στην ενότητα 9.3.2)
μέσω της μελέτης ευστάθειας αυτών των διατάξεων κάτω από
γωνιακές διακυμάνσεις.
Ωστόσο, στην περίπτωση που έχουμε παραμόρφωση και
η υπερσυμμετρία $\cN=4$ έχει ελαττωθεί σε $\cN=1$,
η παρουσία τέτοιου δυναμικού
περιορισμού εν γένει είναι αναμενόμενη \cite{Seiberg}. Αξιοσημείωτα, αυτό
επιβεβαιώνεται από την ανάλυση ευστάθειας που ακολουθεί.\\
\no
Αρχικά παρατηρούμε ότι οι εγκάρσιες και οι διαμήκεις διακυμάνσεις
είναι ευσταθείς, το οποίο προκύπτει από το γεγονός ότι
η συνάρτηση του μήκους δεν έχει ακρότατα. Για τις γωνιακές διακυμάνσεις,
το δυναμικό \en Schr\"odinger \gr ισούται με
\be
\label{11-26}
V_\th = -1 + 4 \s^2(u^2-1)\ ,
\ee
όπου παρατηρούμε την επίδραση της παραμόρφωσης στο
ότι τροποποιήσει το δυναμικό $V_\th$ κατά έναν προσθετικό θετικό
παράγοντα. Είναι προφανές ότι η παρουσία
αυτού του όρου θα τείνει να σταθεροποιήσει τις γωνιακές
διακυμάνσεις. Η συμπεριφορά αυτή περιγράφεται στο Σχήμα~10.3.
Συμπερασματικά, η περιοχή εγκλωβισμού γίνεται όλο και πιο ευσταθής όσο
αυξάνουμε το $\s$.
\begin{figure}[!t]
\begin{center}
\begin{tabular}{cc}
\includegraphics[height=6cm]{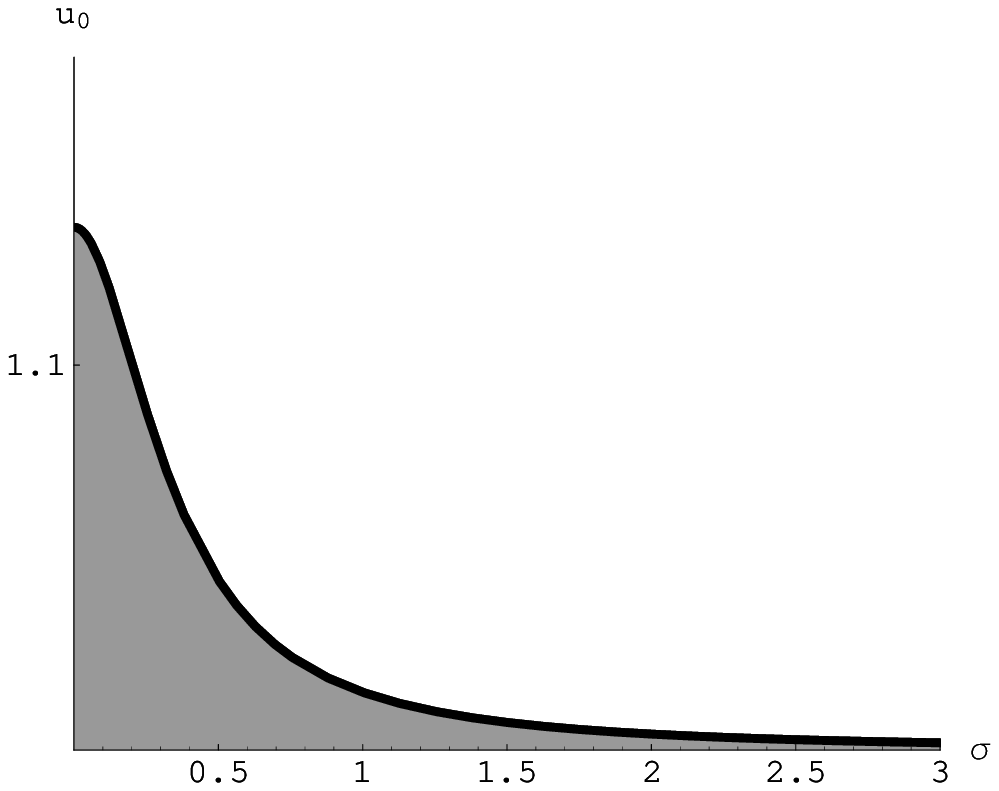}
&\includegraphics[height=6cm]{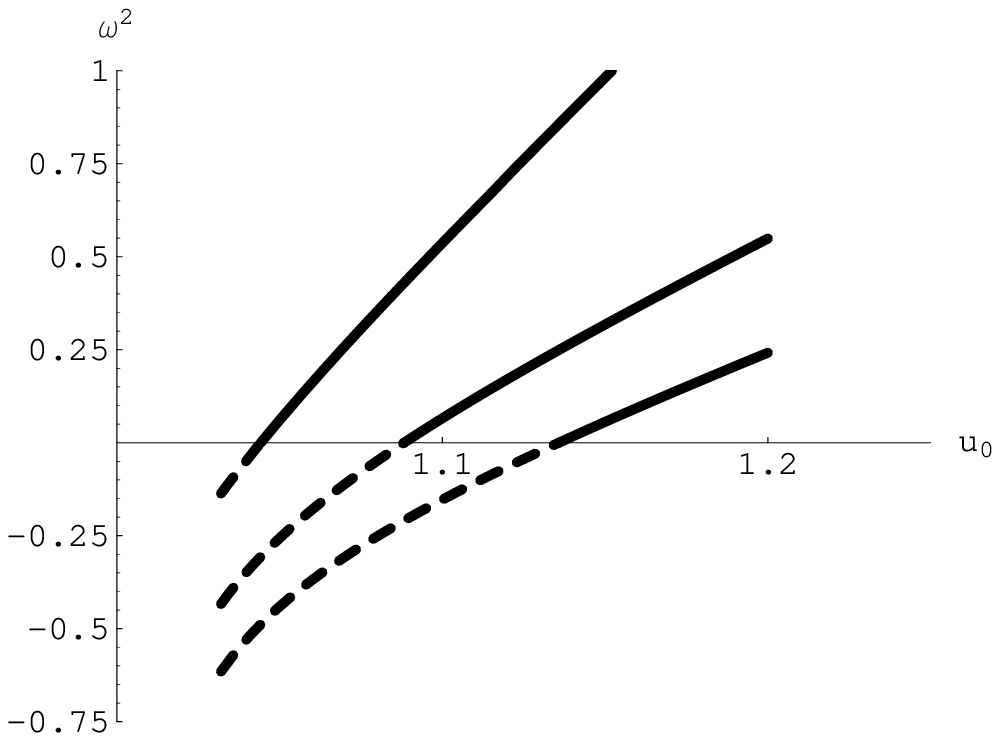}\\
(α) & (β)
\end{tabular}
\end{center}
\vskip -.5 cm \caption{(α) Διάγραμμα ευστάθειας στο επίπεδο $(\s,u_0)$
για την περίπτωση της παραμορφωμένης σφαίρας και για τροχιά
$(\th_0,\psi_0)=(\pi/2,{\textrm{οποιαδήποτε}})$.
Η σκιασμένη περιοχή αντιστοιχεί σε αστάθεις κάτω από γωνιακές διαταραχές.
(β) Γραφική παράσταση της μικρότερης ιδιοτιμής των γωνιακών διακυμάνσεων με το
$u_0$, για $\s=0$, $0.25$ και $0.5$ (από κάτω προς τα πάνω).}
\label{fig6.deformed}
\end{figure}

\section{Ο δίσκος}

Τέλος θα ασχοληθούμε με την παραμόρφωση του υποβάθρου το οποίο αντιστοιχεί
σε κατανομή $D3$ βρανών πάνω σε δίσκο ακτίνας $r_0$, η οποία μπορεί να
βρεθεί από αναλυτική επέκταση $r_0^2\to-r_0^2$. Οι δυνατές τροχιές
στις γωνιακές διευθύνσεις είναι οι ίδιες με την περίπτωση της σφαίρας.

\no $\bullet$ $(\th_0,\psi_0)=(0,\pi/4)$. Οι συναρτήσεις του μήκους και της ενέργειας
δίνονται απο (δες ενότητα 6.2 του \cite{asz2})
\be
\label{11-27} L(u_0,\s) = { 2 u_0 \ov \sqrt{(u_0^2+{1 \ov
1+\s^2})(2 u_0^2+{1 \ov 1+\s^2})} } \left[ \elPi(a^2,k) - \elK(k)
\right]\ ,
\ee
και
\be
\label{11-28} E(u_0,\s) = \sqrt{1+\s^2} \left\{  \sqrt{2 u_0^2+{1
\ov 1+\s^2}} \left[ a^2 \elK(k) - \elE(k) \right] + \elE(c)-{1 \ov
1+\s^2}\elK(c) \right\}\ ,
\ee
όπου
\be
\label{11-29} k = \sqrt{{u_0^2 - {\s^2 \ov 1+\s^2} \ov 2 u_0^2 + {1
\ov 1+\s^2}}}\ ,\qq a = \sqrt{{u_0^2 + {1 \ov 1+\s^2} \ov 2 u_0^2
+ {1 \ov 1+\s^2}}}\ , \qq c={\s \ov \sqrt{1 + \s ^2}}\ .
\ee
Για $u_0\gg 1$ (μικρά $L$) έχουμε την γνωστή συμπεριφορά \en Coulomb \gr
πολλαπλασιαμένη με έναν παράγοντα $\sqrt{1+\s^2}$, ενώ για $u_0\to 0$ έχουμε
την ασυμπτωτική έκφραση

\be
\label{11-30} E \simeq - {(\pi - L)^2 \ov 8 \elE({\rm i} \s)}\ ,
\ee
και η οποία δείχνει ότι υπάρχει πλήρης θωράκιση για μήκη
\be
\label{11-31}
L_{\rm c} = \pi\ ,
\ee
το οποίο παραμένει αναλλοίωτο και στην περίπτωση
που έχουμε παραμορφώσει το υπόβαθρο με την παράμετρο $\s$.\\
\no
Όσον αφορά στην ανάλυση της ευστάθειας, η λύση είναι ευσταθή κάτω από
εγκάρσιες και διαμήκεις διακυμάνσεις. Όσον αφορά στις γωνιακές, τα αντίστοιχα δυναμικά
\en Schr\"odinger \gr δίνονται από πολύπλοκες εκφράσεις, δες Εξ. (Α.16) \cite{ASS2}.
Αρχικά θεωρούμε τις γωνιακές διακυμάνσεις $\d\th$, και χρησιμοποιώντας την προσέγγιση του
απειρόβαθρου πηγαδιού βρίσκουμε ότι ο πυθμένας του πηγαδιού είναι ανυψωμένος
και αυξάνει με το $\s$ και υπάρχει μια κρίσιμη τιμή της παραμέτρου $\s$, $\s^{(\th)}_{\rm
cr}>1/\sqrt{15}$ πάνω από την οποία δεν υπάρχει αστάθεια της κλασική λύσης.
Για τις γωνιακές διακυμάνσεις $\d\psi$, για τον ίδιο  λόγο που προανεφέρθει στην
σύμμορφη περίπτωση για την ίδια τροχιά, υπάρχει μια κρίσιμη τιμή $\s^{(\psi)}_{\rm cr} = 1/\sqrt{15}$
πάνω από την οποία υπάρχει αστάθεια για κάθε τιμή της παραμέτρου $u_0$.

\no $\bullet$ $(\th_0,\psi_0)=(0,0 \hbox{ ή } \pi/2)$.
Σε αυτή την περίπτωση βρίσκουμε ότι η κλασική λύση είναι η ίδια
με την περίπτωση του δίσκου χωρίς παραμόρφωση του υποβάθρου για
$\th=0$ \cite{bs}
\be
\label{11-32} L(u_0) = {2 u_0 k^{\prime} \ov u_0^2 +1} \left[ \elPi
(k^{\prime 2},k) - \elK(k) \right]
\ee
και
\be
\label{11-33} E(u_0) = \sqrt{2u_0^2+1} \left[ k^{\prime 2} \elK (k) -
\elE(k) \right] \ ,
\ee
όπου
\be
\label{11-34} k={u_0 \ov \sqrt{2u_0^2+1}}\ , \qq k^{\prime
}=\sqrt{1-k^2}\ .
\ee
Η συμπεριφορά τουσ είναι ανάλογη με το Σχήμα~10.1a.
Από την γενική μας ανάλυση, η λύση είναι ευσταθής κάτω από εγκάρσιες και διαμήκεις
διακυμάνσεις. Όσον αφορά στις γωνιακές διακυμάνσεις $\d\th$ και $\d\psi$, τα δυναμικά
\en Schr\"odinger \gr δίνονται από πολύπλοκες εκφράσεις, δες Εξ. (Α.17) του \cite{ASS2}
και τα αντίστοιχα δυναμικά διαφέρουν κατά έναν προσθετικό παράγοντα από την
περίπτωση που δεν έχουμε παραμόρφωση. Για τις γωνιακές διακυμάνσεις $\d\th$,
για τις οποίες έχουμε αστάθεια όπως και στην περίπτωση που δεν έχουμε παραμόρφωση,
ο επιπλέον όρος σταθεροποιεί τις διακυμάνσεις και υπάρχει μια κρίσιμη τιμή
$\s^{(\th)}_{\rm cr}$ πάνω από την οποία δεν υπάρχει αστάθεια. Τέλος,
όσον αφορά τις γωνιακές διακυμάνσεις $\d\psi$ είναι ευσταθείς όπως και στην
περίπτωση που δεν έχουμε παραμόρφωση.

\no $\bullet$ $(\th_0,\psi_0)=(\pi/2,{\textrm{οποιαδήποτε}})$. Οι συναρτήσεις του μήκους
και της ενέργειας δίνονται από \cite{bs}

\be
\label{11-35} L(u_0) = {2 u_<^2 \ov \sqrt{u_0^2+u_>^2}}\left[ {\bf
\Pi} \left({u_>^2\ov u_0^2+u_>^2},k\right)-{\bf K}(k)\right ]\ ,
\ee
και
\be
\label{11-36} E(u_0) = {u_0^2\ov \sqrt{u_0^2+u_>^2}} \ {\bf
K}(k) - \sqrt{u_0^2+u_>^2} \ {\bf E}(k) + 1\ ,
\ee
όπου τώρα
\be
\label{11-37} k = \sqrt{{u_>^2 -u_<^2 \ov u_0^2+u_>^2}}\ , \qq
k^{\prime}=\sqrt{1-k^2}\ ,
\ee
και $u_>$ ($u_<$) συμβολίζουν το μεγαλύτερο (μικρότερο) μεταξύ του $u_0$ και
$1$. Η συμπεριφορά τους είναι ανάλογη με το Σχήμα 10.1α, και συγκεκριμένα,
έχουμε ένα θωρακισμένο δυναμικό \en Coulomb \gr με μήκος θωράκισης
\be
\label{11-38} L_{\rm c} = {\pi \ov 2}\ .
\ee
Η κλασική λύση είναι κατά τα γνωστά ευσταθής κάτω από εγκάρσιες και
διαμήκεις διακυμάνσεις, των οποίων η συμπεριφορά δεν εξαρτάται
από την παραμόρφωση του υποβάθρου. Όσον αφορά τις γωνιακές
διαταραχές, το δυναμικό \en Schr\"odinger \gr ισούται με
\ba
\label{11-39} V_\th = 1 + 4 \s^2(u^2+1) > 0\ ,
\ea
και επομένως η λύση είναι ευσταθής κάτω από τις γωνιακές διακυμάνσεις.

\part{Δυόνια}

\chapter{Εισαγωγή}
Η συνένωση τριών ή περισσοτέρων χορδών (\en string junction)  \cite{AhaSoYa,Schwarz}\gr
\ είναι ένα αντικείμενο στην θεωρία χορδών με ιδιαίτερο ενδιαφέρον
για διάφορους λόγους: Διότι είναι καταστάσεις \en BPS \gr και ως εκ τούτου
μπορούν να χρησιμοποιηθούν για την κατασκευή υπερσυμμετρικών καταστάσεων
\cite{Dasgupta,Sen,ReyYee} και στην μελέτη θεωριών βαθμίδας εντός
της θεωρίας χορδών \en SYM \cite{Bergman,Gabe}.
\gr Η σκέδαση των τρόπων ταλάντωσης μιας σύνενωσης χορδών σε επίπεδο χωροχρόνο
έχει μελετηθεί στην εργασία \cite{CaTho}. Συγκεκριμένα, σημαντικές εφαρμογές των συνενώσεων αυτών αποτελεί
το γεγονός ότι συνδέουν όλα τα είδη των χορδών, και μέσω της αντιστοιχίας \en AdS/CFT \gr
μπορεί να υπολογίσει την ενέργεια δυονίων και συγκεκριμένα ενός βαρέου ζεύγους
κουάρκ-μονοπόλου \cite{Minahan}.\\
\no
Σκοπός αυτού του μέρους της διατριβής είναι η κατασκευή συνενώσεων χορδών
κατάλληλων για τον υπολογισμό της ενέργειας αλληλεπίδρασης ισχυρώς συζευγμένων βαριών δυονίων,
εντός της αντιστοιχίας \en AdS/CFT. \gr Η κατασκευή αυτή γίνεται για μια μεγάλη κατηγορία
καμπυλομένων υπόβαθρων, μη σύμμορφων και με μειωμένη ή και καθόλου υπερσυμμετρία.
Έχοντας κατασκευάσει αυτές τις συνενώσεις χορδών δεν σημαίνει ότι
το αντίστοιχο δυναμικό που περιγράφει την αλληλεπίδραση του δυονίου είναι φυσικώς επιτρεπτό.
Θα πρέπει να εξετάσουμε την ευστάθεια των λύσεων τουλάχιστον κάτω από μικρές διακυμάνσεις
και η οποία θα μας περιορίσει τον παραμετρικό χώρο για τον οποίο η λύση είναι ευσταθής.
Τέτοιου είδους μελέτες περιγράφησαν διεξοδικά για τα μεσόνια και σκοπός μας σε αυτό
το μέρος είναι η γενίκευση αυτών των αποτελεσμάτων για την περίπτωση των δυονίων. Σε αντίθεση με
την περίπτωση των μεσονίων θα δούμε ότι η ανάλυση της ευστάθειας των συνενώσεων χορδών
οδηγεί σε αναμενόμενα και μη αποτελέσματα, θα δούμε συγκεκριμένα ότι τα αποτελέσματα
της ανάλυσης της ευστάθειας των λύσεων έρχονται σε αντίθεση με ενεργειακά επιχειρήματα
\cite{Sfetsos:2007nd}.\\
\no
Η οργάνωση αυτού του μέρους της διατριβής είναι η ακόλουθη. Στο 12ο κεφάλαιο θα διατυπώσουμε
το πρόβλημα των συνενώσεων χορδών για μια μεγάλη κατηγορία υποβάθρων. Θα αποδείξουμε γενικές
σχέσεις από τις οποίες μπορούμε να υπολογίσουμε την ενέργεια αλληλεπίδρασης του δυονίου. Στο 13ο
κεφάλαιο θα μελέτησουμε με λεπτομέρεια την ανάλυση της ευστάθειας των συνενώσεων χορδών
κάτω από μικρές διακυμάνσεις με σκοπό να ανακαλύψουμε τις φυσικώς επιτρεπτές περιοχές.
Θα δώσουμε ιδιαίτερη έμφαση στην διατύπωση και ερμηνεία των συνοριακών συνθηκών και θα εξάγουμε
γενικά συμπεράσματα για την ευστάθεια για τα είδη των διακυμάνσεων. Στο 14ο κεφάλαιο, θα παρουσιάσουμε
διάφορα παραδείγματα από συνενώσεις χορδών, όπως οι μελανές $D3$ βράνες και
οι πολυκεντρικές λύσεις $D3$ βρανών στον κλάδο \en Coulomb, \gr οι οποίες λύσεις
εντός της  αντιστοιχίας \en AdS/CFT \gr περιγράφουν την ${\cN}=4$ \en SYM \gr για πεπερασμένη
θερμοκρασία και σε γενικά σημεία του κλαδου \en Coulomb \gr της θεωρίας, αντίστοιχα.
Στο 15ο κεφάλαιο, θα εφαρμόσουμε τα γενικά αποτελέσματα του 13ου κεφαλαίου για την μελέτη των
παραδειγμάτων του 14ου κεφαλαίου.

\chapter{Κλασικές λύσεις}
\begin{figure}[!t]
\begin{center}
\begin{tabular}{cc}
\includegraphics[height=6cm]{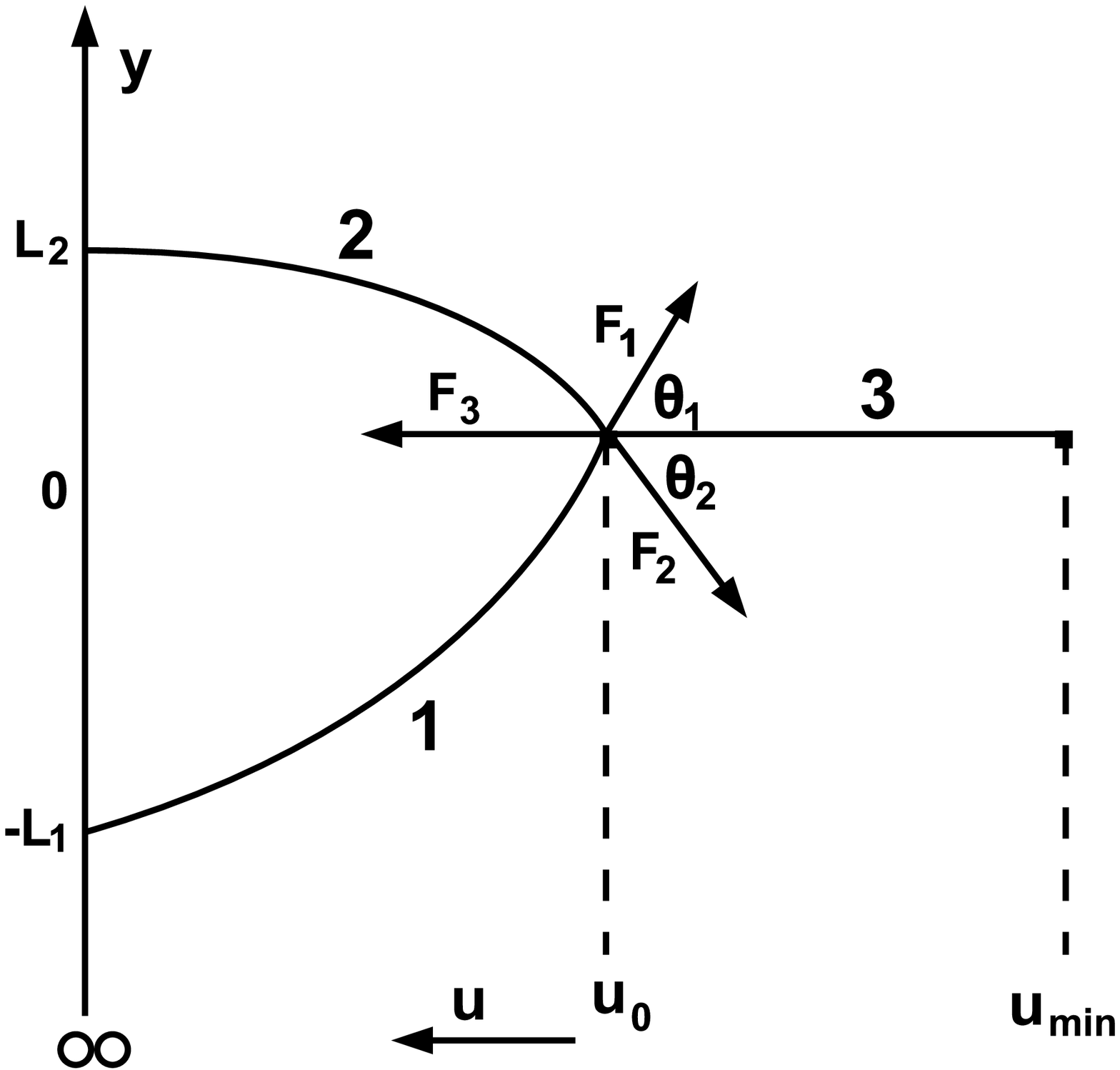}
\end{tabular}
\end{center}
\vskip -.5 cm \caption{Διάταξη συνένωσης: συμβολίζουμε με ${\bf F}_{i}$, $i=1,2,3$ (
${\bf F}_{pq}$ στο κείμενο)
τις δυνάμεις οι οποίες ασκούνται από κάθε χορδή στο σημείο συνένωσης $u_0$.
Τα σημεία στροφής (\en turning points) \gr $u_i$, $i=1,2$
για τις χορδές 1 και 2, δεν συμβολίζονται και βρίσκονται στα δεξιά του $u_0$.
Έχουμε πάντα ότι $u_{\rm min}<u_0$,
αλλά τα $u_i$ μπορεί να είναι και να μην είναι μεγαλύτερα από το $u_{\rm min}$.}
\label{junction}
\end{figure}

Σε αυτό το κεφάλαιο θα αναπτύξουμε την γενική μεθοδολογία
υπολογισμού ενός βαριού δυονίου εντός της αντιστοιχίας \en AdS/CFT
\gr με γενικά φορτία \en Neveu-Schwarz (NS) \gr και \en
Ramond-Ramond (RR) \cite{Schwarz}. \gr Οι υπολογισμοί αυτοί εμπεριέχουν τρείς
χορδές και ως εκ τούτου βασίζονται στα αποτελέσματα που
αναπτύχθηκαν στο μέρος των μεσονίων, και του υπολογισμού της
ενέργειας αλληλεπίδρασης ενός δυονίου στην σύμμορφη περίπτωση
\cite{Minahan}.\\
\no Θεωρούμε ένα υπόβαθρο ενός ασυμπτωτικού χώρου $AdS_5\times S^5$
Εξ.\eqn{7-3},\eqn{7-6} καθώς και τις συναρτήσεις
που ορίστηκαν στις Εξ.\eqn{7-4},\eqn{7-5} με
το σύμμορφο τους όριο να δίνεται από την Εξ.\eqn{7-7}.
Θεωρούμε επίσης τους ανάλογους ορισμούς για τις συντεταγμένες $(x,y,\th)$
ως εγκάρσιες, διαμήκεις και γωνιακές διακυμάνσεις.\\
\no
Μια συνένωση χορδών αποτελείται από τρείς συνεπίπεδες χορδές
οι οποίες ενώνονται όπως φαίνεται στο Σχήμα \ref{junction}. Η χορδή που
συμβολίζεται με $1$ έχει φορτία $(p,q)$ \en NS \gr και \en RR \gr,
η δεύτερη που συμβολίζεται με $2$ έχει φορτία $(p^\prime,q^\prime)$,
και η τρίτη ευθεία χορδή που συμβολίζεται με $3$ έχει φορτία
$(m,n)=(p+p^\prime,q+q^\prime)$ λόγω της διατήρησης φορτίου στο σημείο συνένωσης των χορδών.
Οι θεμελιώδεις χορδές είναι η θεμελιώδης χορδή $F$ με φορτίο $(1,0)$ και η $D$ χορδή με
φορτίο $(0,1)$. Όλες οι άλλες χορδές μπορούν να επιτευχθούν με έναν μετασχηματισμό
$SL(2,\mathbb{R})$ ο οποίος ικανοποιεί την συνθήκη $p\ q^{\prime}-q\ p^{\prime}=\pm 1$ \cite{Gabe}.
Οι δύο πρώτες χορδές (1 και 2) έχουν το ένα άκρο τους στο σύνορο του χωροχρόνου που αντιστοιχεί
στο υπεριώδες $(u\to\infty)$, ενώ η τρίτη ευθεία χορδή έχει το άκρο της στο υπέρυθρο $(u=u_{min})$.
Οι τρείς αυτές χορδές τέμνονται στο σημείο συνένωσης στο $u=u_0$. Εντός της αντιστοιχίας
\en AdS/CFT, \gr η ενέργεια αλληλεπίδρασης του δυονίου $(p,q)$ δίνεται από
\be
\label{13-1}
e^{-{\rm i} E T} = \langle W(C) \rangle = e^{{\rm i} S_{p,q}[C]}\ ,
\ee
όπου
\be
\label{13-2}
S_{p,q}[C] = - {T_{p,q} \ov 2 \pi} \int d \tau d \sigma \sqrt{- \det g_{\a \b} }\ ,
\qq g_{\a\b} = G_{\mu\nu}
\partial_\alpha x^\mu \partial_\b x^\nu \ ,
\ee
είναι η δράση \en Nambu--Goto \gr για μια χορδή η οποία διαδίδεται στο δυικό
υπερβαρυτικό υπόβαθρο και της οποίας τα άκρα βρίσκονται πάνω στον ορθογώνιο παραλληλόγραμμο βρόχο \en Wilson
\gr με μια χωρική και μια χρονική ακμή και όπου $T_{p,q}=\sqrt{p^2+q^2/g_s^2}$.
Αντίστοιχες εκφράσεις ισχύουν για τις χορδές 2 και 3. Τέλος, δεν θα θεωρήσουμε
συνεισφορά από τον όρο \en Wess-Zumino, \gr επειδή είτε δεν υπάρχει είτε
είναι μηδέν στις περιπτώσεις που θα θεωρήσουμε.\\
\no
Στην συνέχεια καθορίζουμε την διδιάστατη παραμετροποίηση επιλέγοντας την βαθμίδα
\be
\label{13-3}
t=\tau \ ,\qq u=\s \ .
\ee
Υποθέτωντας ότι δεν έχουμε εξάρτηση από τον χρόνο, θεωρούμε την παρακάτω εμβάπτιση
\be
\label{13-4}
y = y(u)\ ,\qq x =0\ ,\qq \th = \th_0 = \textrm{σταθερά}\ ,\qq \hbox{Υπόλοιπες} = \textrm{σταθερές}\ ,
\ee
η οποία υπόκεινται στις συνοριακές συνθήκες
\be
\label{13-5}
u \left(-L_1\right)=u \left(L_2\right) = \infty\ ,
\ee
κατάλληλες για ένα δυόνιο $(p,q)$ τοποθετημένο στο $y=-L_1$ και ένα
δυόνιο $(p^\prime,q^\prime)$ τοποθετημένο στο $y=-L_2$. Η χορδή με
φορτία $(m,n)$ είναι ευθεία $y=\textrm{σταθερά}$ και εκτείνεται από το σημείο συνένωσης
μέχρι το $u=u_{min}$. Η σταθερή τιμή στις γωνιακές (μη αγνοήσιμες) συντεταγμένες δίνεται από
τις Εξ.\eqn{7-13}.

\no
Με την παραπάνω υπόθεση, η δράση \en Nambu--Goto \gr γράφεται
\be
\label{13-6}
S_{p,q} = - {T_{p,q}{\cal T}  \ov 2 \pi} \int d u \sqrt{ g(u) + f_y(u) y^{\prime 2}}\ ,
\ee
όπου το ${\cal T}$ συμβολίζει την χρονική ακμή του βρόχου \en Wilson, \gr ο τόνος
συμβολίζει παραγώγιση ως προς την μεταβλητή $u$ ενώ $g(u) \equiv
g(u,\th_0)$ και $f_y(u) \equiv f_y(u,\th_0)$ είναι οι συναρτήσεις
της Εξ.\eqn{7-4} υπολογισμένες στην σταθερά τιμή $\th_0$ της $\th$.
Αντίστοιχες δράσεις μπορούν να γραφούν και για την χορδή $(p^\prime,q^\prime)$,
καθώς και για την ευθεία $(m,n)$. Το γεγονός ότι η πυκνότητα Λαγκρανζιανής
είναι ανεξαρτητη από την μεταβλητή $y$, σημαίνει ότι η αντίστοιχη
γενικευμένη ορμή διατηρείται, ως εκ τούτου
\be
\label{13-7} {f_y y_{\rm cl}^\prime \ov \sqrt{ g + f_y y_{\rm
cl}^{\prime 2}}} = \mp f_{yi}^{1/2}\qq \Longrightarrow \qq y_{\rm
cl}^\prime = \mp {\sqrt{f_{yi} F_i}\ov f_y}\ ,
\ee
όπου $u_i$ είναι οι τιμές της μεταβλητής $u$ στα σημεία στροφής της κάθε χορδής, $f_{yi}
\equiv f_y(u_i)\ ,$ $f_{y0}\equiv f_y(u_0)$ και $y_{\rm cl}$
είναι η κλασική λύση με τα δύο πρόσημα να αντιστοιχούν στην κάτω (χορδή 1) και άνω λύση (χορδή 2).
Το σύμβολο $F_i$ (να μην το συγχύσουμε με τις δυνάμεις στο Σχήμα \ref{junction}) αντιστοιχεί στην συνάρτηση
\be
\label{13-8} F_i = {g f_y \ov f_y - f_{yi}}\ ,\qq i=1,2\ .
\ee
Σημειώστε ότι, για το σημείο διακλάδωσης $u_0$ έχουμε πάντα ότι $u_0 \geq u_{min}$.
Για τα σημεία στροφής έχουμε ότι $u_i\leq u_0$, αλλά δεν είναι απαραιτήτως μεγαλύτερα
από το $u_{min}$, διότι οι χορδές 1 και 2 δεν εκτείνονται μέχρι τα σημεία στροφής. Ολοκληρώνοντας
την Εξ.\eqn{13-8}, βρίσκουμε το διαχωριστικό μήκος
\be
\label{13-9}
L =L_1+L_2=f_{y1}^{1/2}
\int_{u_0}^{\infty} d u {\sqrt{F_1} \ov f_y}+f_{y2}^{1/2} \int_{u_0}^{\infty}
d u {\sqrt{F_2} \ov f_y}\ .
\ee
Τέλος, αντικαθιστώντας την λύση για την $y_{\rm cl}^\prime$ στην Εξ.\eqn{13-6} και
αφαιρώντας τις αποκλίνουσες αυτοενέργειες των δυονίων, γράφουμε την ενέργεια αλληλεπίδρασης
ως
\ba
\label{13-10}
&&E = {1 \ov 2\pi}\left(T_{p,q}{\cal E}_1  +T_{p^\prime,q^\prime} {\cal E}_2
+ T_{m,n} \int_{u_{\rm min}}^{u_0} d u \sqrt{g}\right)\ , \\
&&{\cal E}_i=\int_{u_0}^\infty d u \sqrt{F_i} -
\int_{u_{\rm min}}^\infty d u \sqrt{g}\ ,\qq i=1,2\ \nonumber.
\ea
Στην ιδανική περίπτωση, θα θέλαμε να υπολογίσουμε τα ολοκληρώματα
των εξισώσεων \eqn{13-9} και \eqn{13-10}, να λύσουμε την Εξ.\eqn{13-9}
ως προς την παράμετρο $u_0$ και να αντικαθιστήσουμε στην Εξ.\eqn{13-10}
και να εκφράσουμε την ενέργεια σαν συνάρτηση του μήκους $L$ και των $u_1,u_2$.
Στην συνέχεια θα ελαχιστοποιούσαμε την έκφραση αυτή ως προς τις παραμέτρους
$u_1,u_2$ και θα υπολογίζαμε την ενέργεια αλληλεπίδρασης συναρτήσει του διαχωριστικού μήκους
$L$ για ένα ζεύγος βαριών δυονίων.\\
\no
Ωστόσο, αυτό δεν μπορεί να γίνει ακριβώς, παρά μόνο στην
σύμμορφη περίπτωση, επομένως οι Εξ.\eqn{13-9} και \eqn{13-10} μπορούν να θεωρηθούν
ως παραμετρικές εξισώσεις με παραμέτρους $u_0,u_1,u_2$. Ένας πιο
εύκολος τρόπος για να βρούμε τις παραμέτρους $u_1,u_2$ είναι
απαιτώντας η συνισταμένη δύναμη στο σημείο συνένωσης να ισούται με μηδέν
\cite{Schwarz,Sen,CaTho}, μια προσέγγιση που εφαρμόστηκε για το παρόν πρόβλημα στην
σύμμορφη περίπτωση \cite{Minahan}. Τα απειροστά μήκη σε κάθε χορδή είναι ίσα με
\be
\label{13-11}
d\ell_i^2=(G_{yy} y'^2_{\rm cl} + G_{uu}) du^2 = -{1\ov G_{tt}}F_i du^2\ ,\qq i=1,2\ .
\ee
Επομένως από την δράση Εξ.\eqn{13-6} οι τάσεις των χορδών στο σημείο στροφής $u=u_0$
είναι $ \displaystyle{T_{p,q}\ov 2\pi}\sqrt{-G_{tt}}$ και ομοίως για τις άλλες δύο
χορδές που συναντιούνται στο σημείο στροφής. Οι γωνίες μεταξύ των χορδών και του άξονα $u$
στο σημείο στροφής είναι
\ba
\label{13-12}
{\textrm{Για}}\ u=u_0:\qq \tan \th_i = \Bigg| {\sqrt{G_{yy}} dy\ov \sqrt{G_{uu}} du}\Bigg |
\qq \Longrightarrow \qq
\sin\th_i=\sqrt{{f_{yi}\ov f_{y0}}}\ , \qq i=1,2,
\ea
όπου χρησιμοποιήσαμε την Εξ.\eqn{13-7} και έχουμε λάβει υπ'όψιν πως οι μετατοπίσεις στους άξονες
υπολογίζονται με καμπύλη μετρική. Οι δυνάμεις οι οποίες εξέρχονται από την κάθε χορδή στο επίπεδο
$u-y$ είναι
\ba
\label{13-13}
&&{\bf F}_{p,q}={T_{p,q}\ov 2\pi}\sqrt{-G_{tt}}\ (-\cos\th_1,\sin\th_1)\ ,
\nonumber  \\
&&{\bf F}_{p^{\prime},q^{\prime}} = -{T_{p^{\prime},q^{\prime}}\ov 2\pi}\sqrt{-G_{tt}}\ (\cos\th_2,\sin\th_2)\ ,
\\
&&{\bf F}_{m,n} = {T_{m,n}\ov 2\pi}\sqrt{-G_{tt}}\ (1,0)\ ,
\nonumber
\ea
όπου η πρώτη συνιστώσα είναι η $u$ και η δεύτερη είναι η $y$. Απαιτώντας η συνολική δύναμη να είναι μηδέν
βρίσκουμε ότι οι γωνίες στο σημείο ισορροπίας είναι
\ba
\label{13-14}
\cos\th_1 &=& {T_{m,n}^2+T_{p,q}^2-T_{p^{\prime},q^{\prime}}^2\ov{2 T_{m,n}T_{p,q}}}\ , \nonumber\\
\cos\th_2 &=& {T_{m,n}^2+T_{p^{\prime},q^{\prime}}^2-T_{p,q}^2\ov{2 T_{m,n}T_{p^{\prime},q^{\prime}}}}\ .
\ea
Από τις εκφράσεις αυτές οι γωνίες δίνονται συναρτήσει των φορτίων \en NS \gr και \en RR \gr
των χορδών οπότε στην συνέχεια από την Εξ.\eqn{13-12} μπορούμε να εκφράσουμε τις παραμέτρους
$u_i$ συναρτήσει της παραμέτρου $u_0$ και των φορτίων των χορδών. Συνεπώς απαλείφοντας με αυτόν τον τρόπο
τις παραμέτρους $u_i$, έχουμε μια έκφραση της ενέργειας αλληλεπίδρασης ως συνάρτηση του μήκους
και των φορτίων των χορδών.\\
\no
Μπορούμε να επαληθεύσουμε ότι αυτή η προσέγγιση καθορισμού των παραμέτρων $u_i$ μέσω
της εξισορρόπησης των δυνάμεων στο σημείο συνένωσης είναι ισοδύναμη με την ελαχιστοποίηση της ενέργειας.
Εφόσον δεν μπορούμε να λύσουμε την Εξ.\eqn{13-9} για γενική τιμή της παραμέτrου $u_0$,
θα υποθέσουμε ότι έχουμε μια αναλυτική έκφραση της παραμέτρου $u_0$ συναρτήση των $L,u_1,u_2$,
αντιστρέφοντας την Εξ.\eqn{13-9}. Εφόσον θέλουμε να ελαχιστοποιήσουμε την ενέργεια ως προς
τις παραμέτρους $u_i$, το φυσικό μήκος $L$ δεν θα εξαρτάται από αυτές, δηλαδή
\ba
\label{13-15}
{\partial L\ov\partial u_i}=0\ ,\qq i=1,2\ ,
\ea
\no
το οποίο μπορεί να χρησιμοποιηθεί μαζί με την Εξ.\eqn{13-12}
για να εκφράσουμε την παράγωγο του $u_0$ ως προς $u_i$ σαν
\ba
\label{13-16}
{\partial u_0\ov\partial u_i}={1\ov \tan\th_1+\tan\th_2}\ {f_{yi}^\prime\ov 2}\ \sqrt{f_{y0} \ov g_0f_{yi}}\
\int_{u_0}^\infty du {\sqrt{gf_y}\ov(f_y-f_{yi})^{3/2}}\ ,\qq i=1,2\ .
\ea
Τονίζουμε ότι για την απόδειξη της παραπάνω σχέσης δεν έχουμε χρησιμοποιήσει την εξισορρόπηση δυνάμεων.
Ισοδύναμα, δεν έχουμε θεωρήσει ότι οι γωνίες $\th_i$ είναι σταθερές πέραν του σημείου ισορροπίας. Τέλος,
χρησιμοποιώντας την Εξ.\eqn{13-16} και την εξισορρόπηση δυνάμεων, μετά από προσεκτικούς υπολογισμούς
βρίκουμε το προσδοκώμενο αποτέλεσμα
\be
\label{13-17}
{\partial E\ov\partial u_i}=0\ ,\qq i=1,2 \ .
\ee
Η απόδειξη μας είναι εντελώς γενική για υπόβαθρα της μορφής Εξ.\eqn{7-3} και
δεν χρειάστηκε  να υπολογίσουμε τα ολοκληρώματα που αντιστοιχούν σε συγκεκριμένα
παραδείγματα. ’ν είχαμε ακολουθήσει αυτήν την προσέγγιση θα είχαμε καταλήξει σε περίπλοκες εκφράσεις
και η απόδειξη μας θα στηριζόταν στο να αποδείξουμε διάφορες ταυτότητες, εν γένει,
με ειδικές συναρτήσεις (για παράδειγμα στην σύμμορφη περίπτωση \cite{Minahan}).\\
\no
Αξίζει να σημειώσουμε ότι η έκφραση της ενέργειας αλληλεπίδρασης $E$ ως συνάρτηση του μήκους
$L$ για την διάταξη $(p,q)$ και $(q,p)$ δυονίων είναι αναλλοίωτη κάτω από τον $S$-δυικό μετασχηματισμό
$g_s\to 1/g_s$, όπως και θα έπρεπε. Ωστόσο, αυτό δεν ισχύει
για το μήκος και την ενέργεια ως μεμονομένες συναρτήσεις τις παραμέτρου $u_0$.

\section{Αλληλεπίδραση κουάρκ με μονόπολο}

Η πιο σημαντική εφαρμογή των ανωτέρω είναι στην ενέργεια
αλληλεπίδρασης ενός κουάρκ με ένα μονόπολο με φορτία
$(1,0)$ και $(0,1)$, αντίστοιχα. Σε αυτήν την περίπτωση,
μπορεί εύκολα να δειχθεί από την Εξ.\eqn{13-14} ότι
$\th_1+\th_2=\pi/2$. Στην συνέχεια θα μελετήσουμε το όριο όπου
$g_s\to 1$. Σε αυτήν την περίπτωση
η τάση της χορδής $(0,1)$ που αντιστοιχεί στο μονόπολο γίνεται πολύ μεγάλη,
οπότε περιμένουμε ότι θα παραμείνει σχεδόν ευθεία, ενώ για να υπάρχει εξισορρόπηση
των δυνάμεων, η χορδή $(1,0)$ θα τέμνει κάθετα την ευθεία χορδή. Αυτή
η συμπεριφορά εμφανίζεται στα αναπτύγματα
\ba
\label{13-18}
\th_1 & = & {\pi\ov 2} - g_s + {1\ov 3} g_s^3 + {\cal O}(g_s^5)\ ,
\nonumber\\
\th_2 & = &  g_s - {1\ov 3} g_s^3 + {\cal O}(g_s^5)\ .
\ea
Στην περίπτωση όπου το $g_s\to\infty$, υπάρχει μια ισοδύναμη έκφραση με την εναλλαγή των
$\th_1,\th_2$ και $g_s\to1/g_s$. Στο όριο των μικρών $g_s$ μπορούμε να δούμε από
τις Εξ.\eqn{13-12} και \eqn{13-18} ότι τα σημεία στροφής των χορδών 1 και 2 είναι
\ba
\label{13-19}
u_1 = u_0 + {\cal O}(g_s^2)\ ,\qq u_2 = u_{\textrm{ρίζα}} + {\cal O}(g_s^4)\ ,\qq
\ea
όπου $u_{\textrm{ρίζα}}$ είναι η μεγαλύτερη ρίζα της εξίσωσης $f_{y2}=0$.
Στα παραδείγματα που θα παρουσιάσουμε παρακάτω, $u_{\textrm{ρίζα}}=u_{min}$, εκτός
από μια περίπτωση όπου έχει μικρότερη τιμή. Ως εκ τούτου οι εκφράσεις για το μήκος Εξ.\eqn{13-9}
και την ενέργεια Εξ.\eqn{13-10}, γίνονται
\ba
\label{13-20}
L =f_{y0}^{1/2}
\int_{u_0}^{\infty} d u {\sqrt{F_0} \ov f_{y}} + {\cal O}(g_s) \
\ea
και
\ba
\label{13-21}
E = {1 \ov 2\pi} \left( \int_{u_0}^\infty d u \sqrt{F_0} -
\int_{u_{\rm min}}^\infty d u \sqrt{g}\right) + {\cal O}(g_s)\ ,
\ea
όπου έχουμε ορίσει την συνάρτηση $F_0$ όπως στην Εξ.\eqn{13-8}
με το $f_{yi}$ να αντικαθίσταται με το $f_{y0}\equiv f_y(u_0)$.
Παρατηρούμε ότι οι εκφράσεις
για το μήκος και την ενέργεια είναι μικρότερες κατά έναν παράγοντα
δύο από τις αντίστοιχες εκφράσεις για την περίπτωση του δυναμικού
κουάρκ-αντικουάρκ Εξ.\eqn{7-17} και \eqn{7-18}. Τονίζουμε ότι
αυτό δεν συνεπάγεται ότι το φάσμα των μικρών διακυμάνσεων θα είναι το ίδιο.
Συγκεκριμένα θα δείξουμε ότι συμβαίνει το αντίθετο, και άσχετα από το πόσο
άκαμπτες μπορούν να γίνουν οι δύο χορδές (η ευθεία και μία από τις χορδές 1 και 2),
οι διακυμάνσεις δεν αποσυζεύγνυνται και επηρεάζουν αυτές της τρίτης χορδής.

\chapter{Ανάλυση Ευστάθειας}
Εδώ θα μελετήσουμε την ευστάθεια αυτών των δυονίων,
έχοντας ως σκοπό την εύρεση των φυσικώς αποδεκτών περιοχών στον
παραμετρικό χώρο στις οποίες η κλασική λύση είναι ευσταθής,
τονίζοντας τις ομοιότητες και διαφορές με την περίπτωση
του δυναμικού κουάρκ και αντικουάρκ.
Οι μικρές διακυμάνσεις της κλασικής λύσης Εξ.\eqn{13-4} ανήκουν
στις τρείς κατηγορίες των εγκάρσιων, διαμήκων και γωνιακών διακυμάνσεων.
Το κεφάλαιο αυτό έχει αρκετές ομοιότητες με το 7ο κεφάλαιο αλλά και αρκετές
ουσιώδεις διαφορές, όπως οι συνθήκες συρραφής στο σημείο συνένωσης.

\section{Μικρές διακυμάνσεις}

Θεωρούμε την παρακάτω διαταραχή στην εμβάπτιση

\be
\label{14-1}
x_i = \d x_i (t,u)\ ,\qq y_i = y_{\rm cl,i}(u) + \d y_i (t,u)\ ,\qq \th_i = \th_0 + \d \th_i(t,u)\ ,
\ee
όπου ο δείκτης $i$ αναφέρεται στις τρείς χορδές που δημιουργούν την συνένωση.
Στην συνέχεια υπολογίζουμε την δράση \en Nambu--Goto \gr για αυτήν την περίπτωση και την
αναπτύσουμε σε δυνάμεις των διακυμάνσεων. Ο όρος μηδενικής τάξεως κατά τα γνωστά δίνει
την κλασική δράση και ο όρος πρώτης ταξεως μηδενίζεται χάρη στις κλασικές εξισώσεις κίνησης.\footnote{
Μια προσεκτική μελέτη των όρων πρώτης τάξεως μας οδηγεί στην ύπαρξη επιφανειακών όρων. Απαιτώντας
ότι μηδενίζονται βρίσκουμε τις συνθήκες ισορροπίας στο σημείο συνένωσης. Αυτό μπορεί
να δειχθεί για συνενώσεις σε επίπεδο χωροχρόνο πριν την επιλογή βαθμίδας, δες \cite{CaTho}.
Ωστόσο, για καμπύλους χωροχρόνους έχοντας επιλέξει το $u=\s$ για βαθμίδα, βρίσκουμε
την $y$-συνιστώσα της ισορροπίας δυνάμεων. Αλλάζοντας βαθμίδα σε $y=\s$ βρίσκουμε την
$u$-συνιστώσα.} Το αποτέλεσμα της ανάπτυξης για τους τετραγωνικούς όρους για τις χορδές
$i=1,2$ σε αντιστοιχία με την Εξ.\eqn{8-4} γράφεται ως

\ba
\label{14-2} \!\!\!\!\!\!\!\!\!\!\!\! S^{(i)}_2 &=& - {1 \ov 2\pi} \int
dt du \biggl[ {f_x \ov 2 F_i^{1/2}}
\d x_i^{\prime 2} - {h f_x F_i^{1/2} \ov 2 g f_y} \d \dot{x_i}^2 \nonumber\\
 && \qq\qq\quad\; + {g f_y \ov 2 F_i^{3/2}} \d y_i^{\prime 2} - {h \ov 2 F_i^{1/2}} \d \dot{y_i}^2 \\
&& \qq\qq\quad\; + {f_\th \ov 2 F_i^{1/2}} \d \th_i^{\prime 2} - {h
f_\th F_i^{1/2} \ov 2 g f_y} \d \dot{\th_i}^2 + \left( {1 \ov 4
F_i^{1/2}}
\partial_\th^2 g
+ {f_{y0} F_i^{1/2} \ov 4 f_y^2} \partial_\th^2 f_y \right) \d \th_i^2
\biggr]\ ,
\nonumber
\ea
όπου έχουμε χρησιμοποιήσει και πάλι την Εξ.\eqn{7-13}
ενώ όλες οι συναρτήσεις είναι υπολογισμένες στο σημείο
$\th=\th_0$. Γράφοντας τις εξισώσεις κίνησης για αυτήν την
δράση και κάνοντας μια ανάλυση κατά \en Fourier \gr με συχνότητα $\om$ (\gr δεν έχουμε
εξάρτηση από το χρόνο) βρίσκουμε τις εξισώσεις κίνησης για τις διακυμάνσεις

\ba
\label{14-3}
&&\left[ {d \ov du} \left({f_x \ov F_i^{1/2}} {d \ov du} \right)
+ \omega^2 {h f_x F_i^{1/2} \ov g f_y} \right] \d x_i = 0\ ,
\nonumber\\
&&\left[ {d \ov du} \left( {g f_y \ov F_i^{3/2}} {d \ov du} \right)
+ \omega^2 {h \ov F_i^{1/2}} \right] \d y_i = 0\ ,
\\
&&\left[ {d \ov du} \left({f_\th \ov F_i^{1/2}} {d \ov du} \right)
 + \left( \omega^2 {h f_\th F_i^{1/2} \ov g f_y}
- {1 \ov 2 F_i^{1/2}} \partial_\th^2 g - {f_{y0} F_i^{1/2} \ov 2
f_y^2} \partial_\th^2 f_y \right) \right] \d \th_i = 0\ . \nonumber
\ea

Για την ευθεία χορδή η δράση των τετραγωνικών διακυμάνσεων είναι

\ba
\label{14-4}
S^{(3)}_2 &= & -{1\ov 4\pi}
 \int dt du \biggl[{f_y\ov\sqrt{g}}\d y_3^{\prime 2}+{f_x\ov\sqrt{g}}\d x_3^{\prime 2}+{f_{\th}\ov\sqrt{g}}
\d \th_3^{\prime 2}-
{h\ov\sqrt{g}}\d\dot{y}_3^2-{f_x h\ov \sqrt{g}f_y}\d\dot{x}_3^2
\nonumber\\
&&\phantom{xxxxxxxxxx}
-{f_{\th} h\ov \sqrt{g}f_y}\d\dot{\th}_3^2  +{\del^2_\th g\ov 2 \sqrt{g}} \d\th_3^2\biggr]\ ,
\ea
από την οποία προκύπτουν οι εξισώσεις

\ba
\label{14-5}
&&
{d\ov du}\left({f_x\ov\sqrt{g}}\d x_3^{\prime}\right)+\omega^2{hf_x\ov\sqrt{g}f_y}\d x_3=0\ ,
\nonumber \\
&&{d\ov du}\left({f_y\ov\sqrt{g}}\d y_3^{\prime}\right)+\omega^2{h\ov\sqrt{g}}\d y_3=0\ ,
 \\
&&{d\ov du}\left({f_{\th}\ov\sqrt{g}}\d \th_3^{\prime}\right)
+\omega^2{hf_{\th}\ov\sqrt{g}f_y}\d\th_3
={1\ov 2}{\partial_\th^2 g\ov\sqrt{g}}\d\th_3\ .
\nonumber
\ea
Σημειώστε ότι για την ευθεία χορδή μπορούμε να πάρουμε τις εξισώσεις
ως το όριο όπου\\ $F_i\to g$ ($f_{y0}\to 0$) στην Εξ.\eqn{14-3}.
\no
Επομένως το πρόβλημα εύρεσης της ευστάθειας
ανάγεται στην λύση ενός προβλήματος ιδιοτιμών \en Sturm--Liouville
\gr (Εξ.\eqn{8-7} ) με την διαφορά ότι στην περίπτωση μας υπάρχει εξάρτηση
από το σημείο συνένωσης $u_0$ και τα φορτία \en NS \gr και \en RR
\gr των χορδών. Ο σκοπός μας είναι η εύρεση του εύρους τιμών του $u_0$ για τις οποίες
το $\om^2$ είναι αρνητικό. Η εύρεση αυτού του θα γίνει
καθορίζοντας την μηδενική ιδιοτιμή, το οποίο είναι ένα ευκολότερο πρόβλημα.
Σημειώνουμε, επίσης ότι τα τρία είδη διακυμάνσεων είναι έμμεσα συζευγμένα
λόγω των συνθηκών συρραφής στο σημείο συνένωσης παρότι οι εξισώσεις κίνησης δεν είναι
συζευγμένες.

\section{Συνοριακές συνθήκες}

Για να λύσουμε το πρόβλημα ιδιοτιμών, πρέπει να εφαρμόσουμε συνοριακές συνθήκες
στο υπεριώδες ($u\to\infty$) για τις διακυμάνσεις, στο υπέρυθρο ($u=u_{min}$),
και συνθήκες συρραφής στο σημείο συνένωσης στο $u=u_0$. Η συνοριακή συνθήκη στο υπέρυθρο
αντιπροσωπεύει το γεγονός ότι ότι δεν μετακινούμε τα δυόνια στο σύνορο, δηλαδή

\be
\label{14-6}
\Phi(u) = 0\ ,\qq \textrm{όταν}\quad u\to \infty \ ,
\ee
όπου η $\Phi$ είναι οποιαδήποτε από τις τρείς διακυμάνσεις των χορδών 1 και 2.
Στην περιοχή του υπέρυθρου, στην οποία εκτείνεται η χορδή 3, απαιτούμε η λύση και
η παράγωγος της ως προς $u$ να είναι πεπερασμένες. Αλλιώς, δεν θα ισχύε η προσέγγιση
των μικρών διακυμάνσεων.\\
\no
Απαιτούμε οι εγκάρσιες διακυμάνσεις $\d x$ όπως και οι μεταβολές τους
στην δράση να είναι ίσες στο σημείο συνένωσης, ώστε να μην σπάσει η συνένωση των χορδών στο
$u=u_0$. Επιπλέον, απαιτούμε ο συνοριακός όρος που προκύπτει απο την εύρεση
των εξισώσεων κίνησης μέσω της δράσης των τετραγωνικών διακυμάνσεων $S_2^{(i)}$ να μηδενίζεται.
Οι απαιτήσεις αυτές ισοδυναμούν με

\ba
\label{14-7}
&&\d x_1=\d x_2=\d x_3\ , \qq {\textrm{όταν}} \quad u=u_0\ ,
\nonumber \\
&&T_{p,q}\cos\th_1\d x_1^{\prime}+T_{p^\prime,q^\prime}\cos\th_2\d x_2^{\prime}
-T_{m,n}\d x_3^\prime=0\ , \qq {\textrm{όταν}} \quad u=u_0\ ,
\ea
όπου χρησιμοποιήσαμε την Εξ.\eqn{13-12} για να απλοποιήσουμε την δεύτερη σχέση.\\
\no
Παρόμοιες σχέσεις προκύπτουν επίσης για τις γωνιακές διακυμάνσεις $\d\th$

\ba
\label{14-8}
&&\d \th_1=\d \th_2=\d \th_3\ , \qq {\textrm{όταν}} \quad u=u_0\ ,
\nonumber \\
&&T_{p,q}\cos\th_1\d \th_1^{\prime}+T_{p^\prime,q^\prime}\cos\th_2\d \th_2^{\prime}
-T_{m,n}\d \th_3^\prime=0\ , \qq {\textrm{όταν}} \quad u=u_0\ .
\ea
Για την περίπτωση των διαμήκων διακυμάνσεων, οι συνοριακές συνθήκες, σε αναλογία
με την περίπτωση του μεσονίου, πρέπει να βρεθούν σε ένα σύστημα
στο οποίο δεν θα αλλάζει η κλασική λύση. Αυτό επιτυγχάνεται μέσω της
Εξ.\eqn{8-17}. Στο σημείο συνένωσης οι διακυμάνσεις $\d u$ είναι
συνεχείς και αυτό οδηγεί σε μια ασυνέχεια στις $\d y$ λόγω της
Εξ.\eqn{8-17} για τις χορδές 1 και 2. Όπως πριν, μηδενίζουμε
τον συνοριακό όρο που προκύπτει κατά την εύρεση των συνοριακών συνθηκών από
τις $S_2^{(i)}$. Χρησιμοποιώντας και την Εξ.\eqn{13-12} έχουμε για
τις διαμήκεις διακυμάνσεις $\d y$ τις συνοριακές σχέσεις:
\ba
\label{14-9}
&&\d y_1\cot\th_1+\d y_2\cot\th_2=0\ ,\qq {\textrm{όταν}} \quad u=u_0\ ,
\nonumber \\
&&T_{p,q}\cos^3\th_1\d y_1^{\prime}+T_{p^\prime,q^\prime}\cos^3\th_2\d y_2^{\prime}
-T_{m,n}\d y_3^\prime=0\ ,\qq {\textrm{όταν}} \quad u=u_0\ .
\ea

\section{Μηδενικοί τρόποι}

\subsection{Εγκάρσιοι μηδενικοί τρόποι}

Ο μηδενικός τρόπος για τις εγκάρσιες διακυμάνσεις των χορδών 1 και 2
που ικανοποιεί την Εξ.\eqn{14-3} είναι

\ba
\label{14-10}
\d x_i   =  a_i I_i(u)\ ,
\qq I_i(u)=\int^{\infty}_u {du\ov f_x} \sqrt{gf_y\ov f_y-f_{yi}}\ ,\qq i=1,2\ ,
\ea
όπου τα $a_i$ είναι πολλαπλασιαστικές σταθερές.
Ο μηδενικός τρόπος για την ευθεία χορδή είναι
\ba
\label{14-11}
\d x_3=a_3\int^{u_{0}}_u du {\sqrt{g}\ov f_x}+\textrm{σταθερά}\ .
\ea
Ειδικεύοντας την Εξ.\eqn{14-11} για τα παράδειγματα μας, θα δούμε
ότι για να μην απειρίζεται το $\d x_3$ ή/και $\d x^\prime_3$ όταν
το $u\to u_{min}$, επιλέγουμε $a_3=0$ έτσι ώστε να έχουμε πεπερασμένη λύση.
Οπότε $\d x_3=\textrm{σταθερά}$, και συνεπώς ο μηδενικός αυτός τρόπος
δεν εμφανίζεται στην ανάλυση της ευστάθειας. Από την Εξ.\eqn{14-7} και
την συνθήκη ισορροπίας δυνάμεων, βρίσκουμε ένα γραμμικό σύστημα για
τις παραμέτρους $a_1,a_2$, του οποίου η ορίζουσα πρέπει να μηδενίζεται
ώστε να έχουμε λύση πέρα της μηδενικής. Με τον τρόπο αυτό βρίσκουμε

\ba
\label{14-12}
\sin\th_1 I_1(u_0)+\sin\th_2 I_2(u_0)=0\ .
\ea
Εφόσον $I_i(u_0) > 0$, η συνθήκη αυτή δεν μπορεί να ικανοποιηθεί
και επομένως καταλήγουμε στο συμπέρασμα ότι δεν υπάρχουν μηδενικοί τρόποι
για τις διαμήκεις διακυμάνσεις, με συνέπεια αυτές να
είναι ευσταθείς. Συμπερασματικά, οι εγκάρσιες διακυμάνσεις είναι ευσταθείς
στην περίπτωση του ζεύγους δυονίων όπως ήταν και στο ζεύγος κουάρκ-αντικουάρκ
 (\gr δες ενότητα 7.3.1).

\subsection{Διαμήκεις μηδενικοί τρόποι}

Στην συνέχεια θα θεωρήσουμε τους διαμήκεις μηδενικούς τρόπους.
Ο μηδενικός τρόπος για τις διαμήκεις διακυμάνσεις των χορδών 1 και 2
που ικανοποιεί την Εξ.\eqn{14-3} είναι

\ba
\label{14-13}
\d y_i   = b_i J_i(u)\ ,\qq  J_i(u)=\int^{\infty}_u du {\sqrt{gf_y}\ov (f_y-f_{yi})^{3/2}}\ ,\qq i=1,2 \ ,
\ea
όπου τα $b_i$ είναι πολλαπλασιαστικές σταθερές. Ο μηδενικός τρόπος για την ευθεία χορδή
είναι
\ba
\label{14-14}
\d y_3=b_3\int^{u_0}_u du{\sqrt{g}\ov f_y}+ \textrm{σταθερά}\ .
\ea
Όπως και πριν,  για τα παράδειγματα μας θα δούμε
ότι για να μην απειρίζονται τα $\d y_3$ ή/και $\d y^\prime_3$ όταν
το $u\to u_{min}$, επιλέγουμε το $b_3=0$ ώστε να έχουμε πεπερασμένη λύση.
Κατά συνέπεια ο μηδενικός αυτός τρόπος δεν εμφανίζεται στην ανάλυση της ευστάθειας.
Αντίστοιχα με πριν, απο την Εξ.\eqn{14-9} και την συνθήκη ισορροπίας δυνάμεων
βρίσκουμε ότι
\ba
\label{14-15}
\cos\th_1 J_1(u_0)=\cos\th_2 J_2(u_0)\ .
\ea
Αυτή είναι μια εξίσωση για την παράμετρο $u_0$ συναρτήσει της
σταθεράς ζεύξης της χορδής και των φορτίων των χορδών, η οποία
μπορεί να λυθεί αριθμητικά. Θα συμβολίσουμε την λύση της, όπου αυτή υπάρχει,
με $u_{0c}$. Εν γένει θα έχουμε αναπτύγματα της μορφής
\ba
\label{14-16}
u_{0c}(g_s)& = & u^{(0)}_{0c}+ \sum_{n=1}^\infty u_n^{(0)} g_s^n\ ,
\nonumber\\
u_{0c}(g_s)& = & u^{(\infty)}_{0c}+ \sum_{n=1}^\infty u^{(\infty)}_n g_s^{-n}\ .
\ea
Στην πιο σημαντική περίπτωση της αλληλεπίδρασης ενός κουάρκ και ενός
μονοπόλου, λόγω του $S$-δυικού μετασχηματισμού οι συντελεστές σε κάθε τάξη των αναπτυγμάτων
είναι ίσοι, δηλαδή $ u^{(0)}_{0c}=  u^{(\infty)}_{0c}$ και $ u_n^{(0)}=u^{(\infty)}_n$.\\
\no
Κατά αναλογία με την περίπτωση της ανάλυσης της ευστάθειας του μεσονίου (παράγραφος 7.3.2), τίθεται
η ερώτηση αν η τιμή της παραμέτρου $u_{0c}$ συμπίπτει με τα πιθανά ακρότατα
του μήκους $L(u_0)$, τα οποία στην περίπτωση που υπάρχουν τα συμβολίζουμε με $u_{0m}$ (
ένα τυπικό παράδειγμα παρουσιάζεται στο Σχήμα 13.1)
\begin{figure}[!t]
\begin{center}
\begin{tabular}{cc}
\includegraphics[height=5.2cm]{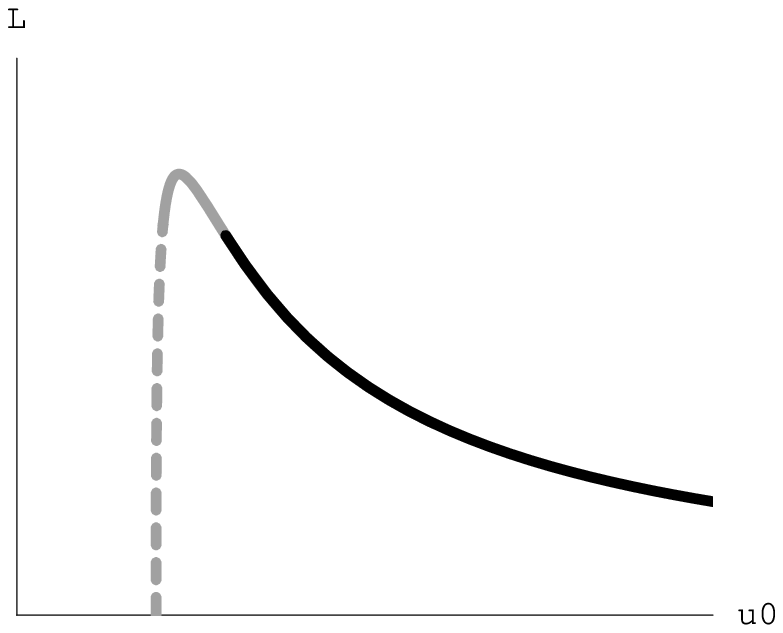}
&\includegraphics[height=5.2cm]{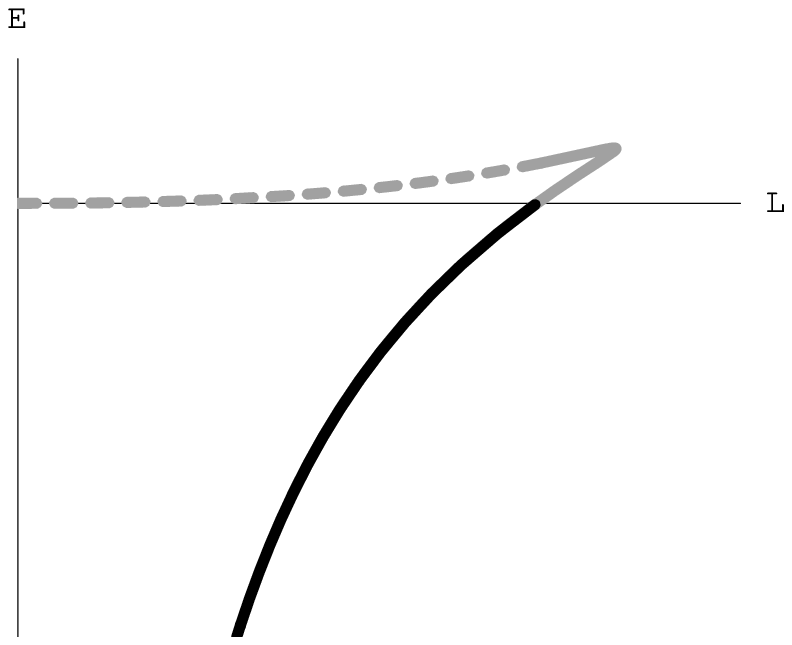}\\
(α) & (β)
\end{tabular}
\end{center}
\vskip -.5 cm \caption{Γραφικές παραστάσεις του $L(u_0)$
και του $E(L)$ για τις περιπτώσεις στην οποίες υπάρχει αστάθεια στις διαμήκεις διακυμάνσεις.
Το διάγραμμα αυτό περιγράφει τις περιπτώσεις της μελανής $D3$ βράνης,
πολυκεντρικών λύσεων $D3$ βρανών σε μία σφαίρα για τροχιές στο $\th_0=0$.
Τα διάφορα είδη γραμμών αντιστοιχούν σε ευσταθής (μαύρη γραμμή)
μετασταθής (γκρί γραμμή) και ασταθής (διακεκομμένη γκρί γραμμή),
των οποίων η ευστάθεια θα καθορίστει απο την ανάλυση του 16ου κεφαλαίου.
}
\label{fig1.dyon}
\end{figure}
Στην περίπτωση του μεσονίου δείξαμε αναλυτικά ότι τα ενεργειακά επιχειρήματα ταυτίζονται με
την ανάλυση της ευστάθειας της κλασική λύσεως, δηλαδή
$u_{0c}=u_{0m}$. Δουλεύοντας ανάλογα με την περίπτωση του μεσονίου
απο τις Εξ.\eqn{13-9} και \eqn{13-10} και χρησιμοποιώντας την Εξ.\eqn{13-12} για
να βρούμε στο σημείο ισορροπίας την παράγωγο των $u_i$ ως προς το $u_0$,
καταλήγουμε στις:
\ba
\label{14-17}
 L^{\prime}(u_0) & = & {f_{y0}^{\prime}\ov\sqrt{f_{y0}}}  \ (\sin\th_1 K_1+\sin\th_2 K_2)\ ,
\nonumber\\
E'(u_0)& = &  {1\ov 2\pi} {f_{y0}^{\prime}} \sin\th_1 T_{p,q} \ (\sin\th_1 K_1+\sin\th_2 K_2)\ ,
\ea
οι οποίες εκφράσεις είναι ανάλογες μεταξύ τους και έχουμε ορίσει
\be
\label{14-18}
 K_i=\int_{u_0}^{\infty}du\ \partial_u
\left({\sqrt{gf_y}\ov f^{\prime}_y}\right){1\ov(f_y-f_{yi})^{1/2}}\ ,\qq i=1,2\ .
\ee
Η τιμή για την οποία η Εξ.\eqn{14-15} ικανοποιείται είναι διαφορετική από την
$u_0=u_{0m}$ όπου και ισχύει $L^\prime(u_{0m})=E^\prime(u_{0m})=0$.
Όπως θα δούμε, στα παραδείγματα, η αστάθεια εμφανίζεται για μικρότερες τιμές ($u_{0c} < u_{0m}$),
συνεπώς τμήμα του άνω κλαδου είναι ευσταθές κάτω από μικρές διακυμάνσεις
παρότι έχει μεγαλύτερη ενέργεια από το αντίστοιχο τμήμα του κάτω κλαδου.
Παρόλα αυτά, η αποκομμένη διάταξη είναι προτιμητέα ενεργειακά για όλες
τις θετικές τιμές της ενέργειας. Οπότε το μέρος του άνω κλάδου το
οποίο ήταν διαταρακτικά ευσταθές είναι στην πραγματικότητα μετασταθές.
Το γεγονός ότι $u_{0c}\neq u_{0m}$ μπορεί να δειχθεί αναλυτικά
για μικρές σταθερές ζεύξης για την πιο σημαντική περίπτωση, την
αλληλεπίδραση κουάρκ και μονόπολου. Μπορούμε να δείξουμε ότι
\be
\label{14-19}
L'(u_0)= {f_{y0}^{\prime}\ov\sqrt{f_{y0}}}  \int_{u_0}^{\infty}du\ \partial_u
\left({\sqrt{gf_y}\ov f^{\prime}_y}\right){1\ov(f_y-f_{y0})^{1/2}}\ + {\cal O}(g_s) \ ,
\ee
(μαζί με μια ανάλογη ανάπτυξη για την ενέργεια $E(u_0)$). Από την έκφραση αυτή μπορούμε
να υπολογίσουμε τον όρο πρώτης τάξεως $u_{0m}^{(0)}$ σε μια ανάπτυση της παραμέτρου
$u_{0m}$ σε αντιστοιχία με την Εξ.\eqn{14-16}. Χρησιμοποιώντας τις Εξ.\eqn{13-18} και
\eqn{13-19} και την ταυτότητα
\be
\label{14-20}
J_i(u_0)=2 K_i(u_0) + 2 {\sqrt{g_0}\ov f'_{y0} \cos\th_i }\ ,
\ee
η οποία μπορεί να αποδειχθεί με ολοκλήρωση κατά παράγοντες, βρίσκουμε
ότι η συνθήκη ύπαρξης μηδενικού τρόπου Εξ.\eqn{14-15} γίνεται
\be
\label{14-21}
2 {\sqrt{g_0}\ov f'_{y0}} - \int_{u_0}^\infty du {\sqrt{g}\ov f_y}= {\cal O}(g_s)\ ,
\ee
από την οποία μπορούμε να βρούμε τον όρο πρώτης τάξεως $u_{0c}^{(0)}$ στο
ανάπτυγμα της\\ Εξ.\eqn{14-16}. Συγκρίνοντας τις Εξ.\eqn{14-19} και \eqn{14-21}
βρίσκουμε ότι εν γένει $u_{0m}^{(0)}\neq u_{0c}^{(0)}$ και κατά συνέπεια
η τιμή της παραμέτρου $u_0$ για την οποία έχουμε μέγιστο μήκος $u_{0m}$ είναι
διαφορετική απο την τιμή για την οποία έχουμε ύπαρξη μηδενικού τρόπου $u_{0c}$ στις
διαμήκεις διακυμάνσεις, ο οποίος συνεπάγεται την αστάθεια του συστήματος.\\
\no
Παρατηρούμε επίσης ότι η Εξ.\eqn{14-21} μπορεί να προκύψει από τις συνοριακές συνθήκες
Εξ.\eqn{14-9} ειδικεύοντας στην περίπτωση της λύσης μηδενικού τρόπου για τις διαμήκεις διακυμάνσεις.
Τονίζουμε ότι στην απόδειξη έχουμε ενσωματώσει ένα παράγοντα $1/g_s$ στην πολλαπλασιαστική σταθερά
της δεύτερης χορδής. Συνεπώς, είναι προφανές ότι παρότι η δεύτερη χορδή είναι άκαμπτη και σχεδόν
ευθεία, οι διακυμάνσεις της είναι σε σύζευξη με τις διακυμάνσεις της πρώτης χορδής.
Αντίστοιχη επιχειρηματολογία υπάρχει και για τις διεγερμένες καταστάσεις. Το παραπάνω
επιχείρημα εξηγεί γιατί στο όριο $g_s\to 0$ (ή $g_s\to\infty$)
οι συνθήκες ύπαρξης μηδενικού τρόπου είναι διαφορετικές με την περίπτωση του μεσονίου παρόλο που
το σημείο στροφής $u_1\to u_0$ (ή $u_2\to u_0$) και κατά συνέπεια οι εξισώσεις των
διακυμάνσεων \eqn{14-3} είναι οι ίδιες με την περίπτωση του μεσονίου.
Αυτό οφείλεται στην ύπαρξη διαφορετικών συνοριακών συνθήκων.\\
\no
Τέλος, θα σχολιάσουμε αν ο υπολογισμός των δυναμικών δυονίου-δυονίου μέσω
της αντιστοιχίας \en AdS-CFT \gr παραβιάζει κάποιες γενικές αρχές σε
σχέση με το πρόσημο και την μονοτονία της δύναμης. Για παράδειγμα,
όπως προανεφέρθει στην περίπτωση του μεσονίου η δύναμη πρέπει
να είναι ελκτική και να είναι μια αύξουσα συνάρτηση της απόστασης \cite{concavity}.
Αυτή η συνθήκη κυρτότητας μας οδηγεί στο συμπέρασμα ότι ο άνω κλάδος
του διαγράμματος $E-L$ είναι ασταθής και το γεγονός αυτό επιβεβαιώθηκε
από την ανάλυση της ευστάθειας της κλασικής λύσεως.
Στην περίπτωση μας, χρησιμοποιώντας την Εξ.\eqn{14-17}, βρίσκουμε
\ba
\label{14-22}
{dE\ov d L}& = & {T_{p,q} \sin\th_1\ov 2\pi} \sqrt{f_{y0}}\ ,
\nonumber\\
{d^2E\ov d L^2} & = & {T_{p,q} \sin\th_1\ov 4\pi} {f'_{y0}\ov \sqrt{f_{y0}} L'(u_0)}\ ,
\ea
η οποία μπορεί να γραφεί σε συμμετρικό τρόπο για τις χορδές 1 και 2 χρησιμοποιώντας την
$y$ συνιστώσα της συνθήκης εξισορρόπησης δυνάμεων. Το αποτέλεσμα της Εξ.\eqn{14-21} είναι
ανάλογο με αυτό στην περίπτωση του μεσονίου Εξ.\eqn{7-20}, \cite{bs}. Από την άλλη
πλευρά, βρήκαμε ότι μέρος του άνω κλάδου είναι διαταρακτικά ευσταθές, οπότε η συνθήκη κυρτότητας
έρχεται σε αντίθεση με την ανάλυση της ευστάθειας σε αυτήν την περίπτωση. Ωστόσο,
στην περίπτωση μας δεν προκύπτει συνθήκη κυρτότητας από την ανάλυση του βρόχου \en Wilson \gr για
την περίπτωση ενός βαριού ζεύγους κουάρκ και μονοπόλου. Αυτό που μπορούμε να δείξουμε στην
περίπτωση μας είναι ότι
με χρήση κατοπτρισμού και θετικότητας, η ενέργεια του ζεύγους προκύπτει να είναι μεγαλύτερη ή
και ίση από τον μέσο όρο ενεργειών των συστημάτων
κουάρκ-αντικουάρκ και \gr μονοπόλου-αντιμονοπόλου.\footnote{Θα ήθελα να ευχαριστήσω τον
Καθηγητή Κώστα Μπαχά για αυτήν την πληροφορία.}

\subsection{Γωνιακοί μηδενικοί τρόποι}

Για τους γωνιακούς μηδενικούς τρόπους δεν μπορούμε να γράψουμε
την λύση για τον μηδενικό τρόπο λόγω ύπαρξης του όρου μάζας
στις αντίστοιχες εξισώσεις \en Sturm--Liouvlle \gr στην τρίτη γραμμή
των Εξ.\eqn{14-3} και \eqn{14-5}. Για το λόγο αυτό, θα αναπτύξουμε
την αντίστοιχη μεθοδολογία στα παραδείγματα που θα λύσουμε. Η μεθοδολογία
αυτή στηρίζεται στην περιγραφή του προβλήματος μέσω της εξίσωσης
\en Schr\"odinger \gr που αναπτύξαμε στην παράγραφο (7.3.3).

\chapter{Παραδείγματα κλασικών λύσεων}
Σε αυτό το κεφάλαιο, θα μελετήσουμε την συμπεριφορά των δυναμικών
δυονίου-δυονίου τα οποία προκύπτουν από τον υπολογισμό του βρόχου
\en Wilson \gr για υπόβαθρα που αντιστοιχούν σε μελανές $D3$ βράνες
και σε πολυκεντρικές $D3$ βράνες.

\section{Μελανές $D3$ βράνες}

Στην περίπτωση αυτή η μετρική αυτού του υποβάθρου στο όριο της θεωρίας
πεδίου και οι συναρτήσεις ($g,f_y,f_x,f_{\th},h$) δόθηκαν στις
Εξ.\eqn{9-1},\eqn{9-3} και \eqn{9-4}, όπου έχουμε κανονικοποιήσει
το μήκος και την ενέργεια ως
$\displaystyle{L \to {1 \ov \m} L\ ,\quad E \to {\m \ov 2\pi} E}$.
Στην περίπτωση αυτή έχουμε ότι $u_{\rm min}=u_{\textrm{ρίζα}}=1$.
Για το σύστημα δυονίου-δυονίου βρίσκουμε λοιπόν ότι το μήκος του
δίνεται συναρτήσει του σημείου συνένωσης $u_0$ ως εξής:

\ba
\label{15-1}
L &=& R^2 \left( \sqrt{u_1^4-1} \int_{u_0}^\infty {du \ov
 \sqrt{(u^4-1)(u^4-u_1^4)}}+(1\rightarrow 2)\right)=\ L_1+L_2\ ,\nonumber\\
L_i&=& R^2 {\sqrt{u_i^4-1}\ov 3u_0^3}
F_1\left({3\ov 4},{1\ov 2},{1\ov 2},{7\ov 4},{1\ov u_0^4},{u_i^4\ov u_0^4}\right)\ , \qq i=1,2\ ,
\ea
και όπου η ενέργεια δίνεται από
\ba
\label{15-2}
E&=&T_{p,q}{\cal E}_1+T_{p^{\prime},q^{\prime}}{\cal E}_2+T_{m,n}(u_0-1)\ ,
\nonumber\\
{\cal E}_i&=&\int_{u_0}^{\infty}du\left(\sqrt{{u^4-1\ov u^4-u_i^4}}-1\right)-(u_0-1)
\\
&=&
-u_0F_1\left(-{1\ov 4},-{1\ov 2},{1\ov 2},{3\ov 4},{1\ov u_0^4},{u_i^4\ov u_0^4}\right)+1\ ,\qq i=1,2\ .
\nonumber
\ea
Στην παραπάνω έκφραση η $F_1(a,b_1,b_2,c,z_1,z_2)$ είναι η υπεργεωμετρική
συνάρτηση \en Appell \gr και $u_0 \geqslant u_i\geqslant 1$.
Από την Εξ.\eqn{13-12} βρίσκουμε ότι τα σημεία στροφής για τις δύο χορδές
δίνονται από
\be
\label{15-3}
u_i=(u_0^4-(u_0^4-1)\cos^2\th_i)^{1/4}\ ,\qq i=1,2\ ,
\ee
όπου οι γωνίες $\th_i$ δίνονται από την Εξ.\eqn{13-14} σαν συνάρτηση των φορτίων των χορδών.
Εφαρμόζοντας την Εξ.\eqn{14-17} σε αυτήν περίπτωση βρίσκουμε ότι:
\ba
\label{15-4}
&&L^\prime(u_0)= R^2 {4u_0^3\ov\sqrt{u_0^4-1}}\left(\sin\th_1K_1+\sin\th_2K_2\right)\ ,
\nonumber \\
&&K_i={3\ov28u_0^7}F_1\left({7\ov 4},{1\ov 2},{1\ov 2},{11\ov 4},{1\ov u_0^4},{u_i^4\ov u_0^4}\right)
-{1\ov12u_0^3}F_1\left({3\ov 4},{1\ov 2},{1\ov 2},{7\ov 4},{1\ov u_0^4},{u_i^4\ov u_0^4}\right)\ .
\ea
Η συνάρτηση $L(u_0)$ έχει ένα ολικό μέγιστο $u_{0\rm m}$ για κάθε τιμή της σταθεράς ζεύξης της χορδής,
η θέση του οποίου εξαρτάται από την σταθερά αυτή και για το σύστημα
κουάρκ και μονόπολου βρίσκουμε την παρακάτω ανάπτυξη
\be
\label{15-5}
u_{0m}\simeq 1.177 - 0.037 g_s + {\cal O}(g_s^2)\ .
\ee
Για $L > L_{max}$, μόνο η αποκομμένη λύση υπάρχει. Για $L < L_{max}$,
η Εξ.\eqn{15-1} έχει δύο λύσεις για την παράμετρο $u_0$, οι οποίες
αντιστοιχούν στην κοντή και μακρυά χορδή, και η ενέργεια $E$ είναι μια δίτιμη
συνάρτηση του μήκους $L$. Η συμπεριφορά που περιγράψαμε παραπάνω εμφανίζεται
στο Σχήμα \ref{fig1.dyon} και είναι πανομοιότυπη με την συμπεριφορά του μήκους και
της ενέργειας στην περίπτωση του μεσονίου, δες παράγραφο 8.1.
Ωστόσο, η συμπεριφορά τους διαφέρει κάτω από μικρές διακυμάνσεις,
αφού όπως θα δούμε με λεπτομέρεια στο επόμενο κεφάλαιο, μέρος του άνω κλάδου είναι διαταρακτικά ευσταθές.
Η κλασική λύση που περιγράψαμε παραπάνω και η δίτιμη συμπεριφορά της ενέργειας έχει επίσης
περιγραφεί στην εργασία \cite{Park}.

\section{Πολυκεντρικές λύσεις $D3$ βρανών σε τριδιάστατη σφαίρα}

Στην συνέχεια θα μελετήσουμε την ειδική περίπτωση ομοιόμορφων κατανομών
$D3$ βρανών πάνω σε μια σε τριδιάστατη σφαίρα.
Στην περίπτωση αυτή η μετρική αυτού του υποβάθρου\\ στο όριο της θεωρίας
πεδίου και οι συναρτήσεις ($g,f_y,f_x,f_{\th},h$) δόθηκαν στις
Εξ.\eqn{9-23},\\ \eqn{9-24}, \eqn{9-25} και \eqn{9-26}, όπου έχουμε κανονικοποιήσει
το μήκος και την ενέργεια ως\\
$\displaystyle{L \to {1 \ov r_0} L\ , E \to {r_0 \ov 2\pi} E}$.

\subsection{Η τροχιά για $\th_0=0$}

Σε αυτή την περίπτωση έχουμε ότι $u_{\rm min}=u_{\textrm{ρίζα}}=1$ και
τα ολοκληρώματα για το αδιάστατο μήκος και την αδιάστατη ενέργεια δίνονται από:
\ba
\label{15-6}
L &=& R^2 \left( u_1 \sqrt{u_1^2 - 1}
\int_{u_0}^\infty {du \ov u \sqrt{(u^2-1)(u^2-u_1^2)(u^2+u_1^2-1)}} +(1\to 2)\right)\nonumber\\
&=& R^2 \left({ u_1 k_1^{\prime} \ov u_1^2-1}
\left[ \elPi (\nu_1,k_1^{\prime 2},k_1) - \elF(\nu_1,k_1) \right]+(1\to 2)\right)
\ea
και
\ba
\label{15-7}
E &=& T_{p,q}{\cal E}_1+T_{p^{\prime},q^{\prime}}{\cal E}_2+T_{m,n}(u_0-1)\ ,
\nonumber\\
{\cal E}_i&=&\int_{u_0}^\infty du\left[ u \sqrt{u^2-1 \ov (u^2-u_1^2)(u^2+u_1^2-1)} - 1 \right] - (u_0-1)\ ,
\nonumber\\
&=&\sqrt{2u_i^2-1}[k_i^{\prime 2}({\bf K}(k_i)-{\bf F}(\mu_i,k_i))-({\bf E}(k_i)-{\bf E}(\mu_i,k_i))]
\\
&& +\sqrt{{(u_0^2-u_i^2)(u_0^2+u_i^2-1)\ov u_0^2-1}}+1\ ,
\qq i=1,2\ ,
\nonumber
\ea
\no
όπου ${\bf F}(\nu,k),{\bf E}(\nu,k)$ και $\elPi(\nu,\alpha,k)$
είναι τα μη πλήρη ελλειπτικά ολοκληρώματα πρώτου, δεύτερου
και τρίτου είδους αντίστοιχα, ενώ
${\bf K}(k),{\bf E}(k)$ και $\elPi(\alpha,k)$ είναι τα αντίστοιχα πλήρη, με:
\ba
\label{15-8}
&&k_i={u_i \ov \sqrt{2u_i^2-1}}\ ,\quad k_i^{\prime}=\sqrt{1-k_i^2}\ ,\nonumber \\
&&\mu_i=\sin^{-1}\sqrt{{u_0^2-u_i^2\ov u_0^2-1}} ,\quad
\nu_i=\sin^{-1}\sqrt{{2u_i^2-1\ov u_0^2+u_i^2-1}}\ .
\ea
Από την Εξ.\eqn{13-12} βρίσκουμε ότι
\be
\label{15-9}
u_i=\sqrt{{1+\sqrt{\D_i}\ov 2}}\ , \qq \D_i=1+4u_0^2(u_0^2-1)\sin^2\th_i\ ,
\qq i=1,2 \ ,
\ee
όπου οι γωνίες $\th_i$ δίνονται από την Εξ.\eqn{13-14}.
Η συνάρτηση του μήκους $L(u_0)$ έχει ένα ολικό μέγιστο το οποίο εξαρτάται από
την σταθερά ζεύξης της χορδής $g_s$. Η συμπεριφορά είναι ανάλογη με την περίπτωση του μεσονίου
που περιγράψαμε στην παράγραφο 8.2.2 όσο και με την περίπτωσης που περιγράψαμε πριν της συνένωσης χορδών
για μελανές $D3$ βράνες. Η συμπεριφορά αυτή εμφανίζεται στο Σχήμα \ref{fig1.dyon}.

\subsection{Η τροχιά για $\th_0=\pi/2$}

Σε αυτή την περίπτωση έχουμε ότι $u_{\rm min}=1$ και $u_{\textrm{ρίζα}}=1$. Επιπλέον
τα ολοκληρώματα για το αδιάστατο μήκος και την αδιάστατη ενέργεια δίνονται από:

\ba
\label{15-10}
L
&=& R^2  \left(u_1^2 \int_{u_0}^\infty {du \ov u \sqrt{(u^2-1)
(u^4-u_1^4)}}\ +(1\to 2)\right) \nonumber \\
&=& {R^2\ov \sqrt{2}}\left({\elPi(\nu_1,{1\ov 2},k_1)
-{\bf F}(\nu_1,k_1)\ov u_1}+(1\to 2)\right) ,
\ea
και
\ba
\label{15-11}
E &=& T_{p,q}{\cal E}_1+T_{p^{\prime},q^{\prime}}{\cal E}_2+T_{m,n}\sqrt{u_0^2-1} ,\nonumber \\
{\cal E}_i&=&\int_{u_0}^{\infty}{du u\ov\sqrt{u^2-1}}
\left({u^2\ov\sqrt{u^4-u_i^4}}-1\right)-\int_1^{u_0}{du u\ov\sqrt{u^2-1}}\ ,
\\
&=&\sqrt{2}u_i({\bf E}(\mu_i,k_i)-{\bf E}(k_i))+{u_i\ov\sqrt{2}}{\bf F}(\nu_i,k_i)
-\sqrt{(u_0^2-u_i^2)(u_0^2+u_i^2)\ov u_0^2-1}\ ,\nonumber
\ea
\begin{figure}[!t]
\begin{center}
\begin{tabular}{cc}
\includegraphics[height=5.2cm]{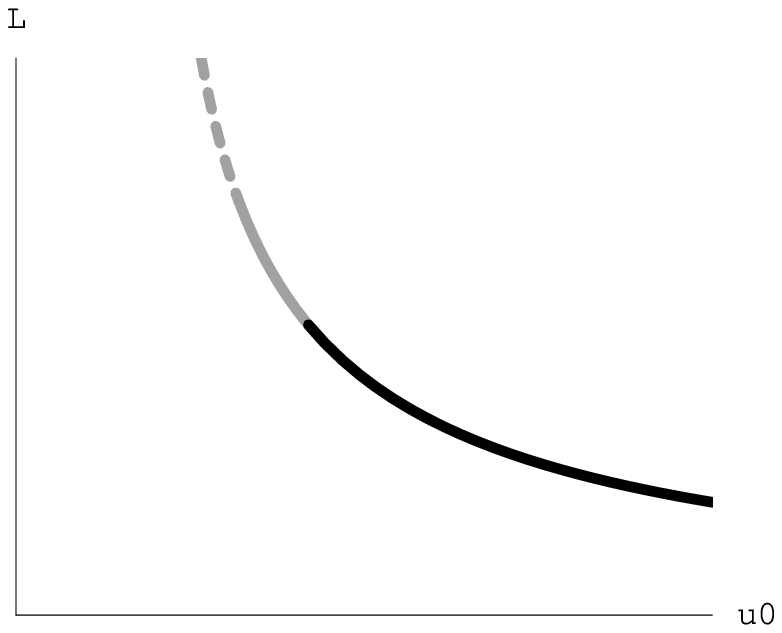}
&\includegraphics[height=5.2cm]{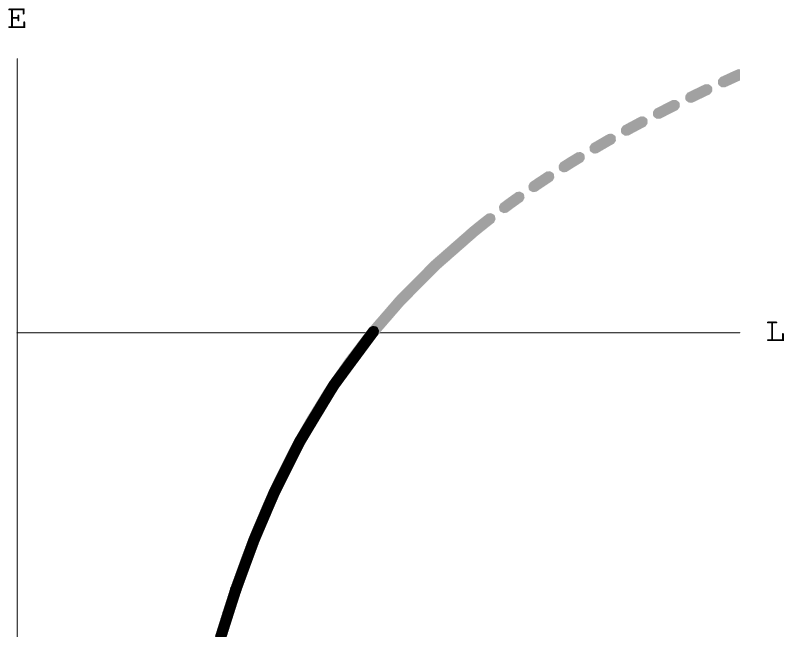}\\
(α) & (β)
\end{tabular}
\end{center}
\vskip -.5 cm \caption{
Γραφικές παραστάσεις των $L(u_0)$ και $E(L)$ για τις περιπτώσεις που πιθανώς υπάρχει αστάθεια
η οποία οφείλεται σε γωνιακές διακυμάνσεις.
Σημειώνουμε την εμφάνιση ενός δυναμικού εγκλωβισμού για τις περιπτώσεις
που υπάρχουν αστάθειες που να οφείλονται σε γωνιακές διακυμάνσεις.
Αυτές οι γραφικές παρασταάσεις καλύπτουν το παράδειγμα μας, για πολυκεντρικές $D3$ βράνες
πάνω σε μια σφαίρα και για τροχιά $\th_0=\pi/2$.
Η σημασία των σκιάσεων είναι όπως στο Σχήμα \ref{fig1.dyon}.}
\label{fig2.dyon}
\end{figure}
\no
όπου τώρα
\ba
\label{15-12}
&&k_i=\sqrt{{u_i^2+1 \ov 2 u_i^2}}\ ,\quad k^{\prime}=\sqrt{1-k^2}\ ,\nonumber\\
&&\mu_i=\sin^{-1}\left(\sqrt{{u_0^2-u_i^2\ov u_0^2-1}}\right)\ ,\quad
\nu_i=\sin^{-1}\left(\sqrt{{2u_i^2\ov u_0^2+u_i^2}}\right)\ .
\ea
Από την Εξ.\eqn{13-12} βρίσκουμε ότι
\be
\label{15-13}
u_i=\displaystyle u_0\sqrt{\sin\th_i}\ , \qq i=1,2 \ ,
\ee
όπου οι γωνίες $\th_i$ δίνονται από την Εξ.\eqn{13-14}.
Σε αυτή την περίπτωση η $L(u_0)$ είναι γνησίως
μονότονη φθίνουσα που απειρίζεται στο υπέρυθρο ($u_0\to 1$)
και μηδενίζεται στο υπεριώδες ($u_0\to\infty$). Από την Εξ.\eqn{14-22}
συμπεραίνουμε ότι η ενέργεια $E$ είναι μία γνησίως αύξουσα συνάρτηση του μήκους $L$.
Η συμπεριφορά αυτή φαίνεται στο Σχήμα \ref{fig2.dyon}. Η συμπεριφορά εγκλωβισμού
είναι ανάλογη με την περίπτωση του μεσονίου, δες παράγραφο 9.3.2 για $\th_0=\pi/2$.

\chapter{Παραδείγματα ανάλυσης ευστάθειας}
Εδώ θα εφαρμόσουμε την ανάλυση ευστάθειας που αναπτύξαμε στο 13ο κεφάλαιο για
τα παραδείγματα του 14ου κεφαλαίου. Θα βρούμε ότι για τις μελανές
$D3$ βράνες αλλά και για την σφαίρα, όταν $\th_0=0$ παρατηρείται
αστάθεια που οφείλεται στις διαμήκεις διακυμάνσεις και αντιστοιχεί
σε μέρος του άνω κλαδου, ενώ όταν $\th_0=\pi/2$ έχουμε αστάθεια
λόγω των γωνιακών διακυμάνσεων στο υπέρυθρο.

\section{Η σύμμορφη περίπτωση}

Για πληρότητα, θα θεωρήσουμε πρώτα την σύμμορφη περίπτωση, η οποία
αντιστοιχεί στο όριο όπου $\mu\to 0$ και $r_0\to 0$ στις παραπάνω λύσεις.
Η κλασική λύση για συνένωση χορδών έχει μελετηθεί στην εργασία \cite{Minahan}.
Εφόσον δεν υπάρχει εξάρτηση από την γωνία $\th$, οι γωνιακές διακυμάνσεις
$\d\th$ είναι ισοδύναμες με τις εγκάρσιες διακυμάνσεις $\d x$ και ως εκ τούτου
είναι ευσταθείς. Από την άλλη πλευρά για τις διαμήκεις διακυμάνσεις πρέπει να λύσουμε
την Εξ.\eqn{14-15}. Από την Εξ.\eqn{13-12} βρίσκουμε ότι
\be
\label{16-1}
u_i=u_0\sqrt{\sin\th_i}\ ,\qq i=1,2 \ ,
\ee
όπου οι γωνίες $\th_i$ δίνονται από την Εξ.\eqn{13-14}.
Επίσης υπολογίζουμε τα ολοκληρώματα $J_i$
\be
\label{16-2}
 J_i(u_0)={1\ov 3u_0^3}\ {_2F_1}\left({3\ov 4},{3\ov 2},{7\ov 4},{u_i^4\ov u_0^4}\right)\ .
\ee
Χρησιμοποιώντας τα παραπάνω, μπορούμε να δούμε ότι η Εξ.\eqn{13-12} δεν έχει λύση για καμία τιμή της
παραμέτρου $g_s$, επομένως οι συνενώσεις χορδών είναι ευσταθείς στην σύμμορφη περίπτωση.

\section{Μελανές $D3$ βράνες}

Στην συνέχεια θα μελετήσουμε την περίπτωση των μελανών $D3$ βρανών,
στην οποία η ενέργεια είναι μια δίτιμη συνάρτηση του μήκους. Όπως και πρίν,
εφόσον δεν υπάρχει εξάρτηση από την γωνία $\th$, οι διακυμάνσεις $\d\th$ είναι
ισοδύναμες με τις $\d x$ και κατά συνέπεια είναι ευσταθείς. Για τις διαμήκεις πρέπει
να λύσουμε την Εξ.\eqn{14-15}, όπου
\ba
\label{16-3}
 J_i(u_0)={1\ov 3u_0^3}\
F_1\left({3\ov 4},-{1\ov 2},{3\ov 2},{7\ov 4},{1\ov u_0^4},{u_i^4\ov u_0^4}\right)\ ,
\qq i=1,2\ .
\ea
Η εξίσωση \eqn{14-15} έχει λύση για κάθε τιμή της σταθεράς ζεύξης της χορδής $g_s$.
Η τιμή $u_{0c}$, είναι μικρότερη από την τιμή $u_{0m}$ για την οποία έχουμε μέγιστο μήκος.
Για μικρές σταθερές ζεύξης και για το σύστημα κουάρκ-μονοπόλου έχουμε δείξει ότι
η Εξ.\eqn{14-15} προσεγγίζεται με την Εξ.\eqn{14-21} και η οποία στην περίπτωση μας
ισούται με
\be
\label{16-4}
{}_2F_1\left({3\ov 4},1,{7\ov 4},{1\ov u_0^4}\right)-{3\ov 2}={\cal O}(g_s)\ .
\ee
Η λύση η οποία δίνει την κρίσιμη τιμή της παραμέτρου $u_0$ για την οποία
έχουμε αστάθεια σε πρώτη τάξη είναι
\be
\label{16-5}
u_{0c}\simeq  1.117 -0.166 g_s + {\cal O}(g_s^2)\ ,
\ee
όπου έχουμε επίσης συμπεριλάβει και την διόρθωση πρώτης τάξεως.
Η ανάπτυξη αυτή είναι διαφορετική από το αντίστοιχο ανάπτυγμα για
την $u_{0m}$ στην Εξ.\eqn{15-5}. Οπότε μέρος του άνω κλάδου είναι
διαταρακτικά ευσταθές, μια συμπεριφορά η οποία δεν αναμενόταν
από ενεργειακής απόψεως. Στην συνέχεια θα συγκρίνουμε το μέγιστο
μήκος για τις περιπτώσεις του ζεύγους κουάρκ-αντικουάρκ και του ζεύγους
κουάρκ-μονοπόλου. Καθώς το $g_s$ μεταβάλλεται από
μικρές σε μεγάλες τιμές βρίσκουμε ότι $L^{\rm qm}_{\rm max}\simeq (0.425\pm 0.005) R^2$,
ενώ για το ζεύγος κουάρκ-αντικουάρκ από την Εξ.\eqn{9-8} έχουμε ότι
$L^{\rm q\bar q}_{\rm max}\simeq 0.869 R^2\simeq 2 L^{\rm qm}_{\rm max}$.
Παρατηρούμε ότι το ζεύγος κουάρκ-μονοπόλου είναι περισσότερο θωρακισμένο
από το ζεύγος κουάρκ-αντικουάρκ κατά έναν παράγοντα δύο.
Η συμπεριφορά αυτή είναι γενική και θα την συναντήσουμε και στα άλλα δύο παραδείγματα.

\section{Πολυκεντρικές $D3$ βράνες πάνω σε σφαίρα}

\subsection{Η τροχία για $\th_0=0$}

Οι γωνιακές διακυμάνσεις είναι ευσταθείς διότι το δυναμικό
\en Schr\"odinger \gr κάθε χορδής είναι θετικό για όλες τις τιμές του $u_i$,
Εξ.\eqn{10-20} και $u_0\to u_i$ για $i=1,2$. Για τις διαμήκεις διακυμάνσεις πρέπει να
λύσουμε την Εξ.\eqn{14-15}, όπου
\ba
\label{16-6}
J_i(u_0)=\int_{u_0}^\infty du{u\sqrt{u^2-1}\ov (u^2(u^2-1)-u_i^2(u_i^2-1))^{3/2}}\ ,
\qq i=1,2\ .
\ea
Η εξίσωση αυτή έχει λύση για κάθε τιμή της σταθεράς ζεύξης της χορδής. Όταν αυτή παίρνει
μικρές τιμές τότε για το σύστημα κουάρκ-μονοπόλου, η Εξ.\eqn{14-15} προσεγγίζεται
από την Εξ.\eqn{14-21} η οποία στην περίπτωση μας ισούται με
\be
\label{16-7}
{2u_0^2\ov 2u_0^2-1} -{u_0\ov 2}\ln\left(u_0+1\ov u_0-1\right) ={\cal O}(g_s)\ .
\ee
Η λύση της Εξ.\eqn{16-7} μας δίνει δίνει σε πρώτη τάξη
την κρίσιμη τιμή της παραμέτρου $u_0$ για την οποία έχουμε αστάθεια
\be
\label{16-8}
u_{0c}\simeq  1.084 + {\cal O}(g_s)\ .
\ee
Συνεπώς, μέρος του άνω κλάδου είναι διαταρακτικά ευσταθές όπως
στην προηγούμενη περίπτωση. Όπως και πριν, θα συγκρίνουμε το μέγιστο
μήκος για τις περιπτώσεις του ζεύγους κουάρκ-αντικουάρκ και του ζεύγους
κουάρκ-μονοπόλου. Καθώς το $g_s$ μεταβάλλεται από
μικρές σε μεγάλες τιμές έχω ότι $L^{\rm qm}_{\rm max}\simeq (0.425\pm 0.050) R^2$,
ενώ για το ζεύγος κουάρκ-αντικουάρκ από την Εξ.\eqn{9-31} έχουμε
$L^{\rm q\bar q}_{\rm max}\simeq 1.002 R^2\simeq 2.3 L^{\rm qm}_{\rm max}$.

\subsection{Η τροχία για $\th_0=\pi/2$}

Οι διαμήκεις διακυμάνσεις είναι ευσταθείς διότι όπως μπορούμε να δείξουμε
η Εξ.\eqn{14-15} δεν έχει λύση. Εναλλακτικά, τα δυναμικά \en Schr\"odinger \gr
του προβλήματος είναι θετικά, Εξ.\eqn{10-23} και $u_0\to u_i$ για $i=1,2$.
Για τις γωνιακές διακυμάνσεις το δυναμικό είναι μια σταθερά
\be
\label{16-9}
V_\th =-1 \ , \qq i=1,2\ .
\ee
Η αλλαγή μεταβλητών για τις δύο χορδές είναι
\be
\label{16-10}
 z_i(u)={1\ov u_i\sqrt{2}}\ {\bf F}(\nu_i,k_i)\ ,\quad  i=1,2\ ,
\ee
όπου $\nu_i$ και $ k_i$ δίνονται από τις:
\ba
\label{16-11}
\nu_i=\sin^{-1}\left(\sqrt{{2u_0^2 \sin\th_i\ov u^2+u_0^2 \sin\th_i}}\right)\ ,
\qq k_i=\sqrt{{u_0^2\sin\th_i+1\ov 2u_0^2\sin\th_i}}\ ,\qq i=1,2\ .
\ea
Η λύση που αντιστοιχεί στην εξίσωσης \en Schr\"odinger \gr για τις χορδές $i=1,2$
και ικανοποιεί την συνοριακή εξίσωση \eqn{14-6} στο υπεριώδες, ισούται με
\ba
\label{16-12}
\d \th_i(z_i)=c_i\sin z_i\ ,\quad z_i\in[0,\zeta_{i}]\ (\textrm{όταν}\ u\in (\infty,u_0])\ ,
\quad \zeta_{i}=z_i(u_0) \ ,
\ea
όπου οι $c_i$ είναι πολλαπλασιαστικές σταθερές. Η ευθεία χορδή παίζει ρόλο
στις συνοριακές συνθήκες, και η αιτία είναι ο όρος μάζας που υπάρχει στην Εξ.\eqn{14-5}.
Για την ευθεία χορδή έχουμε δύο γραμμικές ανεξάρτητες λύσεις $\sin z$ και $\cos z$,
όπου η κατάλληλη αλλαγή μεταβλητών είναι
\be
\label{16-13}
z={\pi\ov 2}-\sin^{-1}{1\ov u} \ .
\ee
Θα μελετήσουμε με λεπτομέρεια μόνο την λύση $\sin z$ και θα σχολιάσουμε στην συνέχεια
την γενική λύση. Από τις συνθήκες συρραφής στο $u=u_0$ Εξ.\eqn{14-8} έχουμε
\ba
\label{16-14}
\sin\th_1\sin \zeta_1 \cos \zeta_2 +\sin\th_2\sin  \zeta_2 \cos \zeta_1
- {\sin(\th_1+\th_2)\ov u_0(u_0^2-1)}\sin \zeta_1\sin \zeta_2=0\ .
\ea
Στην περίπτωση του ζεύγους κουάρκ-μονοπόλου, μπορούμε
να χρησιμοποιήσουμε την σχέση $\th_1+\th_2=\pi/2$ και να απλοποιήσουμε τον παραπάνω τύπο.
Επιπλέον, για μικρές σταθερές ζεύξης η Εξ.\eqn{16-14} απλοποιείται στην:
\be
\label{16-15}
\tan \left({1\ov 2 u_0} + {1\ov 2 \sqrt{u_0^2-1}}\right)-u_0(u_0^2-1) ={\cal O}(g_s)\ .
\ee
Η λύση αυτής μας δίνει σε πρώτη τάξη την κρίσιμη τιμή της παραμέτρου $u_0$ για την οποία
έχουμε αστάθεια από τις γωνιακές διακυμάνσεις
\be
\label{16-16}
u_{0 c}\simeq 1.378 + {\cal O}(g_s)\ .
\ee
Η συμπεριφορά εγκλωβισμού είναι ασταθής διότι η Εξ.\eqn{16-14} έχει λύση για κάθε τιμή της
σταθεράς ζεύξης της χορδής. Καθώς το $g_s$ μεταβάλλεται από
μικρές σε μεγάλες τιμές βρίσκουμε ότι $L^{\rm qm}_{\rm c}\simeq (0.560\pm 0.040) R^2$,
ενώ για το ζεύγος κουάρκ-αντικουάρκ από την Εξ.\eqn{10-26} έχουμε ότι
$L^{\rm q\bar q}_{\rm c}\simeq 1.700 R^2\simeq 3 L^{\rm qm}_{\rm c}$.\\
\no
’ν είχαμε θεωρήσει την γενική λύση της εξίσωσης \en Schr\"odinger \gr για
την ευθεία χορδή, δηλαδή $\cos(z+\phi)$, θα είχαμε καταλήξει σε ανάλογα αποτελέσματα
εκτός από την περίπτωση που αντιστοιχεί σε $\phi=0$ για την οποία δεν έχουμε αστάθεια.
Κατά συνέπεια τα παραπάνω αποτελέσματα είναι τελείως γενικά.\\
\no
Τέλος, αναφέρουμε επιγραμματικά τα αποτελέσματα για την κατανομή $D3$ βρανών πάνω σε
δίσκο. Για $\th_0=\pi/2$ η λύση είναι ευσταθής κάτω από όλα τα είδη διακυμάνσεων,
ενώ για $\th_0=0$ είναι ευσταθής κάτω από διαμήκεις και εγκάρσιες
και ασταθείς κάτω από τις γωνιακές διαταραχές μετά από ένα συγκεκριμένο μήκος.
Αυτή η συμπεριφορά είναι ανάλογη με την περίπτωση του ζεύγους κουάρκ-αντικουάρκ
στην οποία το μήκος θωράκισης ήταν ανεξάρτητο από τον προσανατολισμό της χορδής.

\part{Βαρυόνια}

\chapter{Εισαγωγή}
Στο πλαίσιο της αντιστοιχίας \en AdS/CFT \gr το βαρυόνιο ορίζεται ως η δέσμια
κατάσταση που σχηματίζουν $N$ κουάρκ, τα οποία βρίσκονται
στο σύνορο του $AdS_5$ τοποθετημένα στα άκρα $N$ θεμελιωδών χορδών και οι οποίες
εκτείνονται μέχρι έναν κόμβο (\en vertex)\ \gr στο εσωτερικό του $AdS_5$. Κάθε χορδή
συνεισφέρει μια μονάδα φορτίου, ώστε το ολικό φορτίο να διατηρείται
στον κόμβο. Όπως προτάθηκε στην εργασία \cite{Witten:1998} μια $D5$ βράνη θα
έπρεπε να τυλιχθεί γύρω από την εσωτερική πενταδιάστατη σφαίρα, $S^5$, και
λόγω της παρουσίας της, υπάρχει μια σύζευξη μεταξύ του όρου
\en Wess--Zumino \gr και του ηλεκτρικού πεδίου της \en Dirac--Born--Infeld (DBI)
\gr ούτως ώστε να διατηρείται το συνολικό φορτίο. Αυτή η ιδέα εφαρμόστηκε
για τον υπολογισμό της κλασικής λύσης που αντιστοιχεί σε αυτή την διάταξη
\cite{Brandhuber:1998,Imamura1} και είναι γενίκευση των τεχνικών που αναπτύχθηκαν
στο μέρος των μεσονίων.\\
\no
Σε τεχνικό επίπεδο έχουμε μια διάταξη από $N$ χορδές οι οποίες ενωνόνται
σε έναν κόμβο στο εσωτερικό του $AdS_5$ ο οποίος δεν είναι μαθηματικό
σημείο, όπως αυτή περίπτωση της συνένωσης χορδών, αλλά ένα σολιτονικό
αντικείμενο, όπως μια $D5$ βράνη τυλιγμένη γύρω από την εσωτερική πενταδιάστατη σφαίρα, $S^5$.
Επιπλέον έχουμε διατάξεις με $5N/8 < k < N$ κουάρκ \cite{Brandhuber:1998,Imamura1},
οι οποίες ικανοποιούν τις κλασικές εξισώσεις κίνησης, πέρα από τις αναμενόμενες για
$k=N$. Το γεγονός αυτό έρχεται σε αντίθεση με το ότι οι δέσμιες καταστάσεις κουάρκ
και αντικουάρκ είναι άχρωμες καταστάσεις της ομάδας βαθμίδας. Παρότι
η θεωρία η οποία μελετούμε είναι μέγιστα υπερσυμμετρική, θα θέλαμε να δείξουμε
ότι υπάρχει ένα καλύτερο κάτω φράγμα για το $k/N$, αν όχι $k=N$ ακριβώς \cite{Sfetsos:2008yr}.\\
\no
Η οργάνωση αυτού του μέρους της διατριβής είναι η εξής: Στο 17ο κεφάλαιο θα ορίσουμε τα
βαρυόνια για μια μεγάλη κατηγορία υποβάθρων. Ειδικότερα, θα βρούμε γενικές σχέσεις για την
δέσμια ενέργεια του βαρυονίου στην περίπτωση που τα κουάρκ κατά τον σχηματισμό
του βαρυονίου είναι ομοιόμορφα κατανεμημένα πάνω σε ένα σφαιρικό κέλυφος. Στο 18ο κεφάλαιο
θα μελετήσουμε την ευστάθεια των λύσεων κάτω από μικρές διακυμάνσεις και θα βρούμε τις
εξισώσεις κίνησης όπως και τις συνοριακές συνθήκες και συνθήκες συρραφής στο σύνορο
του $AdS_5$ και στον κόμβο του βαρυονίου. Στο 19ο κεφάλαιο θα παρουσιάσουμε παραδείγματα
βαρυονικών διατάξεων στα πλαίσια της αντιστοιχίας \en AdS/CFT, \gr για σύμμορφα
και πεπερασμένης θερμοκρασίας υπόβαθρα. Θα δείξουμε ότι η ευστάθεια περιορίζει
την τιμή $k$ να είναι μεγαλύτερη από μια κρίσιμη τιμή ακόμα και για την σύμμορφη
περίπτωση. Στην περίπτωση των πλειότιμων δυναμικών, ακόμα και μέρος του ενεργειακά
προτιμητέου κλάδου είναι διαταρακτικά ασταθές.

\chapter{Κλασικές λύσεις}
\begin{figure}[!t]
\begin{center}
\begin{tabular}{cc}
\includegraphics[height=10cm]{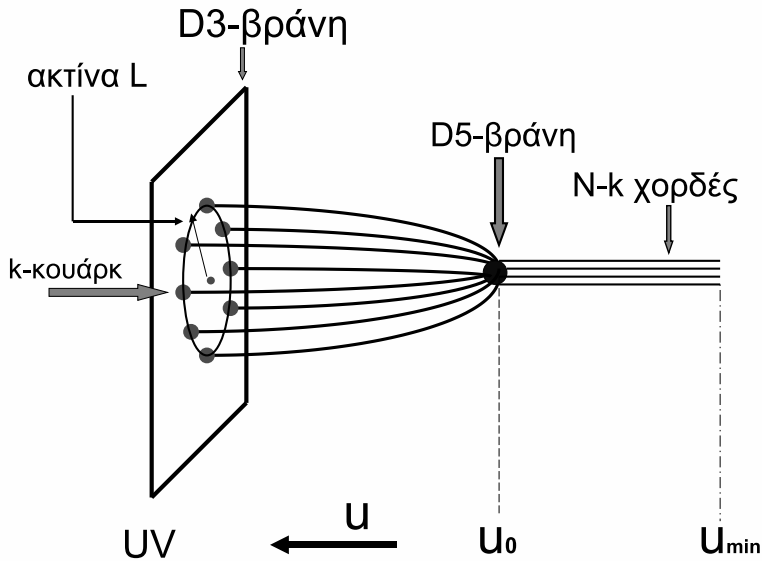}
\end{tabular}
\end{center}
\vskip -2.5 cm \caption{Η διάταξη του βαρυονίου με
 $k$ εξωτερικά κουάρκ τοποθετημένα πάνω σε ένα σφαιρικό κέλυφος ακτίνας $L$ στο
σύνορο του χώρου \en AdS, \gr κάθε ένα από αυτά είναι συνδεδεμένο με την
$D5$ βράνη η οποία είναι τυλιγμένη πάνω σε μια σφαίρα στο σημείο
$u=u_0$ και $N-k$ ευθείες χορδές με άκρο στο $u_{\rm min}$.}
\label{baryon}
\end{figure}

Σε αυτό το κεφάλαιο θα παρουσιάσουμε τον γενικό φορμαλισμό
στο πλαίσιο της αντιστοιχίας \en AdS/CFT \gr
για τον υπολογισμό της ενέργειας ενός βαρυονίου το οποίο αποτελείται
από βαριά κουάρκ. Εφόσον η δυική θεωρία πεδίου είναι η
$\cN=4$\ \en SYM, \gr οι υπολογισμοί περικλείουν $Ν$
χορδές και μια $D5$ βράνη να είναι τυλιγμένη γύρω από μια
εσωτερική σφαίρα $S^5$. Ο υπολογισμός στηρίζεται στις τεχνικές
που αναπτύχθηκαν στο μέρος του μεσονίου και τον κόμβο του βαρυονίου
\cite{Witten:1998}. Η εύρεση της κλασικής λύσης θα είναι μια γενίκευση
της σύμμορφης περίπτωσης \cite{Brandhuber:1998,Imamura1} σε πιο γενικά υπόβαθρα.
Θα θεωρήσουμε διαγώνιες μετρικές με υπογραφή \en Lorentz \gr της μορφής
\ba
\label{18-1}
ds^2 = G_{tt} dt^2 + G_{xx}(dx^2 + dy^2 + dz^2)+G_{uu}du^2 + R^2d\Om_5^2 \ ,
\ea
όπου $x,y$ και $z$ συμβολίζουν αγνοήσιμες συντεγμένες και $u$ είναι η ακτινική
συντεταγμένη η οποία παίζει τον ρόλο της κλίμακας ενέργειας στην δυική θεωρία πεδίου.
Εκτείνεται από το υπεριώδες ($u\to\infty$) έως το υπέρυθρο σε κάποια ελάχιστη τιμή $u_{min}$
η οποία θα καθορίζεται από την γεωμετρία.\\
\no
Είναι βολικό για την ανάλυση μας να ορίσουμε τις συναρτήσεις
\ba
\label{18-2}
f(u) = - G_{tt}G_{xx}\ ,\qq g(u)=- G_{tt} G_{uu}\ , \qq h(u)=G_{yy}G_{uu} \ .
\ea
Στο σύμμορφο όριο μπορούμε να προσεγγίσουμε την γεωμετρία με τον χώρο
$AdS_5\times S^5$ με ακτίνα ($\a^\prime$=1) $R=(4\pi g_sN)^{1/4}$, όπου η
$g_s$ είναι η σταθερά ζεύξης της χορδής. Στο σύμμορφο όριο οι συναρτήσεις
της εξίσωσης προσεγγίζονται με
\ba
\label{18-3}
 f(u)\simeq u^4\ ,\qq g(u)\simeq 1\ ,\qq h(u)\simeq 1\ .
\ea
Εξ ορισμού το βαρυόνιο είναι μια δέσμια κατάσταση από $N$ κουάρκ τα οποία
σχηματίζουν μια πλήρως αντισυμμετρική αναπαράσταση της ομάδας $SU(N)$. Ωστόσο, στο
πλαίσιο της αντιστοιχίας \en AdS/CFT, \gr μπορούμε να φτιάξουμε μια δέσμια κατάσταση
από $k$ κουάρκ με $k < N$. Αυτή η δέσμια κατάσταση αποτελείται από μια
$D5$ βράνη η οποία βρίσκεται στο εσωτερικό και $k$ εξωτερικά κουάρκ στο σύνορο
του $AdS_5$. Κάθε κουάρκ είναι συνδεδεμένο μέσω μίας χορδής με την $D5$ βράνη.
Υπάρχουν επίσης $N-k$ ευθείες χορδές οι οποίες εκτείνονται από την $D5$ βράνη μέχρι
το $u_{min}$. Η περιγραφή αυτή αποτυπώνεται στο Σχήμα \ref{baryon}. Θα θεωρήσουμε $N,k\gg 1$,
με τον λόγο τους $k/N$ να είναι πεπερασμένος.\\
\no
Θα πρέπει να θεωρήσουμε και μια κατανομή των κουάρκ μέσα στο βαρυόνιο.
Αναμένουμε μια κατανομή με ακτινική συμμετρία και για ευκολία θα θεωρήσουμε
μια ομοιόμορφη πάνω σε ένα σφαιρικό κέλυφος ακτίνας $L$. Επιπλέον, αξίζει να
σημειώσουμε ότι στην προσέγγιση μας τα άκρα των $N$ χορδών είναι ομοιόμορφα
κατανεμημένα στην εσωτερική πενταδιάστατη σφαίρα $S^5$, για να μην
παραμορφωθεί και η προσέγγιση της στρογγυλής σφαίρας να ισχύει (περαιτέρω ανάλυση αυτού του σημείου δίδεται
στην εργασία \cite{Imamura1}. Αυτή η επιλογή σπάει τελείως την υπερσυμμετρία \cite{Imamura2,Callan:1998iq}.)

\no
Στα πλαίσια της αντιστοιχίας βαρύτητας και θεωρίας βαθμίδας, η δέσμια ενέργεια ενός
βαρυονίου δίνεται από
\ba
\label{18-4}
e^{-{\rm i} E T} = e^{{\rm i} S_{\rm cl}}\ ,
\ea
όπου $S_{\rm cl}$ είναι η κλασική δράση της παραπάνω διάταξης. Η δράση αυτή
αποτελείται από τρείς όρους, μια δράση \en Nambu--Goto \gr για τις $N$ χορδές,
μια δράση \en Dirac--Born--Infeld (DBI) \gr για την $D5$ βράνη και έναν όρο
\en Wess--Zumino \gr μεταξύ των $N$ χορδών και της $D5$ βράνης
\ba
\label{18-5}
&&S_{NG}[C] =  {1\ov 2 \pi} \int_C d \tau d \sigma \sqrt{-\det g_{\a \b}}\ ,
\nonumber\\
&& S_{D5}={1\ov (2\pi)^5 g_s}\int_M d^6x \sqrt{-h_{ij}-2 \pi F_{ij}}\ ,
\\
&&S_{WZ}={1\ov (2\pi)^5 g_s}\int_M d^6x\ C^{(4)}\wedge F\ ,\qq M={\mathbb{R}\times S^5}\ ,
\nonumber
\ea
όπου $g_{\a \b},h_{ij},F_{ij}$ είναι οι επαγώμενες μετρικές για την χορδή, την $D5$ βράνη
και του τανυστή πεδίου \en Born--Infeld\ \gr αντίστοιχα. Θα θεωρήσουμε την πιο απλή περιπτωση όπου
η δράση \en DBI \gr δίνεται από την επαγώμενη μετρική και ο όρος
\en Wess--Zumino \gr δρα ως όρος διατήρησης φορτίου \cite{Witten:1998}.
Εντούτις, υπάρχουν περιπτώσεις στις οποίες έχουμε $F_{ij}\neq 0$ \cite{Lozano:2006,Sfetsos:2008yr}.
Τέλος, δεν έχουμε συμπεριλάβει το πεδίο \en Kalb--Ramond \gr διότι δεν υπάρχει στα παραδείγματα μας.\\
\no
Στην συνέχεια, κατά τα γνωστά, θα καθορίσουμε την διδιάστατη παραμετροποίηση
\ba
\label{18-6} t=\tau \ ,\qq u=\s \ .
\ea
Για στατικές λύσεις θα θεωρήσουμε την παρακάτω εμβάπτιση της διδιάστατης σφαίρας $S^2$
πάνω στην $D3$ βράνη σε σφαιρικές συντεταγμένες $(r,\th,\phi)$
\ba
\label{18-7} r = r(u)\ ,\qq \phi, \ \th =\textrm{σταθερές}\ ,\qq \hbox{$S^5$--γωνίες} = \textrm{σταθερές}\ ,
\ea
η οποία υπόκειται στην συνοριακή συνθήκη
\ba
\label{18-8} u \left(L \right)=\infty\ .
\ea
Για αυτήν την υπόθεση, η δράση \en Nambu--Goto \gr γράφεται ως:
\ba
\label{18-9} S = - {T\ov 2 \pi} \int du\sqrt{ g(u) + f(u) r^{\prime 2}}\ ,
\ea
όπου $T$ συμβολίζει τον χρόνο και ο τόνος την παράγωγο ως προς
την μεταβλητή $u$. Η Λαγκρανζιανή του προβλήματος είναι ανεξάρτητη της μεταβλητής
$y$, οπότε η αντίστοιχη γενικευμένη ορμή διατηρείται
\ba
\label{18-10}{f r_{\rm cl}^\prime \ov \sqrt{g+ f r_{\rm cl}^{\prime 2}}}
=  f_1^{1/2}\qq \Longrightarrow
\qq r_{\rm cl}^\prime =  {\sqrt{f_1F}\ov f}\ ,
\ea
όπου $u_1$ είναι η τιμή της παραμέτρου $u$ στο σημείο στροφής της κάθε χορδής,\\
$f_1 \equiv f(u_1),\ f_0\equiv f(u_0)$ και
\ba
\label{18-11} F = {g f \ov f - f_1}\ .
\ea
Οι $N-k$ χορδές οι οποίες εκτείνονται από τον βαρυονικό κόμβο
μέχρι το $u_{min}$ είναι ευθείες, και εφόσον το $r=\textrm{σταθερά}$
είναι πάντα λύση των εξισώσεων (με $f_1=0$). Στα παραδείγματα
που θα ακολουθήσουν $u_{min}\leq u_1\leq u_0$ και $f(u)$ είναι
μια αύξουσα συνάρτηση. Ολοκληρώνοντας την Εξ.\eqn{18-10}, μπορούμε να
εκφράσουμε την ακτίνα του σφαιρικού κέλυφους ως
\ba
\label{18-12} L =\sqrt{f_1}\int_{u_0}^{\infty} d u {\sqrt{F}\ov f}\ .
\ea
Στην συνέχεια θα καθορίσουμε την πενταδιάστατη παραμετροποίηση της
τυλιγμένης $D5$ βράνης επιλέγοντας
\ba
\label{18-13} t=\tau\ ,\qq \th_a=\s_\a\ , \quad \a=1,2,\dots,5\ .
\ea
Για αυτήν την υπόθεση, η δράση \en DBI \gr γράφεται
\ba
\label{18-14}
S_{D5}={TNR\ov 8\pi}\sqrt{-G_{tt}}\Big{|}_{u=u_0}\ ,
\ea
Εδώ χρησιμοποιήσαμε το γεγονός ότι ο όγκος μιας πενταδιάστατης σφαίρα μοναδιαίας ακτίνας
είναι $\pi^3$.\\
\no
Τέλος, αντικαθιστώντας την λύση για την $r_{\rm cl}^\prime$ στην \eqn{18-9},
αφαιρώντας τις αποκλίνουσες μάζες των $k$-κουάρκ και προσθέτοντας
την ενέργεια της $D5$ βράνης Εξ.\eqn{18-14}, βρίσκουμε την δέσμια ενέργεια του
βαρυονίου
\ba
\label{18-15}
E = {k \ov 2\pi}\left\{\int_{u_0}^\infty d u \sqrt{F}
- \int_{u_{\rm min}}^\infty d u \sqrt{g}
+{1-a\ov a} \int_{u_{\rm min}}^{u_0} d u \sqrt{g} +
{R\sqrt{-G_{tt}}\ov 4 a }\ \bigg|_{u=u_0} \right\}\ ,
\ea
όπου
\be
\label{18-16}
a\equiv {k\ov N}\ ,\qq 0< a \leqslant 1\ .
\ee

\no
Οι εκφράσεις του μήκους και της ενέργειας, Εξ.\eqn{18-12} και \eqn{18-15}
εξαρτώνται από την παράμετρο $u_1$ η οποία θα πρέπει να εκφραστεί σαν συνάρτηση
της θέσης του βαρυονικού κόμβου $u_0$. Ο πιο εύκολος τρόπος είναι να απαιτήσουμε
ότι η συνολική δύναμη στον βαρυονικό κόμβο είναι μηδέν. Ακολουθώντας μια ανάλογη διαδικασία
με την περίπτωση της διακλάδωσης χορδών \cite{Sfetsos:2007nd}, αλλάζουμε προσωρινά
βαθμίδα σε $r=\s$ και από τους συνοριακούς όρους που προκύπτουν κατά την εξαγωγή των εξισώσεων κίνησης
για μια ομοιόμορφη κατανομή κουάρκ πάνω σε ένα σφαιρικό κέλυφος, βρίσκουμε
την σχέση (η σύμμορφη περίπτωση έχει μελετηθεί στην εργασία \cite{Brandhuber:1998})
\ba
\label{18-17}
&&\cos\Th = {1-a\ov a} + {R\ov 4a\sqrt{g}}\del_u\sqrt{-G_{tt}}\Big{|}_{u=u_0}\ ,
\nonumber\\
&& \cos\Th =  \sqrt{1-f_1/f_0}\ ,
\ea
όπου $\Th$ είναι η γωνία μεταξύ του καθενός από τα $k$ κουάρκ και του άξονα $u$
στον βαρυονικό κόμβο και καθορίζει την σχέση μεταξύ του $u_1$ και του $u_0$. Η παραπάνω
σχέση μπορεί να ερμηνευτεί και σαν συνθήκη μηδενισμού της $u$ συνιστώσα της δύναμης
\cite{Imamura1}. Ο όρος στο αριστερό μέρος της Εξ.\eqn{18-17} οφείλεται στις χορδές
που προσκολούνται στο σύνορο στο υπεριώδες. Η δύναμη με την οποία έλκουν τον βαρυονικό κόμβο
αντισταθμίζεται από τις ευθείες χορδές και την τάση που έχει η $D5$ βράνη
να κινηθεί προς το $u_{min}$, και αντιπροσωπεύονται απο τον πρώτο και δεύτερο
όρο της Εξ.\eqn{18-17}. Όπως θα δούμε στα παραδείγματα μας, η Εξ.\eqn{18-17} έχει λύση για μια περιοχή
του παραμετρικού χώρου $(a,u_0)$. Ωστόσο, θα πρέπει να μελετήσουμε την ευστάθεια
των κλασικών λύσεων ούτως ώστε να απομονώσουμε τις παραμετρικές
περιοχές που είναι φυσικώς ενδιαφέρουσες.

\chapter{Ανάλυση Ευστάθειας}
Στην συνέχεια θα μελετήσουμε την ευστάθεια των παραπάνω λύσεων,
σκοπεύοντας να απομονώσουμε τις φυσικώς ενδιαφέρουσες παραμετρικές
περιοχές. Η ανάλυση αυτή στηρίζεται στα γενικά αποτελέσματα που
αναλύθηκαν διεξοδικά στο μέρος των μεσονίων και βαρυονίων. Ωστόσο,
τα ποιοτικά αποτελέσματα θα είναι εντελώς διαφορετικά, λόγω των
διαφορετικών συνοριακών συνθηκών που θα επιβάλουμε.

\section{Μικρές διακυμάνσεις}

Θεωρούμε την παρακάτω διαταραχή στην εμβάπτιση της χορδής
\ba
\label{19-1}
r = r_{\rm cl}(u) + \d r (t,u)\ ,\qq \th =\th_0 + \d\th (t,u)\ ,\qq \phi = \phi_0+\d\phi (t,u)\ .
\ea
\no
όπου έχουμε κρατήσει την επιλογή βαθμίδας αναλλοίωτη Εξ.\eqn{18-6}. Στην συνέχεια
θα υπολογίσουμε  την δράση \en Nambu--Goto \gr για αύτην την υπόθεση και θα την
αναπτύξουμε σε δυνάμεις των διακυμάνσεων. Η ανάπτυξη σε δεύτερη τάξη για τις διακυμάνσεις
των $k$ χορδών γράφεται ως
\ba
\label{19-2}
\!\!\!\!\!\!\!\!\!\!\!\! S_2 &=& - {1 \ov 2\pi} \int
dt du \biggl[ {g f \ov 2 F^{3/2}}\ \d r^{\prime 2} - {h \ov 2 F^{1/2}}\
\d \dot{r}^2+{f r_{\rm cl}^2 \ov 2 F^{1/2}}
\ \d\th^{\prime 2} - {h r_{\rm cl}^2 F^{1/2} \ov 2 g}\  \d \dot{\th}^2 \nonumber\\
&&\qq\qq\qq + {f r_{\rm cl}^2\sin^2\th_0 \ov 2 F^{1/2}}\
\d\phi^{\prime 2} - {h r_{\rm cl}^2\sin^2\th_0  F^{1/2} \ov 2 g}\ \d \dot{\phi}^2\biggr]\ .
\ea
Χρησιμοποιώντας τις εξισώσεις \en Euler--Lagrange \gr και την ανάπτυξη \en Fourier \gr
\ba
\label{19-3}
\d x^\m (t,u) = \d x^\m (u) e^{-\imath \om t}\ ,\qq x^\m=r,\th,\phi\ ,
\ea
βρίσκουμε τις γραμμικές διαφορικές εξισώσεις για
τις διαμήκεις και εγκάρσιες διακυμάνσεις, για κάθε μια απο τις $k$ χορδές
\ba
\label{19-4}
&&\left[ {d \ov du} \left( {g f \ov F^{3/2}} {d \ov du} \right)
+ \omega^2 {h \ov F^{1/2}} \right] \d r = 0\ ,\nonumber \\
&&\left[ {d \ov du} \left({f r_{\rm cl}^2 \ov F^{1/2}} {d \ov du} \right)
+ \omega^2 {h r_{\rm cl}^2 F^{1/2} \ov g } \right] \d\th = 0\ ,\\
&&\left[ {d \ov du} \left({f r_{\rm cl}^2 \ov F^{1/2}} {d \ov du} \right)
+ \omega^2 {h r_{\rm cl}^2 F^{1/2} \ov g } \right] \d\phi = 0\ .\nonumber
\ea
Για τις ευθείες χορδές οι εκφράσεις για την δράση και τις εξισώσεις κίνησης
δόθηκαν στην παράγραφο 13.1, Εξ.\eqn{14-4} και \eqn{14-5}. Οπως αναφέρθηκε
αναλυτικά στα μέρη των μεσονίων και των δυονίων, η ύπαρξη ασταθειών
σχετίζεται με την παρουσία ενός κανονικοποιημένου μηδενικού τρόπου για
κάποια συγκεκριμένη τιμή της παραμέτρου $u_0$, την οποία συμβολίζουμε
με $u_{0\rm c}$. Οι τιμές αυτές είναι εν γένει διαφορετικές από τις
πιθανές τιμές για τις οποίες το μήκος είναι μέγιστο, τις οποίες συμβολίζουμε με
$u_{0\rm m}$ και οι οποίες βρίσκονται λύνοντας την εξίσωση $L^\prime(u_0)$.
Ο σκοπός μας είναι να βρούμε την κρίσιμη καμπύλη στον παραμετρικό χώρο
των $(\a,u_0)$ η οποία χωρίζει τις ευσταθείς από τις ασταθείς περιοχές.
Τέλος, θα μελετήσουμε τις διακυμάνσεις της $D5$ βράνης και θα θεωρήσουμε
την παρακάτω διαταραχή
\ba
\label{19-5}
x^\m=\d x^\m(t,\th_\a)\ , \qq u=u_0\ ,\qq x^\m = x,y,z\ ,
\ea
αφήνοντας την θέση της $D5$ βράνης αναλλοίωτη στο σημείο $u=u_0$ λόγω
της επιλογής βαθμίδας $u=\s$ για τις χορδές. Σε δεύτερη τάξη στις διακυμάνσεις,
ο όρος ανάπτυξης για την $D5$ βράνη ισούται με
\ba
\label{19-6}
S={NR\ov 8\pi^4}\int dtd\Om_5\sqrt{\g}\sqrt{-G_{tt}}
\bigg\{1+
{G_{\m\n}\ov 2}\g^{\a\b}\del_\a\d x^\m\del_\b\d x^\n
+{G_{\m\n}\ov 2G_{tt}}\d\dot{x}^\m\d\dot{x}^\n\bigg\} \ ,
\ea
όπου $\g_{\a\b}$ είναι η μετρική της $S^5$ και η δράση έχει
υπολογιστεί στο σημείο $u=u_0$. Οι δείκτες $\a,\mu$ αναφέρονται
στις γωνίες της $S^5$ και στις συντεταγμένες $x,y,z$ αντίστοιχα.
Αναπτύσουμε τις διακυμάνσεις σε όρους σφαιρικών αρμονικών $\Psi_\ell$ της
πενταδιάστατης σφαίρας $S^5$, οι οποίες ικανοποιούν
την εξίσωση ιδιοτιμών
$\nabla_{\g}^2\Psi_\ell=-\ell(\ell+4)\Psi_\ell$, $\ell=0,1,2,\dots$,
ως
\be
\label{19-7}
\d x^\m(t,\th_\a) = \d x^\m(t) \Psi_\ell(\th_\a)\ .
\ee
Τότε, από τις εξισώσεις \en Euler--Lagrange \gr για την δράση βρίσκουμε ότι
\ba
\label{19-8}
{d^2\d x^\m\ov dt^2}+\Om_\ell^2\d x^\m=0\ , \quad \Om_\ell^2=-G_{tt}(u_0)\ell(\ell+4)>0\ .
\ea
Οπότε η κλασική λύση είναι ευσταθής κάτω από τις διακυμάνσεις $\d x^\mu$.
Παρατηρούμε ότι δεν υπάρχουν συνοριακές συνθήκες για αυτές τις διακυμάνσεις,
λόγω του ότι ο χώρος $\mathbb{R}\times S^5$ δεν έχει σύνορο.

\section{Συνοριακές συνθήκες}

Για να καθορίσουμε πλήρως το πρόβλημα ιδιοτιμών μας, θα πρέπει να επιβάλουμε
συνοριακές συνθήκες στο υπεριώδες, $u\to\infty$ και στο υπέρυθρο, $u=u_{min}$,\
όπως και τις συνθήκες συρραφής στο βαρυονικό κόμβο, $u=u_0$.
Οι συνοριακές συνθήκες στο υπεριώδες είναι τέτοιες ώστε να μην διαταράξουμε
τα κουάρκ, δηλαδή
\ba
\Phi(u) = 0\ ,\qq {\textrm{όταν}}\quad u\to \infty \ ,
\label{19-9}
\ea
όπου η $\Phi(u)$ είναι οποιαδήποτε από τις διακυμάνσεις των $k$
κουάρκ. Στο σημείο $u=u_{min}$ απαιτούμε η λύση, καθώς και
η παράγωγος της ως προς $u$ για τις $N-k$
χορδές να είναι πεπερασμένη, ώστε να ισχύει η διαταρακτική προσέγγιση.
Όπως και στην περίπτωση του δυονίου, για τον μηδενικό τρόπο
αυτές οι διακυμάνσεις απεμπλέκονται από την
ευστάθεια της λύσης. Εφόσον θεωρούμε μια ομοιογενή κατανομή από
κουάρκ πάνω σε ένα σφαιρικό κέλυφος, η $D5$ βράνη βρίσκεται
στο σημείο $x=y=z=0$. Για κάθε μια από τις $k$ χορδές, η $\d r$ είναι η διαμήκης
διακύμανση της και οι $\d\th,\d\phi$ είναι οι εγκάρσιες διακυμάνσεις της.\\
\no
Για να βρούμε τις συνθήκες συρραφής για τις διαμήκεις διακυμάνσεις $\d r$
στον βαρυονικό κόμβο, είναι πιο βολικό να αλλάξουμε βαθμίδα
σε $r=\s$, αντί για $u=\s$ και στην συνέχεια να εκφράσουμε το αποτέλεσμα
στην νέα βαθμίδα όπου οι εξισώσεις κίνησης είναι πιο εύκολες. Δουλεύοντας ανάλογα
με τις περιπτώσεις των μεσονίων και δυονίων, κάνουμε την αλλαγή συντεταγμένων
\ba
\label{19-10}
u=\overline{u}+\d u(t,u)\ ,\qq \d u(t,u)=-{\d r(t,u)\ov r_{\rm cl}^{\prime}}\ .
\ea
\no
Οπότε, η συνοριακή συνθήκη στον βαρυονικό κόμβο μπορεί να βρεθεί, αν αλλάξουμε
προσωρινά βαθμίδα σε $r=\s$ και εφαρμόσουμε μικρές διακυμάνσεις για την μεταβλητή
$\overline{u}$. Με την απαίτηση να έχουμε ένα σωστά ορισμένο συνοριακό
πρόβλημα για τις διακυμάνσεις $\d u$ βρίσκουμε
\ba
\label{19-11}
{d\d u\ov d r}\bigg{|}_{u=u_0}=0\ .
\ea
Από τις Εξ.\eqn{18-10},\eqn{19-10} και \eqn{19-11}
βρίσκουμε ότι οι διακυμάνσεις $\d r$ για κάθε χορδή ικανοποιούν
την συνοριακή συνθήκη
\ba
\label{19-12}
u=u_0:\qq 2(f-f_1)\d r^{\prime}
+\d r\left(2f^{\prime}-{f^{\prime}\ov f}f_1-{g^{\prime}\ov g}(f-f_1)\right)=0\ .
\ea
Αυτή είναι μια εξίσωση για την παράμετρο $u_0$ συναρτήσει του $a$ και των παραμέτρων
του συγκεκριμένου προβλήματος. Θα συμβολίσουμε την λύση της, όπου αυτή υπάρχει με $u_{0\rm c}$.
Δεν θα υποθέσουμε σχέσεις συνέχειας μεταξύ των διακυμάνσεων της $D5$ βράνης και των $N$ χορδών.\\
\no
Γενικά, η τιμή της παραμέτρου $u_{0\rm c}$ δεν συμπίπτει με τα πιθανά ακρότατα της ακτίνας $L(u_0)$,
τα οποία συμβολίζουμε με $u_{0m}$. Όπως είδαμε για τα μεσόνια οι τιμές αυτές συνέπιπταν,
αλλά διαφέρανε στην περίπτωση των δυονίων. Στην περίπτωση μας χρησιμοποιώντας τις
Εξ.\eqn{18-12},\eqn{18-15} και \eqn{18-17} βρίσκουμε ότι
\ba
\label{19-13}
&&{dE\ov du_0}={k\sqrt{f_1}\ov2\pi}{dL\ov du_0}\ , \nonumber \\
&&{dL\ov du_0}= - {\sqrt{f_1F}\ov f}+{\del u_1\ov\del u_0}\ {f_1^{\prime}\ov2\sqrt{f_1}}
\int_{u_0}^\infty du{\sqrt{gf}\ov(f-f_1)^{3/2}}\ ,
\ea
και από την Εξ.\eqn{18-17} υπολογίζουμε ότι
\be
\label{19-14}
f_1' {\del u_1\ov\del u_0} = \sin^2\Th f_0'-{R\ov 2 a} f_0 \cos \Th
\del_u\left({\del_u \sqrt{-G_{tt}}\ov \sqrt{g}}\right)\! \bigg |_{u=u_0}\ .
\ee
Μια ικανή συνθήκη για να υπάρχει $u_{0m}$ είναι το δεξί μέλος της Εξ.\eqn{19-13}
να είναι θετικό, το οποίο όντως συμβαίνει στην περίπτωση των μελανών $D3$ βρανών
που θα μελετήσουμε στο επόμενο κεφάλαιο. Επίσης τα ακρότατα του μήκους και της
ενέργειας στο $u_0=u_{0m}$ ταυτίζονται όπως και στις περιπτώσεις του μεσονίου
και δυονίου. Από την Εξ.\eqn{19-14} μπορούμε εύκολα να δείξουμε ότι
\ba
\label{19-15}
&&{dE\ov dL}={k\ov2\pi}\sqrt{f_1}\ ,\nonumber \\
&&{d^2E\ov dL^2}={k\ov4\pi}{f_1^{\prime}\ov\sqrt{f_1}}\ {1\ov L^{\prime}(u_0)}\
{\del u_1\ov\del u_0}\ .
\ea
Εφόσον $dE/dL>0$ η δύναμη είναι ελκτική. Από την άλλη πλευρά
η ένταση της μπορεί να αυξάνεται ή να μειώνεται ανάλογα με το πρόσημο
του δεξιού μέλους της έκφρασης $d^2E/dL^2$. Τονίζουμε ότι αυτό
δεν έρχεται σε αντίθεση με τα αναμενόμενα αποτελέσματα από
την θεωρία βαθμίδας εφόσον δεν υπάρχει ανάλογη συνθήκη για την δέσμια ενέργεια
των βαρυονίων με την συνθήκη κοιλότητας για την περίπτωση των μεσονίων \cite{concavity}.

\section{Μηδενικοί τρόποι}

Στην συνέχεια θα μελετήσουμε το πρόβλημα του μηδενικού τρόπου για
μια κατανομή σφαιρικού κέλυφους και θα βρούμε την κρίσιμη καμπύλη
στον παραμετρικό χώρο $(a,u_0)$.
Οι διαμήκεις και εγκάρσιες διακυμάνσεις κάθε χορδής, που
ικανοποιούν τις συνοριακές στο υπεριώδες, είναι
\ba
\label{19-16}
&&\d r=A\ J(u)\ ,\qq \d\th=B\ K(u)\ ,\qq \d\phi=C\ K(u)\ ,
\nonumber\\
&&J(u)=\int^{\infty}_u du {\sqrt{g f}\ov (f-f_1)^{3/2}}\ ,
\qq K(u)=\int^{\infty}_u {du\ov f r_{\rm cl}^2}\ \sqrt{gf\ov f-f_1}\ ,
\ea
όπου $A,B$ και $C$ είναι σταθερές ολοκλήρωσης. Από τις Εξ.\eqn{18-3} και
\eqn{18-10} βρίσκουμε ότι το $K(u)$ αποκλίνει, άρα πρέπει να θεωρήσουμε
$B=C=0$. Ως εκ τούτου, οι εγκάρσιες διακυμάνσεις είνα ευσταθείς, όπως
ήταν και στις περιπτώσεις του μεσονίου και δυονίου.\\
\no
Για τις διαμήκεις διακυμάνσεις θα πρέπει να αντικαταστήσουμε
την παραπάνω έκφραση για το $\d r$ στην Εξ.\eqn{19-12} και να βρούμε
μια υπερβατική εξίσωση που θα καθορίσει την κρίσιμη καμπύλη η οποία
διαχωρίζει τις ευσταθείς από τις ασταθείς περιοχές.

\chapter{Παραδείγματα}
Σε αυτό το κεφάλαιο θα κάνουμε μια ανασκόπηση στην κατασκευή ενός
βαρυονίου στην σύμμορφη περίπτωση \cite{Brandhuber:1998,Imamura1}
και θα μελετήσουμε την ευστάθεια του. Στην συνέχεια θα παρουσιάσουμε
την αντίστοιχη ανάλυση στην περίπτωση της πεπερασμένης θερμοκρασίας.

\section{Σύμμορφη περίπτωση}

\subsection{Κλασική λύση}

Αρχικά θα θεωρήσουμε την σύμμορφη περίπτωση, στην οποία η μετρική
του χώρου $AdS_5\times S^5$ έχει την μορφή της Εξ.\eqn{19-1}, με
\be
\label{20-1}
-G_{tt}=G_{xx}= G_{uu}^{-1}={u^2\ov R^2}\ .
\ee
Η ακτίνα και η ενέργεια που αντιστοιχούν σε μια ομοιόμορφη κατανομή σφαιρικού κελύφους
από κουάρκ συναρτήσει της θέσης της $D5$ βράνης $u_0$ και του σημείου στροφής $u_1$
της κάθε χορδής είναι:
\ba
\label{20-2}
L={R^2u_1^2\ov 3u_0^3}\ {\cal I}\ ,\qq E={k u_0\ov 2\pi}\left(- {\cal J}+{5-4a\ov 4a }\right)\ ,
\ea
με
\be
\label{20-3}
{\cal I}= {_2F_1}\left({1\ov 2},{3\ov 4},{7\ov 4};{u_1^4\ov u_0^4}\right)\ ,\qq
{\cal J}={_2F_1}\left(-{1\ov 4},{1\ov 2},{3\ov 4};{u_1^4\ov u_0^4}\right)\
\ee
και όπου ${_2F_1}(a,b,c;z)$ είναι η υπεργεωμετρική συνάρτηση και $u_0\geqslant u_1\geqslant 0$.
Από την Εξ.\eqn{18-17} βρίσκουμε την συνθήκη μηδενισμού της δύναμης στην $u$ διεύθυνση
\ba
\label{20-4}
u_1=u_0(1-\lambda^2)^{1/4}\ ,\qq \l={5-4a\ov 4a}\ .
\ea
Παρατηρούμε ότι η Εξ.\eqn{20-4} δεν περιορίζει το εύρος του $u_0$,
αλλά αυτό δεν θα ισχύει στην περίπτωση της πεπερασμένης θερμοκρασίας.
Εφόσον $\l<1$, η διάταξη του βαρυονίου υπάρχει για $a>a_<$, $a_<=5/8=0.625$.
Η τιμή αυτή βρέθηκε στις εργασίες \cite{Brandhuber:1998,Imamura1}. Χρησιμοποιώντας τις Εξ.\eqn{20-2}
και \eqn{20-4} βρίσκουμε την έκφραση την δέσμιας ενέργειας συναρτήσει του φυσικού μήκους
\ba
\label{20-5}
E=-{R^2\ov2\pi L}{k\sqrt{1-\l^2}\ov 3}\left({\cal J}-{5-4a\ov 4 a}\right){\cal I}\ .
\ea
Από την έκφραση αυτή είναι προφανές γιατί το $a=5/8$ ($\l=1$) απορρίπτεται.
Η ενέργεια του βαρυονίου είναι ανάλογη του $1/L$ όπως αυτό αναμενόταν από
την σύμμορφη συμμετρία και έχει την γνωστή μη διαταρακτική εξάρτηση από
την σταθερά ζεύξης \en 't Hooft.\gr

\subsection{Ανάλυση ευστάθειας}

Στην συνέχεια θα μελετήσουμε την ευστάθεια της σύμμορφης περίπτωσης.
Για της διαμήκεις διακήμάνσεις πρέπει να λύσουμε την Εξ.\eqn{19-12},
όπου
\ba
\label{20-6}
\d r(u)= A \int_u^\infty du{u^2\ov(u^4-u_1^4)^{3/2}}={A\ov 3u^3}\ {_2F_1}
\left({3\ov 4},{3\ov 2},{7\ov 4};{u_1^4\ov u^4}\right)\ .
\ea
Αντικαθιστώντας τις Εξ.\eqn{20-4} και \eqn{20-6} στην \eqn{19-12} βρίσκουμε
την υπερβατική εξίσωση για τον μηδενικό τρόπο
\ba
\label{20-7}
{_2F_1}\left({3\ov4},{3\ov2},{7\ov4};1-\l^2\right)={3\ov2\l(1+\l^2)}\ .
\ea
Χρησιμοποιώντας τις Εξ.\eqn{20-4} και \eqn{20-7} βρίσκουμε αριθμητικά την κρίσιμη τιμή
για το $a$, η οποία είναι $a_c\simeq 0.813$.\\
\no
Οπότε, η ανάλυση της ευστάθειας θέτει ένα κάτω όριο για το $a$ το οποίο είναι μικρότερο
από την μονάδα. Ωστόσο αυτό είναι αξιοσημείωτο, δεδομένου ότι αντιστοιχεί στο σύμμορφο σημείο
της μέγιστα υπερσυμμετρικής $\cN=4$ \en SYM. \gr Η μόνη άλλη περίπτωση στην οποία
υπάρχει αστάθεια στο σύμμορφο σημείο είναι στην περίπτωση των μεσονίων που μελετήσαμε στο 10ο
κεφάλαιο για $\b$ παραμορφωμένες $\cN=1$ θεωρίες \cite{LS}, και του οποίου το δυικό
υπερβαρυτικό υπόβαθρο μελετήθηκε στο \cite{LM}.
Σε αυτήν την περίπτωση, όπως είδαμε στο 10ο κεφάλαιο υπάρχει αστάθεια για τιμές της παραμέτρου
διαμόρφωσης $\s$ οι οποίες είναι μεγαλύτερες από μια κρίσιμη τιμή.

\section{Μελανές $D3$ βράνες}

\subsection{Κλασική λύση}

\begin{figure}[!t]
\begin{center}
\begin{tabular}{cc}
\includegraphics[height=5.0cm]{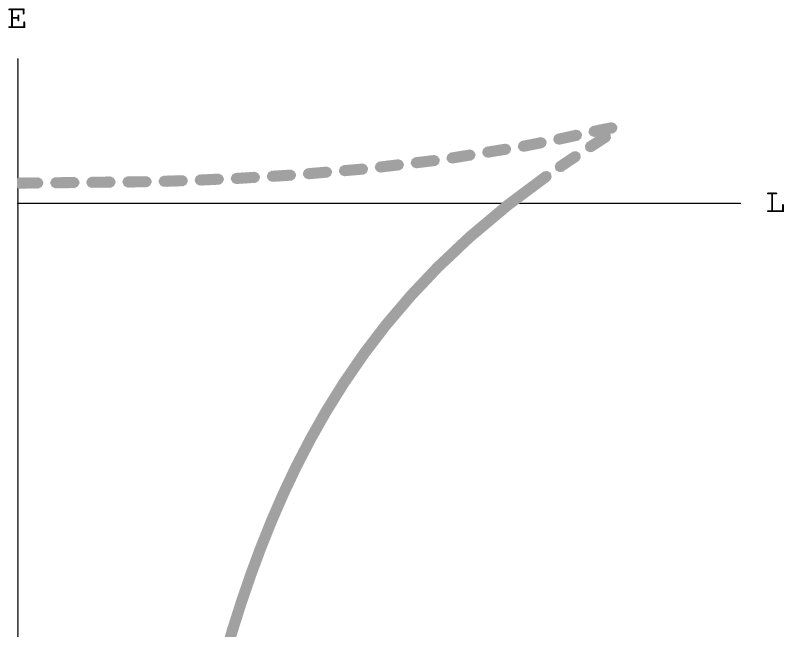}
&\includegraphics[height=6.6cm]{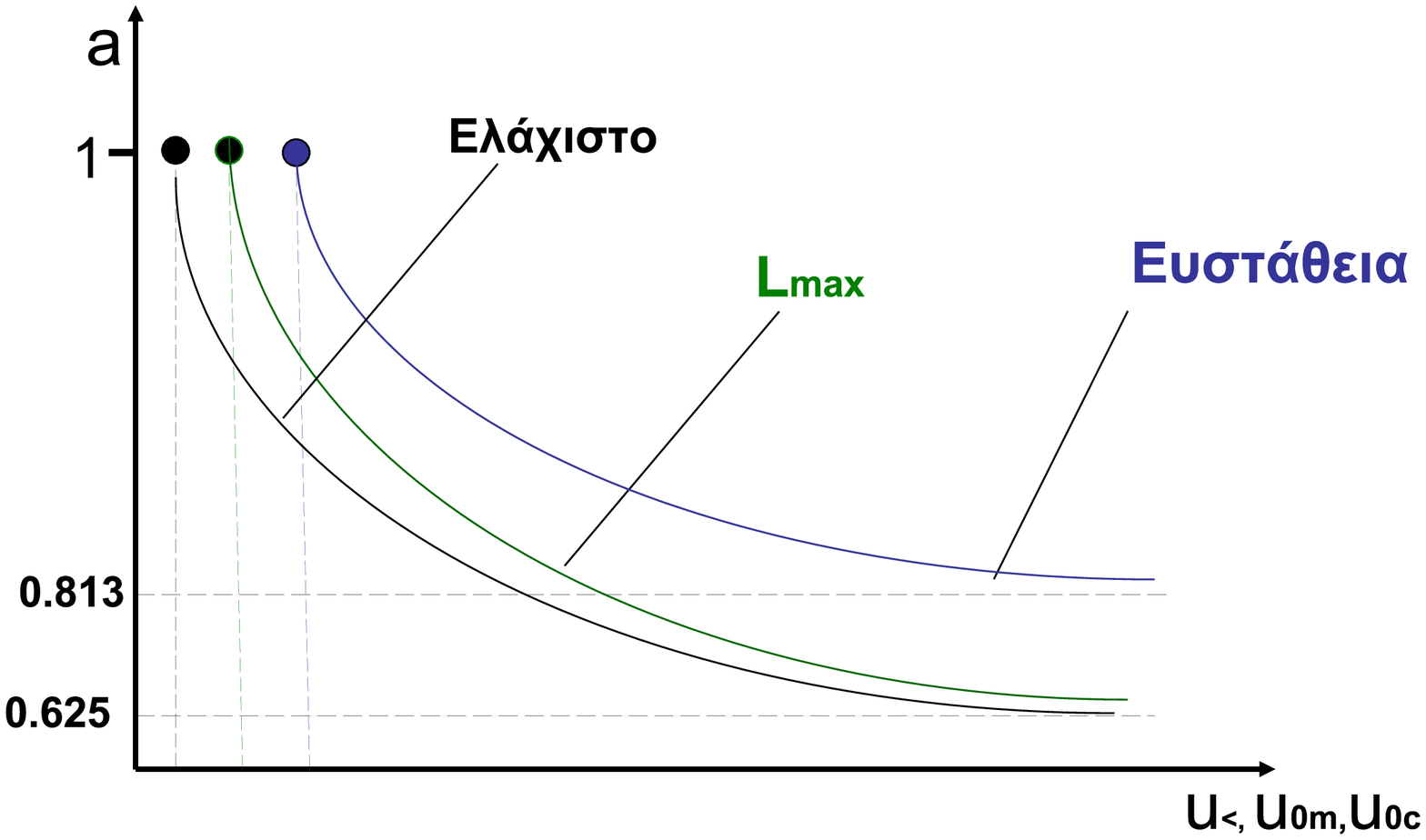}\\
(α) & (β)
\end{tabular}
\end{center}
\vskip -.4 cm \caption{(α) Γραφική παράσταση της
$E(L)$ για μελανές $D3$ βράνες. Τα διάφορα είδη γραμμών αντιστοιχούν σε ευσταθείς (γκρί γραμμή)
και ασταθείς (διακεκομμένη γκρί γραμμή). Όπως το $a$ μειώνεται το ασταθές τμήμα εκτείνεται επίσης και στον
κάτω κλάδο. Η συμπεριφορά αυτή περιγράφεται στην ενότητα (19.2.2).
(β) Από αριστερά προς τα δεξιά 1) Γραφική παράσταση του $a$ με το $u_<$, 2)
Γραφική παράσταση του $a$ με το $u_{min}$, 3) Γραφική παράσταση του $a$ με το $u_{0c}$.
Οι περιοχές στα δεξιά και αριστερά της κριτικής καμπύλης είναι ευσταθείς και ασταθείς, αντίστοιχα.
}
\label{D3blackbaryon}
\end{figure}

Σε αυτήν την ενότητα θα θεωρήσουμε ένα σύνολο από $N$ μελανές $D3$ βράνες. Στο όριο της θεωρίας πεδίου
η μετρική έχει την μορφή της Εξ.\eqn{18-1} όπου
\ba
\label{20-8}
G_{tt}= -G_{uu}^{-1}=-{u^2\ov R^2}\left(1-{\mu^4\ov u^4}\right)\ ,\qq G_{xx}={u^2\ov R^2}\ .
\ea
Η αντίστοιχη θερμοκρασία \en Hawking \gr είναι $T_H=\m/(\pi R^2)$. Σε ότι ακολουθεί, θα τα
κανονικοποιήσουμε όλα σε μονάδες $\mu$, όπου θέτουμε $\mu=1$. Η ακτίνα και η ενέργεια είναι ίσες με
\ba
\label{20-9}
&&L=R^2{\sqrt{u_1^4-1}\ov 3u_0^3}
F_1\left({3\ov 4},{1\ov 2},{1\ov 2},{7\ov 4};{1\ov u_0^4},{u_1^4\ov u_0^4}\right)\ ,
\nonumber \\
&&E={k\ov 2\pi}\left\{{\cal E}+{u_0\ov 4a}\sqrt{1-{1\ov u_0^4}}+{1-a\ov a}(u_0-1)\right\}\ ,
\\
&&{\cal E}=-u_0 F_1\left(-{1\ov 4},-{1\ov 2},{1\ov 2},{3\ov 4};{1\ov u_0^4},{u_1^4\ov u_0^4}\right)
+1\ ,
\nonumber
\ea
όπου $F_1(a,b_1,b_2,c;z_1,z_2)$
είναι η υπεργεωμετρική συνάρτηση του \en Appell \gr και $u_0\geqslant u_1\geqslant 1$. Από την Εξ.\eqn{18-17}
βρίσκουμε την συνθήκη μηδενισμού της $u$ διεύθυνσης της δύναμης
\ba
\label{20-10}
\cos\Th={1-a\ov a}+{1+1/u_0^4\ov 4a\sqrt{1-1/u_0^4}}\ ,
\qq \cos\Th=\sqrt{{u_0^4-u_1^4\ov u_0^4-1}}\ .
\ea
Χρησιμοποιώντας το γεγονός ότι $\cos\Th\leqslant1$ στην Εξ.\eqn{20-10}
καταλήγουμε με μια πολυωνυμική ανισότητα δευτέρου βαθμού η οποία έχει λύση για
$a \geq 5/8$, την
\ba
\label{20-11}
&&u_0\geqslant u_<(a)\ ,\qq u_<(a)=1/\l^{1/4}\ ,
\nonumber \\
&&\l=-9-32a(a-1)+4\sqrt{2}(2a-1)\sqrt{3-8 a(1-a)}\ ,
\\
&&a\in[5/8,1]\ ,\quad \l\in[0,-9+4\sqrt{6}]\ ,\quad u_0\in[\infty, u_<(1)]\ ,
\quad u_<(1)\simeq 1.058\ .
\nonumber
\ea
Ωστόσο, για $a=5/8$ δεν υπάρχει κλασική λύση εφόσον $u_0\to\infty$.
Για $a > 5/8$ η λύση της Εξ.\eqn{20-10} είναι
\ba
\label{20-12}
u_1=u_0(1-\L^2(u_0))^{1/4}\ ,\qq \L(u_0)={1-a\ov a}\sqrt{1-1/u_0^4}+{1+1/u_0^4\ov 4a}\ .
\ea
\no
Σε αυτή την περίπτωση δεν μπορούμε να εκφράσουμε την ενέργεια συναρτήσει του μήκους.
Στην πραγματικότητα η ενέργεια είναι μια δίτιμη συνάρτηση του μήκους με ενεργειακά
προτιμητέους και μη κλάδους, όπως φαίνεται στο σχήμα \ref{D3blackbaryon}.
Αυτή η συμπεριφορά είναι ανάλογη με το δυναμικό ενός μεσονίου για πεπερασμένη θερμοκρασία
\cite{wilsonloopTemp}.

\subsection{Ανάλυση ευστάθειας}

Στην συνέχεια θα μελετήσουμε την ευστάθεια των μελανών $D3$ βρανών. Για τις διαμήκεις διακυμάνσεις
πρέπει να λύσουμε την Εξ.\eqn{19-12} όπου
\ba
\label{20-13} \d r(u)=A\int_u^\infty du
{\sqrt{u^4-1}\ov(u^4-u_1^4)^{3/2}}= {A\ov 3u^3}F_1\left({3\ov
4},-{1\ov 2},{3\ov 2},{7\ov 4};{1\ov u^4},{u_1^4\ov u^4}\right)\ .
\ea
Από τις Εξ.\eqn{19-12} και \eqn{20-13} βρίσκουμε ότι η συνθήκη ύπαρξης μηδενικού τρόπου δίνει
την υπερβατική εξίσωση
\ba
\label{20-14} {3\ov
2\sqrt{u_0^4-u_1^4}}={2u_0^4-u_1^4-1\ov(u_0^4-1)^{3/2}}
F_1\left({3\ov 4},-{1\ov 2},{3\ov 2},{7\ov 4};{1\ov u_0^4},{u_1^4\ov u_0^4}\right)\ .
\ea
Με απλές αριθμητικές μεθόδους βρίσκουμε ότι υπάρχει λύση για κάθε τιμή του $a$ η οποία είναι
μεγαλύτερη από την κρίσιμη τιμή που βρήκαμε στο σύμμορφο όριο $a_c\simeq 0.813$, ενώ υπάρχει
επίσης ένας μηδενικός τρόπος στον ενεργειακά προτιμητέο κλάδο, ο οποίος είναι $u_{0c}>u_{0m}$.
Η λύση έχει την ακόλουθη ανάπτυξη
\be
\label{20-15} a_c\simeq 0.813+{0.679\ov u_{0c}^4}+{0.778\ov
u_{0c}^8}+{\cal{O}}\left(1\ov  u_{0c}^{12}\right)\ .
\ee
Όταν το $a_c$ μειώνεται τότε αυξάνεται το $u_{0c}$ και για $a_c\simeq 0.813$
ακόμα και το σύμμορφο τμήμα του διαγράμματος γίνεται ασταθές. Τα αποτελέσματα της ανάλυσης
ευστάθειας συνοψίζονται στο Σχήμα \ref{D3blackbaryon}.\\

\begin{table}
\begin{center}
\begin{tabular}{| c || c | c | c |}
\hline
{\bf Τιμές για $a=1$} &{\bf Ελάχιστο} & {\bf Κρίσιμη} & {\bf Μέγιστο} \\
\hline\hline  $u_0$ & 1.0581 & 1.4292 & 1.3437 \\
\hline   $L/R^2$ & 0 & 0.2920 & 0.2951 \\
\hline  $E/N$ & 0.0097 & 0.0266 & 0.0273 \\
\hline
\end{tabular}
\end{center}
\caption{Το μήκος και η ενέργεια για την ελάχιστη, κρίσιμη και μέγιστη τιμή $u_0$ για $a = 1$.}
\end{table}

\newpage

\Huge
\addcontentsline{toc}{chapter}{Επίλογος}
\textbf{Επίλογος}

\normalsize

\vspace{2cm}
\no
Σε αυτή την διδακτορική διατριβή μελετήσαμε την ευστάθεια δέσμιων καταστάσεων
κουάρκ και μονοπόλων εντός της αντιστοιχίας \en AdS/CFT.
\gr Αρχικά θεωρήσαμε τις δυικές διατάξεις χορδών ενός ζεύγους
κουάρκ-αντικουάρκ στη θεωρία $\cN=4$ \en SYM \gr σε
πεπερασμένη θερμοκρασία και στον κλάδο \en Coulomb.\ \gr Το κίνητρο για την μελέτη
μας ήταν ότι τα δυναμικά των μεσονίων όπως αυτά υπολογίζονται εντός
της \en AdS/CFT \gr έχουν συμπεριφορές οι οποίες σε μερικές περιπτώσεις είναι σε αντίθεση με τα
προσδοκώμενα από την θεωρία πεδίου αποτελέσματα. Συγκεκριμένα συναντούμε: (1) πλειότιμα δυναμικά,
(2) δυναμικό του οποίου το μήκος θωράκισης εξαρτάται ισχυρώς από τον προσανατολισμό στο χώρο της κλασικής
λύσης της χορδής σε σχέση με τον άξονα του ζεύγους και (3) συμπεριφορές εγκλωβισμού των κουάρκ στην
$\cN=4$ \en SYM \gr. Η ανάλυση της ευστάθειας των κλασικών λύσεων επιλύει αυτές τις
αντιφάσεις δείχνοντας ότι οι διατάξεις αυτές είναι ασταθείς και
κατά συνέπεια είναι φυσικώς μη αποδεκτές.
Επιπλέον, μελετήσαμε υπόβαθρα $D3$ βρανών τα οποία είναι δυικά σε
θεωρίες βαθμίδας με $\cN=1$ υπερσυμμετρία. Όπως είδαμε για τιμές
της παράμετρου παραμόρφωσης μεγαλύτερες από μια οριακή τιμή
ακόμα και στο σύμμορφο όριο υπάρχει αστάθεια. Επίσης βρήκαμε
ότι για ορισμένες περιοχές της παραμέτρου παραμόρφωσης
το γραμμικό δυναμικό εγκλωβισμού γίνεται ευσταθές όπως αναμένεται απο επιχειρήματα
της θεωρίας πεδίου.\\
\no
Στην συνέχεια κατασκευάσαμε συνενώσεις χορδών και
τις χρησιμοποιήσαμε για να υπολογίσουμε την ενέργεια
αλληλεπίδρασης βαριών δυονίων σε ισχυρή ζεύξη στο
πλαίσιο της αντιστοιχίας \en AdS/CFT.\ \gr Η πιο σημαντική
εφαρμογή είναι αυτή του κουάρκ και μονοπόλου,
την οποία μελετήσαμε με λεπτομέρεια. Στην συνέχεια
εφαρμόσαμε μικρές διακυμάνσεις για να καθορίσουμε
τις φυσικώς αποδεκτές παραμετρικές περιοχές. Βρήκαμε ότι τμήματα
ενεργειακά μη προτιμητέων κλάδων είναι ευσταθή κάτω από
μικρές διακυμάνσεις. Ο λόγος για αυτή την συμπεριφορά
έγκειται στο ότι οι συνοριακές συνθήκες για τις χορδές στο σημείο
διακλάδωσης ενισχύουν την ευστάθεια της. Η συμπεριφορά
αυτή όπως τονίσαμε δεν έρχεται σε αντίθεση με τα αναμενόμενα
αποτελέσματα απο την θεωρία πεδίου. Διότι,
δεν υπάρχει συνθήκη κυρτότητας για το δυναμικό κουάρκ
και μονοπόλου όπως στα μεσόνια.
Βρήκαμε επίσης ότι η συμπεριφορά εγκλωβισμού
στο δυναμικό του δυονίου στην περίπτωση του κλάδου \en Coulomb \gr για
ομοιόμορφη κατανομή $D3$ βρανών πάνω σε σφαίρα, είναι ασταθής
και είναι κάτι το οποίο δεν προβλέπεται από την θεωρία βαθμίδας.\\
\no
Τέλος θεωρήσαμε την κατασκευή βαρυονίου μέσω χορδών και μιας $D5$ βράνης
εντός την αντιστοιχίας \en AdS/CFT \gr
για μια μεγάλη κατηγορία υποβάθρων. Μελετήσαμε διεξοδικά την $\cN=4$ \en SYM\ \gr
στο σύμμορφο σημείο και για πεπερασμένη
θερμοκρασία. Η ανάλυση της ευστάθειας, βελτίωσε το κάτω φράγμα του αριθμού των
κουάρκ για τον οποίο μπορεί να υπάρξει βαρυόνιο.
Επιπλέον, βρήκαμε ότι μέρος του ενεργειακά προτιμητέου κλάδου στην περίπτωση της πεπερασμένης
θερμοκρασίας είναι διαταρακτικά ασταθές, ακόμα και για $k=N$. \\
\no
Είναι ενδιαφέρον για την περίπτωση του βαρυονίου
να θεωρήσουμε μια κατανομή από κουάρκ
και να καθορίσουμε με λογισμό των μεταβολών την
κατανομή των κουάρκ με την ελάχιστη ενέργεια.
Εν συνεχεία να μελετήσουμε την ευστάθεια της  αυτής
κάτω από μικρές διακυμάνσεις, έχοντας ως σκοπό την ύπαρξη ενός ευσταθούς
βαρυονίου με $k=N$.

\end{document}